\documentclass[11pt]{report}
\usepackage{graphicx,subfigure}
\usepackage{bm,bbm}
\usepackage{cite}
\usepackage{amsmath,amssymb}

\newcommand{\Z}{\bm{\mathbbm{Z}}_2}
\newcommand{\be}{\begin{equation}}
\newcommand{\ee}{\end{equation}}
\newcommand{\bea}{\begin{eqnarray}}
\newcommand{\eea}{\end{eqnarray}}
\newcommand{\N}{\mathcal{N}}
\renewcommand{\b}[1]{\bar{#1}}
\newcommand{\Del}{\nabla}
\newcommand{\del}{\partial}
\newcommand{\bi}{\bar{\imath}}
\newcommand{\bj}{\bar{\jmath}}
\newcommand{\ap}{\alpha^\prime}
\newcommand{\ket}[1]{| #1 \rangle}
\newcommand{\tr}{\mathnormal{tr}}

\newcommand{\im}{\mathnormal{Im}\,}
\newcommand{\comment}[1]{}
\newcommand{\bd}{\ensuremath{\overline{\textnormal{D3}}}}
\renewcommand{\k}{\kappa_4}
\newcommand{\kten}{\kappa_{10}}
\newcommand{\ot}{\overline{\Theta}}
\newcommand{\tg}{\tilde{g}}
\renewcommand{\t}[1]{\tilde{#1}}
\newcommand{\wo}{W_0}

\title{Warped Strings: Self-dual Flux and Contemporary Compactifications}
\author{Andrew R. Frey\\ Department of Physics\\ University of California,
Santa Barbara}
\date{August 16, 2003}

\begin{document}

\maketitle

\begin{abstract}
String theory is a leading candidate for the quantum theory of gravitation
and is the only serious unified theory of all the forces of physics.
Additionally, string theory seems to be mathematically unique, and it is
possible that all the parameters of physics could be determined by
the dynamics of string theory.  However, there are two intertwined problems.
First is that string theory requires ten dimensions for mathematical 
consistency, so six must be compact or otherwise decoupled from the physics
we observe.  Also, there exist many different ground
states of string theory, so it is difficult to make predictions that are
generic to string theory.  Therefore, it is crucial to gain a broad
understanding of all the string theory vacua, particularly those 
compactified to four dimensions.

In recent years, our understanding of string theory 
compactifications has expanded greatly to include many new classes of
solutions.  In this dissertation, I discuss
type IIB string compactifications in which the three-form field strengths 
satisfy a self-duality condition on the internal manifold.  I begin with
an overview of the models, giving preliminary formulae and several 
points of view from which they can be understood.  Then I will describe
windows into the small radius behavior of the compactifications, which is
more complicated than compactifications without fluxes, which are more
well-known.  I will describe
details of the flux-generated potential and nonperturbative corrections to
it.  These nonperturbative corrections allow a discussion of the cosmological
constant and possible mechanisms for the universe to decay from one energy
state to another.  I conclude with comments on related topics and interesting
directions for future study.  

As this document is a dissertation, 
I will indicate my own contributions to the
subject.  However, for the purpose of completeness and in hopes of helping
the readers, I will include a fairly comprehensive 
review of related literature.  The early part of the dissertation is
much more a literature review because there are many results upon which
my contributions have been founded.
\end{abstract}

\newpage
\begin{center}\textbf{Dedication}
\ \\ \ \\ \ \\
To Rebecca With All My Love
\end{center}

\newpage
\begin{center}\textbf{Acknowledgements}\end{center}

First of all, I thank the faculty who have influenced my education,
including Omer Blaes and Claudio Campagnari for various forms of support.
Particularly I thank the string theory faculty, Steven Giddings, 
David Gross, Gary Horowitz, and
Joseph Polchinski, for encouragement and instruction.  Especially I 
thank my advisor, Joseph Polchinski, for guidance.  
Not only has he taught me a great deal of physics, but he has
also ensured that I had the necessary support to get through difficult
times. I am very grateful to have had him for a mentor.

I would also like to thank the many people with whom I have had 
interesting discussions.  Specifically, I have benefited greatly 
from UCSB graduate students and postdocs, including
I. Bena, O. deWolfe, H. Elvang, T. Erler, S. Hellerman, 
V. Hubeny, A. Maharana, and N. Mann.  Outside of UCSB, I have benefited
from discussions with E. D'Hoker, 
S. Ferrara, S. Kachru, M. Schulz, and E. Silverstein.
I would especially like to thank my collaborators, A. Buchel, M. Gra\~na,
M. Lippert, A. Mazumdar, and B. Williams.

For financial support, I am grateful to the National Science Foundation
and the UCSB Physics Department and Graduate Division.  
Part of my work has been supported through UCSB fellowships and an NSF
Graduate Research Fellowship.

Finally, I thank my family for much support throughout my entire education
and life.  Most importantly, I thank my wife, Rebecca Danos, whose
love and support have helped me through every aspect of my life.

\newpage

\tableofcontents\listoftables\listoffigures

\chapter{General Theory}\label{c:general}
This chapter gives some motivational words along with an introduction to
the compactifications that are the subject of this dissertation.

\section{Introduction}\label{s:intro}

The Standard Model of particle physics is, by all accounts, remarkably 
successful in that it is very well corroborated by experiment (with the
exception of neutrino mixing), so much so that it is almost universally 
capitalized in spelling and pronunciation.  However, theorists have long
been frustrated by the number of parameters in the Standard Model, such as
gauge couplings and the Yukawa couplings that determine fermion masses.  
Naturally, there have been many attempts to extend the Standard Model to
a theory that predicts those parameters.  

Possibly the most promising of the attempts has been string theory, a 
reformulation of particle physics in terms of extended one dimensional 
objects, or strings.  One of the great advantages of string theory is
that all of its parameters are dynamical variables, so presumably all the
parameters of the Standard Model can be determined by finding the ground
state of all those variables in string theory.  Unfortunately, there are
an incredible number of ground states of string theory, and, through the
turn of the century, the best understood of these vacuum states allow many
of the parameters, or moduli, to take on a continuum of values.  This problem
of free moduli is related to the number of ways to compactify string theory
from ten dimensions (necessary for the mathematical consistency of the theory)
to the four dimensions in which we live.  

Starting just before the new millenium, though, string theorists have 
developed mechanisms for fixing the values of moduli in a number of different
contexts (as it turns out, some of these mechanisms are related).
In this document, I review the dimensional reduction of
type IIB string theory from ten to four dimensions, with an emphasis on
the use of rank 3 field strengths to freeze moduli.  A particular feature
of these models is that the field strengths must satisfy a self-duality 
condition on the compact space, so I will often call them self-dual fluxes.  
This first chapter will
give the general theory of such compactifications, including the solution
of type IIB supergravity, and the second will find the 
effective four dimensional theory from several perspectives.
The discussion of the first two chapters 
will mainly be valid at large volumes for the
compact six dimensions; in a subsequent chapter, 
I will consider the small volume regime.  
Later, I will discuss the application
of these models to cosmological issues (chapter \ref{c:potential})
and some open questions (chapter \ref{c:future}).

Throughout, I will give particular attention to work to which I have made 
a contribution in the course of graduate study.  The papers to which I have
contributed relevant to this review are 
\cite{Frey:2002hf,Grana:2002ti,Frey:2002qc,Frey:2003jq,Frey:2003dm,Frey:2003sd}.  
Results from them appear as follows:
\begin{itemize}
\item Results from \cite{Frey:2002hf} are scattered throughout the
dissertation, in sections \ref{s:eom}, \ref{sss:n3sol}, \ref{s:moduli}, 
\ref{ss:describe}, and \ref{ss:udual}.  Some results which were derived
in preparation for \cite{Frey:2002hf} but not previously published 
also appear in section \ref{s:eom}.
\item The counting of BPS states from \cite{Grana:2002ti} appears in
section \ref{ss:counting}.
\item The cosmology discussed in section \ref{s:fluxcosmo} is from
\cite{Frey:2002qc}.
\item The discussion of \cite{Frey:2003jq} regarding string corrections 
and symmetry breaking is included, in condensed form, in section 
\ref{ss:analogy}.
\item The explicit description of de Sitter vacua given in \cite{Frey:2003dm}
appears in section \ref{sss:parameters}, while the calculations of
decay times from that paper appear in \ref{ss:dsdecays}.
\item The supersymmetric solutions that will appear in \cite{Frey:2003sd} are given
in section \ref{ss:susy}.
\end{itemize}

Appendix \ref{a:conventions} discusses the conventions and notation that I 
use.  Actions and equations of motion are given in appendix 
\ref{a:actionsetc}.  Some derivations that are too detailed for the main 
text are collected
in appendix \ref{a:ancillary}.

\section{Phenomenological Motivation}\label{s:warped}

To start, it is probably useful to review the basic phenomenological 
models that inspired some of the recent developments in string 
compactifications.  This section is mainly intended as a short motivation,
along with the idea of moduli stabilation discussed above, for the more
specific results developed later.

The first major development was the implication that the Standard Model
might be confined to a brane, or domain wall, so that only gravity could move
through some number of extra dimensions; since gravity is very poorly 
measured in comparison to the Standard Model, the extra dimensions can be
fairly large before giving problems due to graviton Kaluza-Klein modes
\cite{Arkani-Hamed:1998rs,Rubakov:1983bb,Akama:1982jy}.  
An exciting feature of these
models is that the large extra dimensions can allow the 4D Planck scale
to be much higher than the extra dimensional Planck scale; taking the
metric on the internal compact space to be fixed, the reduction of
the Einstein-Hilbert action is (up to convention dependent factors of $2\pi$,
etc)
\be\label{planck1} 
M_D^{D-2} \int d^D x \sqrt{-g_D} R = \left( M_D^{D-2} V_{D-4}\right) 
\int d^4x\sqrt{-g_4} R \Rightarrow M_4^2 = M_D^{D-2} V_{D-4}\ .\ee
(We will also use the gravitational coupling, $\kappa_D^2 = 1/M_D^{D-2}$,
to represent the Planck scale.)
Branes are very natural objects in string
theory, and the Standard Model braneworld was identified with D-branes
of string theory \cite{Antoniadis:1998ig}.  As long as the compactification
length scale is large
compared to the fundamental length scale of the higher dimensional theory,
the 4D Planck scale can be quite large.

The next discovery was the importance of non-factorizable
geometry; the traditional compactification ansatz is that the metric is just
a direct product of the internal space and the noncompact spacetime.  
However, that is not the most general ansatz for a Poincar\'e invariant 
ground state; the metric
\be\label{warp1}
ds^2 = e^{2A(x^m)}\eta_{\mu\nu} dx^\mu dx^\nu + g_{mn} dx^m dx^n
\ee
is fully consistent with the Poincar\'e group.  Here and henceforth, we
use Greek indices $\mu,\nu,\cdots$ for the noncompact spacetime,
and Latin $m,n,\cdots$ for the compact dimensions.  In two simple models
(with one and two branes), \cite{Randall:1999ee,Randall:1999vf} showed that
the warp factor modifies the Kaluza-Klein decomposition, localizing
the graviton zero-mode near one of the branes.  In fact, 
\cite{Randall:1999vf} showed that infinitely large 
extra dimensions can still give
rise to four dimensional effective theories of gravity due to this effect.
These simplest models are five dimensional, compactified on $S^1/\Z$ with
one brane at each fixed point, and a negative bulk cosmological constant that
generates the warp factor.  The focus of this document is on string 
compactifications that reproduce the physics of this simple two-brane 
model \cite{Verlinde:1999fy,Chan:2000ms}.

What exactly does the four dimensional graviton look like?  As it turns out,
for the simple models of \cite{Randall:1999ee,Randall:1999vf}, the zero
mode graviton is just $\delta g_{\mu\nu}=e^{2A(x^m)}h_{\mu\nu}(x^\mu)$.  
In fact,
the graviton takes precisely this form for a wide class of stress tensors
\cite{Csaki:2000fc}, and it is known to have this form for the self-dual 
flux compactifications which I discuss in the rest of this review
\cite{Greene:2000gh}; section \ref{s:eom} will give a demonstration.
In terms of the effective theory of the light modes, then, the Planck
constant (\ref{planck1}) essentially picks up an additional $e^{2A}$ due
to the rescaling of the 4D metric.

To put things more precisely, consider a compactification of string theory. 
The gravitational action of all string theories is, in the string frame,
\bea
S&=& \frac{1}{2\kappa_{10}^2} \int d^{10}x \sqrt{-g_{10}}e^{-2\phi}R_{10}
= \frac{1}{2\kappa_{10}^2} \int d^{10}x \sqrt{-g_4}
e^{2A}e^{6u} e^{-2\phi}R_{4}\ \ \label{stringframe}\\
 ds^2&=&e^{2A}g_{4,\mu\nu}dx^\mu dx^\nu+\cdots\label{metric4D}
\eea
where $e^{6u}$ gives the volume of the compact metric (other metric 
perturbations preserve volume) and where 
$2\kappa_{10}^2 = (2\pi)^7\ap{}^4$ (see, for example, 
\cite{Polchinski:1998rr})\footnote{Also, much of the literature uses the
Einstein frame in 10D, along with other conventions for the supergravity
fields. See \ref{aa:frames} for a summary of different conventions.  I
will use string frame in 10D unless otherwise specified.}.  Clearly,
equation (\ref{stringframe}) does not give the usual Einstein-Hilbert term
for $g_{4,\mu\nu}$; instead, we must convert to the 4D Einstein frame
metric 
\be\label{einstein4d}
g_{E,\mu\nu}= e^{6u} e^{-2\phi}g_{4,\mu\nu}\ee
to decouple the dilaton and the overall volume of the compactification.
Then the effective action is the usual
\be\label{planck2}
S=\frac{1}{2\k^2}\int d^4x\sqrt{-g_E} R_E\ , \ \
\frac{1}{\k^2} = \frac{V_w}{\kappa_{10}^2}\ ,\ \ 
V_w = \int d^6x \sqrt{\det g_{mn}}\, e^{2A}e^{-6u}\ .\ee
Note that now there can be a large hierarchy between 10D and 4D Planck 
constants due to a large warp factor.  This particular form of the 
relation between Planck constants is due to \cite{DeWolfe:2002nn}.

What does this mean for the gauge/gravity hierarchy?  Consider a Higgs
scalar that lives on, for example, a D3-brane, with some string frame
mass $m$ (ie, a fundamental mass $m$),
much as in \cite{Randall:1999ee,DeWolfe:2002nn}.  In the string frame,
the action (rescaling the D3 tension into the scalar to get canonical 
normalization) is
\be\label{higgs1}
S=-\frac{1}{2}\int d^4x e^{-\phi}\sqrt{-g_4} e^{4A} \left[ 
e^{-2A}\del_\mu H\del^\mu H +m^2 H^2\right]\ .\ee
We can already see that the mass as measured by the 4D effective theory
is modified by a factor of $e^{2A}$, and going to the Einstein frame
gives
\be\label{higgs2}
S=-\frac{1}{2}\int d^4x \sqrt{-g_E} e^{\phi-6u+2A}\left[
\del_\mu H\del^\mu H + e^{2\phi-6u+2A}m^2H^2\right]\ .\ee
So the Higgs mass, compared to the Planck mass,
measured by an observer on the brane is
\be\label{higgs3}
\frac{m_H^2}{M_4^2} = e^{2\phi-6u+2A}(\textnormal{SM})
\frac{m^2}{M_{10}^8 V_w}\ .\ee
So we can get a large hierarchy in a number of ways: 
\begin{enumerate}
\item Weakly coupled strings
\item A large compactification volume ($u$ large), as in equation
(\ref{planck1})
\item A small warp factor $e^A$ at the brane
\item $m\ll M_{10}$ for some reason such as supersymmetry
\item $m\sim M_{10}$, but large warped volume compared to $1/M_{10}^6$
\end{enumerate}
or some combination of those.  In fact, some of these reasons are
essentially identical.  For example,  
the first reason is not often discussed because it does not appear in the 
10D Einstein frame; the relations (\ref{einstein10d},\ref{einsteinmass}) 
mean that the string 
coupling rescales the compactification volume and the fundamental mass.
Also, its always possible to scale
the coordinates so the warp factor is $1$ at the Standard Model brane,
so the third and fifth reasons are identical.  I will use the convention
that, at large compactification radii, $A\sim 0$ away from special points
such as a Standard Model brane.
An important question is how the warp factor can be given a fixed
small value near the Standard Model brane (in other words, how does the
geometry develop a large hierarchy?), and the first phenomenological
answer was given in \cite{Goldberger:1999uk,Goldberger:1999un}.

Similar considerations apply to other mass scales 
\cite{DeWolfe:2002nn}; an observer on 
the Standard Model brane always sees a mass rescaled by the warp factor
relative to the warp factor on the brane (similar considerations would
apply to a varying dilaton).  For example, a Kaluza-Klein mode could
be localized in some region of the bulk, or supersymmetry could be
broken on a different brane.  Then 
\be\label{otherscales}
m^2(\textnormal{SM observed}) = 
\frac{e^{2A}(\textnormal{KK/SUSY/etc.})}{e^{2A}(\textnormal{SM})}
m^2(\textnormal{fundamental})\ .
\ee

A number of warped compactifications of string theory were discussed in
\cite{Mayr:2000zd} from the point of view of using the warp factor to
generate large hierarchies.  In the rest of this disseration, I will 
discuss another type of string compactification, type IIB strings with 
3-form fluxes.  In some instances, the fluxes can give large ratios of
warp factors, and they always freeze many of the moduli.

\section{Ten Dimensional Supergravity}\label{s:tend}

Based on the supersymmetry variation of the gravitino in M-theory, 
\cite{Becker:1996gj} first studied compactifications of 11D supergravity
to 3D with nonvanishing 4-form field strengths.  Then, 
\cite{Dasgupta:1999ss} explored duals of these 
compactifications in type IIB and heterotic string theory, mainly focusing
on some simple examples, and \cite{Greene:2000gh} described the 
compactifications with fluxes as warped compactifications, paying special
attention to the effect of the warp factor and fluxes on the dimensional
reduction.  In fact, \cite{Verlinde:1999fy,Chan:2000ms} 
argued that similar string
compactifications could implement the models of \cite{Randall:1999ee};
the Standard Model would live on D-branes in a region of low warp factor,
and the 6D compactification manifold would play the role of the second
brane.  More complicated type IIB compactifications appeared in 
\cite{Giddings:2001yu}, which also showed how to stabilize the warp
factor at small values near special points.

In the next subsection, I will describe the full ten dimensional
solution of type IIB 
string theory with self-dual flux, 
mainly following the discussion in \cite{Giddings:2001yu}.
Then, in \ref{ss:susy}, I review the fermion supersymmetry variations of
type IIB supergravity and show which of the self-dual flux backgrounds are 
supersymmetric (at tree level), as well as exploring other relevant 
solutions.

\subsubsection{Historical Note}\label{sss:historical}

Before launching into the actual compactifications that we will study,
it might be useful to note a few early studies of fluxes in string 
compactifications \cite{Polchinski:1996sm,Michelson:1997pn,Taylor:1999ii,
Curio:2000sc,Curio:2001ae,Dall'Agata:2001zh}.  These papers introduced
fluxes to Calabi-Yau compactifications, but most of them did not account
for the importance of string theoretic objects such as orientifold planes,
which we will develop below.  However, they did develop an understanding of
the role of fluxes in compactifications and provided motivation for the
self-dual flux compactifications we review here.

In brief, \cite{Polchinski:1996sm} introduced fluxes in type IIA 
compactifications, studying the flux-induced scalar potential using the 
language of 4D $\N =2$ supergravity, and \cite{Michelson:1997pn} extended
this work to type IIB supergravity.  The role of fluxes in string dualities
was studied in \cite{Curio:2001ae}; 
\cite{Taylor:1999ii,Curio:2000sc,Dall'Agata:2001zh} detailed various 
aspects of the potential. It is worth noting that \cite{Dall'Agata:2001zh}
found that fluxes create a runaway potential toward infinite volume on
the compactification manifold; presumably this effect arises because of the
need for negative tension objects such as orientifold planes.

\subsection{Solution to Equations of Motion}\label{ss:background}

Type IIB string theory reduces in the low energy limit to 10D type IIB
supergravity (SUGRA).  The fields of IIB SUGRA are the metric, a complex
scalar, two 3-form field strengths, and a self-dual 5-form field strength.
The dilaton-axion scalar $\tau=C_0+ie^{-\phi}$ combines the RR scalar with
the string coupling; thre is a
$SL(2;\bm{\mathbbm{Z}})$ S-duality that takes $\tau\to (a\tau+b)/(c\tau +d)$.
It is often convenient to combine the R-R ($F_3=dC_2$) and NS-NS 
($H_3=dB_2$) 3-forms 
into a single complex field $G_3=F_3-\tau H_3$ that transforms as
$G_3 \to G_3/(c\tau+d)$
\footnote{Sometimes it is also convenient to write $\t F_3=F_3-CH_3.$}.  
The 5-form $\tilde F_5 = dC_4 -C_2 H_3$  is 
neutral under S-duality, and self-duality $\star \tilde F_5=\tilde F_5$
is imposed as an equation of motion.  I work in the 10D string frame; much
of the literature uses 10D Einstein frame (see appendix \ref{aa:frames}).

As in equation (\ref{warp1}), we will take the most general Poincar\'e
invariant metric, although we find it convenient to scale the warp factor
out of the compact metric, as well.  We take as a fairly general ansatz
\bea
ds^2 &=& e^{2A} \eta_{\mu\nu}dx^\mu dx^\nu + e^{-2A} \tg_{mn}dx^m dx^n
\ \textnormal{and}\label{warp2}\\
\tilde F_5 &=& e^{-4A}\epsilon_{(4)} dF -e^{-4A}\star_{(6)} dF \ ,
\label{fiveform1}
\eea
with $F_3,H_3$ harmonic and $\tau$ satisfying its equations of motion.
The warp factor in the 5-form ensures that $\tilde F_{\mu\nu\lambda\rho m}$
is closed, as required by the Bianchi identity (\ref{bianchi5}) given in
appendix \ref{aa:frames} along with the other equations of motion.
As is consistent with Poincar\'e invarience, no fields vary in the spacetime.

\subsubsection{A SUGRA No-Go Theorem and a Stringy Evasion}\label{sss:nogo}

In examining the equations of motion for the metric, we will find the 
following  $D$ dimensional identities very helpful:
\bea
\b g_{MN}&=& e^{2\omega} g_{MN} \Rightarrow\nonumber\\
\b R_{MN} &=& R_{MN} - (D-2)\Del_M\Del_N\omega+(D-2)\Del_M\omega \Del_N\omega
\nonumber\\
&&-g_{MN}\Del^2\omega-(D-2)g_{MN}\left(\Del\omega\right)^2\label{riccirescale}\\
\b\Del^2\psi &=& e^{-2\omega}\Del^2\psi+(D-2)e^{-2\omega}\Del^P\omega\Del_P\psi
\, , \ \psi\ \textnormal{scalar}\label{laplacerescale}\ .\eea
In particular, for the metric (\ref{warp2}), the rescaling 
(\ref{laplacerescale}) means that the 10D Laplacian $\Del^2$ is equal to the
6D Laplacian $e^{2A}\t\Del^2$ acting on scalars independent of $x^\mu$.  
To see that, remember that 
$\Del^2$ has a term $-g^{\mu\nu} \Gamma^m_{\mu\nu}\del_m$ and that
$\Gamma^m_{\mu\nu}=-g_{\mu\nu}\Del^m A$.

The spacetime components of the Ricci tensor of equation (\ref{warp2})
are
\be\label{ricci1}
R_{\mu\nu} = -\eta_{\mu\nu} e^{4A} \tilde\Del^2 A\ , \ee
where $\tilde\Del$ is the covariant derivative with respect to the 
\textit{six dimensional} metric $\tg_{mn}$ \cite{Giddings:2001yu}. 
One way to find equation (\ref{ricci1}) is to use (\ref{riccirescale}) 
with metric (\ref{warp2}); take $\b R_{\mu\nu}$ from (\ref{warp2}),
and use $e^{-2A}$ times (\ref{warp2}) on the right-hand side.  That metric has 
$\Gamma^m_{\mu\nu}=0,R_{\mu\nu}=0$, 
and rescaling the Laplacian back gets rid of the extra terms in 
(\ref{riccirescale}).  

The spacetime components of
Einstein's equation (\ref{einstein}) can be written as
\bea
R_{\mu\nu} &=& -\frac{1}{4}g_{\mu\nu}\Del^2\phi +\frac{1}{2}g_{\mu\nu}
\Del_m\phi\Del^m\phi -2\Del_\mu\Del_\nu\phi
-g_{\mu\nu} \frac{e^{2\phi}}{48}G_{mnp}\b G^{mnp} 
\nonumber\\
&&+\frac{1}{24} e^{2\phi} \tilde F_{\mu\lambda\rho\sigma\tau m}
\tilde F_\nu{}^{\lambda\rho\sigma\tau m}+\kten^2 e^{2\phi}
\left( T^\prime_{\mu\nu}-\frac{1}{8}
g_{\mu\nu}T^{\prime M}_M\right)\ .\label{einsteineqn1}
\eea
The local contributions, denoted by primes, 
represent string theoretic objects, such as D-branes
and orientifold planes; note that the stress tensor contribution is
multiplied by $e^{2\phi}$ due to the prefactor of the Hilbert term in
the string frame action.  The $\Del_\mu\Del_\nu\phi$ terms is nonvanishing 
because of the nontrivial
Christoffel symbol $\Gamma^M_{\mu\nu}$.
Remembering that the gradient on the dilaton is the full 10D derivative,
we can rewrite (\ref{einsteineqn1}) to 
\bea
e^{4A}\tilde\Del^2\left( A-\frac{\phi}{4}\right)\!\! &= \!\!& 
-\frac{1}{2} e^{4A}(\t\Del\phi)^{\t 2}+2e^{4A} \t\Del^m A\t\Del_m\phi
+ \frac{e^{2\phi}}{48}e^{2A}G_{mnp}\b G^{mnp} \nonumber\\
\!\!&\!\!& +\frac{e^{2\phi-6A}}{4} \del_m F\del^{m}\! F +\frac{\kten^2}{8}
e^{2A+2\phi}\left( T^{\prime m}_m-T^{\prime \mu}_\mu\right)
\label{einsteineqn2}
\eea
after tracing by $\eta^{\mu\nu}$.  (Here $T^\mu_\mu$ has indices raised
by the full string frame metric including warp factor.) Rescaling to a metric 
$\hat g_{mn}=e^{-\phi}\tg_{mn}$ gives
\bea
e^{4A-\phi}\hat\Del^2\left( A-\frac{\phi}{4}\right) &=&
\frac{e^{2\phi}}{48}e^{2A}G_{mnp}\b G^{mnp}
+\frac{e^{2\phi}}{4} e^{-6A}\del_m F\del^{m} F\nonumber\\
&&+\frac{\kten^2}{8}e^{2A+2\phi}
\left( T^{\prime m}_m-T^{\prime \mu}_\mu\right)
\ .\label{einsteineqn3}\eea
As was noted in \cite{Giddings:2001yu}, this equation demonstrates the
SUGRA no-go theorem of \cite{deWit:1987xg,Maldacena:2000mw} directly in IIB
supergravity; without the (purely stringy) local contributions, the right-hand 
side is positive definite, while the left-hand side integrates to zero
on a compact manifold.

In fact, for a $p$-brane extendend through the entire spacetime and wrapped
on a submanifold $\Sigma$ of the compact space, 
the local contribution to the stress tensor
gives
\be\label{local}
\left( T^m_m-T^\mu_\mu\right)= (7-p) T_p \delta(\Sigma)\ ,\ee
where $T_p$ is the $p$-brane tension \cite{Giddings:2001yu}.  Therefore,
to get a negative contribution to the right-hand side of (\ref{einsteineqn3}),
there must either be branes with $p>7$ or negative tension objects.  
In type IIB string theory, both D9-branes and negative tension
orientifold planes are stable objects.  From here on, we will focus on the
orientifolds.

\subsubsection{Stringy Constraints}\label{sss:constraints}

Gauss's law assures us that the charges for all the field strengths must
integrate to zero over the compact manifold.  We are particularly interested
in D3-brane charge, where Gauss's law comes from the 5-form Bianchi
identity (\ref{bianchi5}).  In fact, we can obtain both local and global
constraints on the compactification from the Bianchi identity.

\paragraph{Local Constraints}\label{p:localconst}

Consider first the derivative side of the Bianchi identity.  The components
with spacetime indices are trivial, so we consider the part with
only internal indices.  If we Hodge
dualize on it using the string frame metric in 6D, the derivative part 
becomes
\bea
\star_{(6)} d\t F_5 &=& -\star_{(6)} d\left( e^{-4A} \star_{(6)} dF \right)=
-e^{6A-3\phi} \hat\star_{(6)} d\left( e^{-8A+2\phi} \hat \star_{(6)} dF
\right)\nonumber\\
&=& e^{-2A-\phi} \hat\Del^2 F -2e^{-2A-\phi}\del^{\hat m} \left(4A-\phi\right)
\del_m F
\label{starbianchi1}\eea
Then the entire Bianchi identity becomes
\bea
\hat\Del^2 F &=& \frac{i}{12} e^{2A+2\phi} G_{mnp}\left(\star_{(6)}\b G
\right)^{mnp} +2\kten^2 \mu_3 e^{2A+\phi}\star_{(6)}\rho_3\nonumber\\
&&+2e^{-6A+2\phi} \del^m F \del_m\left( e^{4A-\phi}\right)\label{starbianchi2}
\eea
where $\rho_3$ is the charge density 6-form of D3-branes (normalized to $1$
for a single brane)
\cite{Giddings:2001yu}.

Now rewrite equation (\ref{einsteineqn3}) as an equation for the
second derivative of $e^{4A-\phi}$ and subtract equation (\ref{starbianchi2}).
With some algebraic rearrangement,
\bea
\hat\Del^2\left( e^{4A-\phi}-F\right) &=& \frac{e^{2A+2\phi}}{2} 
\left| iG-\star_{(6)} G\right|^2 +e^{-6A+2\phi} \left| \del\left(
e^{4A-\phi}-F\right)\right|^2\nonumber\\
&& +\frac{1}{2}\kten^2 e^{2A+2\phi}\left( T^{\prime m}_m-T^{\prime \mu}_\mu
-4\mu_3 e^{-\phi} \star_{(6)}\rho_3\right)\ .\label{stringnogo}\eea
Consider the last term of this equation.  According to equation
(\ref{local}), this term vanishes for a D3-brane, since the tension and
charge for D-branes are related by $T_p = \mu_3 e^{-\phi}$ (and the
number density is clearly given by the delta function).
As in \cite{Giddings:2001yu}, make the assumption that all the local
sources satisfy the inequality
\be\label{d3ineq}
T^{\prime m}_m-T^{\prime \mu}_\mu \geq 4\mu_3 e^{-\phi} \star_{(6)}\rho_3
\ .\ee
This inequality is valid for D3-branes and \bd-branes, as well as O3-planes
and fractional D3-branes.  
Of these, only \bd-branes do not saturate it because
they have a negative D3-brane charge density.  7-branes can have an
induced D3-brane charge due to worldvolume curvature and field strengths
(from $\ap$ corrections to the action, discussed more fully in SECTION),
and \cite{Giddings:2001yu} argued that they in fact saturate 
(\ref{d3ineq}).  We are, however, ruling out D5-branes and O5-planes; those
must appear in some other class of solutions.

With the inequality (\ref{d3ineq}), the three terms on the right hand side
of (\ref{stringnogo}) are positive definite, while the left hand side
integrates to zero.  This is the no-go theorem of \cite{Giddings:2001yu}:
for solutions with only local objects that obey (\ref{d3ineq}),
the 3-form is imaginary self-dual
\be\label{imsd}
iG_3=\star_{(6)} G_3\Rightarrow e^{-\phi}\star_{(6)}H_3 = -\t F_3\, ,\
\star_{(6)} \t F_3 = e^{-\phi} H_3
\ , \ee
the potential energy of a D3-brane cancels between the DBI and 
WZ parts (D3-branes feel no force)
\be\label{d3noforce}
F=e^{4A-\phi}\ ,\ee
and the inequality (\ref{d3ineq}) is saturated.  Because the D3-branes
feel no forces, we might expect that these solutions are related to
supersymmetric solutions, and, indeed, all the localized sources that 
saturate the bound preserve the same SUSYs.  We will see in the next
section, \ref{ss:susy}, which fluxes also preserve SUSY.

We should also mention that the constraints and equation (\ref{starbianchi2})
lead to Poisson's equation for the warp factor:
\be\label{warppoisson}
-\hat\Del^2 e^{-(4A-\phi)} = \frac{1}{12} e^\phi G_{mnp}\b G^{\widehat{mnp}}
+2\kten^2 \mu_3 \hat\star_{(6)} \rho_3\ .\ee
This equation tells us that the warp factor becomes trivial in the large radius
limit (except at special points), as follows.
If we scale $\hat g_{mn}\to e^{2u}\hat g_{mn}$, then the left hand side
goes as $e^{-2u}$, while the right hand side goes as $e^{-6u}$.
Therefore, the warp factor (combined with the dilaton) becomes a constant,
which we take to be unity.  So in a case where self-duality of the 
flux (\ref{imsd}) freezes the dilaton, $A\sim e^{-4u}$.

With all of these conditions satisfied, we only need to discuss the
metric $\tg_{mn}$, 3-form, and dilaton-axion.  The simplest model is
that the dilaton and axion are constant, in which case $\tg_{mn}$ is a
(orientifolded) CY 3-fold.  Then the 3-form equations of motion 
(\ref{3formeom}) are
satisfied by closed 3-forms as long as equations (\ref{imsd},\ref{d3noforce})
hold.  
The case of nonconstant
$\tau$ is more complicated.  It has been shown in 
\cite{Giddings:2001yu} that these form a compactification of F-theory
(given here in the IIB string frame) on a CY 4-fold.  
In F-theory, $\tau$ represents
the complex structure of a $T^2$, and the degeneration points of the 
torus are $(p,q)$7-branes.  These 7-branes can be described in string
perturbation theory when particular combinations of them coincide.  This
is the ``orientifold limit'' of F-theory in which each O7-plane is
coincident with 4 D7-branes, precisely enough so that the D7-brane
charge cancels locally \cite{Sen:1996vd}.  In the orientifold limit,
$\tau$ becomes a constant once again.

\paragraph{Global Constraints}\label{p:globalconst}

Constraints global on the compactification manifold arise from 
integrating the Bianchi identity (\ref{bianchi5}).  If the 
F-theory 4-fold has Euler number $\chi$, the integrated Bianchi identity
becomes \cite{Becker:1996gj,Sethi:1996es}
\be\label{intbianchi}
\frac{\chi}{24} = N_{\textnormal{D3}} -\frac{1}{4} N_{\textnormal{O3}}
+\frac{1}{4} N_{\widetilde{\textnormal{O3}}}
+\frac{1}{2\kten^2 \mu_3}\int H_3\wedge F_3 
\ .\ee
Here $N_{\textnormal{O3}}$ is the number of ``regular'' O3-planes and
$N_{\widetilde{\textnormal{O3}}}$ the number of O3-planes with flux
discussed below under the quantization of fluxes.  Both
\cite{Sethi:1996es,Giddings:2001yu} argued that $-\chi/24$ is the complete
D3-brane number induced on 7-branes through $\ap$ corrections to F-theory 
(or, rather, the D-brane action in string theory).\footnote{Technically
speaking, the O3 charges should also be included in $\chi$, but since we
always look at the O3- and 07-plane cases separately, it is useful to 
separate them.} We have left off
\bd-brane number because anti-branes do not saturate the inequality
(\ref{d3ineq}).

Let us also note that this equation is a nonperturbative generalization of
worldsheet tadpole cancellation conditions, which are necessary for 
consistency of the string theory (just as the integrated Bianchi is a 
geometric consistency condition).  Hence we will sometimes refer to 
the integrated Bianchi in terms of tadpole cancellation.

\subsubsection{Orientifold Projections}\label{sss:orientproj}

There are two types of orientifold planes that could appear in the 
self-dual flux compactifications, as discussed above.

\paragraph{O3-planes}\label{p:orient3}

If the compactification manifold has a reflection symmetry of all
compact coordinates $x^m$,
\begin{equation}
R: \quad (x^4, x^5, x^6, x^7, x^8, x^9) \to  (-x^4, -x^5, -x^6, -x^7, -x^8,
-x^9)\ ,
\end{equation}
it can be orientifolded by placing O3-planes at the fixed points.

The action of the orientifold
$\Z$ can be derived by using $T$-duality to the type I theory, where
$g_{MN}$, $C_{2}$, and
$\phi$ are even under world-sheet parity $\Omega$ and $B_{2}$,
$C_{0}$, and $C_{4}$ are odd.
Alternately, one may derive it by noting that the orientifold $\Z$
must include a factor of $(-1)^{\bm{F}_L}$, where $\bm{F}_L$ is the
spacetime fermion number carried by the left-movers:
$\mathcal{R} \equiv \Omega R (-1)^{\bm{F}_L}$ 
\cite{Dabholkar:1996pc,Sen:1996vd}.  This is necessary in order that
it square to unity,
\begin{equation}
\mathcal{R}^2 =
\Omega^2 R^2 (-1)^{\bm{F}_L + \bm{F}_R} =1\ .
\end{equation}
Note that $\Omega^2 = 1$, as $\Omega$ acts as $\pm 1$ on all fields.
$R$ is equivalent to a rotation by $\pi$ in each of three planes, so $R^2$
is a rotation by $2\pi$ in an odd number of planes and therefore equal to
$(-1)^{\bm{F}}$.

By either means one finds that $\Z$ acts on the various fields as
follows:
\begin{eqnarray}
\textnormal{even:}&&
g_{\mu\nu}\ ,\ g_{mn}\ ,\ B_{\mu m}\ ,\ C_{\mu m}\ , \
C_{mnpq}\ ,\ C_{\mu\nu mn}\ ,\ C_{\mu \nu\lambda\rho}\ ,\
\phi\ ,\  C \ ;\nonumber\\
\textnormal{odd:}&& g_{\mu m}\ ,\ B_{\mu \nu}\ ,\ B_{mn}\ ,\ C_{\mu \nu} \
,\ C_{mn}\ ,\  C_{\mu mnp}\ ,\ C_{\mu\nu\lambda m}\ .
\label{reflect}
\end{eqnarray}
It follows that the fluxes $H_{mnp}$ and $F_{mnp}$ are even, and so
harmonic three-form fluxes are allowed.

\paragraph{O7-planes}\label{p:orient7}

In F-theory compactifications, there is a limit in which the $(p,q)$7-branes
come together to form familiar objects from string perturbation theory, 
namely an O7-plane with 4 coincident D7-branes\cite{Sen:1996vd}.  
The boundary conditions of the SUGRA fields at the O7-planes can be
determined just as those at O3-planes.  In fact, the results are given
in (\ref{reflect}) with $\mu,\nu$ going to indices parallel to the O7 and
$m,n$ going to indices perpendicular.  In particular, $H_3,F_3$ must always
have one leg orthogonal to the O7, and $\tilde F_5$ has either 0 or 2
legs orthogonal to the O7.  This is as in
\cite{Dasgupta:1999ss,Tripathy:2002qw,Becker:2002sx}.

\subsubsection{Quantization of Flux}\label{sss:fluxquant}

The three-form fluxes must be appropriately quantized
by a generalization of the argument for monopoles.  The usual
quantization conditions are
\begin{equation}
\frac{1}{2\pi \alpha'} \int_{C} H_{3} \in 2\pi \bm{\mathbbm{Z}}\ ,\quad
\frac{1}{2\pi \alpha'} \int_{C} F_{3} \in 2\pi \bm{\mathbbm{Z}}
\label{gquant}
\end{equation}
for every three-cycle $C$.  (Note that $\t F_5$ integrates to zero.)
However, orientifolds can present subtleties.
To understand the issues, we can look at the simplest possibility of 
compactification on $T^6/\Z$ with O3-planes\footnote{This presentation
was given in \cite{Frey:2002hf,Grana:2002ti}.}.

Consider first $T^6$ compactification without the orientifold.  
Letting $C$ run over all $T^3$'s, one finds that constant fluxes
\begin{equation}
H_{mnp} = \frac{\alpha'}{2\pi R_1R_2R_3} h_{mnp}\ ,\quad F_{mnp} =
\frac{\alpha'}{2\pi R_1 R_2R_3} f_{mnp}\ ;
\quad h_{mnp}\ ,\ f_{mnp} \in \bm{\mathbbm{Z}}
\label{3quant}
\end{equation}
are allowed, where the three coordinates on the $T^3$ have periodicities
$2\pi R_i$.  Any cycle on the covering space $T^6$ descends to a cycle on
$T^6/\Z$, so the
conditions (\ref{3quant}) are still necessary.  In addition, there are new
3-cycles on the coset space, such as
\begin{equation}
0 \leq x^4 \leq 2\pi R_1\ ,\
0 \leq x^5 \leq 2\pi R_2\ ,\
0 \leq x^6 \leq \pi R_3\ ,\
x^7 = x^8 = x^9 = 0\ . \label{halfc}
\end{equation}
The conditions (\ref{gquant}) on this cycle\footnote
{The cycle (\ref{halfc}) is unoriented, but the three-form fluxes can be
integrated on it because they have odd intrinsic parity.}
would appear to
require that
$h_{456}$ and $f_{456}$ be even, and similarly for all other components.
However, we claim that $h_{mnp}$ and $f_{mnp}$ can still be arbitrary odd
or even integers.

To understand this, consider first the reduced problem of a charge moving
in a constant magnetic field $F_{12}=B$ on a square 
torus $0 \leq x^{1,2} \leq 2\pi R$. Let us work in the gauge
\begin{equation}
A_1 = -\frac{1}{2}Bx^2\ ,\quad A_2 = \frac{1}{2} B x^1\ .
\end{equation}
The gauge field is periodic up to a gauge transformation,
\begin{equation}\label{gaugetr}
A_m (x^1 + 2\pi R, x^2) = A_m (x^1 , x^2) + \partial_m \lambda_1\ ,
\
A_m (x^1, x^2 + 2\pi R) = A_m (x^1 , x^2) + \partial_m \lambda_2\ ,
\end{equation}
with $\lambda_1 = \pi R B x^2$ and $\lambda_2 = -B\pi R x^1$.  
Similarly a field of charge $e$ satisfies
\begin{equation}
\psi (x^1 + 2\pi R, x^2) = e^{i e \lambda_1} \psi (x^1 , x^1) \ ,
\
\psi (x^1, x^2 + 2\pi R) = e^{i e \lambda_2} \psi (x^1 , x^2)\ .
\label{gtran}
\end{equation}
Now suppose that we translate the charged particle around the path
$x^1=0\to 2\pi R$, $x^2=0\to 2\pi R$, $x^1=2\pi R\to 0$, then
$x^2=2\pi R\to 0$.  The combined gauge transformations give a phase of
$\exp [-iB(2\pi R)^2]$.  Since we have described a contractible path,
any nonsingular wavefunction must be single-valued, so
\be\label{Bquant}
B=\frac{n}{2\pi e R^2}\ ,\ n\in \bm{\mathbbm{Z}}\ .\ee

Now let us form the orbifold $T^2 /\Z$ (this is topologically $S^2$)
by identifying
$(x^1,x^2)$ with $(-x^1,-x^2)$.  For any value of
$n$ we can define the quantum mechanics for the charged particle on the
coset space simply by restricting to wavefunctions such 
that 
\begin{equation}
\psi (-x^1, -x^2) = + \psi (x^1 , x^2) \ . \label{gref}
\end{equation}
We have chosen a gauge in which $A_m$ is explicitly 
$\Z$ symmetric, so no gauge transformation is needed.
However, the integral of $F_{12}$ over the orbifold 
is half of the integral over $T^2$, so for $n$ odd the flux
is not quantized.

To see how this can make sense, note that there are four fixed points
$(x^1,x^2) = (0,0),\ (\pi R,0),\ (0,\pi R),\ (\pi R,\pi R)$.  
At the first three, the
periodicities (\ref{gtran}) and (\ref{gref}) are compatible, but at
$(\pi R,\pi R)$ they are incompatible and the wavefunction must
vanish.  If we circle this fixed point, from
$(\pi R-\epsilon,\pi R)$ to the identified point
$(\pi R+\epsilon,\pi R)$, the wavefunction is required to change sign:
there is a half-unit of magnetic flux at the fixed point
$(\pi R,\pi R)$.  Thus the Dirac quantization condition is in
fact satisfied.
Of course, the fixed point $(\pi R,\pi R)$ is not special: the quantization
condition is satisfied if there is a half-unit of flux at any one fixed
point, or at any three.  Similarly for $n$ even there can be
half-integer flux at zero, two, or four fixed points.  In each case
there are eight configurations, which can be obtained in the orbifold
construction by including discrete Wilson lines on the torus, and a
discrete gauge transformation in the orientifold projection.

This analysis extends directly to the quantum mechanics of an F-string or
D-string wrapped in the 4-direction, moving in the fluxes $H_{456}$ and
$F_{456}$.  This is consistent for any integers $h_{456}$ and $f_{456}$,
but if either of these is odd then there must NS-NS or R-R flux at
some fixed points, for example all those with $x^4 = x^5 =
x^6 =\pi R_{1,2,3}$.
Indeed, there are four kinds of O3-plane, distinguished by the presence or
absence of discrete NS-NS and R-R fluxes \cite{Witten:1998xy}; for
recent reviews see \cite{Hanany:2000fq,Bergman:2001rp}.  The
cycle(\ref{halfc}), and each of the others obtained from it by a rotation
of the torus, contains four fixed points.  If the NS-NS flux through the
cycle is even (odd) then an even (odd) number of the fixed points must have
discrete NS-NS flux, and correspondingly for the R-R flux.   An O3-plane
with no discrete fluxes has a charge and tension equal to $-1/4$ that of a 
D3-brane, but the O3-planes with any discrete fluxes have $+1/4$ D3-brane
charge and tension.

There are also multiple types of O7-planes \cite{Witten:1998bs}.  
However, since we are
mainly interested in the orientifold limit of F-theory, which requires
negative tension O7-planes, I will not discuss them further.

\subsection{Solution to Supersymmetry Conditions}\label{ss:susy}

For supersymmetric solutions, it is often both convenient and illustrative
to solve the supersymmetry constraints.  This strategy will also happen
to demonstrate which of our classical backgrounds are supersymmetric.
As it turns out, 
there are several well-known classes of supersymmetric solutions to IIB
SUGRA.  In this section, I will describe three of them in some detail by
finding a class of solutions that interpolates between them.  This work
is to reported in \cite{Frey:2003sd}.  Similar solutions of M-theory are
described in \cite{Martelli:2003ki}.

\subsubsection{Nomenclature of IIB Solutions}\label{sss:nomenclature}

We start by giving a brief overview of the three types of solutions:

\begin{enumerate}

\item In 1986, A. Strominger \cite{Strominger:1986uh} gave a comprehensive
description of supersymmetric solutions of type I supergravity from the
perspective of the heterotic string.  The solutions involve only the
NS-NS fields in the bosonic sector, so they equally well describe IIB 
SUGRA in the NS-NS sector, including the NS5-brane solution.  These type A
(light-heartedly named for A. Strominger) include the NS5-brane background
and notably the AdS/CFT background of \cite{Maldacena:2000yy}.  While the
compactifications we study in this review are not type A solutions,
we will see in section \ref{ss:tdual} that they are related.  A recent
exposition of type A compactificatoins in the heterotic string is
\cite{Becker:2003yv}.

\item Type B solutions are named for K. \& M. Becker \cite{Becker:1996gj},
who discovered them in M-theory.  In type IIB, they are generalized D3-brane
backgrounds, as described in \cite{Grana:2000jj,Gubser:2000vg}.  In 
particular, the spacetime warp factor and the 4-form potential are simply
related, so that they cancel in the D3-brane action.  As in section
\ref{ss:background}, the 3-forms satisfy imaginary self-duality 
$\star_6 G_3=iG_3$.  Type B solutions can also be generalized to include
a nontrivial dilaton-axion, as in F-theory
\cite{Kehagias:1998gn,Grana:2001xn}.  Indeed, the supersymmetric
cases of the backgrounds described above are type B solutions.

\item We call the third class of solutions type C solutions to continue
the pattern.  These solutions are the S dual of type A solutions and 
therefore include D5-brane solutions, and they are also related to 
type B solutions by duality, as discussed in section \ref{ss:tdual}.
They are very well understood because of their simple relation to type A
solutions.

\end{enumerate}

We will give below a class of solutions that interpolates between type B
and C solutions, since we will discuss both types of solution.  Solutions
that interpolating between types A and B can be obtained easily by S duality.

\subsubsection{SUSY Variations and Ansatz}\label{sss:susyvar}

In a supersymmetric solution, the supersymmetry variations of all fields
in the background should vanish.  Since expectation values of an odd number
of fermions must vanish classically, the SUSY variations of bosonic fields
automatically vanish.  Consider then a background with only bosonic 
components.  The fermion SUSY variations in the string frame and in our
conventions for the fields are 
\cite{Schwarz:1983wa,Schwarz:1983qr,Bergshoeff:1999bx,Hassan:1999bv}
\begin{eqnarray}
\delta\lambda&=& \frac{1}{2} \Gamma^M \partial_M \phi \varepsilon  
-\frac{1}{24}
\Gamma^{M}\del_{M}C\, (i\sigma^2)\, \varepsilon - \frac{e^\phi}{24}
\Gamma^{MNP}\t F_{MNP}\, \sigma^1\, \varepsilon \nonumber\\
&&-\frac{1}{24}\Gamma^{MNP}H_{MNP}\, \sigma^3\,\varepsilon \nonumber\\
\delta\psi_{M}&=&  D_{M} \varepsilon +\frac{e^\phi}{8} \,\Gamma^N \Gamma_M\,
\del_{N} C (i \sigma ^2) \varepsilon -\frac{1}{8}
\Gamma^{NP} H_{MNP}\, \sigma^3 \varepsilon  \nonumber\\
&&+\frac{e^\phi}{48} 
\Gamma^{NPQ} \Gamma_M  \t F_{NPQ} \sigma^1 \varepsilon
\! +\! \frac{e^\phi}{16 \cdot 5!} 
\Gamma^{NPQRS} \t F_{NPQRS} \Gamma_M (i\sigma^2) \varepsilon
\label{susyvars}
\end{eqnarray}
with $\t F_3 = F_3-CH_3$.  The spinors all represent two Majorana-Weyl
spinors; for example, the superparameter is
\be\label{vareps}
\varepsilon =\left[ \begin{array}{c} \varepsilon_1 \\ \varepsilon_2
\end{array}\right] ,\ \Gamma_{(\widehat{10})}\varepsilon =-\varepsilon\ .\ee
The Pauli matrices $\sigma^i$ mix the two spinors $\varepsilon_{1,2}$.\
Finally, $D_M$ is a covariant derivative containing the spin connection
and a $U(1)$ connection related to the construction of the 
dilaton-axion system as a coset (the $U(1)$ connection vanishes when the R-R
scalar is uniform).

For the background, we take essentially the metric as in equation 
(\ref{warp1}) and the 5-form as in (\ref{fiveform1});
this does not specify a particular warp factor on the internal manifold.
For the spinors, we take 
\be\label{spinoransatz}
\varepsilon= e^{A/2} \left[ \begin{array}{c} 
\zeta\otimes \chi + \zeta^* \otimes \chi^* \\
i e ^{i\alpha} \zeta\otimes \chi - i  e ^{-i\alpha} 
\zeta^* \otimes \chi^*\end{array}\right]\ ,
\ee
where $\gamma_{(\hat 4)}\zeta=\zeta, \gamma_{(\hat 6)} \chi=-\chi$.
We will see that we recover the results of section \ref{ss:background}
and type B solutions in the $\alpha\to 0$ limit, while type C has
$\alpha\to \pi/2$.  For convenience, we will set the R-R scalar $C=0$
throughout this section.  The results should generalize to allow
type B F-theory compactifications.

\subsubsection{Gravitino Equations}\label{sss:gravitino}

From $\delta \psi_{\mu}$, we find that
\be
\gamma^{\mu} \gamma_{(\hat 4)}
\left(\frac{1}{2} \gamma^m \partial_m A  +\frac{i}{8}
e^{-4A} e^{\phi} \gamma^m \partial_m F \gamma_{(\hat 4)} i\sigma^2 -
\frac{e^\phi}{48} F_{mnp}\gamma^{mnp} \sigma^1 \right) \varepsilon=0
\ee
which means by 4D Lorentz invariance
\be
\left(\frac{1}{2} \gamma^m \partial_m A  +\frac{i}{8}
e^{-4A} e^{\phi} \gamma^m \partial_m F \gamma_{( \hat 4)} i\sigma^2
-\frac{1}{48}e^{\phi} F_{mnp}\gamma^{mnp} \sigma^1\right)
 \varepsilon=0\ .
\label{gravitino0}
\ee
From  $\delta \psi_{m}$ we get
\bea
0&=&\Del_m \varepsilon -\frac{1}{8}H_{mnp} \gamma^{np} \sigma^3 \varepsilon 
+ \frac{e^\phi}{8} F_{mnp}\gamma^{np} \sigma^1 \varepsilon 
-\frac{e^\phi}{48}  \gamma_m F_{npq}\gamma^{npq} \sigma^1 \varepsilon
\nonumber\\
&& +\frac{i}{8}
e^{-4A} e^{\phi}\gamma^n \gamma_m \partial_n F \gamma_{(\hat 4)} 
i\sigma^2 \varepsilon 
\eea
The last two terms, using $\gamma^n \gamma_m=2\gamma^n\,_m+\gamma_m 
\gamma^n$, can be written in terms of derivative of the warp factor
using the $\delta \psi_{\mu}$ equation (\ref{gravitino0}), 
and we are left with
\bea
0&=& \left(\Del_m -\frac{1}{2} \partial_m A - \frac{1}{2} \gamma_m{}^n 
\partial_n A  - 
\frac{ie^\phi}{4}e^{-4A} \gamma_m{}^n \partial_n F \gamma_{(\hat 4)} 
i \sigma^2 
-\frac{1}{8} H_{mnp}\gamma^{np} \sigma^3 \right.\nonumber\\
&&\left. + \frac{1}{8} F_{mnp} \gamma^{np}\sigma^1 \right)\varepsilon\ .
\label{gravitino1}
\eea

Inserting the ansatz (\ref{spinoransatz}) 
for the spinors in (\ref{gravitino1}) and using
(\ref{relhA}) below\footnote{We don't need to use (\ref{relhA}) here, but 
it makes things look a little simpler.},  we get
\bea
0\!\!&=\!\!\!&\left(\Del_m +\frac{e ^{2i \alpha}}{2} 
\gamma_m{}^n \partial_n A   
-\frac{1}{8} H_{mnp}\gamma^{np}  + \frac{ie ^{i \alpha+\phi}}{8} F_{mnp}
\gamma^{np}\right)
\chi\ , \label{grav1}\\
0\!\!&=\!\!\!&
\left(\!\Del_m \! +\! \frac{e ^{-2i \alpha}}{2}  \gamma_m{}^n \partial_n A   
+\!\frac{1}{8} H_{mnp}\gamma^{np}\!  -\! \frac{ie ^{-i \alpha+\phi}}{8} 
F_{mnp}\gamma^{np}
+\!i\del_m\alpha\right)\!\! \chi\! .
\label{grav2}
\eea
The complex conjugate equations also follow from the $\chi^*$ part of
$\varepsilon$.
Now add up (\ref{grav1},\ref{grav2}) to get
\be
\left(\Del_m +\frac{1}{2}\cos(2\alpha)\del_n A\gamma_m{}^n
-\frac{1}{8} \sin(\alpha) e ^{\phi} F_{mnp}\gamma^{np} 
+\frac{i}{2}\del_m\alpha
\right) \chi =0\ .
\label{grav3}
\ee
Now take $\chi= e^{-i\alpha/2}\chi^\prime$.
If we assume for now that 
\be\label{defAprime}
\cos(2\alpha)\del_m A = \del_m A^\prime\ee is a 
total derivative, then we can rescale the internal metric
\be\label{rescale}
ds^2_6 = e^{-2A^\prime} d\tilde{s}_6^2\ \ee
to eliminate the warp factor terms,
which clearly agrees with the type B metric (\ref{warp2}) at $\alpha=0$.
Note that results below will show that there is such an $A^\prime$ as
long as there is a complex structure.
Then we have   
\be\label{cov+torsion}
\left(\tilde{\Del}_m -\frac{1}{8} \sin(\alpha) e ^{\phi+2A^\prime} F_{mnp}
\tilde{\gamma}^{np}\right) \chi^\prime =0\ ,\ee
which means that our manifold has torsion
and that there is a normalized spinor that is covariantly
constant with respect to this torsion, with which we can build a
complex structure \cite{Strominger:1986uh}. Equation (\ref{cov+torsion})
leads to many more conclusions \cite{Strominger:1986uh}, but we will leave
those to section \ref{sss:complexstructure} below.  For now, we just need
that the manifold has a complex structure, and we can without loss
of generality take $\gamma^{\bi} \chi=0$ because of the chirality.

Inserting (\ref{spinoransatz}) into equation (\ref{gravitino0}) gives us
\bea
\left(\frac{1}{2} \gamma^m \partial_m A  -\frac{1}{8}e^{-4A} e^{\phi} 
e ^{i\alpha}
 \gamma^m \partial_m F
-\frac{i}{48} e ^{i\alpha} e^{\phi} F_{mnp}\gamma^{mnp} \right)
 \chi&=&0 \nonumber\\
\left(\frac{1}{2} i e ^{i\alpha} \gamma^m \partial_m A  
-\frac{i}{8} e^{-4A} e^{\phi} \gamma^m \partial_m F
-\frac{1}{48} e^{\phi} F_{mnp}\gamma^{mnp} \right)
 \chi&=&0\ . \label{gravitino01}
\eea
Multiplying the second by $-ie ^{i\alpha} $ and adding both up, we get
\be
\del_m F = 4e^{4A} e^{-\phi}\cos\alpha\del_m A
\label{relhA}
\ee
because the $\gamma^i$ terms do not annihilate $\chi$ automatically
(and the conjugate equation also holds from (\ref{gravitino0})).
Note that this means the right hand side of (\ref{relhA}) is a total
derivative, at least in patches.
We also see  
from (\ref{gravitino01}) that  
\be
F_{ijk}=F_{\bi\bj\b k}=0
\ee
(and hence from the dilatino equations (\ref{dilatino1},\ref{dilatino2})
that $H_{ijk}=H_{\bi\bj\b k}=0$, so there are no (3,0) and (0,3)
pieces).
The other equation that we get from equations (\ref{gravitino01}) is
\be\label{FA1}
e^\phi F_{ij}{}^j = 4\sin\alpha \del_i A\ .\ee
Again, the complex conjugate follows from the equations for $\chi^*$.

Substituting (\ref{grav3}) into (\ref{grav1},\ref{grav2}), we get
\be
\left(\frac{i}{2} \sin(2\alpha)   \gamma_m{}^n \partial_n A   
-\frac{1}{8} H_{mnp}\gamma^{np}   + \frac{i}{8}\cos(\alpha)  e^{\phi} 
F_{mnp}\gamma^{np} -\frac{i}{2}\del_m\alpha\right)\chi=0
\ .\label{grav4}
\ee
For $m=i$, we end up with
\be\label{grav5}
\left(H-i\cos\alpha e^\phi F\right)_{ij}{}^j = 2i\del_i\alpha-
2i\sin(2\alpha)\del_i A\ ,\ee
and, for $m=\bi$, we get
\bea
\left(H-i\cos\alpha e^\phi F\right)_{\bi\bj}{}^{\bj} = 2i\del_{\bi}\alpha
+2i\sin(2\alpha)\del_{\bi} A\label{grav6}\\
\left[-\left(H-i\cos\alpha e^\phi F\right)_{\bi jk}\gamma^{jk}
+4i\sin(2\alpha)\del_j A\gamma_{\bi}{}^j\right]\chi =0\ .\label{grav7}
\eea

Then we end up getting
\be\label{final}
H_{ij}{}^j=-2i\sin(2\alpha)\del_i A=-i\cos(\alpha)e^\phi F_{ij}{}^j 
=2i\del_i \alpha \ee
and
\be\label{21HF}
\left(\cos(\alpha) F -ie^{-\phi} H\right)_{i \bj \b k}
=-4 e^{-\phi} \sin(2\alpha)g_{i[\bj }\partial_{\b k]}A\ .
\ee
The second equality in equation (\ref{final}) follows from equation 
(\ref{FA1}).

\subsubsection{Dilatino Variation}\label{sss:dilatinovar}

Now, let's turn to the two equations coming from $\delta \lambda=0$:
\bea
\frac{1}{2}\gamma^m \partial_m \phi \chi -\frac{1}{24}H_{mnp}\gamma^{mnp} \chi
-\frac{i}{24} e ^{i\alpha} e ^{\phi} F_{mnp}\gamma^{mnp} \chi =0 
\ ,\label{dilatino1} \\
 \frac{1}{2}\gamma ^m \partial_m \phi \chi + 
\frac{1}{24}H_{mnp}\gamma^{mnp} \chi
+\frac{i}{24}  e ^{-i\alpha} e ^{\phi} F_{mnp}\gamma^{mnp} \chi =0\ .
 \label{dilatino2} 
\eea
Substracting (\ref{dilatino2}) from (\ref{dilatino1}), we get
\be
\left(H + i \cos(\alpha) e ^{\phi} F\right)_{mnp}\gamma^{mnp} \chi =0
\ee
from which
\be
\cos(\alpha) F_{ij}{}^j -i e ^{-\phi} H_{ij}{}^j =0
\label{tracesFH2}
\ee
(the (3,0) pieces of both fluxes are zero) and the conjugate equation.

Finally, adding  (\ref{dilatino2}) and (\ref{dilatino1}), we get the equation 
\be
\frac{1}{2}\gamma ^m \partial_m \phi \chi +
\frac{1}{24}  \sin(\alpha) e ^{\phi} F_{mnp}\gamma^{mnp} \chi =0 
\ ,\label{dilatinouseful} 
\ee
yielding
\be
\sin(\alpha) e ^{\phi} F_{ij}{}^j = 2 \partial_i \phi
=4\sin^2\alpha\del_i A\ .
\label{traceF}
\ee
Here we have used equation (\ref{FA1}) to write the last equality.

\subsubsection{Scalar Relations}\label{sss:scalarrelate}

We now have many relations among $h,A,A^\prime,\phi,\alpha$, which are
in fact sufficient to determine them all in terms of a single function.
Equations (\ref{defAprime},\ref{relhA},\ref{final},\ref{traceF}) lead to
\bea
A&=& -\frac{1}{2}\ln\tan\alpha +A_0\label{Aalpha}\\
h&=&  e^{4A_0} e^{-\phi_0} \cot^2 (\alpha) +h_0\label{halpha}\\
\phi&=& \ln\cos\alpha +\phi_0\label{phialpha}\\
A^\prime&=& -\frac{1}{2}\ln\sin(2\alpha) +A^\prime_0\ .\label{Apalpha}
\eea
The subscript $0$ denotes an integration constant that is uniform over
the entire compact space, while $\alpha$ is a function of the internal
coordinates.
Note that these scalar solutions appear to be singular 
when $\alpha\to 0,\pi/2$, which I have claimed are the
type B and C limits.  
Let us make the type B and C limits precise.
We can include the type B and C cases if we allow the integration
constants to be infinite, so that they cancel the
divergences due to the dependence on $\alpha$.  

\paragraph{Type B}\label{p:typeB} 

Let us consider first
the type B limit, where $\alpha(x)=\delta\beta(x)$ with $\delta\to 0$ a 
constant.  Then the scalar equations (\ref{Aalpha},\ref{halpha},
\ref{phialpha},\ref{Apalpha}) become
\bea
A&=& -\frac{1}{2}\ln\beta +\hat A_0, \ A_0 =\hat A_0+\frac{1}{2}\ln\delta 
\nonumber\\
h&=&  e^{4\hat A_0} e^{-\phi_0} \beta^{-2} +h_0=-e^{-\phi}e^{4A}+h_0
\nonumber\\
\phi&=& \phi_0=\mathnormal{constant}\nonumber\\
A^\prime&=& -\frac{1}{2}\ln\beta +\hat A^\prime_0,\ A^\prime_0 
=\hat{A}^\prime_0+\frac{1}{2}\ln(2\delta),\ A^\prime = A+
\mathnormal{constant}\ .\label{renormB}
\eea
The relations (\ref{renormB}) are indeed what we expect from
section \ref{ss:background} for type B solutions without including
F-theoretical variations of the dilaton-axion
(the constant in the relation between $A^\prime$ and $A$ is usually set to
zero).  Additionally, as $\delta\to 0$, eqn (\ref{21HF}) shows that
the $(1,2)$ part of $G_3$ vanishes (we already know that the $(3,0)$ part
also vanishes).  Something new we find is that also the the $(0,3)$ part of
$G_3$ vanishes, so not all off the compactifications of section
\ref{ss:background} are supersymmetric.  
Finally, eqn (\ref{tracesFH2}) 
also shows that the contraction of $G_3$ with the metric vanishes ($G_3$
is primitive), so we we have the well-known result
that $G$ should be a primitive $(2,1)$ form for type B solutions.

\paragraph{Type C}\label{p:typeC}

For type C, let $\alpha(x^m) = \pi/2 -\delta\beta(x^m)$, where again
$\delta\to 0$ is a constant.  Then the scalars become
\bea
A&=& \frac{1}{2}\ln\beta +\hat{A}_0,\ A_0 =\hat{A}_0-\frac{1}{2}\ln\delta
\nonumber\\
h&=& e^{4\hat{A}_0}e^{-\hat\phi_0} \delta\beta^2 +h_0\to h_0=
\mathnormal{constant}\nonumber\\
\phi&=& \ln\beta +\hat\phi_0,\ \phi_0=\hat\phi_0-\ln\delta,\ 
\phi=2A+\mathnormal{constant}\nonumber\\
A^\prime &=& -\frac{1}{2}\ln\beta +\hat A^\prime_0,\ A^\prime_0 
=\hat{A}^\prime_0+\frac{1}{2}\ln(2\delta),\ A^\prime = 
-A+\mathnormal{constant}\ .\label{renormC}
\eea
These are, in fact, the expected relations among the scalars for type C.
In the familiar type A solution, the warp factor vanishes in the string 
frame, and by S duality we should have $g_{mn}(\textnormal{A})=
e^{-\phi(\textnormal{\scriptsize C})}g_{mn}(\textnormal{C})=
\tg_{mn}(\textnormal{C})$.  Additionally, the 
NS-NS 3-form vanishes because of 
(\ref{21HF}) and its conjugate (along with the fact that no $(0,3),(3,0)$
fluxes are allowed).  This is just what we would expect for a D5-brane
type background.

\comment{
Solutions that go to the type B or C limit at some position pose an
interesting problem.  In some cases, the divergent scalars give what we
expect due to the presence of a source, such as the dilaton in the presence
of a D5-brane.  Then we would not want to renormalize out the divergence.
We will examine this case in more detail in the example of the D5/D3 bound
state.
In cases where the solution approaches a limit at infinity,
the situation is much more complicated, depending on the expected solution.
For example, if the solution should go to type B AdS at infinity, the
warp factor is expected to diverge in a certain way, so, again, we might
not have to renormalize out that divergence.}

\subsubsection{Relations to Complex Structure}\label{sss:complexstructure}

The Killing spinor equation (\ref{cov+torsion}) implies that the 
supersymmetry parameters feel a torsion in the metric $d\tilde s_6^2$;
the torsion is equal to 
\be\label{torsion}
T=-\frac{1}{2} \sin (\alpha) e^{\phi+2A'} F=-\frac{1}{4}e^{\phi_0+2A'_0} F
\ .\ee
Clearly, if we go to the type B limit, the torsion vanishes, which means
that $d\tilde s^2_6$ is a Calabi-Yau manifold, which we also expect.  
Note that the results following equation 
(\ref{acs}) do not apply to type B solutions 
because of division
by zero problems.  Therefore,
this section is a bit out of the main line of this dissertation, but does
elucidate features of type C solutions.  The results presented here
were derived in \cite{Strominger:1986uh}\footnote{I follow 
the preprint version of \cite{Buchel:2001qi} to get signs and factors in 
the type IIB string theory conventions I use.}.

First of all, equation (\ref{cov+torsion}) implies that there is an almost
complex structure
\be\label{acs}
\tilde J_m{}^n = i\chi^{\prime \dagger} \tilde\gamma_m{}^n\chi^\prime\ .\ee
The complex structure is covariantly constant with respect to the
connection with the torsion given in (\ref{torsion}). 
Therefore it is possible to show that $\tilde J$ is an integrable complex
structure and satisfies $\tilde J = i\tilde g_{i\bj}dz^i d\b z^{\bj}$.
Additionally, there is an expression for the R-R 3-form flux in terms of the
fundamental two form, as in \cite{Strominger:1986uh}:
\be
e ^{\phi_0+2A'_0} F= -2 i (\bar{\del} - \del) \tilde J
\label{Ffromtorsion}
\ee
Again, in the type B solutions, the left hand side of (\ref{Ffromtorsion}) 
vanishes due to the divergence of $A'_0$, so we cannot divide by zero
and use the following results.

With this $F_{3}$, let's plug into eqn (\ref{21HF}).  Using
(\ref{Aalpha}) we get
\be\label{H21}
H_{(2,1)}=-2 e ^{-2A'_0} \left(\cos^2(\alpha) \del \tilde{J}-
\sin(2\alpha) \del\alpha\wedge\tilde{J}  \right)
\ee
and
\be\label{H12}
H_{(1,2)}=-2 e ^{-2A'_0} \left(\cos^2(\alpha) \b\del \tilde{J}-
\sin(2\alpha) \tilde{J}\wedge\b\del\alpha  \right)\ .
\ee
Being very careful with the ordering of basis forms lets us write
\be\label{Hfromtorsion}
H=-2 e ^{-2A'_0} \left(\cos^2(\alpha) d\tilde{J}-
\sin(2\alpha) \tilde{J}\wedge d\alpha  \right)\ .
\ee

Also, it is possible to write the dilaton and $\alpha$ in terms of the 
holomorphic $(3,0)$ form of the metric $\tg_{mn}$.  We will not
need this result.

\subsubsection{Bianchi Identities}\label{sss:susybianchi}

The Bianchi identities will be just as in section \ref{sss:constraints},
except that now we will also allow D5-brane charge, meaning that the
Bianchi for $F_3$ will have sources.  We do not allow sources for $H_3$ 
because they should not appear in either type B or type C solutions;
only D3-branes and D5-branes will.

These Bianchi identities may still place severe constraints on any
compactifications.  Just as in \ref{sss:constraints} above, the integrated
Bianchis require that the sources on the compact manifold add to zero.
However, it is unknown how to combine string theoretic sources that 
include both signs of D3- and D5-brane charges in a supersymmetric way.
It may be that the only solutions
in this class that can be compactified are the type B and C limits or
else compactifications with no D5-brane charges at all.

Still, noncompact solutions interpolating between types B and C 
are interesting in their own right, as in the context of 
the AdS/CFT correspondence.  They also would include D-brane solutions;
\cite{Frey:2003sd} 
will show explicitly that the D5/D3-brane bound state solution
is in this interpolating class.

\comment{
I'M HAVING A LITTLE TROUBLE GETTING THIS BUT AM WORKING ON IT.
To be consistent with  equation  (\ref{traceF}) for the trace of $F$ we need
\be
d\alpha= -2 e ^{4A'_0} cotg (\alpha) {\tilde J} \odot d{\tilde J}
\label{dalpha}
\ee 
(THIS IS J CONTRACTED WITH dJ) ACTUALLY, I GET
\be
d\alpha= \frac{1}{2}\cot\alpha \left[(\del\tilde J)_{mnp}-
(\b\del \tilde J)_{mnp}\right] \tilde J^{np} dx^m
\ ,\ee
WHICH I DON'T THINK IS THE SAME.  IN FACT, THIS SEEMS TO GIVE US AN
EQN FOR THE DILATON (BC THE LHS WOULD BECOME $\tan\alpha d\alpha$).
THIS IS A STEP IN GETTING THE (3,0) FORM BIT I MENTION BELOW.

Now we turn to Bianchi identities
\be 
dF=2\kappa_{10}^2 \mu_5 \rho_5,\ dH=0, \ dF_{5}= 
H \wedge F+2\kappa_{10}^2 \mu_3 \rho_3\ .
\ee
Note that we do not include any NS5-brane sources because they should lie 
outside this ansatz.

The Bianchi identity for $F$ gives
\be
e ^{-\phi_0 -2A'_0} \bar{\del}\del \tilde J=2\kappa_{10}^2 \rho_5
\ ,\label{delbardelJ}
\ee
which is just as we would get following Strominger.
The corresponding equation for $H$ gives
\be
2 \sin(2\alpha) e ^{-2A'_0} \left(d {\tilde J}\wedge d\alpha+d\alpha
\wedge d {\tilde J}\right)=0
\ee
which is satisfied automatically because of the wedge product.
The Bianchi identity for $F_{5}$
leads to the equation
\be
-\tilde \Del^2 h=
\ee
THIS HAS TO BE CHANGED BC THE FORM OF H CHANGED.  I DON'T THINK IT LOOKS
VERY INFORMATIVE IN TERMS OF J, ETC.

THERE IS ALSO THE DILATON IN TERMS OF THE (3,0) FORM OF THE MANIFOLD.  DO YOU
THINK WE SHOULD BOTHER WITH IT?}

\chapter{4D Effective Theory and Moduli Fixing}\label{c:effective}
Now that we have the full ten dimensional structure for compactifications
with self-dual flux (see section \ref{s:tend}), we can start to
understand the four dimensional effective theory, what an observer
would see at low energies.  In this chapter, I will first describe the
Kaluza-Klein ansatz that arises from the full equations of motion. 
Then, recognizing that the effective theory should be a 4D SUGRA, I
review the effective superpotential and K\"ahler potential.  Section
\ref{s:special} will give specific, detailed examples for compactifications
that are geometrically $T^6/\Z$ and $K3\times T^2/\Z$, as well as 
briefly mention how these special cases are related to gauged supergravities
in 4D.  Understanding the effective theory completely requires a 
combination of these techniques.

In the introductory section \ref{s:intro}, I related the problem of 
moduli in string theoretic compactifications.  This chapter will return
to that problem: based purely on geometry, the solutions of section
\ref{s:tend}, with and without SUSY, have many classical moduli. 
Throughout this chapter, I will pay particular attention to how the
fluxes freeze the moduli of the compactification at \textit{tree level}.
An important point is that the naive moduli get masses that are
parametrically lower than any other mass scale, so (classically at least)
we are able to ignore Kaluza-Klein modes, string excitations, and winding
states in the effective theory.  I will develop this hierarchy in
section \ref{s:eom}.

The reader should be aware that much of this chapter is a review of
the literature, but I will denote my own work.

\section{Reduction of Equations of Motion}\label{s:eom}

The dimensional reduction of a theory requires the decomposition of
the higher dimensional fields into a tower of Kaluza-Klein modes.  
We find the modes by studying the equations of motion at linear order
(since particles are in fact perturbations of the background); on a 
torus, solving the equations of motion is trivial.  Life becomes 
complicated very rapidly, however, in the presence of a warp factor.
We will see below that, even at linear order in the perturbations,
the warp factor mixes Kaluza-Klein modes as well as different 10D fields.

All those effects come on top of the effects from the 3-form fluxes.
Therefore, in this section, we will work in the large radius limit, ignoring
the warp factor, for a first approximation.  In that approximation, we will
discuss how the 3-forms freeze some moduli and show that the mass scale is
hierarchically below the Kaluza-Klein and Planck scales.  Then we will
turn on the warp factor and examine the effect on the dimensional reduction.

\subsection{Dimensional Reduction with Fluxes Only}\label{ss:fluxonly}

Here we want to isolate the effect of the 3-form flux on perturbations
around the background, so we consider the large radius limit, where
the warp factor becomes negligible (see the discussion following equation
(\ref{warppoisson}) at least away from special points.  Our main intent
is to see that the fluxes introduce a new mass scale parametrically lower
than the others\footnote{The results of this subsection were obtained
in preparation for \cite{Frey:2002hf} in the special case of compactification
on $T^6/\Z$ with the $\N=3$ background of section \ref{sss:n3sol}, 
although most were not included in that publication.}.  Throughout,
if we are in a case with D7-branes, we work in the orientifold limit of 
F-theory for simplicity.  Note that while we ignore the warp factor,
$g_{mn}=\tg_{mn}$.

\subsubsection{Scalars and Mass Hierarchies}\label{sss:scalarmass}

\paragraph{The Dilaton-Axion}\label{p:dil-mass}

Given a fixed background of quantized 3-form flux, equation (\ref{imsd})
generically requires a fixed value of the dilaton and R-R scalar. Therefore
we would expect that perturbations in $\tau$ should be massive.  I will
use the dilaton-axion as an example of our general approach to dimensional
reduction.  For notational convenience, we will take the background
value of the string coupling to be $e^\phi=g_s$.

The linearized form of the equation of motion (\ref{dilaxeom}) is
\bea
\del_\mu\del^\mu \delta\tau &=& -\t\Del^2\delta\tau 
-\frac{ig_s}{12}\delta\phi
G_{mnp}G^{mnp}+\frac{ig_s}{4}\delta g_{mn} G^m{}_{pq}G^{npq}\nonumber\\
&&-\frac{ig_s}{6} G_{mnp}\delta G^{mnp}\ .
\label{taumass1}\eea
Since we are most interested in the light modes, we should drop the 
Kaluza-Klein excitations.  This not only drops the $\t\Del^2$ term, but it
also removes fluctuations $\delta F_3,\delta H_3$ as the constant fluxes
are quantized.  Additionally, the second term on the right vanishes for
self-dual flux.  Therefore, writing $H_3 =ig_s(G_3-\b G_3)/2$ gives
\be\label{taumass2}
\del_\mu\del^\mu \delta\tau = \frac{ig_s}{4}\delta g_{mn} G^m{}_{pq}G^{npq}
+\frac{g_s^2}{2} \delta\tau |G_3|^2\ .\ee
The final term is clearly a mass for $\tau$.  The first term on the
right represents potential mixing of the dilaton with geometrical
moduli; this is just a non-diagonal mass matrix.  Indeed, some simple
examples show this effect (see section \ref{ss:potential}).

For simplicity, ignore mixing with other moduli and look carefully at 
the mass term.  Take the coordinate ``radii'' to be string length and
the metric to scale as $g_{mn}\propto e^{2u}$.  Then the mass in the
10D string frame is
\be\label{taumass4}
m^2 = \frac{g_s^2}{2} e^{-6u} \frac{n_f}{\ap}
\ ,\ee
where $n_f$ is the square of the numbers of flux quanta (divided by $6$).
This sets up a clear hierarchy of scales
\be\label{fluxhierarchy}
\begin{array}{cccccc}
&\textnormal{massless}&\textnormal{flux}&\textnormal{KK}&
\textnormal{string}/M_{10}
&\textnormal{winding}\\
\ap m^2 \sim &0& e^{-6u}& e^{-2u} & 1& e^{2u}\end{array}\ee
at large radius (where supergravity is valid).  Since the dilaton is fixed,
we assume it is order $1$.
In the 4D Einstein frame, the mass is rescaled; for the flux-induced masses,
we find
\be\label{taumass5}
m^2_E \sim g_s^4 e^{-12u} M_4^2\ee
since the 4D Planck mass is string mass in this frame.
One consequence of this hierarchy is that any supersymmetry breaking
caused by the fluxes must be spontaneous.

\paragraph{Other Perturbations}\label{p:metricmasses}

Finding which metric components are massive and which massless directly
from the equations of motion is a complicated and unenlightening process.
However, there is a short-cut, outlined by \cite{Greene:2000gh} and
used by \cite{Frey:2002hf}.  Any perturbations that preserve the nature of
the fluxes should be massless; in a non-supersymmetric compactification, 
this means that (\ref{imsd}) should still be obeyed.  For a supersymmetric
compactification, complex structure deformations that leave $G_3$ a $(2,1)$
form and K\"ahler deformations that leave $G_3$ primitive are massless.
Clearly, the overall scale modulus is massless by this reasoning.
However, at least for the simplest case of an $\N=3$ compactification on
$T^6/\Z$, it is possible to check this ansatz explicitly by linearizing
the equation of motion (\ref{einstein}) for internal 
components\footnote{This was unpublished work done for \cite{Frey:2002hf}.}.

Similarly, nothing constrains the position of
D3-branes, so we expect their positions to be moduli; it is in fact 
straightforward to see that no other Kaluza-Klein zero modes enter
their equations of motion at linear order.  D7-branes are another story.
Moving D7-branes away from O7-planes requires $\tau$ to vary on the
compact space, which is impossible if it is massive, so we expect the 
D7-brane scalars to be massive also.  This fact has not been checked
explicitly to my knowledge.

\subsubsection{Form Perturbations and Higgsing}\label{sss:forms}

Perturbations of the $p$-form potentials have some complications, so
it is important to discuss them carefully.  

\paragraph{Periodicities}\label{p:cperiod}

The first subtlety with the form fields is that they may only be periodic
up to gauge transformations on cycles.  Let us look at the $T^6/\Z$ case
for specificity.  This argument was given first in \cite{Frey:2002hf}.

The gauge transformations of the various potentials are
\begin{eqnarray}
\delta C_{2} &=& d\lambda_C \nonumber\\
\delta B_{2} &=& d\lambda_B \nonumber\\
\delta C_{4} &=& d\chi + \lambda_C \wedge H_{3}\ ,
\end{eqnarray}
in terms of one-forms $\lambda_C$ and $\lambda_B$ and three-form $\chi$.
The forms have periodicities
\begin{eqnarray}
C_{2}(x+e^{m}) &=& C_{2}(x) + d\lambda^{m}_C(x) \nonumber\\
B_{2}(x+e^{m}) &=& B_{2}(x) + d\lambda^{m}_B(x) \nonumber\\
C_{4}(x+e^{m}) &=& C_{4}(x) + d\chi^{m}(x) + \lambda^{m}_C(x)
\wedge H_{3}(x)\ .
\end{eqnarray}
Here $e^{m}$ is the lattice vector in the $m$-direction, $(e^m)^n = 2\pi
\sqrt{\ap}\delta^{mn}$, and $\lambda^{m}_C$, $\lambda^{m}_B$,
and $\chi^{m}$ are specified gauge transformations.  To analyze these it
is convenient to write each field as its background value plus a shift,
for example $C_{4}(x) = \hat C_{4}(x) + \delta C_{4}(x)$.

The three-form
flux backgrounds are constant, and so for the corresponding potentials we
can choose a gauge
\begin{equation}
\hat C_{mn} = \frac{1}{3} \hat F_{mnp} x^p\ ,\quad \hat B_{mn} = \frac{1}{3}
\hat H_{mnp} x^p\ .
\end{equation}
It follows that
\begin{equation}
\lambda^m_C = \frac{1}{6} \hat F_{mnp} x^n dx^p\ ,\quad \lambda^m_B =
\frac{1}{6} \hat H_{mnp} x^n dx^p\ .
\end{equation}
The quantized fluxes cannot fluctuate, and so the $\lambda^{m}_{B,C}$ are
fixed.  It then follows that the two-form fluctuations are periodic, so
any scalars that descend from them follow the usual story for orientifold
compactifications.

The four-form must satisfy a more complicated boundary condition.  This
can be deduced from the condition that $C_{4}(x + e^m + e^n)$ be
consistently defined.
It takes only a few lines of algebra to see that
\begin{equation}
\tilde c_{4} = \delta c_{4} - \hat C_{2} \wedge \delta B_{2}\ ,
\end{equation}
has periodicity\footnote{Strictly speaking, this is only the linearized
form; the full form is given in \cite{Frey:2002hf}.}
\begin{equation}
\tilde c_{4}(x+e^m) = \tilde c_{4}(x) + d\tilde\chi^m(x)\ ,
\end{equation}
where $\t \chi^m=\chi^m -\lambda^m_C\delta B_2$.
We can set $\t \chi^m$ to zero by a gauge transformation, since a nontrivial
gauge transformation of $\t c_4$ would imply a quantized 5-form, which 
is inconsistent with the orientifold condition.
The 5-form fluctuation to linear order is then
\begin{equation}
\delta\t F_{5} = d \tilde c_{4} + \hat F_{3}
\wedge \delta B_{2}
- \delta C_{2} \wedge \hat H_{3} \ .
\label{f5pert}\end{equation}

The gauge variation of $\tilde c_{4}$ is, after some algebra,
\be
\delta \tilde c_{4} = 
d\tilde\chi + \frac{i}{2\im(\tau)}
\left( \b{\lambda}_A \wedge G_{3}  - \lambda_A \wedge
\b G_{3}  \right) 
 \ , \label{c4var}
\ee
where $\tilde\chi = \chi - C_{2} \wedge \lambda_B$
and $\lambda_A = \lambda_C - \tau \lambda_B$.
We are now dropping hats from the background for notational convenience.
(Note that the
background is defined to be fixed, so the gauge transformation goes
entirely into the fluctuation).  The gauge transformation $\tilde\chi$
must be periodic. 

In more general geometries, the story should work out similarly.  For a 
CY 3-fold compactification, the scalars (and potentially vectors) from 
$\t c_4$ descend as normal.

\paragraph{4-Form Scalars and Gauge Fields}\label{p:c4-gauge}

Some interesting physics arises when the internal manifold has 1-cycles,
since $B_2,C_2$ can lead to spacetime gauge fields.  Using the same sort of
general reasoning for the moduli, we expect gauge transformations
that alter the background to correspond to broken vector fields.

We can make this expectation more precise.  We define the complex
field $A_2 = \delta C_2 -\tau \delta B_2$, so $\delta G_3=dA_2$ with $\tau$
fixed.  Then the linearized 5-form (\ref{f5pert}) becomes 
\cite{Frey:2002hf}
\begin{equation}
\delta \t F_{5} = d \tilde c_{4} + \frac{i}{2 \im\tau}\left( A_{2} \wedge 
\b G_{3} - \b A_{2} \wedge G_{3} \right)\ . \label{elhiggs}
\end{equation}
($G_3$ is here the background field only.)
The relevant component for the spacetime vectors is $\delta \t F_{\mu mnpq}$,
so the scalar is $\t c_{mnpq}$.
Comparing with the nonlinear Higgs covariant derivative
$\partial_\mu \phi - A_\mu$, we see that the background 3-form serves
as a charge coupling the 4-form to the perturbative gauge potential.

In fact, the equation of motion for $\t c_4$, after dropping Kaluza-Klein
modes, is just
\be\label{goldstone}
\del^\mu\left[\del_\mu \t c_{mnpq} +2ig_s \left( A_{\mu[m}\b G_{npq]}-
\b A_{\mu[m}G_{npq]}\right)\right]=0\ .\ee
This is precisely the equation of motion for a Goldstone boson
for those $\t c_{mnpq}$ such that the wedge products
do not vanish.  We find this by Hodge dualizing the Bianchi identity
(\ref{bianchi5}) and realizing that the 3-form contribution cannot give a 
source at the linear level.  The vectors have a linearized 
equation of motion (from Hodge dualizing (\ref{3formeom}) and using
self-duality of $\t F$)
\be\label{brokenvector}
\left(\star d\star \delta G_3\right)_{\mu m} =-\frac{i g_s}{6} \delta \t 
F_{\mu mnpq}G^{npq}\sim \frac{g_s^2}{2}|G_3|^2 A_{\mu m}\ .\ee
The $\sim$ represents going into the unitary gauge and schematically 
illustrates the flux contraction with the different terms of the wedge
products.

We should note also that each D3-brane also contributes a $U(1)$ gauge 
field (assuming they are not coincident).  These remain massless for the
simple reason that there are no charged scalars for those $U(1)$ factors.

\subsection{Complications from the Warp Factor}\label{ss:warpeom}

In Scherk-Schwarz type compactifications, it is well known that the 
Kaluza-Klein zero modes are not trivial \cite{Scherk:1979zr}.
We are about to see that the warp factor also modifies the Kaluza-Klein
decomposition.  In fact, the 10D equations of motion in the presence
of the warp factor are unsolved, although there is some recent progress
\cite{deAlwis:2003sn,gm}\footnote{I thank the authors of \cite{gm} for 
useful discussions
of their work prior to its publication.}.

\subsubsection{Graviton Reduction}\label{sss:4dmetric}

As we mentioned in section \ref{s:warped}, the graviton zero mode
is known even in the presence of a warp factor.  It was briefly argued
that the zero mode should be $\delta g_{\mu\nu} = e^{2A}h_{\mu\nu}(x^\mu)$
in \cite{Greene:2000gh}, which also gave the eigenvalue equation for the
Kaluza-Klein masses.

Here we can give a slight amplification of the argument
in \cite{Greene:2000gh}.  Consider a 10D metric with a general 4D 
line element
\be\label{newwarp}
ds^2=e^{2A} g_{4,\mu\nu}dx^\mu dx^\nu + e^{-2A}\tg_{mn}dx^mdx^n\ee
and set the fluctuations of all other fields to zero.
Then we can just follow the analysis of section \ref{sss:nogo};
the only difference is that the left hand side of the Einstein equation 
now contains a term $R_{\mu\nu}(g_4)$.  The warp factor cancels with
everything on the right hand side as before, so we find
\be\label{einsteinfluct}
R_{\mu\nu}\left(g_{4,\mu\nu}\right) =0\ .\ee
Since $g_{4,\mu\nu}$ satisfies the vacuum Einstein
equations when we have turned off all other fluctuations, it is clearly the
zero mode of the graviton.  In addition, it is possible to see the
appropriate kinetic terms from the conversion to 4D Einstein frame if we
allow the internal metric to fluctuate, but we will not go into that
detail here.

\subsubsection{Massive Modes and Kaluza-Klein Mixing}\label{sss:kkmixing}

Unfortunately, most of the time, the Kaluza-Klein decomposition is not as
straightforward as that for the metric.  One complication is that the
light but massive fields -- moduli frozen by fluxes -- no longer have 
simple Kaluza-Klein modes, despite the fact that the warp factor does not
appear in their covariant derivatives (remember from \ref{ss:background}
that the 10D Laplacian is related by scaling to the 6D Laplacian $\t\Del^2$)
\cite{Greene:2000gh}.  

To see this effect, look at the dilaton equation 
of motion (\ref{taumass1}), neglecting fluctuations of other fields but
including the warp factor.  We have
\be\label{tauwarp}
\eta^{\mu\nu}\del_\mu\del_{\nu}\delta \tau = -e^{4A}\widetilde{\Del^2}
\delta\tau +\frac{g_s^2}{2}e^{8A}\widetilde{|G_3|^2} \delta \tau\ .\ee
Because of the warp factor on the mass term, the zero mode clearly cannot
be the usual harmonic 0-form.  Solving this mass equation is difficult
because the warp factor is not a simple function (see \cite{Greene:2000gh}
for the solution on a toroidal compactification).  However, for a simple
geometry in the large radius limit, it is possible to treat the warp factor
as a perturbation mixing the usual Kaluza-Klein modes.

\subsubsection{Constraints for Massless Fields}\label{sss:masslessconstrain}

Another complication that occurs in the presence of the warp factor is
related to the mixing of Kaluza-Klein modes discussed above.  Just above,
we discussed mixing due to the appearance of the warp factor in the
mass terms.  In this section, we examine mixing that occurs because, with
the warp factor, the massless modes appear in more components of the
10D equations of motion.  

A particularly problematic instance of this effect is the introduction of
metric deformations $\delta g_{mn}$ into the Einstein equation for 
$R_{\mu m}$.  Following the logic that $x^\mu$ independent 
deformations which lead to a valid background in 10D should lead to 
massless fields, we might expect $\delta g_{mn}=e^{-2A}\delta\tg_{mn}
-2\delta A e^{-2A}\tg_{mn}$, where $\delta A$ is the change in the
warp factor due to the linearization of (\ref{warppoisson}), along with
an appropriate $\delta \t F_5$, to be massless.  However, it has so
far proven difficult to find the Kaluza-Klein zero mode for the
internal metric.  This is the focus of \cite{gm} and is essentially the 
issue raised in \cite{deAlwis:2003sn}.

To illustrate this type of issue, we can look at the 2-form gauge
fields compactified on $T^6/\Z$, which were considered in \cite{Frey:2002hf}.
We take as an ansatz that the only nontrivial components
of the fluctuations are $\delta G_{\mu\nu m}$ and $\delta \t F_{\mu\nu mnp}$. 
We further take
\begin{eqnarray}
\delta G_{\mu\nu m} (x^\mu,x^m) &=& f_{\mu\nu}(x^\mu) u_m (x^m) 
+ (\star_{(4)} f)_{\mu\nu}(x^\mu)
v_m(x^m)\ ,\nonumber\\
\delta \t F_{\mu\nu mnp} &=& f_{\mu\nu} \gamma_{mnp} (x^m) +
(\star_{(4)} f)_{\mu\nu} (\star_{(6)} \gamma)_{mnp} (x^m)\ . \label{ansatz}
\end{eqnarray}
Here $u_m$ and $v_m$ are complex, and $\gamma_m$ and $f_{\mu\nu}$
are real.  
Note that on two-forms $\star_{(4)}$ is the same as the
flat spacetime Hodge star.

Inserting this ansatz into the field
equations gives the usual 4D equations
\begin{equation}
d \star_{(4)} f_{2} = d f_{2} = 0
\end{equation}
and the internal equations
\begin{eqnarray}
d u_{1} &=& d v_{1} = 0 ,\nonumber\\
d \star_{(6)} u_{1} &=& - i g_{s} v_{1} \wedge \tilde
F_{5} - i g_{s} G_{3} \wedge \star_{(6)} \gamma_{3}\ ,
\nonumber\\
d \star_{(6)} v_{1} &=& i g_{s} u_{1} \wedge \t
F_{5} + i g_{s} G_{3} \wedge \gamma_{3}\ ,
\nonumber\\
d \gamma_{3} &=& \frac{i g_{s}}{2} (G_{3} \wedge
\b u_{1} + u_{1} \wedge \b G_{3})
\ ,
\nonumber\\
d \star_{(6)} \gamma_{3} &=& \frac{i g_{s}}{2} (G_{3} \wedge
\b v_{1} + v_{1} \wedge \b G_{3})
\ .
\end{eqnarray}
The Bianchi identities for $u_{1}$ and $v_{1}$ are solved by
\begin{equation}
u_{1} = \omega_{1} + da\ ,\ v_{1} =
\nu_{1} + db
\end{equation}
where $\omega_{1}$ and $\nu_{1}$ are harmonic on the
internal space and
$a$ and
$b$ are periodic scalars.
The equations for $\gamma_{3}$ are then solved by
\begin{equation}
\gamma_{3} = \frac{ig_{s}}{2} (a \b G_{3} - \b a
G_{3})\ ,
\end{equation}
if
\begin{equation}\label{wedgezero}
b = - ia\ ,\ \omega\wedge \b G_3 = \nu\wedge \b G_3 = 0\ .
\end{equation}
Finally, the field equations for $u_{1}$ and $v_{1}$ both
become
\begin{equation}
e^{-4A} \t\Del^2 a + 2 \del^{\t m} e^{-4A} \partial_m a + \del^{\t m}e^{-4A}
(\omega_m + i
\nu_m) = \frac{g_{s}^2}{12} a G_{mnp} \b G_{mnp}\ ,
\end{equation}
where all contractions are with the flat internal metric.  There are then
two solutions for each $\omega_1,\nu_1$ satisfying (\ref{wedgezero}):
\begin{eqnarray}
\omega_{1} &=& - i \nu_{1} \ ,\ a = \gamma_{3} = 0\ ;
\
v_{1} = iu_{1}\ ,\nonumber\\
\omega_{1} &=& i \nu_{1} \ ,\ a\neq 0\ ,\ \gamma_{3}
\neq 0 \ ;\
v_{1} = -iu_{1}\ .
\end{eqnarray}
There is no known closed form for the second solution, but it can be seen
to exist by considering the large radius limit in which $A\to 0$.
This is a very interesting situation; fluxes with $\star_{(4)}f_2=-if_2$ 
have a trivial Kaluza-Klein zero mode, but those with 
$\star_{(4)}f_2=if_2$ do not.  This situation persists even without fluxes.
Note that we must beware of overcounting 
because the ansatz is invariant under $u \to v$, $v \to
-u$, $f_{2} \to \star_{(4)} f_{2}$.

\section{Flux Superpotential and K\"ahler Potential}\label{s:gvw}

We have seen from the equations of motion that
the fluxes give masses parametrically lower than the Planck scale,
even without considering warping, so we should be able to put self-dual
flux compactifications into the context of 4D supergravity.  In fact,
the 3-form generates a superpotential, which can break SUSY, and
the associated potential will fix the values of many moduli.  

\subsection{Superpotential from Fluxes}\label{ss:superpot}

In the presence of fluxes, the classical 4D effective $\N=1$
superpotential was argued to be \cite{Gukov:1999ya,Gukov:1999gr} 
\be
\label{superpotential}
W = \frac{1}{\k^8}\int_M G_3 \wedge \Omega\ , 
\ee
where $\Omega$ is the holomorphic (3,0) form on the compactification
manifold $M$.  $W$  then 
is given by the $(0,3)$ part of
the $G_3$ flux which, because the fluxes are quantized, can only be
tuned discretely.  This superpotential was rederived by dimensional
reduction of the type IIB SUGRA action in \cite{Giddings:2001yu} 
ignoring the
warp factor and in \cite{DeWolfe:2002nn} with the warp factor.

To avoid the details of a derivation, let us recap the argument of 
\cite{Gukov:1999ya,Gukov:1999gr} 
to justify the superpotential (\ref{superpotential}).
First, $\Omega$ is, loosely speaking, a function of the complex moduli
of $M$\footnote{Strictly speaking, a superpotential is supposed to be
a section of a $U(1)$ bundle over the scalar manifold, which this
is \cite{Gukov:1999ya}.}.  It is also
independent of K\"ahler moduli.  This is indeed
as expected.  Second, we know that all $(2,1)$ contributions to the 
$G_3$ flux leave supersymmetry unbroken.  Despite the fact that the
flux components are quantized, it would be rather surprising if the 
supersymmetry conditions could be satisfied for all $(2,1)$ fluxes if
the superpotential depended on them.  Indeed, in combination with the
K\"ahler potential for the moduli given below, this superpotential 
requires that the $(0,3)$ flux vanish in supersymmetric compactifications.
Finally, we could imagine domain walls constructed of D5-branes or
NS5-branes wrapping 3-cycles in the compact manifold.  These were, in
fact, considered in detail by \cite{Gukov:1999ya,Kachru:2002ns} (along
with their instantonic counterparts in \cite{Frey:2003dm} and section
\ref{sss:ns5decay}).
In the limit that gravity decouples, BPS domain walls have a tension
proportional to the change in superpotential.  Indeed, $\Omega$ integrated
over the cycle that the 5-brane wraps is indeed proportional to the 
volume of that cycle and therefore the domain wall tension.  Meanwhile,
the change in superpotential is also proportional to the change in the
flux, which jumps by as many units as there are 5-branes.

\subsection{K\"ahler Potential}\label{ss:kahlerpot}

To define the K\"ahler potential, we must first be precise about the
definition of the fields in the 4D effective theory in terms of 10D 
variables; the only way to get the K\"ahler potential is through dimensional
reduction of the type IIB kinetic terms.  Since the Kaluza-Klein
ansatz in self-dual flux compactifications is not yet fully understood,
as reviewed above, the K\"ahler potential is not completely known.
However, we can proceed in stages.  First, we can ignore
the warp factor completely.  As a next step, we can treat the warp factor
as a fixed 10D quantity independent of the 4D fluctuations\footnote{I thank 
O. deWolfe and S. Giddings for clarifying some of this point.}.  A goal
for future work would be to incorporate the dependence of the warp factor
on the perturbations of the compactified metric and other light fields.

To define the fluctuations of the 4D theory, we start with the 10D metric
in the form (\ref{warp2}) with the warp factor scaled out of the
compact metric.  Then define the scale $u$ of the metric $\tg_{mn}$ 
so that the compactification volume in the 
rescaled metric is approximately $e^{6u}\ap{}^3$ (roughly, $e^{6u}$ is the
determinant of $\tg_{mn}$).  To be precise, all the other fluctuations
in $\tg_{mn}$ are defined to leave the compactification volume invariant at
lowest order.
Then, to the level that
we are working, the Einstein frame in 4D would be as in equation
(\ref{einstein4d}).  (The 4D Einstein frame should really be defined by
using the full form of the massless volume modulus, but we are working
in an approximation with a fixed warp factor.)

We start by ignoring the warp factor.  As shown by \cite{Giddings:2001yu},
the tree-level K\"ahler potential is
\be
\label{kahlerkklt}
\mathcal{K} = -3
\log(-i(\rho-\b\rho)) - \log(-i(\tau-\b\tau)) - \log\left(-\frac{i}{\k^6}  
\int_M \Omega \wedge \b\Omega\right) \ ,
\ee 
where the volume modulus is defined by 
\be\label{rhodef}
\im\rho \equiv \sigma = e^{4u-\phi}\ee
(with the real part an axion from the 4-form potential).  This form
assumes only a single K\"ahler modulus; others can be added similarly.
The last
term includes the holomorphic $(3,0)$ form $\Omega$ and includes
all the complex structure moduli; this form was found for CY 3-folds
by \cite{Candelas:1991pi}.  In the large volume limit, the warp factor
$A\sim e^{-4u}$, so this K\"ahler potential is probably sufficient for
moduli that describe deformations near special points, such as conifolds.
We normalize $\Omega$ so that $\int\Omega$ on a 3-cycle is proportional to 
$(\int d^6x e^{-6u}\sqrt{\tg})^{1/2}$.

Now we can include the warp factor, ignoring its dependence on the 
metric fluctuations.  The K\"ahler metric becomes
\cite{DeWolfe:2002nn}
\bea \mathcal{K} &=& -3
\log(-i(\rho-\b\rho)) - \log(-i(\tau-\b\tau)) -\log\left(
-\frac{i}{\k^6} \int_M e^{-4A}\Omega\wedge \b\Omega
\right)\nonumber\\
&& -\log\left[-\frac{i}{\k^6} \int d^6x e^{-4A-6u} 
\sqrt{\det \left(\tg_{mn}+\delta\tg_{mn}(x^\mu,x^m)\right)}\right]  \ ,  
\label{kahlerwarp}\eea
where the $\delta \tg_{mn}$ include all fluctuations but $\rho$ (possibly
other K\"ahler moduli).  To get this in detail, one has to carry out the 
dimensional reduction, as in \cite{DeWolfe:2002nn}, but we can get the
factors of the warp factor easily. It just comes from the fact that
the compact part of the Einstein-Hilbert term has a prefactor of $e^{-4A}$
when written in terms of the 4D Einstein frame and $\tg_{mn}$ (there
is a factor of $e^{4A}$ from the 4D metric determinant, $e^{-6A}$ from
the 6D metric determinant, and $e^{-2A}$ from contracting the 4D derivatives).

So far, we have assumed that the volume modulus is the only K\"ahler 
modulus.  If there are any free D3-branes, their positions $Z$ are also
K\"ahler moduli.  It is argued in \cite{DeWolfe:2002nn,Kachru:2003sx}
that including the D3-brane moduli is accomplished by taking the $\rho$
term to 
\be\label{d3kahler}
-3\log\left(-i(\rho-\b\rho)+k(Z,\b Z)\right)\ ,\ee
where $\mathcal{k}$ is the K\"ahler potential for the Calabi-Yau metric
(at least in the large-volume, orientifold limit).  The volume is no longer
simply related to the $\rho$ modulus; it is rather \cite{Kachru:2003sx}
\be\label{volumemodulus}
e^{4u-\phi} = \im\rho +\frac{1}{2}k(Z,\b Z)\ .\ee

\subsection{No-Scale Potential}\label{ss:noscale}

The supergravity potential written in terms of the K\"ahler and 
superpotentials is
\be\label{potential1}
V = \k^2 e^{\mathcal{K}}\left[\sum_{i,\bj} 
\mathcal{K}^{i\bj} D_iW \b D_{\bj} \b W -3|W|^2\right] \ ,
\ee
where $i,\bj$ sum over the moduli, $\mathcal{K}_{i\bj}
= \del_i \del_{\bj} \mathcal{K}$ is the K\"ahler metric, and $D_i =
\del_i +\del_i \mathcal{K}$ is the  K\"ahler covariant derivative.
However, with the specific form of the K\"ahler potential (either
(\ref{kahlerkklt}) or (\ref{kahlerwarp}) because the form for $\rho$
does not change) and the fact that the superpotential is independent
of the compactification volume, we get the no-scale potential
\be\label{noscale}
V = \k^2 e^{\mathcal{K}} \sum_{i,\bj} \mathcal{K}^{i\bj} D_iW \b D_{\bj} \b W
\ee
where now $i, j$ sum over all moduli but $\rho$.  
This potential generically
fixes all complex moduli (and $\tau$) such that $G_3$ is imaginary
self-dual.  Some of the K\"ahler moduli can be fixed by the requirement
that $G_3$ be primitive, but $\rho$ is never fixed since it is an overall
scale of the metric.  Thus, these are ``no-scale'' compactifications,
as noticed in \cite{Giddings:2001yu}.  No-scale SUGRAs and string
compactifications have been studied before, notably in 
\cite{Cremmer:1983bf,Ellis:1984sf,Witten:1985xb,Dine:1985rz}.
Except in cases with enough supersymmetry, this structure should not
survive beyond the approximation of classical supergravity (see,
for example, sections \ref{s:dS} and \ref{s:phenomenology}).

We can see how the masses are related to the Planck scale from this potential
very easily.  Ignoring the warp factor for comparison to the results
listed above, we find
\be\label{masssquaredgvw}
V \propto \k^2 \left(\k^{-8}\right)^2 \left( \ap{}^{5/2}\right)^2
e^{\mathcal{K}} \psi^2 \sim M_4^4 g_s^4e^{-12u} \psi^2 \ee
for some dimensionless modulus $\psi$.  Normalizing the scalar removes
a factor of $M_4^2$, so we get the same result as equation (\ref{taumass5}).

We can also see how SUSY demands that $G_3$ be $(2,1)$; a vacuum is
only supersymmetric if $D_i W=0$ for all moduli.  Since all moduli
but $\rho$ satisfy this equation at the vacuum state, we only need
to consider the $\rho$ condition any further.
As shown in 
\cite{Gukov:1999ya,Giddings:2001yu}, $D_\rho W= \del_\rho \mathcal{K} W=0$ 
is satisfied only when $W = 0$, which
implies  that in supersymmetric vacua $G_3$ is a $(2,1)$ form.

\subsection{Fixing a Complex Structure Modulus}\label{ss:deformation}

The geometry of $M$ is, of course, very complicated but is accurately
described  near conifold points by the Klebanov-Strassler 
solution \cite{Klebanov:2000hb}.   Wrapped  fluxes warp and deform the
conifold; at the tip $y=0$,  the metric is
\be
\label{tipmetric}
ds^2 = e^{2A} \eta_{\mu\nu} dx^\mu dx^\nu +b g_s M\ap \left(
dy^2 +d\Omega_3^2+y^2 d\Omega_2^2\right)
\ee
where $b\sim 1$ is a numerical constant and $e^{u}$ is the
compactification length scale (here we use 10D string frame).   Notice
that the $S^3$ at the tip has a fixed proper size depending only on
the fluxes.  Also, the $S^2$ is nontrivially fibered over the $S^3$.  
We can also find the warp factor of the tip in terms of the deformation
parameter $z$ of the conifold as in \cite{Giddings:2001yu}. The 
argument is that th ewarp factor satisfies roughly 
$e^{-4A} = g_s N \ap{}^2/r^4$
for a D3-brane vacuum, where $r$ is the distance from the branes.  
In the warped deformed conifold, the branes are absent, but there is
an effective number of D3-branes $g_s N_{\textnormal{eff}}=
(g_s M)^2$, which is reflected below 
in equation (\ref{throatwarp}) for the
warp factor away from the tip \cite{Klebanov:2000hb}.  Also, there is
a length scale at the tip of the conifold; the radius of the $S^3$,
which also translates to an effective distance from the D3-branes
(imagine the deformed conifold continuing to the singular conifold; this
is the distance to the singular tip) \cite{Candelas:1990js}.  Since
the volume of the $A$ cycle is given by $e^{3u}\ap{}^{3/2} z$, we get
\be\label{tipwarp} e^{4A}(y=0) \approx 
g_s\sigma\frac{|z|^{4/3}}{a^2(g_s M)^2}\ ,\ a=2^{4/3}3^{1/3} b\ee
where $b$ is the same constant as in equation (\ref{tipmetric}).  This
form of the warp factor, along with the constants, was calculated in
\cite{Herzog:2001xk}.  It is this particular form of the warp factor
that gives the $A$ cycle a fixed proper size at the tip.\footnote{Prior
to \cite{Kachru:2003sx}, the factor of $\sigma$ was usually left out
of the warp factor at the tip.  This did not introduce any qualitative
errors into those papers, however.  See section \ref{s:dS}.}

Away from the tip, the throat has approximately a warped
conifold metric  
\be 
\label{conifold} ds^2\approx e^{2A} \eta_{\mu
\nu} dx^\mu dx^\nu + e^{-2A}(dr^2 + r^2 ds^2_{T^{1,1}}) 
\ee
where $ds^2_{T^{1,1}}$ is the metric on the base $T^{1,1}$.  In this region,
the warp factor is approximately \cite{Klebanov:2000nc}
\be\label{throatwarp} 
e^{-4A} = 1 + (L^4/r^4) \log(r/r_s)\ee 
with the length scale $L^4 = \frac{81}{8}(g_s M \ap)^2$.  
Here, the radial coordinates
$r$ and $y$ are functions of each other, and the tip is at
$y=0, r=\t r$.  For the undeformed conifold, the singular tip is
located at  $r=r_s=\t r e^{-1/4}$.  Splitting the conifold into
the  tip and throat in this manner is described in
\cite{Herzog:2001xk} and references therein. 
For a given conifold throat, the quantized fluxes have the form 
\be 
M = \frac{1}{4\pi^2 \ap} \int_A F_3 \ ,\ \ K =  -\frac{1}{4\pi^2 \ap} 
\int_B H_3\label{fluxquant} 
\ee
where the $A$ cycle is the $S^3$ which stays finite at the tip and the
$B$ cycle is the six-dimensional dual of $A$.   

It is easy to determine how much the conifold tip is deformed by looking
at the fluxes \cite{Giddings:2001yu}.  We start by noting the definitions
\bea 
\int_A \Omega &=& z \left(\int d^6x e^{-6u}\sqrt{\tg}\right)^{1/2}
\approx \ap{}^{3/2}z\
 ,\ \ \int_B \Omega \approx \ap{}^{3/2}\mathcal{G}(z)\nonumber\\
\mathcal{G} &=& \frac{1}{2\pi i} z\log z
+\mathnormal{holomorphic}\label{omegadefs}\eea 
(this form of $\mathcal{G}$ was given in \cite{Strominger:1995cz}).  Then
\cite{Giddings:2001yu}, the superpotential is
\be\label{zsuperpot}
W\approx \frac{(2\pi)^2 \ap{}^{5/2}}{\kappa_4^8} \left( M
\mathcal{G}(z)-i\frac{K}{g_s} z\right)
\ .\ee
We are mainly interested in $z$ small, which we will see requires 
$K\gg g_s M$, which simplifies the calculation immensely.
Then we have
\be\label{dzW}
\del_z W \sim \frac{1}{2\pi i} M \log z -K\tau\ .\ee

Because the metric perturbation associated with $z$ is localized at
the base of the conifold, it feels the warping appreciably.
Including warping and
bunching the  K\"ahler potential for all other complex  moduli
together into $\mathcal{K}_c$ (that is, integrals over other cycles),
eqn (\ref{kahlerwarp}) becomes 
\bea 
\mathcal{K}(\mathnormal{complex}) &=&-\log\left[
e^{-\mathcal{K}_c} -\frac{i}{\k^6} \left(\int_A\Omega\int_B \b\Omega e^{-4A}
-\int_A\b\Omega\int_B\Omega e^{-4A}\right) \right]\nonumber\\ 
&\approx& \mathcal{K}_c -e^{\mathcal{K}_c} \frac{\ap{}^3 (g_s M)^2}{2\pi\k^6}
|z|^{2/3}\log |z|^2\ .
\label{kahlerwarp2}\eea
To calculate this, we use the trick of \cite{Strominger:1995cz} that
the cycles have a monodromy $B\to B+A$ around $z=0$ and take the warp
factor (\ref{tipwarp}) at the tip.
Actually, there will be other terms in  the $B$ cycle integral, but it
is reasonable to believe that, as in the unwarped case, this is the
leading term that depends on $z$.
We can see that $\del_z\mathcal{K}$ grows less rapidly than $1/z$ at small
$z$, so we can ignore the K\"ahler potential contribution to $D_z W$.

Putting it all together, then, the approximate deformation parameter of 
the conifold must be (we take here $\tau=i/g_s$) 
\be\label{deform} z= \exp \left[- \frac{2\pi
K}{g_s M}\right]\ .  \ee 
This is a nice example of moduli fixing in the superpotential formalism
and an important one.  The fixed deformation of the conifold essentially
fixes its length, so \cite{Giddings:2001yu} argued that the 3-forms 
parly stabilize the radial modulus in the reduction of the compactification to
a 5D braneworld as in \cite{Chan:2000ms}.  Additionally, with the warp
factor tied to the deformation parameter as in equation (\ref{tipwarp}), 
we see immediately that it is easy to get large hierarchies without large 
numbers of flux quanta.  This is not quite a solution to radius stabilization
because the D3-branes do not need to sit at the conifold tip; we would
need to add \bd-branes and break SUSY for that (see section \ref{s:dS}).

\section{Special Cases}\label{s:special}

In this section, we will discuss two special geometries, the 
$T^6/\Z$ orientifold by O3-planes, and $K3\times T^2/\Z$ orientifold
by O7-planes.  Since the geometry is simpler in these cases than in
general CY 3-folds, we can give specific examples.  Some early
examples of these compactifications were studied in 
\cite{Dasgupta:1999ss,Greene:2000gh}, and these two examples were 
explored in detail in \cite{Kachru:2002he} and \cite{Tripathy:2002qw}
respectively.

\subsection{$T^6/\Z$, O3-plane Action}\label{ss:t6o3} 

There are many solutions based on the $T^6/\Z$ orientifold,
distinguished by the three-form flux quanta and the discrete fluxes at
orientifold points.  Even with vanishing three-form fluxes there are many
solutions to the tadpole cancellation condition (\ref{intbianchi}).  
One extreme is to have 16
D3-branes and no discrete flux \cite{Verlinde:1999fy}, which is the familiar
T dual to the type I theory on $T^6$.  The other extreme is to have no
D3-branes and 32 fixed points with discrete flux.  For example, the
configuration with discrete R-R flux at all fixed points in the plane $x^4 =
0$ satisfies the quantization conditions and is T dual to a type I
compactification without vector structure \cite{Witten:1998bs}.
In these cases the supersymmetry is
$D=4$, $\N = 4$.

The superpotential formalism is very useful in finding the supersymmetric
values of moduli given a set of fluxes (assuming that they are compatible
with supersymmetry); \cite{Kachru:2002he} explained the supersymmetry
conditions in terms of the torus variables in great detail.  We outline
some of their argument here.  

Start by taking a Hermitean metric $g_{i\bj}dz^i d\b z^{\bj}$ and now
define the complex coordinates by
\be\label{complexgeneral}
z^i = \frac{1}{\sqrt{2}}\left(x^{i+3}+t^{ij} x^{j+6}\right)\ , \ i,j=1,2,3\ .
\ee
(The symmetric part of $t^{ij}$ is the pure holomorphic part of the metric
in a coordinate system with fixed complex coordinates; the antisymmetric
part should come from the NS-NS 2-form.)  A simple example is
that of $T^6=(T^2)^3$ with all the tori square, $t^{ij}=i\delta^{ij}$.
By writing the fluxes and holomorphic $(3,0)$ form out in terms of
a basis of (real) 3-forms, it is straightforward, but tedious, to
write out the superpotential in terms of the flux components, $\tau$, and
$t^{ij}$.  From section \ref{ss:noscale}, we know that supersymmetric
solutions have $W=\del_\tau W = \del_{t^{ij}} W =0$.  These give coupled
cubic equations in the complex moduli and $\tau$.  In fact, these are
generically overdetermined, so all the moduli will be fixed 
(\cite{Tripathy:2002qw} elaborated on conditions necessary to solve
these equations).  In particular, when $\tau$ is fixed near unity, these are
nonperturbative backgrounds of string theory and can be studied reliably
only through low-energy SUGRA unless there is a high degree of SUSY.
Some of the Hermitean
components of the metric might be fixed also; these are found by
considering primitivity.

It is easy to make some comments about the special case that the 
complex structure is proportional to the identity, $t^{ij} = t\delta^{ij}$.
Then two of the three equations are just cubic in $t$ with integer
coefficients, and the coefficients of the third (which includes also 
$\tau$ are determined by the first two equations. This implies that the
two cubics in $t$ have a common factor, so it is particularly easy to solve 
for the complex structure and then for $\tau$.  

\subsubsection{Solution Generation}\label{sss:generate}

There is a clear group action of $GL(2,\bm{\mathbbm{Z}})\times
GL(6,\bm{\mathbbm{Z}})$ on $T^6/\Z$ backgrounds which can be used to
generate new solutions  \cite{Kachru:2002he}.  

In particular, the $GL(2,\bm{\mathbbm{Z}})$ is the immediate generalization
of the $SL(2,\bm{\mathbbm{Z}})$ of type IIB string theory; the only
difference is that now the determinant of the matrix may not be unity,
so $G_3\to (ad-bc)G_3/(c\tau+d)$ ($\tau$ transforms as before, see
section \ref{ss:background}) \cite{Kachru:2002he}.  
This group includes simple rescalings of
the fluxes, which do not affect the vacuum values of the moduli.

The $GL(6,\bm{\mathbbm{Z}})$ acts on the fluxes and moduli like a 
large coordinate transformation, rotating the flux indices as well as
altering the complex structure.  The transformation of the complex 
structure is slightly complicated because the $x^{4,5,6}$ coordinates
can be mixed with $x^{7,8,9}$; this fact requires the redefinition of 
the complex coordinates $z$ back to the form (\ref{complexgeneral}) 
\cite{Kachru:2002he}.
Again, under these transformations, we arrive at a new solution.  The
only caveat (for either group) is that the new fluxes must not become
too large to satisfy the integrated Bianchi identity (\ref{intbianchi}).

As we observed, the $SL(2,\bm{\mathbbm{Z}})$ subgroup is actually
the S duality group of type IIB string theory; similarly, 
$SL(6,\bm{\mathbbm{Z}})\subset GL(6,\bm{\mathbbm{Z}})$ is the group of
large coordinate transformations that leave the torus invariant.  These
transformations, therefore, give equivalent backgrounds, still of
the self-dual flux type. 
Some transformations in 
$SL(2,\bm{\mathbbm{Z}})\times SL(6,\bm{\mathbbm{Z}})$ leave the background
invariant, in fact.  All of these transformations are therefore 
dualities of the 4D effective theory; those that leave the background
unchanged are U dualities.  We discuss the distinction in more detail
in section \ref{ss:udual}.

\subsubsection{An $\N =2$ Example}\label{sss:n2examp}

In section \ref{ss:tdual}, we will find it convenient to work with a 
simple example that preserves 4D $\N=2$ SUSY. We will work with this
example because it has few flux
components.  This example appeared extensively in 
\cite{Kachru:2002ns,Kachru:2002sk}.  The nonzero fluxes are
\be\label{n2fluxes}
f_{459}=f_{789}=h_{456}=h_{786}=2\ .\ee
Using \cite{Kachru:2002he}, it is possible to see that this vacuum is
$\N=2$ and that the complex structure moduli obey \cite{Kachru:2002sk}
\be\label{t-taufix2}
t^{ij}=\mathnormal{diag}(t_1,t_2,t_3)\, ,\ \tau t_1=-1\, ,\ 
t_2 t_3=-1\ .\ee
For simplicity, when we work with this background, we will assume that
$\tau, t_i$ are all imaginary.  The argument that this has $\N=2$ SUSY
is similar to that given below for $\N=3$ solutions.

\subsubsection{$\N=3$ Solutions}\label{sss:n3sol}

In much of this document, we will find it convenient to work in backgrounds
with a high degree of supersymmetry because of many simplifications.
Therefore, consider backgrounds
where the nonzero fluxes are
\begin{eqnarray}
h_{456} = -h_{489} = -h_{759} = -h_{786} &\equiv& h_1\ ,\nonumber\\
f_{456} = -f_{489} = -f_{759} = -f_{786} &\equiv& f_1\ ,\nonumber\\
h_{789} = -h_{756} = -h_{486} = -h_{459} &\equiv& h_2\ ,\nonumber\\
f_{789} = -f_{756} = -f_{486} = -f_{459} &\equiv& f_2\ , \label{n3fluxes}
\end{eqnarray}
and $f_{1,2}$ and $h_{1,2}$ are integers
(see the quantization condition (\ref{3quant}). 
These fluxes
implies that the
$T^6$ is the product of three square $T^2$s and fix the string coupling,
\begin{equation}
t^{ij}=i\delta^{ij}\ ,\ 
\tau = \frac{f_2 - i f_1}{h_2 - i h_1}\ . \label{t-taufix}
\end{equation}
The tadpole cancellation condition is
\begin{equation}
N_{\textnormal{D3}} + \frac{1}{2} N_{\widetilde{\textnormal{O3}}} 
= 16 - 2(h_1 f_2 - h_2 f_1)
\leq 16.
\end{equation}
the last inequality follows from the duality condition (\ref{t-taufix}).

This configuration of fluxes has the simple feature that in terms of the
complex coordinates (\ref{complexgeneral}) with complex structure
(\ref{t-taufix})
there is a single component
\begin{equation}
G_{\bar 1 \bar 2 \bar 3} = \frac{\sqrt 2\ap}{\pi R_1 R_2 R_3} 
(f_1 - \tau h_1)\ .
\label{03}
\end{equation}
That is, $G_{mnp}$ is a $(0,3)$-form.
As a background metric, we take
$\tilde{g}_{i\bj} = R_{i}^2 \delta_{i\bj}$,
although the K\"ahler moduli space is larger than just the three independent
radii (see section \ref{s:moduli}).  These backgrounds are discussed in
detail in \cite{Frey:2002hf} along with a listing of possible fluxes
and the associated moduli.

The $\N=3$ supersymmetry can be understood simply, as follows
\cite{Frey:2002hf}.  Remember that $G_3$ should be $(2,1)$ and primitive
for a supersymmetric solution.  If we choose the coordinates
\begin{equation}
(z^1,z^2,z^3)^\prime = (z^1,\b z^{\b 2},\b z^{\b 3}) \label{z1}
\end{equation}
then the nonzero flux $G_{\b{1} 2 3}$ is indeed $(2,1)$ and
primitive.  There are obviously two other such choices,
\begin{equation}
(z^1,z^2,z^3)'' = (\bar z^{\bar 1},z^2,\bar z^{\bar 3})\ ,\quad
(z^1,z^2,z^3)''' = (\b z^{\bar 1},\bar z^{\bar 2},z^{3})\ . \label{z2}
\end{equation}
Each of these three complex structures leads to an unbroken supersymmetry.

$\N=3$ supersymmetry is unfamiliar because the globally supersymmetric
cases have the same multiplets (with $CPT$ conjugates included).  
However, the gravity multiplets are distinct (see \ref{ss:superhiggs}).  
Previous
examples have been constructed as asymmetric orbifolds in type II
theory \cite{Ferrara:1989nm}, breaking half of the supersymmetry on one
side and three-fourths on the other, and the $\N=3$ supergravity was
constructed in \cite{Castellani:1986ka}.  
The moduli space is completely determined, and we review it in more detail
in section \ref{s:moduli}.

\subsubsection{Non-BPS Domain Walls and Varying $\N$}\label{sss:susychange}

A remarkable feature of self-dual flux compactifications on $T^6/\Z$ is
the variety of possible $\N$, the number of preserved supersymmetries in
the 4D effective theory.  Depending on the fluxes, supersymmetry can be
completely broken or be as high as $\N=4$ (no fluxes; this is the SUSY
preserved by a D3-brane).  However, as was indicated in 
\cite{Gukov:1999ya}, there are domain walls in the 4D effective theory 
over which the fluxes jump; the domain walls are just D5-branes and 
NS5-branes wrapping 3-cycles dual to the fluxes that change.
Following that lead, \cite{Kachru:2002ns} constructed domain walls for
$T^6/\Z$ compactifications that interpolate between solutions with 
different supersymmetries (the entire configuration including the 5-brane,
however, breaks all SUSY).  

Because the domain walls themselves break supersymmetry, \cite{Kachru:2002ns}
considered domain wall bubbles that expand before reaching a maximum radius
in the spacetime and then recontract. 
In a large enough compactification volume, the domain wall tension is small
compared to the Planck scale, and they can also remain outside their
Schwarzschild radius for an arbitrarily long time.  This shows that there
is a real sense in which string theory vacua with different SUSY are
connected, even though there is a large potential barrier between them
\cite{Kachru:2002ns}.  Let us note that these domain walls will also appear
in $K3\times T^2/\Z$ and CY 3-fold compactifications, but there is not
the same range of $\N$ in those compactifications.
We will see a variation on this idea in section \ref{s:dS}.

\subsection{$K3\times T^2/\Z$, O7-plane Action}\label{ss:k3t2}

While models of flux compactifications with O7-planes have appeared
throughout the literature 
(see \cite{Dasgupta:1999ss,Greene:2000gh,Becker:2002sx}, for example),
the most complete analysis of the relation between fluxes and moduli
on $K3\times T^2/\Z$ compactifications appeared in \cite{Tripathy:2002qw}.
Here I will give a brief review of their mathematical formalism and 
the particularly simple example of section 4.1, \cite{Tripathy:2002qw}.

The most important point to remember is that the 3-forms both must have
precisely one leg along the $T^2$, which we take in the $x^6,x^9$
directions.  Thus, we can write the fluxes as
\be\label{k3flux}
F_3 = \frac{2\pi\ap}{R_3}\left( f_6 dx^6 +f_9 dx^9 \right)\ ,\ 
H_3 = \frac{2\pi\ap}{R_3}\left( h_6 dx^6 +h_9 dx^9 \right)\ ,\ee
where $f_{6,9},h_{6,9}$ are in the \textit{integer} cohomology 
$H^2(K3,\bm{\mathbbm{Z}})$ of $K3$.    $H^2(K3,\bm{\mathbbm{Z}})$
inherits an inner product
\be\label{k3inner} 
e_1\cdot e_2 = \int_{K3} e_1\wedge e_2 \ee
of signature $(3,19)$ from $H^2(K3,\bm{\mathbbm{R}})$.  Using this
inner product, the tadpole cancellation condition (\ref{intbianchi})
becomes
\be\label{k3tadpole}
N_{\textnormal{D3}} +N_{\textnormal{inst}}= 24-\frac{1}{2}\left(
f_6\cdot h_9-f_9\cdot h_6\right)\ee
assuming there are no O7-planes with flux.  $N_{\textnormal{inst}}$ is
the number of gauge theory instantons in the $K3$ submanifold of the 
D7-brane worldvolume, which have been considered in 
\cite{Greene:2000gh}; from now on, we set $N_{\textnormal{inst}}=0$.
The $24$ negative D3-brane charges come from the curvature of the
$K3$ part of the D7-brane and O7-plane worldvolumes.

To determine the complex structure of the $K3$ and $T^2$ as well as $\tau$,
\cite{Tripathy:2002qw} use Torelli's theorem, which states that the
complex structure of $K3$ is given entirely by the holmorphic $(2,0)$
form $\omega$.  In particular, the possible values of $\omega$ span a 
2D spacelike subspace of $H^2(K3,\bm{\mathbbm{R}})$.  
They then get a number of conditions on 
$H^2(K3,\bm{\mathbbm{R}})$ inner products by demanding
that $G_3$ be $(2,1)$, which end up reducing to polynomials in $\tau$ and
the complex structure $t$ of $T^2$.  This is very similar to the 
$T^6/\Z$ case.

\subsubsection{A Simple Example}\label{sss:simpleK3}

As discussed in section 4.1 of \cite{Tripathy:2002qw}, we can use
equation (\ref{k3flux}) to write
\be\label{k3gflux}
G_3 = \frac{\sqrt{2}2\pi\ap}{R_3(\b t-t)} 
\left( G_z dz^3+ G_{\b z} d\b z^{\b 3}\right)\ .\ee
For $G_3$ to be $(2,1)$, $G_z$ must be $(1,1)$, and $G_{\b z}$ must be
$(2,0)$.  Since $K3$ has only one $(2,0)$ form, $G_{\b z}\propto \omega$.
Thus the complex structure of $K3$ is determined completely by the
flux $G_{\b z}$ (assuming $G_{\b z}\neq 0$).

Now suppose for this example that $f_{6,9},h_{6,9}$ span a 
two dimensional subspace of $H^2(K3,\bm{\mathbbm{R}})$ (they are not
all linearly independent).  Because the real and imaginary parts of 
$\omega$ span a 2D
subspace also, this must be the same as the flux subspace.  Additionally,
because $G_z$ is $(1,1)$, $G_z\cdot\omega =G_z\cdot \b\omega =0$; since
the 2-plane is spacelike, $G_z=0$.

In terms of the flux components, 
\be\label{gz0}
G_z=0\ \Rightarrow\ (f_6-\tau h_6)\b t  =(f_9-\tau h_9)\ .\ee
But also $G_{\b z}\cdot G_{\b z} =0$ because there are no $(4,0)$ forms.
Using (\ref{gz0}), this requirement gives
\be\label{nx2}
G_{\b z}^2 \propto \im t (f_6-\tau h_6)^2 =0 \ .\ee
Since $\im t =0$ is a singular $T^2$, this is a quadratic equation for
$\tau$.  With the known fixed value of $\tau$, we can then use
equation (\ref{gz0}) to find the fixed value of $t$:
\be\label{k3tfix}
t = \frac{(f_6-\tau h_6)\cdot (f_9-\b\tau h_9)}{|f_6-\tau h_6|^2}\ .\ee

\section{Gauged Supergravity}\label{s:gauged}

In this section, we discuss another perspective on the 4D effective theory,
that of gauged supergravities.  They have been applied to 
the $T^6/\Z$ and $K3\times T^2/\Z$ cases\footnote{Although there are many
similarities in the structure of the SUGRA, warped CY 3-fold 
compactifications cannot be described by gauged supergravity because there
are no gauge fields; there are no 1-cycles on the CY to reduce $B_2,C_2$
to vectors. This of course assumes that the warp factor does not change
dimensional reduction in a radical way, which seems impossible given
the large radius limit (see \ref{s:eom}).}
of the last section in an
extensive literature \cite{Tsokur:1996gr,Andrianopoli:2002mf,
Andrianopoli:2002aq,D'Auria:2002tc,Ferrara:2002bt,D'Auria:2002th,
Andrianopoli:2002vy,Andrianopoli:2002vq,Andrianopoli:2003jf,
Andrianopoli:2003sa,D'Auria:2003jk,Angelantonj:2003rq}.  A nice 
review of the pure supergravity is given in \cite{Ferrara:2002hb}, so
we will be brief.  Our main focus will be to show how the gauged 
supergravity appears in the string compactification, and we work
within the framework of $T^6/\Z$ compactifications for simplicity.

\subsection{SuperHiggs Interpretation}\label{ss:superhiggs}

Consider the $\N=4$ states which become massive due to the
fluxes.  For the $\N=3$ background of section \ref{sss:n3sol}, 
these include one gravitino, so we must have a massive spin 3/2
multiplet.  This must be a large representation because these supergravity
states are all neutral under the $U(1)$ central charges.  Therefore
the supersymmetry breaking (in the bulk fields) is from the $\N=4$ 
gravity multiplet and $6$ matter multiplets
with helicities
\bea
\textnormal{gravity:}&&  
2,\ \textnormal{$\frac{3}{2}$}^4,\ 
1^6,\ \textnormal{$\frac{1}{2}$}^4,\ 0^2,\ -\textnormal{$\frac{1}{2}$}^4,\
- 1^6,\ -\textnormal{$\frac{3}{2}$}^4,\ -2
\label{n4spin2}\\
\textnormal{matter:}&& 1,\ \textnormal{$\frac{1}{2}$}^4,\ 0^6,\ 
-\textnormal{$\frac{1}{2}$}^4,\ -1
\label{n4matter}\eea
to the $\N =3$ gravity, spin $3/2$, and $3$ matter multiplets, whose
helicities are
\bea
\textnormal{gravity:}&&2,\ \textnormal{$\frac{3}{2}$}^3,\ 
1^3,\ \textnormal{$\frac{1}{2}$},\ -\textnormal{$\frac{1}{2}$},\
- 1^3,\ -\textnormal{$\frac{3}{2}$}^3,\ -2\label{n3spin2}\\
\textnormal{spin $3/2$:}&&\textnormal{$\frac{3}{2}$},\ 
1^6,\ \textnormal{$\frac{1}{2}$}^{15},\ 0^{20},\ 
-\textnormal{$\frac{1}{2}$}^{15},\ -1^6,\ -\textnormal{$\frac{3}{2}$}\ .
\label{spin32}
\eea
(The matter multiplets have the same helicities.)

Because there are now six massive vectors, supersymmetry breaking must
be accompanied by symmetry breaking, the SuperHiggs effect.
We see in section \ref{ss:describe} that six gauge symmetries are indeed 
broken.   The
twenty spin-zero components are the dilaton-axion, the six zero-helicity
components of the massive vectors (from $C_{4}$), and the twelve
real components of
$g_{ij}$.  Remember that at large radius these states, with masses-squared
$1/(\ap e^{6u})$, lie parametrically below the Kaluza-Klein scale of
$1/(\ap e^{2u})$.  
Thus we can truncate to an effective field theory in which only
these and the massless states survive.  Since the mass scale is
parametrically below the Planck scale as well, the  SUSY breaking from
$\N=4$ to $\N=3$ must be spontaneous.  In fact, the same story holds for
other degrees of supersymmetry.  There has been some discussion of
such breaking in
supergravity \cite{deRoo:1986yw,Wagemans:1988zy,Tsokur:1996gr} and more
recently in the gauged supergravity literature 
(see especially \cite{Andrianopoli:2001zh,Andrianopoli:2002rm}).

\subsection{The Gauging and Symmetry Breaking}\label{ss:gauging}

Our goal here is in essence to give a definition of gauged supergravity
through some simple string theory.  
Remember from section \ref{p:c4-gauge} that the linearized 5-form 
(\ref{elhiggs}) is a Higgsed gauge covariant derivative, and that
some components of $c_4$ become Goldstone bosons.  In fact, it is just 
the number necessary for the multiplet (\ref{spin32}) to work out
correctly.  In the classical type IIB SUGRA on $T^6/\Z$, continuous
shifts of the axions $\t c_4$ are in fact a symmetry, and gauged 
supergravity just consists of using an existing vector to gauge
a symmetry of the scalars.

We can be a little more formal, following for example the discussion
of \cite{Ferrara:2002hb}.  In an extended supergravity theory, the 
scalars form a manifold, which may have a number of symmetries.  These
are the supergravity U duality group.  To gauge any of the symmetries 
of the scalar manifold, we need vectors that transform in the adjoint of
that symmetry group.  It has long been known \cite{Gaillard:1981rj}
that $n$ field strengths $F$  and their magnetic duals $G$ form a vector
under the most general duality transformation $Sp(2n,\bm{\mathbbm{R}})$:
\be\label{symplectic}
\begin{bmatrix} F'\\ G'\end{bmatrix} = \begin{bmatrix} A & B\\ C& D
\end{bmatrix}\begin{bmatrix} F\\ G\end{bmatrix},\  
 \begin{bmatrix} A & B\\ C& D
\end{bmatrix}\in Sp(2n,\bm{\mathbbm{R}})
.\ee
Therefore, any group we gauge must be embedded in the symplectic group
such that the set of electric field strengths is left invariant, $B=0$,
and such that the action on the field strengths is the adjoint.
In terms of the Lie algebra of $Sp(2n,\bm{\mathbbm{R}})$, the generator
must be 
\be\label{sympgen}
\begin{bmatrix} a & b=0\\ c & -a^T\end{bmatrix}\ .\ee

This is actually the low-energy description of the Scherk-Schwarz
mechanism \cite{Scherk:1979zr} for a non-Abelian group (see
\cite{Ferrara:2002hb} and references therein).  However, we are interested
in an easier example.  For IIB strings on $T^6/\Z$ with no fluxes,
the scalar manifold has symmetry group $SO(6,6)\times SL(2,\bm{\mathbbm{R}})$
(ignoring D3-brane fields).  
There are multiple ways to embed this
in $Sp(24,\bm{\mathbbm{R}})$ for the action on the 12 bulk vectors.
In the type IIB string theory, taking $B_{\mu m},C_{\mu m}$ as the electric
vectors, the Lie algebra of $SO(6,6)$ contains 15 translations.  These
15 translations in fact embed into the matrix $c$ of the symplectic
group generators.  This is an abelian group, so the action on the field
strengths is trivial.

\subsection{Potential from Gauging}\label{ss:gaugepot}

In the SUGRA, the gauging modifies the gravitino and gaugino 
supersymmetry variations, which is why some of the supersymmetries are
broken.  In fact, this is easy to see from the 10D perspective; having
nonzero fluxes clearly changes the SUSY variations (\ref{susyvars}).
While in section \ref{ss:susy} we were intent on finding some unbroken
supersymmetry, here we take the reverse viewpoint. 
There is 4D $\N =4$ SUSY for the $T^6/\Z$ compactification 
with no fluxes, so it is the flux that breaks supersymmetry.

In fact, the modification of the SUSY variations generates a potential
for the moduli (again, see \cite{Ferrara:2002hb} and references therein).
In a very sketchy way, we can write 
\be\label{gaugepot}
V = |\delta \lambda^I |^2 -3 |\delta \psi_{A\mu}|^2 \ ,\ee
where $\lambda^I$ are the gaugini and $\psi_{A\mu}$ the gravitini.  It has
been shown that the potential for the bulk scalars matches exactly 
the prediction from dimensional reduction (say from the $\N=1$ SUGRA
formalism of section \ref{s:gvw}).  In fact, the potential for the brane
scalars (at the classical level, at least) can be derived to a higher
order in the scalars than from the non-Abelian D-brane action
\cite{Myers:1999ps}; this derivation is carried out in detail in
\cite{D'Auria:2002th}.

\chapter{Windows On Small Volume}\label{c:small}
So far, I have reviewed self-dual flux compactifications from the
point of view of supergravity.  There are several reasons why I did not
use string perturbation theory to describe even the simple compactifications
of section \ref{s:special} 
(as would be analogous to \cite{Narain:1986jj,Narain:1987am}).  
One reason is practical: the worldsheet theory in general R-R backgrounds
is not understood.  Another is fundamental: the fluxes generically 
fix the value of the string coupling $g_s\sim 1$, so string perturbation
theory is simply not valid.  Since we have been 
mostly interested in the low
energy physics, having only the tool of supergravity has been acceptable.

However, supergravity is only valid at large compactification volumes. Some
reasons are true in all string compactifications. For example, 
at small volume, we
expect large curvature invariants to make $\ap$ corrections nonnegligible.
Similarly, small volume means that there would be small cycles and light
winding modes of strings and D-branes that start to dominate the effective
theory.  In the warped backgrounds described in section {\ref{s:tend},
the problem is even worse.  In the SUGRA description at large volume,
I smoothed over the fact that the square $e^{4A}$ of the warp factor
becomes negative near O3-planes (and other objects with negative D3-charge),
saying that stringy corrections should take over in those regions
\cite{Chan:2000ms,Greene:2000gh}\footnote{S. Sethi has also emphasized
the importance of the negative warp factor to several authors.}.  At
small radii, those regions fill the entire internal space excepting
various throats, so we lose our understanding of the physics.

In this chapter, I discuss three handles we can use to get a qualitative
hold on
small volume compactifications with self-dual flux.  The first is finding
the moduli spaces in theories with large numbers of supercharges; we study
the case of $\N =3$ SUGRA because the local form of moduli space is
protected from all corrections.  Next, we look at the BPS spectrum and
count the number of multiplets; for similar reasons, this is also protected
from corrections.  Finally, we discuss dualities, both U-dualities of the
compactification and T-dualities to other compactifications, both of which
can lead to new, safe, SUGRA descriptions of the physics.
All of these methods rely on
extended supersymmetry in the 4D effective theory, but they allow us at
least a qualitative understanding of what happens in less supersymmetric
cases.

\section{Moduli Spaces}\label{s:moduli}

When we have addressed the scalars of string compactifications until now,
our emphasis has mostly been on the fact that self-dual fluxes can freeze
moduli.  In this section, we want to address compactifications in which
the fluxes are special, leaving more than just a single radial modulus
massless.  The classical moduli spaces are fairly straightforward to 
determine; if there are no free D-branes, the moduli spaces are
$(U(1,1)/U(1)^2)^3$, $(U(1,1)/U(1)^2)\times U(2,2)/(U(2)\times U(2))$,
and $U(3,3)/(U(3)\times U(3))$ for $T^6/\Z$ orientifolds with 4D $\N =1,2,3$
supersymmetry respectively 
\cite{Andrianopoli:2002rm,Andrianopoli:2002mf,Andrianopoli:2002aq}.  
Strictly speaking, these are only the local
forms at large radius, but in the $\N= 3$ case there is enough 
supersymmetry to extrapolate to the full moduli space.  For 
$\N\leq 2$ it would be very
difficult to analyze the full moduli space.

Therefore,
we analyze the massless scalars of the $\N =3$ models on $T^6/\Z$, 
to verify the structure required by $\N = 3$
supergravity: with the supergravity multiplet plus $n$ matter multiplets
(that is, $n-3$ free D3-branes),
there must be $6n$ moduli and $n+3$ vectors.
We verify, in the large-radius limit, that the metric
on moduli space has the expected form \cite{Castellani:1986ka}
\be\label{metmod}
\frac{U(3,n)}{U(3)\times U(n)}\ .\ee
I will discuss the vectors in the next section, \ref{ss:describe}.
The following results were originally reported in \cite{Frey:2002hf}.

\subsection{Identifying $\N =3$ Moduli}\label{ss:identmod}

The massless scalars arise from the zero modes of the $\Z$-even
scalars in (\ref{reflect}), namely $g_{mn}$, $C_{mnpq}$, $\Phi$ and $C$.
However, not all of these are moduli, as the
fluxes lift some of the directions of moduli
space \cite{Dasgupta:1999ss,Gukov:1999ya,Greene:2000gh}.  For example,
we have already seen that the dilaton and R-R scalar are fixed. Their
potential arises from the three-form flux and the resulting mass-squared 
in the 10D string frame is of order
\begin{equation}
G_{mnp} \b G^{mnp} \sim \frac{1}{\ap}e^{-6u} \sim \frac{\ap{}^2}{V_6}\ .
\label{fluxmass}
\end{equation}
In this rough estimate, we have assumed that all the proper radii are similar
and have used the quantization
conditions (\ref{gquant}).

Now consider the scalars $g_{mn}$.  These are partly fixed by the
self-duality condition (\ref{imsd}), through the dependence of the
$\epsilon$-tensor on $g_{mn}$.  The zero mode of the three-form flux is
fixed by the quantization conditions, so $G_{mnp}$ remains an
$(0,3)$-form in the $z$ coordinates.  The metric $g_{mn}$ must therefore be
Hermitean in these coordinates, or else there will be nonzero components
$\epsilon_{\bar\imath\bar\jmath k}{}^{\bar\imath'\bar\jmath' \bar k'}$.
The self-duality condition is satisfied for any Hermitean metric
$g_{i\bar\jmath}$.  Thus, in terms of the $z$ coordinates, the complex
structure moduli are frozen while the K\"ahler moduli remain free.  In
terms of any of the supersymmetric complex structures (\ref{z1}, \ref{z2})
these are a mix of K\"ahler and complex structure moduli 
\cite{Dasgupta:1999ss,Greene:2000gh}.

The remaining bulk scalars are those from the four-form
potential
$C_{mnpq}$.  We remember from section \ref{s:eom} that the periodicity
of these forms is complicated; the important 
conclusion is that there is a field $\tilde c_{mnpq}$ which is periodic
and which appears in the field strength only through its exterior
derivative.  A constant shift of this field is then a new solution to the
equations of motion.  However, some of these are gauge-equivalent to the
unshifted solution.  We saw in equation (\ref{c4var}) 
that the gauge variation around a given background is
\begin{equation}
\delta \tilde c_{4} = d\tilde\chi +i
(\b{\lambda}_A \wedge \hat G_{3} - \lambda_A \wedge
\hat{\b{G}}_{3})/2\,\im (\tau)\ , \label{c4varx}
\end{equation}
with $\tilde \chi$ periodic and $\lambda_A$ a complex
one-form.  Since the background $\hat G_{3}$ is a
$(0,3)$ form, the $(1,3)$ and $(3,1)$ parts of $\tilde c_{4}$ can be
gauged away.  The $(2,2)$ parts $\tilde c_{ij\bar k\bar l}$ are the
moduli.

Finally, there is no restriction on the positions of any D3-branes that
might be present, so their world-volume scalars are also
moduli.  It will be convenient to write these in complex form,
as $Z_I^i$, $\b{Z}_I^{\bj}$ 
where $I$ labels the D3-brane (perturbatively
speaking, it would be a Chan-Paton factor diagonal on the two endpoints).
When some D3-branes become coincident, we expect that the index $I$ should
become an index in the adjoint of the enhanced gauge group.  We will 
discuss enhanced symmetries later, in section \ref{ss:udual}.

\subsection{Metric on $\N =3$ Moduli Space}\label{ss:modmetric}

Here I will show that the low-energy action for the scalars 
takes the form of a $U(3,n)/U(3) \times U(n)$ coset.  We
only consider the large-radius limit, where the warp factor $A$ vanishes
as discussed in section \ref{sss:constraints}.
Thus, for convenience, I will drop the tildes
on the internal metric in this section.  
As usual,``E'' denotes four-dimensional Einstein frame while a subscript
``4'' is the 4D part of the string frame metric;
internal indices will always be raised with the string metric.
Since the dilaton-axion is frozen, we will often write $g_s=e^\phi$.

\subsubsection{$U(3,3)/(U(3)\times U(3))$ Metric}\label{sss:u33}

Here we consider only the moduli from the bulk, those we would have in
the absence of D3-branes.

Let us first find the action for the metric moduli.  The dimensional
reduction of the ten-dimensional string frame Hilbert action gives
\be\label{metricmod1}
S_{\mathrm{g}} = \frac{1}{4\pi \ap g_{s}^2} \int d^4x \sqrt{-g_4}\,
e^{6u} \left[ R_4 +e^{-12 u}\del_\mu e^{6u}\del^\mu e^{6u}
-\frac{1}{2} g^{\bj i}g^{\b{l} k} \del_\mu g_{k\bj} \del^\mu g_{i\b{l}}
\right]\ee
where $e^{6u} = \sqrt{\det g_{mn}}=\det g_{i\bj} $.
The dimensional reduction includes a factor $\frac{1}{2}(2\pi)^6\ap{}^3$ 
from the
volume of $T^6/\Z$.  Switching to the
four-dimensional Einstein frame, equation (\ref{einstein4d}),
\bea
S_{\mathrm{g}} &=& \frac{1}{4\pi \ap} \int d^4x \sqrt{-g_{\rm E}}
\left[ R_{\rm E} -\frac{1}{2}e^{-12u} \del_\mu e^{6u}
\del^\mu e^{6u}
-\frac{1}{2} g^{\bj i}g^{\b{l} k} \del_\mu g_{k\bj} \del^\mu g_{i\b{l}}
\right]\nonumber\\
&=& \frac{1}{4\pi \ap} \int d^4x \sqrt{-g_{\rm E}}
\left[ R_{\rm E} -\frac{1}{2} \gamma^{\bj i} \gamma^{\b{l} k} \del_\mu
\gamma_{k\bj}
\del^\mu \gamma_{i\b{l}} \right]\ , \label{metricmod2}\eea
where all spacetime indices are
raised with the Einstein metric.  We have
defined
\be\label{moddensity}
\gamma_{i\bj} = 2 g_{s} e^{-6u} g_{i\bj}
\ee
in order to eliminate double trace terms from the derivatives of $e^{6u}$.
To see how to get rid of the traces, the identity
\be
\label{trick}
\gamma^{mn}\del_\mu\gamma_{mn} = -\gamma_{mn}\del_\mu\gamma^{mn} 
=6\del_\mu\phi-24\del_\mu u
\ee
is useful.

The other bulk moduli are the R-R scalars, contained in the field strength
fluctuation $\tilde f_{(5)}$.  The moduli kinetic terms arise from $\tilde
f_{\mu npqr}$ and in Hodge dual form from $\tilde
f_{\mu \nu \lambda qr}$; in order to avoid the problems of self-dual
actions we include only the former, in terms of which
\begin{equation}
S_{\mathrm{RR}} = -\frac{ g_{s}^2}{16\pi \ap}
\int d^4x \sqrt{-g_{E}}\,| \tilde
f_{(5)} |^2 \ .
\end{equation}
In the absence of D3-branes, we have $f_{\mu ij \bar k \bar l} =
\partial_\mu c_{ij \bar k \bar l}$, and the action is simply
\begin{equation}
S_{\mathrm{RR}} = -\frac{g_{s}^2}{64\pi \ap}
\int d^4x \sqrt{-g_{E}}\, g^{i\bi'} g^{j\bj'} g^{k\bar k'} g^{l\bar l'}
\partial_\mu c_{ij\b{k}\b{l}}
\partial^\mu c_{\bi'\bj'k'l'}   \ .
\end{equation}
To exhibit the coset
structure we put these moduli in a two-index
form,
\be\label{defc_hat}
c_{ij\b k \b l} = 2 e^{-6u} \epsilon_{ij\b k \b l a \b b}
\beta^{a \b b}\ .
\ee
The action for all the bulk supergravity moduli is then
\be\label{bulkmoduli}
S_{\textnormal{\scriptsize mod}} =
- \frac{1}{8\pi\ap } \int d^4x
\sqrt{-g_{\rm E}}\, \gamma_{k\bar\jmath} \gamma_{i \bar l}
( \partial_\mu \gamma^{i\bj} \partial^\mu \gamma^{k\b{l}}
- \partial_\mu \beta^{i\bj} \partial^\mu \beta^{k\b{l}} )
\ .
\ee
This is just the $U(3,3)/U(3)\times U(3)$ moduli space metric, familiar
from the untwisted moduli of the $\bm{\mathbbm{Z}}_3$ orbifold 
\cite{Ferrara:1986qn,Ovrut:1989eq},
with upper and lower indices exchanged.  The ``wrong-sign'' kinetic term
on $\beta$ corrects for the fact that $\beta^{i\bj}$ is anti-Hermitean.

\subsubsection{With D3-Branes}\label{sss:withd3}

We now consider D3-branes.  Expanding the DBI action gives the kinetic term
\be\label{dbimod}
S_{\textnormal{\scriptsize DBI}} = -\frac{1}{2(2\pi)^3 \ap{}^2}
\int d^4x \sqrt{-g_{E}}\,
\gamma_{i\bj} \del_\mu Z_I^i \del^\mu \b Z_I^{\bj} \ ,
\ee
with an implicit sum on $I$.  In addition there is a dependence on the
collective coordinates from the coupling of the D3-brane to $C_{(4)}$, which
appears through a nontrivial five-form Bianchi identity.  In the D3-brane
rest frame,
\begin{equation}
d\tilde F_{(5)} = (2\pi)^4 \ap{}^2 \delta^6(x^m) d^6 x \to
\frac{\ap{}^2}{2\pi^2}  d^6 x\ ,
\end{equation}
where we have projected onto the zero mode; we omit the flux term in the
Bianchi identity, which makes no contribution to the moduli kinetic terms.
Boosting this gives
\bea
(d\tilde f)_{\mu\nu ij \bar k \bar l} &=& \frac{e^{-6u}}{2\pi^2 \ap}
\epsilon_{ij\bar k \bar l a \bar b} ( \del_{\mu}
\b Z_I^{\bar b} \del_{\nu} Z_I^a - \del_{\mu} Z_I^a  \del_{\nu}
\b z_I^{\bar b} )\ , \nonumber\\
f_{\nu ij \bar k \bar l} &=& \partial_\nu c_{ij \bar k \bar l}
+ \frac{e^{-6u}}{4\pi^2 \ap}
\epsilon_{ij\bar k \bar l a \bar b} (
\b Z_I^{\bar b} \del_{\nu} Z_I^a - Z_I^a  \del_{\nu} \b Z_I^{\bar b} )\ .
\eea
The moduli space action is then
\bea
S_{\textnormal{\scriptsize mod}} &=&
- \frac{1}{8\pi \ap} \int d^4x
\sqrt{-g_{\rm E}}\, \left\{
\gamma_{k\bar\jmath} \gamma_{i \bar l}
( \partial_\mu \gamma^{i\bj} \partial^\mu \gamma^{k\b{l}}
- \mathcal{D}_\mu\beta^{i\bj} \mathcal{D}^{\mu}\beta^{ k\b{l}} )\right.
\nonumber\\
&&\left.
+ \frac{1}{2\pi^2} \gamma_{i\bj}  \del_\mu Z_I^i \del^\mu \b Z_I^{\bj}
\right\}
\ ,\label{allmoduli}
\eea
where
\be
\mathcal{D}_\mu\beta^{i\bj} = \partial_\mu \beta^{i\bj} + \frac{1}{8\pi^2\ap} 
(
\b Z_I^{\bj} \del_{\mu} Z_I^i - Z_I^i \del_{\mu} \b Z_I^{\bj} ) \ .
\ee

\subsubsection{Coset Form}\label{sss:coset}

With a bit of algebra, it is possible to show that the entire action on
moduli space takes the form
\be\label{modaction}
S = \frac{1}{4} \frac{1}{4\pi\ap } \int d^4x \sqrt{-g_{\rm E}}
\,\tr \left( \del_\mu M \eta \del^\mu M \eta\right)
\ee
where $\eta$ is the $U(3,n)$ invariant metric ($\eta =\Omega^\dagger
\eta \Omega$) and $M$ is a Hermitean
$U(3,n)$ matrix that behaves as $M\rightarrow \Omega M \Omega^\dagger$
under $\Omega\in U(3,n)$.  We work in a basis with block diagonal form
\be\label{matrices}
\eta =\! \left[ \begin{array}{ccc} &\, I_3\, &\\ \, I_3\, &&\\&&\, I_N\, 
\end{array}
\right]\, ,\, M\! =\! \left[ \begin{array}{ccc}\gamma^{-1} &
-\gamma^{-1}\mathcal{B}&  -\gamma^{-1} \alpha^\dagger\\
-\mathcal{B}^\dagger \gamma^{-1} &\ \gamma +\mathcal{B}^\dagger\gamma^{-1}
\mathcal{B} +\alpha^\dagger \alpha\
& \mathcal{B}^\dagger\gamma^{-1} \alpha^\dagger +\alpha^\dagger\\
-\alpha \gamma^{-1} & \alpha\gamma^{-1} \mathcal{B}+\alpha&
I_N+\alpha\gamma^{-1} \alpha^\dagger\end{array}
\right]
\ee
with matrix notation $\gamma = \gamma^{\bj i}$,
$\alpha = Z^i_I/2\pi\sqrt{\ap}$, and
$\mathcal{B}=\beta + (1/2)\alpha^\dagger \alpha$.  Here we have used the
fact that the number of matter multiplets $n$ and number of D3-branes $N$
are related by $n=N+3$.  To verify that this takes
the appropriate coset form, note that we can write
\be\label{coset}
M=V^\dagger V\ ,\quad V = \left[\begin{array}{ccc} e& -e\mathcal{B}&
-e\alpha^\dagger\\
0&e^{-1}& 0\\0&\alpha&I_N\end{array}\right]
\ee
where $e$ is the vielbein $e^\dagger e = \gamma^{-1}$.  Following
\cite{Maharana:1993my}, we see that $M$ indeed belongs to the coset
$U(3,3+N)/U(3)\times U(3+N)$, precisely as we expected based on $\N=3$
supersymmetry.

\subsection{Comparison to $\N=4$ heterotic string}\label{ss:heterotic}

The results of \ref{ss:modmetric} are notably similar to work done
by Maharana and Schwarz on the $O(6,22)$ duality of the heterotic string
on $T^6$ \cite{Maharana:1993my}.  This is not an accident.  Starting from
the heterotic string, $S$-duality maps to type I strings, and a further
$T$-duality on all six internal 
dimensions takes the theory to the IIB model of
\cite{Verlinde:1999fy}.  Our $\N=3$ models are then obtained by
nonperturbatively transforming D3-branes into self-dual $G_{3}$ flux,
so we expect that our moduli space should simply be a subspace of the
heterotic moduli space.

To make this more precise, we can follow the action of the $S$- and
$T$-dualities on the moduli of the heterotic theory. 
We will also choose duality conventions such that $\ap$ 
is the same in the heterotic, type I, and type IIB string theories.  To
get the normalization correct including numerical factors, we must be
careful (see \cite{Gimon:1996rq} for some factors in the type I theory,
for example).  Also, please note that these duality mappings are for
the case without fluxes; with flux, see section \ref{ss:tdual} below
and \cite{Dasgupta:1999ss,Greene:2000gh,Becker:2002sx,Kachru:2002sk}.

\subsubsection{Duality Map}\label{sss:hetIIBmap}
We start by considering the heterotic--type I S-duality.  Under this
duality, the heterotic fundamental string maps to the type I D-string; 
in particular the actions must be equal.  Since the D-string tension
and charge are reduced by a factor of $\sqrt{2}$ by the orientifold 
projection in the type I theory, we therefore must have
\bea
\frac{1}{2\pi\ap \sqrt{2}} \int d^2\xi e^{-\phi}(\mathrm{I}) 
\sqrt{-\det g(\mathrm{I})} &=& \frac{1}{2\pi\ap} \int d^2\xi
\sqrt{-\det g(\mathrm{het})}\nonumber\\
\Rightarrow g_{MN}(\mathrm{het}) &=&
\frac{e^{-\phi}(\mathrm{I})}{\sqrt{2}} g_{MN}(\mathrm{I}) \label{hetI}
\eea
and likewise $B_2(\mathrm{het}) = C_2(\mathrm{I})/\sqrt{2}$.
The 10D supergravity actions then map into each other if we take the 
gauge theory potentials to be equal.

In the T-duality between type I on $T^6$ and IIB on 
$T^6/\Z$, the dilaton picks up a well-known factor of
$\sqrt{2}$ \cite{Gimon:1996rq}, so the T-duality is
\be\label{IIIBdilg}
e^{\phi}(\mathrm{I}) = \frac{\sqrt{2}}{\det^{1/2} g_{mn}}
e^{\phi}(\mathrm{IIB})\ ,\
g_{mn}(\mathrm{I}) = g^{mn}(\mathrm{IIB}) \ ,\
g_{\mu\nu}(\mathrm{I}) = g_{\mu\nu}(\mathrm{IIB})\ .
\ee
There is an additional factor in the R-R sector, as follows.  Taking the
prefactor of the 10D action to be the same in the two theories, 
T-duality tells us that we should have the same dimensionally reduced
actions, or
\bea
\frac{(2\pi)^6\ap{}^3}{2\cdot 2}\int d^4x\sqrt{-g_4}e^{6u}
\del_\mu C_{mn} \del^\mu C^{mn} (\mathrm{I})
&=& \nonumber\\
\frac{(2\pi)^6\ap{}^3}{2\cdot 2\cdot 4!}\int d^4x\sqrt{-g_4}
e^{6u} \del_\mu C_{mnpq} \del^\mu C^{mnpq} (\mathrm{IIB})&&
\label{IIIBRRaction}
\eea
for the moduli.  Again, $e^{6u}=\det^{1/2}g_{mn}$ and $g_4$ is the 
string frame metric in the spacetime.  
The additional factor of 2 in the IIB case 
again comes from the volume.  This equality holds if we take
\be\label{IIIBRR}
C_{mn}(\mathrm{I}) = \frac{1}{\sqrt{2}\cdot 4!} e^{6u} \epsilon^{mnpqrs}
C_{pqrs}\ .\ee

Then the heterotic moduli 
(using the notation of \cite{Maharana:1993my}) map
to the IIB $\N=4$ moduli as follows:
\be
\label{STmap}
 g_{mn}\rightarrow
\gamma^{-1}{}^{mn}\ ,\ B_{mn}\rightarrow \beta^{mn}\ ,
\ a_m^I\rightarrow \alpha^m_I\ ,
\ee
following the notation of section \ref{ss:modmetric} for the IIB side.
The $\N=3$ moduli are then clearly the (anti-)Hermitean subset of the
gravitational and R-R moduli along with all the D-brane positions in
complex form.  

Also, the spacetime metrics map from the heterotic variables to IIB
as
\be\label{STmetricmap}
g_{\mu\nu}\to \frac{e^\phi}{2} g_{E,\mu\nu}\ .\ee
As it turns out, \cite{Sen:1994fa,Schwarz:1993mg} gave a ``canonical''
spacetime metric in the heterotic theory as an invariant under 
the U-dualities; up to a factor of $2$, this canonical metric corresponds
to our type IIB Einstein frame metric.

\subsubsection{$\N =4$ Moduli Space Metric}\label{sss:n4metric}

In our IIB notation, then, the $\N =4$ moduli space (and Einstein-Hilbert)
action has the form
\bea
S &=& \frac{M_4^2}{2} \int d^4 x \sqrt{-g_E} \left[ R_E+
\frac{1}{4}\del_\mu\gamma_{mn}\del^\mu\gamma^{mn} +\frac{1}{4}
\gamma_{mp}\gamma_{nq} D_\mu\beta^{mn} D^\mu\beta^{qp}\right.\nonumber\\
&&\left.-\frac{1}{2}
\gamma_{mn} \del_\mu \alpha^m_I \del^\mu \alpha^n_I 
-\frac{1}{2} \del_\mu \phi\del^\mu\phi -\frac{1}{2}e^{2\phi}
\del_\mu C\del^\mu C\right]\ ,\label{allmoduli2}
\eea
where $M_4=(2\pi\ap)^{-1/2}$ is the reduced Planck mass.  This
is the dimensionally reduced action for the heterotic theory of
\cite{Maharana:1993my,Sen:1994fa}, as one might expect. 
As an alternative to the above discussion, we could have arrived at
the moduli space metric (\ref{allmoduli}) by working out equation
(\ref{allmoduli2}) and then projecting onto the massless fields.

Note that there is an additional complex modulus
in the $\N=4$ case which corresponds on the heterotic side to the
four-dimensional dilaton and $B_{\mu\nu}$ axion, and on the IIB side to the
ten-dimensional dilaton and R-R scalar.  In the $\N=3$ theories this
modulus is fixed.  In models with less SUSY, the dilaton-axion is generically
fixed also, but there are special values of the fluxes that only fix
linear combinations of the dilaton with complex structure moduli
(see section \ref{ss:potential} and \cite{Kachru:2002he,Frey:2002qc}).

\section{BPS States}\label{s:bps}

One of the arguments for establishing dualities in string theory,
particularly strong-weak coupling dualities, has been matching the
Bogomol'nyi-Prasad-Sommerfeld (BPS) spectrum
(for example, \cite{Sen:1994fa}).
Since BPS states 
preserve some supercharges, they form short multiplets of the superalgebra.
Therefore the number of BPS states is protected from quantum
corrections, so dual theories should have matching BPS spectra.  Here
I describe the BPS states of flux compactifications, 
emphasizing the $\N =3$ compactification on $T^6/\Z$, and then count the
number of multiplets of BPS electric charges in the same compactification.
In the next section, we will see how these results relate to dualities
of the self-dual flux compactifications.

\subsection{Description of BPS States}\label{ss:describe}

BPS states, those which leave some supercharges unbroken, arise from
cancellations between mass and central charge terms in the superalgebra.
Since all symmetries in string theory are gauged, any central charges
not broken by the orientifolds or fluxes must correspond to gauge symmetries.
Therefore, we start by finding the massless vectors in the $\N =3$ 
backgrounds (the process would be similar in $\N =2$, which also has
BPS charges).  There are also extended BPS objects and BPS instantons in
the spacetime corresponding to tensorial central charges; these are related
to the moduli dicussed in section \ref{s:moduli}.  The $\N =3$ BPS states
were described in \cite{Frey:2002hf}.

\subsubsection{Gauge fields}\label{sss:gauge}

The bulk vector fields that survive the orientifold
projection (\ref{reflect}) are $C_{\mu n}$ and $B_{\mu n}$.  Form the
complex linear combinations
\begin{equation}
A_{\mu m} = C_{\mu m} - \tau B_{\mu m}\ .
\end{equation}
(Since $\tau$ is frozen, we have $G_{\mu\nu m} = (dA)_{\mu\nu m}$.)
The gauge transformation is
$\delta A_{\mu m} = \partial_\mu \lambda_{A m}$,
where the one-form gauge parameter $\lambda_A$ is
as in equations (\ref{c4var}, \ref{c4varx}).  It follows from the
transformation~(\ref{c4varx}) that the
$(1,0)$ parts of $\lambda_A$ leave the background invariant, so the unbroken
gauge fields are $A_{\mu i}$.  This is also evident from the linearized
gauge field strength (\ref{elhiggs}).
The field $A_{\mu \bar\imath}$ appears in the $\mu \bar\imath jkl $
component, so we see that $A_{\mu \bi}$ is Higgsed by
$\t c_{\bar\imath jkl}$, leaving $A_{\mu i}$ and $\t c_{\bi \bj kl}$
as massless fields.

The real and imaginary parts of $A_{\mu i}$ give six
gauge fields; for example when $\tau = i$, these are
\begin{equation}
\frac{C_{\mu 4} - B_{\mu 7}}{\sqrt{2}},\ 
\frac{B_{\mu 4} + C_{\mu 7}}{\sqrt{2}}, \
\frac{C_{\mu 5} - B_{\mu 8}}{\sqrt{2}}, \ 
\frac{B_{\mu 5} + C_{\mu 8}}{\sqrt{2}}, \ 
\frac{C_{\mu 6} - B_{\mu 9}}{\sqrt{2}}, \ 
\frac{B_{\mu 6} + C_{\mu 9}}{\sqrt{2}}\ . \label{realvects}
\end{equation}
In addition each D3-brane adds a $U(1)$ gauge field, for total gauge group
$U(1)^{6 + N}$.  Without getting into the details of the equations of motion,
we see that the brane gauge groups are unbroken because there are no
scalars charged under them.

To summarize, we have $6+N$ vector fields in the $\N =3$ backgrounds.
The total number of moduli is nine from the metric, nine
from $\tilde c_{4}$, and $6N$ from the D3-branes, for $6(3+N)$ in all.
The counting matches $\N = 3$ supergravity with $n=3+N$ matter
multiplets; note that this agreement requires exactly six of the $U(1)$s
from the bulk SUGRA to be broken.

\subsubsection{Electric and Magnetic Charges}\label{sss:charges}

The electric and magnetic BPS charges are easy to read from the unbroken
vectors (as in equation (\ref{realvects})) because a subset of the 
vectors correspond to central charges in the 10D string theory.  It is 
useful to see first how these descend from the well-known BPS charges of
the $\N = 4$ heterotic theory.

\paragraph{BPS States from Heterotic}\label{p:bpshet}

In the heterotic description,
the BPS electric charges are KK states and winding F-strings.  In the type I
description these become KK states and winding D-strings, and then in type
IIB they become winding F-strings and D5-branes.  The magnetic charges
are KK monopoles and NS5-branes, which are KK monopoles and D5-branes in 
type I and NS5-branes and D-strings in type IIB.

The F-string
is obtained by orientifold projection of closed strings, so it wraps a
complete circle of the torus (see \ref{p:singularity} below for more
details); the NS5-brane is simply the $\Z$
reduction of an NS5-brane solution at $x^4 = 0$ on the type I $T^6$.
The D-string also wraps a full circle, while the D5-brane, like the NS5-brane,
wraps a ``half-cycle.''
For the D1- and D5-branes, these statements are $T$-dual to the fact that
in the type I string the D5-brane has two Chan-Paton values while the
D1-brane has one \cite{Witten:1996gx,Gimon:1996rq}: thus, the IIB D1-brane
can move off the fixed plane, while the D5-brane is fixed.

We can see a reversal of the electric-magnetic status of some of the charges
between the heterotic and IIB descriptions.  For example, the F-strings
of the heterotic theory are electric charges, but they map to D5-branes,
magnetic charges of the $C_{\mu m}$ vectors.  This fact will play an
interesting role in our discussion of the U-dualities, 
section \ref{ss:udual}.

\paragraph{Electric Charges}\label{p:bpselectric}

The $\N =3$ BPS electric charges turn out to be just the $\N =4$ states,
although they carry different central charges.
Note that these do not have a perturbative description, because $g_s$ 
is of order one, but we can study them using the effective low energy
description when the radii are large.  In the $\N = 4$ theory, these
states are invariant under eight supersymmetries; one finds that four of
these supersymmetries lie in the $\N=3$ subalgebra of interest.
Thus  these are ``$1/3$-BPS'' states, in agreement with the result that BPS
particles in $\N=3$ preserve four supersymmetries
\cite{Kounnas:1998hi}.  That these charges preserve four supercharges
is explcitly verified in \ref{sss:susycharges} below.

When the torus is rectangular, the R-R backgrounds
zero, and all D3-brane coincident, the central charges are from the bulk
$U(1)$s $A_{\mu i}$.  For simplicity let us focus on the case that
$g_s = 1$.  The unbroken gauge fields associated with the 4-7
torus are
\be
\frac{B_{\mu 4} + C_{\mu 7}}{\sqrt{2}},\
\frac{C_{\mu 4} - B_{\mu 7}}{\sqrt{2}}\ , \label{unbroken}
\ee
while the broken symmetries are
\be
\frac{B_{\mu 4} - C_{\mu 7}}{\sqrt{2}},\
\frac{C_{\mu 4} + B_{\mu 7}}{\sqrt{2}}\ .
\ee
Thus an F-string in the 4-direction, or a D-string in the 7-direction,
have the same BPS charge, electric charge in the first $U(1)$.
Note that even though these states carry the same charge, they are
distinct states.  Similarly, the charges of the second $U(1)$ are
carried by a D-string in the 4-direction or an F-string wrapped on the
7-direction with the opposite orientation.  See figure \ref{f:bpsstrings}
for a visual example.

\begin{figure}
\centering
\includegraphics{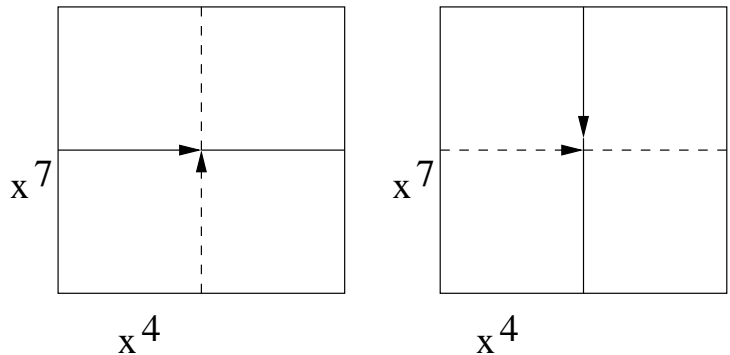}
\caption[Orientation of BPS Strings]{\label{f:bpsstrings} The electric
BPS charges for the two $U(1)$s of equation (\ref{unbroken}).  F-strings
are solid lines, D-strings are dashed.  The first $U(1)$ is the torus
depicted to the left and the second to the right.  The arrows show
the string orientations.}
\end{figure}

\paragraph{Magnetic Charges}\label{p:bpsmagnetic}

Consider again, for example, the $U(1)$s generated by the vectors 
(\ref{unbroken}).  A D5-brane
in the 56789-directions, and an NS5-brane in the 89456-directions, carry
the analogous magnetic charge.\footnote{More generally we can consider
$(p,q)$-strings and 5-branes, at various angles -- a full accounting of
the BPS states is an interesting exercise.}

There is, however, an important subtlety: not all of these states
actually appear in the spectrum.  Each of these objects couples both to a
massless and a massive vector.  The discussion of equation (\ref{elhiggs})
shows that the vector mass arises from electric Higgsing.
For the electrically charged 1-branes the massive charge is screened and
there is no great effect.  However, the 5-branes carry the corresponding
magnetic charge and so must be confined: the Higgsing breaks the symmetry
between these two sets of states.  We can understand this in two other
ways as well.  First, the Higgsing reduces the long-ranged interaction
between the electric and magnetic objects by a factor of two.  Since they
had the minimum relative Dirac quantum in the $\N = 4$ theory, the are no
longer correctly quantized.  Second, the gauge invariant flux on the
D5-brane is
$\mathcal{F}_{2}=F_{2}-B_{2}/2\pi\ap$, which satisfies
\be
d\mathcal{F}_{2} = -\frac{1}{2\pi\ap}H_{3}. \label{dfish}
\ee
The integral of this over any
3-cycle should then vanish, but this is inconsistent because our background
includes at least one of $H_{678}$ or $H_{567}$, among others.
In order that the Bianchi identity be consistent, there must be another
source.  This would be a D3-brane, which is localized in the 3-cycle in
question and extended in the other two compact directions and one
noncompact direction: this is a confining flux tube.

The confinement suggests that there should be a bound state of magnetic
charges that is uncharged under the broken gauge group.  Indeed, a
56789 D5-brane and a 89456 NS5-brane have the same BPS $U(1)$ charge and
the opposite broken charge, and so their bound state is unconfined and is
a BPS state of twice the minimum $\N = 4$ mass.  In a perturbative
description, the D5-brane ends twice on the NS5-brane, as in the
$(p,q)$-5-brane webs of \cite{Aharony:1997ju,Aharony:1998bh}.

\paragraph{BPS Instantons and Cosmic Strings}\label{p:bpsinstanton}

There are two other classes of BPS objects, those that couple to the
R-R moduli $\beta^{\bj i}$.  The first are Euclidean
D3-branes wrapped entirely on the internal torus.  These are instantons
under the unbroken gauge symmetries, and their phases depend on the R-R
moduli.  In field theory, Abelian gauge groups do not have instantons,
so these stringy instantons suggest enhanced symmetries.  Also, these
D3-brane instantons have been discussed before, and they are dual in the
$\N=4$ theory to heterotic worldsheet instantons (see, for example,
\cite{Becker:1995kb,Witten:1996bn,Harvey:1999as,Witten:1999eg}). 
The magnetic analogs to these are spacetime strings,
D3-branes wrapped on the appropriate 2-cycles of the torus and
extended in one direction of the external space; we have already
encountered these above, as confining flux tubes.  
Note that
the instantons wrap enough directions for the identity (\ref{dfish}) to be
relevant, so their spectrum will be subject to restrictions.

There are two physically distinct cases of these instantons and strings.
The simpler case couple to the diagonal $\beta^{\bi i}\ (i=\bi)$ moduli,
as these moduli correspond to a single real component of $\tilde c_{4}$.
For example, $\beta^{\b 1 1}$ couples to an instantonic D3-brane wrapped
on the $5689$ directions and a string D3-brane partially wrapped on
the $47$ directions.  Notice that these instantons do not wrap any 3-cycle
including $H_{3}$ or $F_{3}$ flux.  Additionally, it is easy to check that
these strings preserve supersymmetry; in fact, they preserve 6 supercharges
in common with the background.  The other case correspond to the off-diagonal
moduli, which has real and imaginary parts constructed from two
components of $\tilde c_{4}$ each.  The instantons do wrap 3-cycles
with flux, so they must come in bound states, much as the magnetic BPS
charges discussed above, and the corresponding strings have half as many 
states.  These strings preserve four
supersymmetries in common with the background.

\subsection{Counting of Electric Charges}\label{ss:counting}

Perhaps the most basic of BPS states are the electric charges, and we
have seen that these states are (albeit strongly coupled) F-strings and
D-strings.  I will show here that it is indeed sensible to treat the
electric charges as strings and give their wavefunctions in the fluxes.
These results were presented in \cite{Grana:2002ti}.

Since the electric charges can be either winding D-strings or F-strings, 
our strategy will be to start by finding the $\kappa$-symmetric action
for a D-string wrapped in the $x^4$ direction (without loss of generality).
Then we will find all the supersymmetric states for the D-string;
we obtain the counting for F-string states by S-duality.  It is most
convenient to make the proper periodicities explicit in the coordinate
periodicities $x\simeq x+2\pi R$ with unit metric.

\subsubsection{Example of particle on $T^2$}\label{sss:particle}

The counting of the spectrum of BPS states for the D-string on the
$T^6/\Z$ orientifold with constant fluxes will turn out to be
very similar to the case of spin 1/2 particle on a torus subject to a
constant perpendicular magnetic field, so in this section we review
the latter, simpler, case (\textit{cf.} 
\cite{Fradkin:1991nr} and references therein).  
We work in the context of supersymmetric
quantum mechanics and calculate a Witten index \cite{Witten:1982df}.

The supersymmetric Lagrangian for a spin 1/2 particle with mass $m$
and charge $e$ is 
\be {\mathcal L}= \frac{m}{2}\dot{\vec{x}}^2 + e
\vec{A} \cdot \dot{\vec{x}} + \frac{i}{2} \psi_1 \dot{\psi_1}+
\frac{i}{2} \psi_2 \dot{\psi_2} - i\frac{e}{m} \psi_1 \psi_2 B
\label{lagr}
\ee where $\psi_1$ and $\psi_2$ are two real Grassman variables,
obeying \be\label{anticom} \{\psi_i,\psi_j\}= \delta_{ij}\ . \ee
We should note here that we are using conventions such that 
$(\psi\phi)^\dagger=\phi^\dagger\psi^\dagger$, leading to the real
anticommutator (\ref{anticom}).\footnote{The anticommutator follows from
the theory of systems with constraints, as detailed in
\cite{Casalbuoni:1976tz}; this is important for the D-string, since it
gives constants in the anticommutator.  Note especially that the Dirac 
quantization does not give a factor of $1/2$ in the anticommutator.}
The Lagrangian (\ref{lagr}) is invariant under the 
supersymmetry transformation \be \delta x^i=i \psi_i \epsilon \qquad
\delta \psi_i= -m \dot{x}^i \epsilon
\label{susytransf}
\ee
up to a total derivative, as can be verified by a simple calculation.

We find it convenient to work in the operator formalism, with
Hamiltonian \be\label{ham} H=
\frac{1}{2m}\left(\vec{p}-e\vec{A}\right)^2 +  i\frac{e}{m} \psi_1
\psi_2 B\ . \ee
Then the  supersymmetry (\ref{susytransf}) is generated by the supercharge
operator \be Q=i\left(\vec{p}-e\vec{A}\right)\cdot
\psi
\label{supercharge}
\ee satisfying $Q^2=mH$.

In terms of the complex combinations 
\be z=\frac{1}{2} (x^1 + i x^2)\ ,\qquad
\psi_{\pm}=\frac{1}{2}(\psi_1 \pm i \psi_2)\ , \label{comp}\ee 
the supercharge
(\ref{supercharge}) can be rewritten as \be Q= i(p-eA)_z \psi_+ +i 
(p-eA)_{\bar{z}} \psi_-
\ .\label{superz}
\ee
Supersymmetric ground states should be annihilated by the
supercharge. If we start with a state $\ket{-}$ that is annihilated by
$\psi_-$, then, from (\ref{superz}), supersymmetric wave functions 
obey 
\be (p - eA)_z \phi_- = 0
\ .\label{eqphi}
\ee

As discussed in section \ref{sss:fluxquant}, the magnetic field is quantized
due to single-valuedness of the particle wavefunction, and the
quantization is given in equation (\ref{Bquant}).   Again,
if there is a $\Z$ orbifold, even or odd quanta 
are still allowed, but some
of the fixed points will carry flux, also.  These special fixed points 
will not affect our discussion below, since the translations and reflections
automatically give the wavefunction the correct boundary conditions at
those fixed points.

From (\ref{eqphi}), the wave function should satisfy \be
\frac{\partial \phi_-}{\partial z}= eB \bar{z}\phi_- \ee whose
solution is \be \phi_-=e^{eB |z|^2} F(\bar{z})
\ .\label{sol1}
\ee The periodicity conditions (\ref{gtran}), written in complex
coordinates, are 
\bea 
\phi(z+\pi R, \bar{z}+ \pi R)&=&e^{e B \pi
(z-\bar{z})}\phi(z,\bar{z}), \nonumber\\
\phi(\bar{z}+i\pi R, \bar{z}-i \pi
R)&=&e^{-i e B \pi (z+\bar{z})}\phi(z,\bar{z})
\ .\label{periodzcond}
\eea
Inserting (\ref{sol1}), we get from the first condition in
(\ref{periodzcond}) 
\be F(\bar{z}+ \pi R)= e^{-eB2 \pi R 
\bar{z}-eB\pi^2R^2}F(\bar{z})=e^{-\frac{n}{R}\bar{z}-\frac{n\pi}{2}}
F(\bar{z})
\ ,\label{periodF}
\ee 
where in the last equality we used the quantization condition
(\ref{Bquant}). Defining 
\be \label{defG} F(\bar{z})= e^{-\frac{n}{\pi
R^2}\bar{z}^2}G(\bar{z})\ , \ee 
the condition (\ref{periodF}) implies
that $G(\bar{z})$ is periodic with period $\pi R$. Writing
$G(\bar{z})$ as a sum of Fourier modes, we get for the wave function
\be \phi_-= \exp\left[\frac{n}{2\pi R^2} |z|^2-\frac{n}{2\pi R^2}\bar{z}^2
\right]
\sum_{-\infty}^{\infty} C_m e^{\frac{2i}{R}m \bar{z}}\ .
\label{finsol}
\ee 
The second periodicity condition in (\ref{periodzcond}) implies
the recursion relation \be\label{recurs-} C_{m+n}=C_m e^{\pi (n+2m)}\ . \ee 
So only $n$ of the coefficients $C_m$ are
free. With this recursion relation, the series in (\ref{finsol}) converges if $n<0$. So, for a
magnetic field in the negative $3$ direction, there are $n$ ground
states.

If instead we had started with a ground state annihilated by $\psi_+$,
then the wave function would have been 
\be \phi_+= \exp\left[-\frac{n}{2\pi R^2} |z|^2+\frac{n}{2\pi R^2}z^2\right] 
\sum_{-\infty}^{\infty} C_m e^{\frac{2i}{R}mz}\ ,
\label{finsol+}
\ee 
and the constraint on the components $C_m$ turns out to be 
\be
C_{m+n}=C_m e^{-\pi (n+2m)}\ , \label{recurs+}\ee 
so in this case the sum in (\ref{finsol+})
converges for positive $n$.

Let's take a minute to note the relation of the wavefunctions to the
well-known theta functions on the torus (see chapter 7 of 
\cite{Polchinski:1998rq} for a review).  For example, with $n>0$, the
recursion relation (\ref{recurs+}) has solution $C_m = D e^{-\pi m^2/n}$
with constant $D$,
so the wavefunction (\ref{finsol+}) is a Gaussian times the 
theta function
\be\label{thetafunct}
\phi_+=  \sum_{k=0}^{n-1} D_k \exp\left[-\frac{n}{2\pi
R^2} |z|^2+\frac{n}{2\pi R^2}z^2\right]
\vartheta\left[\begin{array}{c}k/n\\0\end{array}\right] 
\left(\frac{nz}{\pi R},
in\right) \ .\ee
(We use the notation of \cite{Polchinski:1998rq}.)  The sum now has $n$
independent coefficients $D_k$.

A spin $1/2$ particle in a constant perpendicular magnetic field on a
torus has then a finite number of ground states, given by the number
of units of magnetic flux.  On a $\Z$ orbifold, we must have $z\simeq -z$,
or $C_m=C_{-m}$, which force relationships between the $D_k$ coefficients.  
Thus the number of supersymmetric ground states becomes 
$(n+1)/2$ for $n$ odd and $n/2 +1$ for $n$ even.\footnote{It seems naively
that the $T^2/\Z$ orbifold can be smoothly deformed to $S^2$, but in that
case an even number of flux units would give $n/2$ states.  Apparently
the orbifold limit of the sphere is singular; we leave the resolution of
this problem as an exercise for the reader.}
If we think of the states
$\ket{\pm}$ as respectively bosonic and fermionic, the number of ground 
states is therefore the Witten index $\tr (-1)^F$.

\subsubsection{D-string Quantum Mechanics}\label{sss:dqm}
In this section, we find the Lagrangian and Hamiltonian for a D-string
extended along one direction of a $T^6/\Z$ orientifold in imaginary 
self-dual 3-form flux, including the fermionic degrees of freedom.  
For simplicity, we will work with the $\N =3$ backgrounds given in
section \ref{sss:n3sol}, equation (\ref{n3fluxes}), 
although the analysis would be very
similar for an $\N = 2$ background.  Also for simplicity, we work with
the R-R scalar $C=0$.  Without loss of generality, we 
will consider that the D-string winds the $x^4$ direction.

Because the number of BPS states is stable under small
perturbations (in the 4D $\N =3$ case we consider, for example, a long
multiplet is 4 times the size of the BPS multiplet), we will work in the 
large radius limit of the compactification and ignore the warp factor.  
Because we are studying only one type
of electric charge in the 4D theory, the D-string will not be wound in
any other direction or carry any dissolved F-strings.
Also, because motion along the string is quantized by the compactification,
we will set those derivatives to zero.  Finally, D3-branes in the 
compactification appear as singularities on the torus; we ignore D3-branes
and associated 1-3 strings here and explain how to treat D3-branes
in section \ref{sss:extension}.
We will estimate the corrections to
our simple model 
in appendix \ref{aa:assumptions}.

The supersymmetric action for a D1-brane in background fluxes was
worked out  in \cite{Cederwall:1997ri,Bergshoeff:1997tu}. It looks like the
bosonic Dirac-Born-Infeld and Wess-Zumino actions, but the spacetime
fields live in superspace.
\begin{equation}
S=-\frac{1}{2\pi\ap}\int d^2\zeta e^{- \bm{\phi}}\sqrt{-\det \left(
\bm{g}_{\alpha\beta} 
+ \bm{\mathcal{F}}_{\alpha\beta}\right)}+\frac{1}{2\pi\ap}\int 
e^{\bm{\mathcal{F}}} \wedge\bm{C}\ .
\label{susyaction}
\end{equation}
The fields in boldface are superfields; as usual,
\begin{equation}
\bm{\mathcal{F}}=2\pi\ap F-\bm{B}\ ,\ \textnormal{and}\ 
\bm{C}=\oplus_n \bm{C}_{(n)}
\label{somedefs}\end{equation}
is the collection of all RR potentials pulled back to the
world-volume.  The expansions of the superfields in terms of
components fields was developed in \cite{deWit:1998tk}, \cite{Millar:2000ib}, 
and \cite{Grana:2002tu} for 11-dimensional, IIA, and IIB  supergravities
respectively, using a method known as gauge completion. The expansions
of the fields that we will need, as well as our conventions, are listed
in appendix \ref{aa:kappaexpand}. 

Without getting into algebraic details, the action (\ref{susyaction}) 
for a D-string in our background is
\bea
S&=&-\frac{1}{2\pi \ap g_s}\int d^2\zeta \left[ \left(1-\mathcal{F}_{04}^2
\right)^{1/2}-\frac{1}{2}  \left(1-\mathcal{F}_{04}^2\right)^{-1/2}
(\dot{x}^m)^2-g_s C_{m4}\dot{x}^m\right.\nonumber\\
&&\left.+i\frac{g_s^{1/2}}{2} \ot \Gamma^0 \dot{\Theta} 
+i\frac{g_s^{3/2}}{16} \ot \Gamma_0{}^{mn} \Theta F_{4mn} 
+i\frac{g_s^{1/2}}{48} \ot \Gamma^{mnp} \Theta H_{mnp}\right.\nonumber\\
&&\left. -i\frac{g_s^{1/2}}{16}  
\ot \Gamma^{4mn} \Theta H_{4mn}\right]\ ,
\label{action1}
\eea
where we work to second order in the world-volume coordinates and 
fermions\footnote{$F_{4mn}$ in the second fermionic term would be 
$F'_{4mn} = F_{4mn}-CH_{4mn}$, and also the Chern-Simons term would couple
the velocity $\dot{x}^m$ to $C_{m4}+C\mathcal{F}_{m4}$, 
if the R-R scalar were nonvanishing.}.
As noted above, there are corrections to this action, arising from the 
expansion of the D-brane action; we consider these
in appendix \ref{aa:assumptions}.

As discussed in appendix \ref{aa:kappaexpand}, the fermion $\Theta$ is a
10D (Majorana-Weyl) superspace coordinate; 
let us now expand it in terms of 2D spinors.
We do so by noting that
the 10-dimensional gamma matrices can be decomposed into  
$SO(1,1)\otimes SO(8)$ pieces as
\be\label{gammas}
\Gamma^{\parallel}=\gamma^{\parallel}\otimes 1 , \qquad \Gamma^{\perp}= 
\gamma_{(\hat 2)}\otimes \gamma^{\perp}\ ,
\ee
where ``$\parallel$'' and ``$\perp$'' mean along the D-string and
perpendicular to it; $\Gamma$ is a 32x32 Dirac matrix,
$\gamma^{\parallel}$ and $\gamma^{\perp}$ are its $2\times 2$ and 
$16\times 16$
blocks, and $\gamma_{(\hat 2)}$ is the chirality matrix in $SO(1,1)$.
Therefore, a  Majorana-Weyl spinor $\Theta$ can be decomposed into 
\be
\Theta= \ket- \otimes \psi^{a} u_{a} \oplus \ket+ \otimes 
\phi^{\dot{a}} v_{\dot a}
\ ,\label{decomp}
\ee
where $\ket{+}$ and $\ket{-}$ are the eigenfunctions of $\gamma_{(\hat 2)}$, 
and $a$ and 
$\dot a$ 
are indices in the ${\bf 8}$ and ${\bf 8'}$ representations of $SO(8)$
(see appendix \ref{aa:spinors}).
The $\psi$s and $\phi$s are 2D Majorana-Weyl fermions. 
The spin raising (lowering) gamma matrices are the (anti)holomorphic
gamma matrices defined with respect to the coordinates
\bea
y^{0\pm}=\frac{1}{2} (\pm x^0+x^4)&,& y^{1}=\frac{1}{2} 
(x^1+ ix^2)\ , \quad y^{2}=\frac{1}{2} (x^3+ ix^7)\ ,\nonumber\\
\quad y^{3}=\frac{1}{2} (x^5+ ix^8)&,&y^{4}=
\frac{1}{2} (x^6+ ix^9)\ .\label{zcomplex}
\eea
As usual, I label complex coordinates with indices $i,j,\ldots$.

Using the basis (\ref{so8basis}) for the Majorana-Weyl spinors 
and integrating along the string, 
the fermionic Lagrangian from eq (\ref{action1}) can be written
\bea
L_f &=& - i\frac{R_1}{2\ap g_s^{1/2}}\left[
\psi^{a}\dot{\psi}^{a} +  \phi^{\dot a}\dot{\phi}^{\dot a}
+ \frac{g_s\ap}{2\pi R_1R_2R_3}f_2
\left(\psi^1\psi^4+\psi^2\psi^3+\phi^4\phi^1+\phi^3\phi^2\right.\right.
\nonumber\\
&&\left.\left.+\psi^1\phi^3+
\phi^4\psi^2+\phi^1 \psi^3+\psi^4\phi^2\right)-\frac{g_s\ap}{2\pi R_1R_2R_3}
f_1\left(\psi^3\psi^1+\psi^2\psi^4+\phi^1\phi^3\right.\right.\nonumber\\
&&\left.\left.+\phi^4\phi^2
+\psi^1\phi4+\psi^2\phi^3+\phi^2\psi^3+\phi^1\psi^4\right)\right]\ .
\label{fermions}\eea
As well as some algebraic simplification, arriving at 
this result requires 3-form self-duality to relate the NS-NS and R-R 
fluxes as in (\ref{imsd}).
We should note that the fermions $\psi^a$ and $\phi^{\dot a}$ with 
$a,\dot a =5,6,7,8$ enter only through their kinetic terms.

Now to convert to the Hamiltonian formalism, we start by finding the canonical
momentum for the world-volume gauge field $F=dA$.  Following the discussion
in \cite{Witten:1982df}, the Wilson lines are periodic variables, so the
momentum 
\be\label{gaugemomentum}
p_A = \frac{2\pi R_1}{g_s}
\frac{\mathcal{F}_{04}}{[1-\mathcal{F}_{04}^2]^{1/2}}\ee
is quantized in units of $2\pi R_1$.  Further, it is this canonical 
momentum (up to constants) which couples the D-string to $B$, so $p_A$
and therefore $\mathcal{F}$,
\emph{not} the gauge field strength $F$, vanish for a D-string with no
F-string charge.  This issue is somewhat more complicated in the presence
of the R-R scalar $C$, but we have now removed the NS-NS flux from the 
problem.

The canonical momenta for the collective coordinates now simplify to
\be\label{momenta}
p_m = \frac{R_1}{\ap g_s}\dot x_m +\frac{R_1}{\ap} C_{m4}\ , \ee
as in the usual quantum mechanics with a gauge field 
proportional to $C_{m4}$.  Thus, the total Hamiltonian is a
constant mass $m=R_1/\ap g_s$ and a dynamical Hamiltonian
\bea
H&=& \frac{1}{2m}(\vec{p}-\vec{A})^2 + i\frac{C}{2}f_2
\left(\psi^1\psi^4+\psi^2\psi^3+\phi^4\phi^1+\phi^3\phi^2\right.\nonumber\\
&&\left.+\psi^1\phi^3+
\phi^4\psi^2+\phi^1 \psi^3+\psi^4\phi^2\right) - i\frac{C}{2}f_1
\left(\psi^3\psi^1+\psi^2\psi^4+\phi^1\phi^3\right.\nonumber\\
&&\left.+\phi^4\phi^2
+\psi^1\phi4+\psi^2\phi^3+\phi^2\psi^3+\phi^1\psi^4\right)\label{hamil}
\eea
with $C=g_s^{1/2}/(2\pi R_2R_3)$.  The gauge field is defined as
$A_m = (R_1/\ap)C_{m4}$.

\subsubsection{Supercharges}\label{sss:susycharges}

Now we demonstrate that the Hamiltonian (\ref{hamil}) is, in
fact, supersymmetric, and we identify the 4 supercharges that belong
to the unbroken supersymmetries of the $\N =3$ 4D effective theory.  We 
will proceed by first finding the spacetime supersymmetries that leave both
the background (see section \ref{sss:n3sol}) and the BPS states of the string
invariant by starting with the 10D theory.  
We will then relate those to world-volume supercharges,
which should be of the form given in eq. (\ref{supercharge}), 
$Q\sim i(p-A)\psi$.  Finally, we will check that these do actually commute 
with the Hamiltonian.

To start, recall the discussion of section \ref{ss:susy} that described the
spinor parameters of the supersymmetries preserved by the ISD flux
backgrounds.  The key result was that the SUSY spinors satisfy 
$\varepsilon_1 = i\gamma_{(\hat 4)} \varepsilon_2$, 
where $\gamma_{(\hat 4)}$
is the chirality on the noncompact spacetime.  

In fact, the supersymmetries preserved by the $\N =3$ background 
(\ref{n3fluxes}) 
are expressed conveniently by $SO(3,1)\otimes SO(6)$ decomposition 
$\varepsilon_1^A = \zeta\otimes \chi^A+\zeta^*\otimes \chi^{A*}$, 
$\varepsilon_2^A = i\zeta\otimes \chi^A-i\zeta^*\otimes \chi^{A*}$, 
where $\chi^A$ are 3 of the 4 negative
chirality spinors in 6D (those without all three spins parallel)  
\cite{Frey:2002hf}.  
We can re-write these in the $SO(1,1)\otimes SO(8)$ basis
(\ref{so8basis}) as
\bea
\varepsilon^1&=& \epsilon^1_1 \left[ \ket{-}(u_1-iu_2)
-\ket{+}(v_1-iv_2)\right] +\epsilon^1_3 \left[ \ket{-}
(u_3-iu_4)+\ket{+} (v_3-iv_4)\right]\ ,\nonumber\\
\varepsilon^2&=& \epsilon^2_5 \left[ \ket{-} (u_5-iu_6)
-\ket{+} (v_5-iv_6)\right] +\epsilon^2_7 \left[ \ket{-}
(u_7+iu_8)+\ket{+} (v_7+iv_8)\right]\ ,\nonumber\\
\varepsilon^3&=& \epsilon^3_5 \left[ \ket{-} (u_5+iu_6)
+\ket{+} (v_5+iv_6)\right]
+\epsilon^3_7 \left[ \ket{-} (u_7-iu_8)\right.\nonumber\\
&&\left.-\ket{+} (v_7-iv_8)\right]\qquad\label{n3susy}
\eea
with complex Grassman numbers $\epsilon^A_a$.  To save space,
we have written $\varepsilon^A=\varepsilon^A_1-i\varepsilon^A_2$.
Please note that $\varepsilon^A$ are spinors, while
$\epsilon^A_{a}$ are their components.

Now, let's intersect these with the supersymmetries preserved by the 
D-string. Using the superalgebra with central charges, it is easy to 
find that a
D-string wrapped on $x^4$ has supersymmetries $\varepsilon_1=\gamma_{(\hat 2)}
\varepsilon_2$ \cite{Polchinski:1998rr} ($\gamma_{(\hat 2)}$ is the 2D 
worldsheet chirality).  We can obtain spinors that 
satisfy this constraint by taking linear combinations of the latter two
spinors (\ref{n3susy}) such that $\epsilon^2_5=\pm\epsilon^3_5$ and
$\epsilon^2_7=\pm\epsilon^3_7$.  In the end, we find 4 different one 
component spinors $\varepsilon^A_1$ with
\bea
\varepsilon^1_1=\epsilon^1(\ket{-}u_5-\ket{+} v_6)&,&
\varepsilon^2_1=\epsilon^2(\ket{-} u_6+\ket{+} v_5)\ ,
\nonumber\\
\varepsilon^3_1=\epsilon^1(\ket{-}u_7-\ket{+} v_8)&,&
\varepsilon^4_1=\epsilon^4(\ket{-}u_8+\ket{+}v_7)\ .
\label{n3-d4susy}\eea
The coefficients $\epsilon^A$ are now real Grassman numbers.  The spinors
$\varepsilon^A_2$ are just given by the relation for D-strings.

Now we can actually find the worldvolume supercharges.  As discussed in
\cite{Bergshoeff:1997kr,Kallosh:1998ky}, the worldvolume 
supersymmetries are not given simply by their 
action on the spacetime fields (including $\Theta$ as a superspace coordinate)
because that transformation would in general change the $\kappa$-symmetry
gauge.  The supersymmetry transformations of the worldvolume fields,
including a $\kappa$ transformation to keep the same gauge, were found
in \cite{Kallosh:1998ky}.  However, we will not follow this approach.  
Instead, we will take the ansatz
\be\label{Qansatz}
-\frac{1}{\sqrt{2}} Q^A\epsilon^A = i(p-A)_i \b\Theta \Gamma^i 
\varepsilon^A_1
+i(p-A)_{\bi} \b\Theta \Gamma^{\bi}\varepsilon^A_1
\ee
for the supercharges.  We follow 
\cite{Bergshoeff:1997kr} in using the spacetime supersymmetry
parameters, and we know to use the $\varepsilon_1$ because of our
$\kappa$-symmetry gauging, equation (\ref{kappastring}).  
The prefactors are included for convenience.

Using this ansatz, we can write the supercharges in terms of the 2D fermions
$\psi,\phi$ and complex coordinates $y$ as
\bea
Q^1&=& ip_1 (-i\psi_7-\psi_8+\phi_7-i\phi_8)
+i p_{\b 1} (i\psi_7-\psi_8+\phi_7+
i\phi_8)\nonumber\\
&&+ip_2(-i\psi_5+\psi_6-\phi_5+i\phi_6)
+ip_{\b 2}(i\psi_5+\psi_6-\phi_5-i\phi_6)\nonumber\\
&&+i(p-A)_3(-i\psi_3-\psi_4-\phi_3
+i\phi_4)+i(p-A)_{\b 3}(i\psi_3-\psi_4-\phi_3-i\phi_4)\nonumber\\
&&+i(p-A)_4(i\psi_1+\psi_2+\phi_1
-i\phi_2)+i(p-A)_{\b 4}(-i\psi_1+\psi_2+\phi_1+i\phi_2)\ ,\nonumber\\
Q^2&=& ip_1 (-\psi_7+i\psi_8-i\phi_7-\phi_8)+ip_{\b 1} 
(-\psi_7-i\psi_8+i\phi_7-\phi_8)\nonumber\\
&&+ip_2(-\psi_5-i\psi_6-i\phi_5-\phi_6)+ip_{\b 2}
(-\psi_5+i\psi_6+i\phi_5-\phi_6)\nonumber\\
&&+i(p-A)_3(-\psi_3+i\psi_4+i\phi_3
+\phi_4)+i(p-A)_{\b 3}(-\psi_3-i\psi_4-i\phi_3+\phi_4)\nonumber\\
&&+i(p-A)_4(-\psi_1+i\psi_2+i\phi_1
+\phi_2)+i(p-A)_{\b 4}(-\psi_1-i\psi_2-i\phi_1+\phi_2)\ ,\nonumber\\
Q^3&=& ip_1 (i\psi_5+\psi_6-\phi_5+i\phi_6)+ip_{\b 1} (-i\psi_5+\psi_6-\phi_5-
i\phi_6)\nonumber\\
&&+ip_2(-i\psi_7+\psi_8-\phi_7+i\phi_8)
+ip_{\b 2}(i\psi_5+\psi_8-\phi_7-i\phi_8)\nonumber\\
&&+i(p-A)_3(-i\psi_1-\psi_2-\phi_1
+i\phi_2)+i(p-A)_{\b 3}(i\psi_1-\psi_2-\phi_1-i\phi_2)\nonumber\\
&&+i(p-A)_4(-i\psi_3-\psi_4-\phi_3
+i\phi_4)+i(p-A)_{\b 4}(i\psi_3-\psi_4-\phi_3-i\phi_4)\ ,\nonumber\\
Q^4&=& ip_1 (\psi_5-i\psi_6+i\phi_5+\phi_6)+ip_{\b 1} (\psi_5+i\psi_6-i\phi_5
+\phi_6)\nonumber\\
&&+ip_2(-\psi_7-i\psi_8-i\phi_4-\phi_8)+ip_{\b 2}
(-\psi_7+i\psi_8+i\phi_7-\phi_8)\nonumber\\
&&+i(p-A)_3(\psi_1-i\psi_2-i\phi_1
-\phi_2)+i(p-A)_{\b 3}(\psi_1+i\psi_2+i\phi_1-\phi_2)\nonumber\\
&&+i(p-A)_4(-\psi_3+i\psi_4+i\phi_3
+\phi_4)+i(p-A)_{\b 4}\left(-\psi_3-i\psi_4\right.\nonumber\\
&&\left.-i\phi_3+\phi_4\right)\ .\label{wvsusy}
\eea
It is a straightforward but tedious calculation to show that each of these
commute with the Hamiltonian (\ref{hamil}).  We need to note that canonical
quantization gives the anticommutators $\{\psi_a,\psi_b\}=
\{\phi_{\dot a},\phi_{\dot b}\}=\delta_{ab}/(mg_s^{1/2})$
\cite{Casalbuoni:1976tz}.
The magnetic field $F=dA$ (where $A$ was defined below equation
(\ref{hamil})) 
in the $y$ coordinates is also necessary:
\be\label{zmag}
F_{34}=-\frac{1}{\pi R_2R_3 }(f_1+if_2)\ ,\qquad
F_{\b 3\b 4}=-\frac{1}{\pi R_2R_3 }(-f_1+if_2)\ .\ee
This arises in $[Q,H]$ from commutators $[p,A]$; there are no commutators
mixing holomorphic and antiholomorphic indices.

\subsubsection{Supersymmetric Ground States}\label{sss:wavefunction}

To find the states annihilated by the supercharges (\ref{wvsusy}), 
it's easier to 
work in the complex basis
\be
w^1=\frac{1}{2}(\tilde{x}^5+i \tilde{x}^6) , \ w^2=\frac{1}{2}
(\tilde{x}^8+i \tilde{x}^9)\ ,
\label{basis}
\ee
where 
\bea
\tilde{x}^5&=&\frac{1}{\sqrt{2f(f+f_1)}}\left(f_2 x^5+(f+f_1)x^8\right)
\ ,\nonumber\\
\tilde{x}^6&=&\frac{1}{\sqrt{2f(f+f_1)}}\left(f_2 x^6+(f+f_1)x^9\right)
\ ,\nonumber\\
\tilde{x}^8&=&\frac{1}{\sqrt{2f(f+f_1)}}\left( (f+f_1) x^5-f_2x^8\right)
\ ,\nonumber\\
\tilde{x}^9&=&\frac{1}{\sqrt{2}f(f+f_1)}\left((f+f_1) x^6-f_2 x^9\right)
\label{xtilde}
\eea
and $f=\sqrt{f_1^2+f_2^2}$. In the basis (\ref{basis}), the nonzero 
components of the magnetic field are
\be\label{wmagfield}
F_{w^1\bar{w}^1}=-\frac{1}{\pi R_2 R_3}ih\, , \qquad 
F_{w^2\bar{w}^2}=\frac{1}{\pi R_2 R_3}if\ee
We can use a gauge where the potential is
\bea
A_{w^1}=\frac{1}{2 \pi R_2 R_3} if \bar{w}^1 &,& 
A_{\bar{w}^1}=-\frac{1}{2 \pi R_2 R_3}if w^1 \nonumber \\
A_{w^2}=-\frac{1}{2 \pi R_2 R_3} if \bar{w}^2 &,& 
A_{\bar{w}^2}=\frac{1}{2 \pi R_2 R_3}if w^2 \ .
\label{potw}
\eea

The supercharges (\ref{wvsusy}) can be rewritten (up to sign)
\bea
Q^1 &=&(p-A)_{w^1}(\lambda_1-\lambda_4)+ (p-
A)_{\bar{w}^1}(-\bar{\lambda}_1+\bar{\lambda}_4)\nonumber\\
&&+(p-A)_{w^2}(\lambda_2+\lambda_3)+ (p-A)_{\bar{w}^2}
(-\bar{\lambda}_2+\bar{\lambda}_3)+... \nonumber\\
Q^2&=&(p-A)_{w^1}(\bar{\lambda}_2+\bar{\lambda}_3)+ (p
-A)_{\bar{w}^1}(-\lambda_2-\lambda_3)\nonumber\\
&&+(p-A)_{w^2}(-\bar{\lambda}_1+\bar{\lambda}_4)+ (p
-A)_{\bar{w}^2}(\lambda_1-\lambda_4)+... \\
Q^3&=& i(p-A)_{w^1}(\lambda_1+\lambda_4)+ i(p
-A)_{\bar{w}^1}(-\bar{\lambda}_1+\bar{\lambda}_4)\nonumber\\
&&+i(p-A)_{w^2}(-\lambda_2+\lambda_3)+ i(p
-A)_{\bar{w}^2}(-\bar{\lambda}_2-\bar{\lambda}_3)+... \nonumber\\
Q^4&=&i(p-A)_{w^1}(-\bar{\lambda}_2-\bar{\lambda}_3)+ i(p-
A)_{\bar{w}^1}(-\lambda_2-\lambda_3)\nonumber\\
&&+i(p-A)_{w^2}(\bar{\lambda}_1-\bar{\lambda}_4)+ i(p-
A)_{\bar{w}^2}(\lambda_1-\lambda_4)+... 
\label{Qs}
\eea
where $+...$ are terms involving momenta in the noncompact and $x^7$ 
directions, which will give zero when acting on the ground state 
wave-functions. The 
fermions $\lambda_\alpha$ in (\ref{Qs}) are defined in terms of the 
fermions in (\ref{decomp}) as
\bea
\lambda_1&=&\frac{1}{\sqrt{2f(f+f_1)}}\left(f_2 (\psi_2+i\psi_4)+(f+f_1)
(\psi_1+i\psi_3)\right)\nonumber\\
\lambda_2&=&\frac{1}{\sqrt{2f(f+f_1)}}\left((f+f_1) (\psi_2+i\psi_4)-f_2
(\psi_1+i\psi_3)\right)\nonumber\\
\lambda_3&=&\frac{1}{\sqrt{2f(f+f_1)}}\left(f_2 (\phi_2+i\phi_4)+(f+f_1)
(\phi_1+i\phi_3)\right)\nonumber\\
\lambda_4&=&\frac{1}{\sqrt{2f(f+f_1)}}\left((f+f_1) (\phi_2+i\phi_4)-f_2
(\phi_1+i\phi_3)\right)\ .\label{lambdaferms}
\eea
These spinors satisfy the oscillator algebra
\be\label{osc}
\{\lambda_a,\lambda_b\}=\{\bar{\lambda}_a, 
\bar{\lambda_b}\}=0, \ 
\{\lambda_a, \bar{\lambda}_b\}=k \delta_{ab}
\ee
where $k$ is a real constant that can be absorbed in the definition of the 
spinors. So $\lambda$, $\bar{\lambda}$ are raising and lowering operators.

As in the $T^2$ example, to build the wave functions, we start with a state 
that is annihilated by half of the fermionic operators appearing in the 
supercharges (\ref{Qs}). There are only two possible states, $\ket{+}$ and 
$\ket{-}$, satisfying
\bea
\bar{\lambda}_1\ket{+}= \bar{\lambda}_4\ket{+}=\lambda_2\ket{+}=\lambda_3
\ket{+}&=&0,\nonumber\\
\lambda_1\ket{-}= \lambda_4\ket{-}=\bar{\lambda}_2\ket{-}=\bar{\lambda}_3
\ket{-}&=&0\ .
\label{choices}
\eea
It turns out that it is impossible for any other spin choices to preserve
all four supersymmetries; the wave function would have to satisfy 
incompatible differential equations.  From the quantum
mechanics superalgebra, partial supersymmetry breaking is not allowed
\cite{Witten:1982df}, so we have only these two spins.

From (\ref{Qs}), the wave function corresponding to the first state must 
satisfy
\be
(p-A)_{w^1} \phi_+=(p-A)_{\bar{w}^2}\phi_+=0\ .
\ee
Using (\ref{potw}) for the potential, the solution is
\be\label{wave1}
\phi_+=\exp\left[-\frac{f}{2 \pi R_2 R_3}\left(|w^1|^2+|w^2|^2\right)\right] 
F(\bar{w}^1,w^2)\ .
\ee
As in the $T^2$ example, when going around the torus, the wave function picks 
up a phase given by the gauge transformations ({\it cf}. Eqs (\ref{gaugetr}) 
and 
(\ref{gtran})). When $x^5\rightarrow x^5+2\pi R_2$, we get the following 
condition on the function $F(\bar{w}^1,w^2)$
\be\label{Fgauge}
 F\left(\bar{w}^1+\frac{f_2\pi R_2}{\sqrt{2f(f+f_1)}} ,w^2+
\frac{(f+f_1)\pi R_2}{\sqrt{2f(f+f_1)}}\right)
=\exp\left[-f\frac{2v+\pi R_2}{2R_3}\right]\,  F(\bar{w}^1,w^2)
\ ,\ee
where
\be
v=\frac{\left(f_2 \bar{w}^1+(f+f_1) w^2\right)}{\sqrt{2f(f+f_1)}}, 
\ u=\frac{\left((f+f_1)\bar{w}^1-f_2 w^2\right)}{\sqrt{2f(f+f_1)}}
\label{uvdef}
\ee
(we defined $u$ for future use).  Note that $u,v$ bear a striking resemblance
to the original complex coordinates $y$ (modulo some conjugation); this is
because those are the coordinates in which the periodicities of the torus
are simple.

In a similar fashion to the $T^2$ example, we define the function $G$ as
\be
F(\bar{w}^1,w^2)=\exp\left[\frac{f}{2\pi R_2 R_3}
((\bar{w}^1)^2+(w^2)^2)\right] G(\bar{w}^1,w^2)
\ee
which should be periodic when $\bar{w}^1\rightarrow \b w^1+
f_2\pi R_2/\sqrt{2f(f+f_1)}$ and $w^2\rightarrow
(f+f_1)\pi R_2/\sqrt{2f(f+f_1)}$, or, using the variables $u$ and 
$v$ in (\ref{uvdef}), $v\rightarrow v+\pi R_2$ and $u\rightarrow u$. So we 
can decompose $G$ in terms of Fourier modes. Putting  everything together, our 
wave function is
\be
\phi_+=\exp\left[-\frac{f}{2 \pi R_2 R_3}\left(|w^1|^2+|w^2|^2-
(\bar{w}^1)^2-(w^2)^2\right)\right]
\sum_{m} e^{2im\frac{v}{R_2}} g_m(u)
\label{solphi+}
\ee
where $g_m(u)$ is to be determined from the other periodicity conditions. 

When going around the torus in the $x^8$ direction, i.e. when 
$x^8\rightarrow x^8 + 2 \pi R_2$, the condition on the wave function forces 
$g_m(u)$ to have periodicity $\pi R_2$. So our Fourier mode decomposition in 
(\ref{solphi+}) is a double sum, i.e.
\be
\phi_+=\exp\left[-\frac{f}{2 \pi R_2 R_3}\left(|w^1|^2+|w^2|^2-
(\bar{w}^1)^2-(w^2)^2\right)\right]\sum_{m,n}C_{m,n} 
e^{2i(mv+nu)/R_2} \ .
\label{sol2phi+}
\ee
Finally, similar to the $T^2$ case, going around the torus in the $x^6$ 
and $x^9$ gives us two recursive relations for the coefficients $C_{m,n}$, 
of the form
\bea
C_{m-f_1, n+f_2}=C_{m,n} \exp\left[-\pi\frac{R_3}{R_2}\left( 
f+2\frac{(-f_1 m+f_2 n)}{f}\right)\right] \nonumber\\
C_{m+f_2, n+f_1}=C_{m,n} \exp\left[-\pi\frac{R_3}{R_2}\left( 
f+2\frac{(f_2 m+f_1 n)}{f}\right)\right]\ .
\label{recursion}
\eea
The series defined by these recursive relations converges for any sign of 
$f_1$ and $f_2$ and factorizes into theta functions as
\bea
\phi_+&=&\sum_{k,l}D_{k,l}
\exp\left[-\frac{f}{2 \pi R_2 R_3}\left(|w^1|^2+|w^2|^2
-(\bar{w}^1)^2-(w^2)^2\right)\right] \nonumber\\
&&\times
\vartheta\left[\begin{array}{c} \frac{ -f_1 k+f_2 l}{f^2}\\0\end{array}
\right] \left( \frac{-f_1 v+f_2 u}{\pi R_2 },\frac{ifR_3}{2 R_2}
\right) \nonumber\\
&&\times\vartheta\left[\begin{array}{c} \frac{ f_2 k+f_1 l}{f^2}\\0
\end{array}
\right] \left( \frac{f_2 v+f_1 u}{\pi R_2},\frac{ifR_3}{2 R_2}
\right) \label{theta4}\eea
after solving the recursion, where the sum on $k,l$ is over points in the
unit cell for $m,n$ as in (\ref{sol2phi+}) (see figure \ref{f:fund}).

If instead of working with the first choice of ground state in 
(\ref{choices}) we start with the second possibility, we get a similar 
solution to (\ref{sol2phi+}), but the recursive relations are such that the 
series doesn't converge for any sign of $f_1$ and $f_2$.  At first glance,
the fact that the $\ket{+}$ state is normalizable for any magnetic field,
while the $\ket{-}$ is not, appears contradictory with the results for a
particle on $T^2$.  Physically, we expect that, as in the particle case,
a change of sign of the magnetic field would be compensated by a change of
spin.  This is indeed still the case in the current scenario; our 
change of basis (\ref{lambdaferms}) reverses the physical spin of the 
string (given by the $\psi$ and $\phi$ variables) as the field is reversed.

The number of ground states is given by the number of independent 
coefficients. This number is equal to $f^2=f_1^2+f_2^2$, 
as can be seen from figure 
\ref{f:fund}.

\begin{figure}[t]
\centering
\includegraphics[scale=0.7]{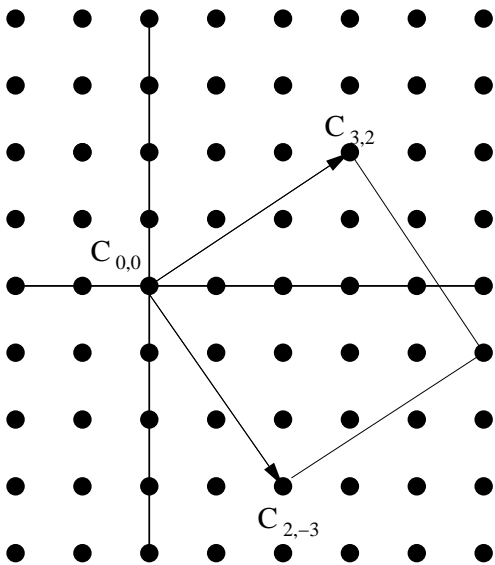}
\caption[BPS State Wavefunctions]{\label{f:fund} Lattice of 
coefficients $C_{m,n}$. For $f_2=2$, 
$f_1=-3$, the independent coefficients are those that lie inside 
the fundamental cell. The number of BPS states in this case is 
$13=f^2$.}
\end{figure}

The orientifold projection forces $C_{m,n}=C_{-m,-n}$. Then, some of the 
elements in the fundamental lattice are related to one another. For the case 
$f_2=2$, $f_1=-3$, we indicate in figure \ref{f:orb} 
all those 
coefficients that are related after the orientifold projection and a 
translation along the lattice basis vectors. In this case, the number of 
independent coefficients is $7$ ($6$ interior points, plus the origin).  

\begin{figure}[t]
\centering
\includegraphics[scale=0.7]{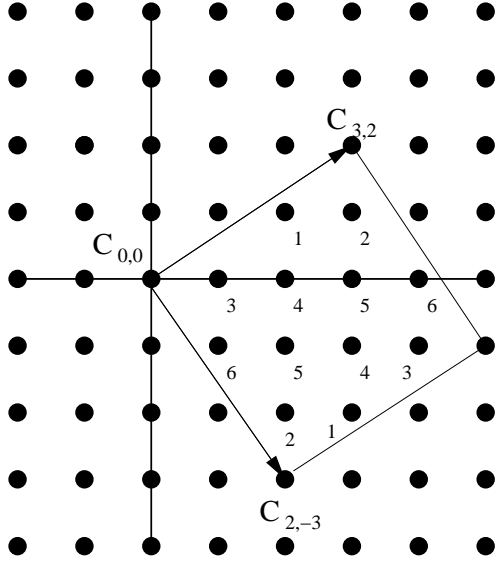}
\caption[BPS State Wavefunctions in Orientifold]{\label{f:orb}  
The lattice points inside the unit cell for
the same case as in figure \ref{f:fund}; numbered points are
identified under the orientifold reflection followed by lattice
translations (equivalent to identifying point related by a reflection
with respect to the center of the lattice). }
\end{figure}

There is no general formula for the number of ground states that
survive after the projection, but nevertheless there are only a finite
number of cases to consider in the string compactification. 
Because 3-form flux carries D3-brane charge,
tadpole cancellation then imposes conditions on this
flux of the form \cite{Frey:2002hf} $g_s f^2\le 8$ 
(see equation (\ref{intbianchi})). 
Besides, the self duality condition on the 3-form flux 
requires $g_s \ge 1/\min f_{1,2}$, leaving only the possibilities
considered in table \ref{t:counting} (the number of ground states for the pair
$\left(f_1, f_2\right)=(1,0)$, for example, is
equal to that for $\left(f_1,f_1\right)=(0,1)$).

\begin{table}[t]
\begin{center}
\begin{tabular}{|c|c|c|}
\hline
$\left(f_1, f_2\right)$ & \# indep. $C_{m,n}$  
&  \# indep. $C_{m,n}$  \\
 & before orientifold & after orientifold\\
\hline
$(0,1)$ & 1 & 1 \\	
$(0,2)$ & 4 & 4 \\
$(0,3)$ & 9 & 5 \\
$(0,4)$ & 16 & 10 \\
$(1,1)$ & 2 & 2 \\
$(1,2)$ & 5 & 3 \\
$(2,3)$ & 13 & 7 \\
$(3,3)$ & 18 & 10 \\
$(4,4)$ & 32 & 17 \\
\hline
\end{tabular}
\end{center}
\caption[Numbers of D-string BPS States]{\label{t:counting} 
Possible combinations of 3-form fluxes, and the number of ground states 
obtained before and after the orientifold projection.}
\end{table}

There is one final issue to consider; in some cases, a point $(m,n)$
maps to itself under the $\Z$ projection followed by several recursion
relations among the Fourier coefficients.  In order to verify that those 
points truly survive the orientifold, we should check that the 
necessary recursion relations truly give $C_{-m,-n}=C_{m,n}$.  First, we note
that, in this case, $\pm (m,n)$ are separated by integer numbers of the
recursion shift vectors:
\be\label{orbshifts}
(m,n) -q(-f_1,f_2)-p(f_2,f_1) = (-m,-n)\ ,\quad
p,q\in \bm{\mathbbm{Z}}\ .\ee
Since the recursion relations (\ref{recursion}) are independent of 
position along the orthogonal shift vector, we have then
\bea
C_{-m,-n} &=&\exp\left\{-\frac{\pi R_3}{R_2} \left[(p+q)f
+\frac{2f_1}{f} \sum_{i=0}^{q-1} 
\left( m+if_1\right)
-\frac{2f_2}{f} \sum_{i=0}^{q-1} 
\left( n-if_2\right)\right.\right.\nonumber\\
&&\left.\left.-\frac{2f_2}{f} \sum_{j=0}^{p-1} 
\left( m-jf_2\right)
-\frac{2f_1}{f} \sum_{j=0}^{p-1} 
\left( n-jf_1\right)\right]\right\} C_{m,n}\ .\label{check}
\eea
Carrying out the arithmetic sums and using (\ref{orbshifts}) shows
easily that the exponent indeed vanishes.

\subsubsection{Other BPS States}\label{sss:extension}

There are other BPS states that we can consider.  In this section, we 
will find BPS states localized at singularities of the D-string moduli
space, show how to count the states of F-strings, and discuss briefly
the problem of threshhold bound states.

\paragraph{Singularities in Moduli Space}\label{p:singularity}
As the D-string moves on the torus, it encounters two types of singularities.
First, at $x^5=\cdots =x^9=0$, it passes through two orientifold planes
and doubles back on itself.  (Note that the D-string does not end on an 
O3-plane; this is because, with vanishing flux, it is T-dual to a type I
D5-brane, which always has 2 CP indices.)
Also, if tadpole cancellation so requires, the
string could intersect a D3-brane (which we specifically ignored in our
calculation).  These are direct analogs of the singularities in the
moduli space of type I and heterotic 5-branes mentioned in 
\cite{Witten:1996gx}.

To count states localized at the singularities, we should be able to 
ignore the 3-form
background.  The reason is the localized states do not have zero modes that
can move through the fluxes.  Therefore, the counting should be the same as
without 3-forms, which is dual to the $SO(32)$ heterotic 
5-brane.\footnote{For an $E_8$ application, see \cite{Keurentjes:2002dc}.}  
Translating the results of \cite{Porrati:1996vj,Sethi:1997kj} to IIB 
language, we find that there are no localized states at the O3-planes and 
that there is one BPS multiplet attached to each D3-brane.

\paragraph{Counting for an F-string}\label{p:duality}

We can obtain the number of BPS states for an F-string by starting with
the Green-Schwarz action for superstrings and accounting for the background
fields; because of supersymmetry, we expect it to suffice for the 
calculation of our index.  However, it is easier to proceed 
by applying S-duality and rotating the background fields,
and then using the results from the previous section. We want to find
the number of BPS states for an F-string wrapped in the 7-direction,
so we will S-dualize the background, and then rotate in the 4-7 plane.

Under an $SL(2,\bm{\mathbbm{Z}})$ transformation, 
\be\label{Sdual} \tau \rightarrow
\frac{a\tau+b}{c\tau+d}, \  \begin{bmatrix} H_{3} \\
F_{3}\end{bmatrix} \rightarrow \begin{bmatrix} \ d & c \cr b & a\ \cr
\end{bmatrix}
\begin{bmatrix}H_{3} \cr F_{3}\cr\end{bmatrix}\ .\ee 
An S-duality is $\tau
\rightarrow -\frac{1}{\tau}$. Then,  $H_{(3)} \rightarrow F_{(3)}$ and
$F_{(3)} \rightarrow -H_{(3)}$.  So $f_1$ and $f_2$ from the previous
sections get mapped to $h_1$ and  $h_2$ respectively. 

A rotation in the 4-7 plane interchanges $h_1$ with $h_2$. From
(\ref{n3fluxes}), we see that a rotation of $\pi /2$ gives $h_1 \rightarrow
-h_2$ and $h_2 \rightarrow h_1$. A rotation of $- \pi /2$ does the
same thing with opposite signs.
Therefore, the
number of BPS states  of the rotated F-string is just
$h_1^2+h_2^2$ plus one state at each D3-brane before taking into account
the orientifold. 

\paragraph{Bound States}\label{p:bound}

So far we have considered only states of  one D- or 
F-string at a time, which are the states of minimal BPS electric charge.
However, it would be very interesting to consider the BPS spectrum of 
multiple strings, since there could be threshhold bound states.  We will
largely leave this question for the future, but we can make some comments.

Consider, for example, the case of two BPS charges.  If they are widely
separated in the noncompact dimensions, then the spectrum should just be
the direct product of the spectra of the individual charges (appropriately
symmetrized).  As the two charges become coincident, we expect that there
would, in addition, be BPS multiplets associated with threshhold bound states.
The counting of multiplets breaks down according to the nature of the 
strings.  A bound state of two D-strings, for example, would have twice the
charge but would be otherwise identical to our previous analysis, so there
would be four times as many nonlocalized states.  A bound state of a
D-string and F-string seems more complicated in that the two strings feel
different fluxes.

Another question that we leave to future work is the nature of the bound
states.  For two D-strings, we really should take into account the 
non-Abelian nature of the worldvolume theory; we can do this using the
D-brane action of \cite{Myers:1999ps}, and supersymmetrizing.  
It is possible that the bound
state is a ``polarized'' configuration, or it is possible that there are
polarized and unpolarized bound states.  We should also mention that
we could start with, instead of the action of \cite{Myers:1999ps}, 
super-Yang-Mills with a superpotential due to the
3-forms along with the appropriate velocity coupling to the vector potential. 
The problem is to guess how the superpotential depends on the 3-form flux.  

\section{Dualities}\label{s:dual}

There are generally two types of dualities in string theory: U dualities,
which leave the string background fixed up to motion in the space of
moduli, and ``string-string'' dualities, which change the string background
but leave the 4D effective theory the same.  Both of these can potentially
move us from a complicated description of the physics to a simple one,
so I will start by discussing the U dualities of the simplest 
compactification with fluxes, the $\N = 3$ cases.

\subsection{U Duality}\label{ss:udual}

In this section, we discuss the stringy duality group of these
compactifications.  In particular, we are interested in the dual
description that governs the physics when the radii become small.
The U duality group of $\N =3$ self-dual flux compactifications was
first discussed in \cite{Frey:2002hf}.  While I will focus on $\N=3$ 
dualities, much of the same discussion should apply for $\N=2$ backgrounds.

\subsubsection{Dualities of the $\N=4$ theory with 16 D3-branes}
\label{sss:hetudual}

As a warmup, let us first consider the dualities of the $\N=4$ theory with
16 D3-branes, which is the T dual of type I on $T^6$ and the TS dual
of the heterotic theory on $T^6$.  The duality of the latter theory is
$SO(22,6,\bm{\mathbbm{Z}}) \times SU(1,1,\bm{\mathbbm{Z}})$
\cite{Schwarz:1993mg,Sen:1994fa}.
Consider first the  perturbative $SO(22,6,\bm{\mathbbm{Z}})$ factor.  
This group is
generated by discrete shifts of the Wilson lines, Weyl reflections in the
gauge group, discrete shifts of
$B_{mn}$, large coordinate transformations on the torus, and the inversion
of one or more directions on the torus (this is not meant to be a minimal
set of generators).  We will call this last operation
$R$-duality to distinguish it from the full perturbative T duality.  The
first three operations are manifest in the IIB description, as the
periodicities of the D3-brane collective coordinates, permutations
of the D3-branes, discrete shifts of the $C_{mnpq}$, and large coordinate
transformations respectively.  The
$R$-duality is not manifest in the IIB description.  Note that this is not
the same as IIB
$R$-duality, because it leaves fixed the ten-dimensional IIB coupling and
not the four-dimensional coupling.  Rather, it is the image of the
heterotic $R$-duality; therefore we will henceforth designate it 
$R_{\mathrm{het}}$.

To see $R_{\mathrm{het}}$ in the IIB description it is useful to focus on its
action on the BPS states.  In the heterotic description
$R_{\mathrm{het}}$ interchanges the electric charges, KK states and 
winding F-strings.  
We remember from section \ref{p:bpshet} that in
IIB these are winding F-strings and D5-branes.  Similarly the duality 
interchanges
winding D-strings and NS5-branes.  This is an electric-magnetic duality in
the IIB case.

To analyze the duality carefully we need the masses of these
objects, taking for simplicity a rectangular torus $ds^2 = r_m^2 dx^m
dx^m$, and vanishing R-R backgrounds. We take the F- and D-strings to be
wound in the 4-direction, and the D5- and NS5-branes to be wound in the
56789-directions.  Then (in the 10D string frame)
\bea
m_{\mathrm{F1}} &=& \frac{r_4}{\alpha'} \ ,\
m_{\mathrm{D1}} = \frac{r_4}{\alpha' g_{s}}\ ,
\nonumber\\
m_{\mathrm{D5}} &=& \frac{v}{2 r_4 \ap{}^3 g_{s}}
\ ,\
m_{\mathrm{NS5}} = \frac{v}{2 r_4 \ap{}^3 g^2_{s}}
\ ,
\eea
where $v = \prod_m r_m$.  The factors of 2 come about because the
strings must be wound on cycles of $T^6$, while the 5-branes can be wound
on the fixed cycle $x^4 = 0$ whose volume is halved.  
For future reference let us also give the masses in the type I description,
where $r_m' = \ap/r_m$; the couplings are related by $v' / g'^2_{s}
= v/2g_{s}^2$, the factor of $2$ being from the orientifold volume.  Then
\be
m_{\mathrm{KK'}} = \frac{1}{r_4'} \ ,\
m_{\mathrm{D5'}} = \frac{v' \sqrt{2}}{r_4' \ap{}^3 g'_{s}}\ ,
\
m_{\mathrm{D1'}} = \frac{r'_4}{v' g'_{s} \sqrt{2}}
\ .
\ee
The factors of $\sqrt 2$ are as found in \cite{Gimon:1996rq}.

In units of the four-dimensional Planck mass $m_4 = (v/2)^{1/2}
\alpha'^{-2}g_{\rm s}^{-1}$ (given in the string frame metric 
with fixed moduli) 
the BPS masses are
\bea
\frac{m_{\mathrm{F1}}}{m_4} &=& \frac{r_4 \ap g_{s}
\sqrt{2}}{v^{1/2}}
= \frac{g^{1/2}_{s}}{\rho_4}\ ,\
\frac{m_{\mathrm{D1}}}{m_4} = \frac{r_4 \ap \sqrt{2}}{v^{1/2}}
= \frac{1}{\rho_4 g_{s}^{1/2}}\ , \nonumber\\
\frac{m_{\mathrm{D5}}}{m_4} &=& \frac{v^{1/2}}{r_4 \ap \sqrt{2}}
= \rho_4 g^{1/2}_{s}\ ,\
\frac{m_{\mathrm{NS5}}}{m_4} = \frac{v^{1/2}}{r_4 \ap g_{s}\sqrt{2}}
= \frac{\rho_4 }{g_{s}^{1/2}}\ . \label{n4mass}
\eea
We have defined $\rho_4 = v^{1/2}/r_4 \ap g^{1/2}_{s}
\sqrt{2}$, which is just the radius in the heterotic string picture, in
heterotic string units.  The first and second lines interchange under
inversion of $\rho_4$, as expected.

The $SU(1,1,\bm{\mathbbm{Z}})$ 
of the heterotic theory maps to the $SU(1,1,\bm{\mathbbm{Z}})$
of the ten-dimensional IIB theory.  In particular, $g_{s} \to g_{s}^{-1}$ 
interchanges the states in each line of (\ref{n4mass}).

\subsubsection{Dualities of the $\N = 3$ theories}\label{sss:n3dual}

We expect that the duality group will be an integer version of the
continuous low energy symmetry $U(3,3+N)$.  The simplest guess would be
that it is the intersection of this continuous group with the discrete
symmetry $SO(6,22,\bm{\mathbbm{Z}}) \times SU(1,1,\bm{\mathbbm{Z}})$ 
of the $\N = 4$ theory.
In other words, the fluxes break the duality symmetry to a subgroup, just as
they do with the supersymmetry.  However, we will see that the situation is
much more complicated.

We will start by considering whether the $\N =4$ duality $R_{\mathrm{het}}$
survives with self-dual flux in the $\N =3$ theory.
The remainder of the duality group would be generated by large coordinate
transformations mixing the holomorphic coordinates, periodicities of the
D3-brane coordinates, permutations of the D3-branes, and shifts of the R-R
backgrounds, just as in the $\N=4$ case.  
Therefore, we will then discuss these dualities.

\paragraph{The Fate of $R_{\mathrm{het}}$}\label{p:rhetfate}

Remember that the 
magnetic objects in this theory are bound states of a D5-brane and an 
NS5-brane.  
We recall from section \ref{p:bpsmagnetic} that these bound states have
twice the minimum $\N = 4$ mass for a given magnetic charge.  
It follows that the duality $R_{\mathrm{het}}$ that interchanges the basic 1-
and 5-branes does not survive in the $\N=3$ theory\footnote{Note 
that this duality interchanges electric and magnetic
objects, while the $SO(6,22,\bm{\mathbbm{Z}})$ of the heterotic theory acts
separately on each. This is because the unbroken gauge
fields (\ref{unbroken}) are a linear combination of electric and magnetic
gauge fields in the heterotic picture: the nonlinear Higgs field has both
electric and magnetic charges.\label{magfoot}}.

The simplest conjecture would then be that the duality group interchanges
the objects of minimum electric and magnetic charge.  With the D5-brane
masses (\ref{n4mass}) doubled, this would now mean that $\rho'_m =
1/(2\rho_m)$; it is not clear whether this symmetry could be inherited from
the $\N=4$ theory\footnote{Such a duality does exist in the heterotic
string for a nonzero axion \cite{Witten:1998bs}, but it has not been
determined if
it can be combined with the heterotic strong-weak coupling duality 
\cite{Sen:1994fa} to generate the proper action on the BPS states. 
This possibility also requires that the axion of the ``heterotic''
description of the $\N=3$ theory be shifted by half a unit, and it
is not immediately clear that this is so.}.
To be precise, this symmetry can act independently on any set of paired
indices, 4-7, 5-8, or 6-9: it must preserve equation (\ref{t-taufix}).  This
conjectured symmetry relates rather different objects, and so for example
the total number of BPS states of a D-string in the 4-direction and an
F-string in the 7-direction must equal that of the D5/NS5 bound states.
This emphasizes the importance of the counting of BPS states carried out
in \cite{Grana:2002ti} and
described in section \ref{ss:counting}.

\paragraph{Large Coordinate Transformations}\label{p:largecoord}

When discussing large coordinate transformations on the torus, we should
distinguish between U dualities, which leave the background invariant, 
and string-string dualities, which take one background into
a different but equivalent self-dual flux compactification on $T^6/\Z$.  
The transformations 
that give string-string dualities are discussed
in \cite{Kachru:2002he} and reviewed in section \ref{s:special}; 
here we are interested in finding those that
give U dualities.  One basic constraint is that any 
large coordinate transformation that is a U duality
must be holomorphic so that it does not change the complex structure.

A large coordinate transformation will leave the background $G_{\b 1\b 2\b
3}$ invariant if its determinant is unity.  Nevertheless, the duality
can include elements of nontrivial determinant.  For example, at $\tau=i$,
rotation of a single coordinate
$z^1\to iz^1$ changes the background 3-form $G_{\b 1\b 2\b 3}\to iG_{\b
1\b 2\b 3}$, but this can be undone by one of the broken
$SL(2;\bm{\mathbbm{Z}})$ dualities of the IIB string, $\tau\to -1/\tau$.
Note that this combined operation leaves the background
invariant and so does not act on the moduli space, but it does mix the
BPS states and so is a nontrivial duality.  Also, if the fluxes are
chosen so that $\tau\neq i$, this duality is not a U duality, so we
find that different $\N=3$ backgrounds have slightly different U duality 
groups.

Note that in models with fluxes on the orientifold planes,
we must restrict to transformations that take
O3-planes of a given type into the same type.  If we insist
that all the fixed points map to themselves under dualities, then the
off-diagonal elements of the linear transformation must be even and the
diagonal elements must be unity (or $-1$ with a translation).  Again,
different backgrounds will have different U duality groups.

\paragraph{D3-brane Shifts}\label{p:d3shifts}

Permutations of the D3-branes are trivial dualities, so we focus here
on the shifts of D3-brane positions, especially considering the
effect on BPS charges.

The D3-brane gauge charges do not appear in the IIB superalgebra, and a
zero-length F- or D-string stretched between coincident D3-branes is
massless, giving an enhanced gauge symmetry.  When the D3-branes are
separated the stretched string begins to couple to the bulk gauge fields,
and acquires a BPS mass and charge.  When the D3-branes shift fully around
the 1-cycles of the torus, the attached F- and D-strings acquire integer
winding charges.  Since the electric charges on the D3-branes are the end
points of F-strings, this duality shifts the
bulk electric charges by the D3-brane electric charges.   Note that
since the magnetic D3-brane charges are D-string end points, the shift
also depends on the D3-brane magnetic charges: as noted in
footnote \ref{magfoot}, the duality group is nontrivially embedded in the
low energy electric-magnetic duality group.

\paragraph{Axion Shifts}\label{p:axionshift}

In order to understand the R-R shift dualities in detail one needs to
consider the BPS instantons and spacetime strings of \ref{p:bpsinstanton}.

It is sufficient to study just the case that couples to the diagonal 
part of $\beta$, in which the instantons are just a single wrapped 
Euclidean D3-brane.   In the $\N=4$ theory the wrapped
D3-branes are dual to type I instantonic D-strings.   These
have a single Chan-Paton index, so the minimal D3-brane instantons
wrap the special half-volume 4-cycles.  Their action is given by
\be\label{c4inst}
\frac{1}{(2\pi)^3\ap{}^2} \int \tilde c_{5689} dx^5dx^6dx^8dx^9
= \pi \tilde c_{5689}\ .\ee
This implies that $\tilde c_{5689}$ can shift by even integers 
without changing the path integral.  As this shifts
$\beta^{\b 1 1}$ by $i/2$ times that integer, we see that the shift
duality has been broken by the instantons to $\bm{\mathbbm{Z}}$ for each
axion.

Let us check that this is consistent with the spacetime strings.  A
D3-brane wrapped on $47$ is dual to a D5-brane in the
$\N=4$ type I theory.  Since the type I D5-brane must have
two Chan-Paton indices, these D3-branes can only wrap 2-cycles of volume
$(2\pi)^2\ap$.  Using the relative coefficients of terms in the action,
the 10-dimensional Bianchi identity for the 5-form integrates to
\be\label{5bianchi}
\frac{1}{(2\pi)^7\ap{}^4} \oint_M \tilde F_{(5)} =
\frac{1}{(2\pi)^3\ap{}^2}\ .
\ee
The surface surrounding the string is $M=S^1\times T^4/\Z$.  
Integrating over the torus factor gives
\be
\oint_{S^1} d\tilde
c_{5689} = 2\ ,
\ee
which is the minimum shift consistent with the instanton amplitude.

\paragraph{The Unfinished Picture}\label{p:unfinished}

It would be very interesting to have a more complete understanding of
U dualities in self-dual flux compactifications, but the problem is not
yet solved.  We would then be able to describe completely the global as
well as local structure of the moduli space and possibly make concrete
statements about the behavior at small volume.

It is also possible that the U dualities are part of a more
intricate pattern of string-string dualities, 
in which the various self-dual flux models mix.
For example, the $\N=3$ dualities might involve other types of $\N=3$
construction, such as the free-fermion models 
of \cite{Ferrara:1989nm}, though we have
no particular reason to expect this.  I will review some results 
on dualities to other constructions below.

A related issue is that of enhanced symmetries at special points of
moduli space due to new massless states.  These usually occur at fixed
points of a U duality group, such as when D3-branes become coincident,
which we mentioned above.  However, there are more exotic situations;
in the heterotic string on $T^6$, with zero $B_{mn}$, winding strings with
other charges become massless when the proper radii 
of the compactification are string length.  This is in fact the fixed
point of the $R_{\mathrm{het}}$ duality.  In the IIB picture, these
are the wrapped D5-branes; it is very difficult to see how to find
a massless state of the 5-brane bound states necessary for the 
compactifications with flux, however.  This problem is clearly tied to
the survival of $R_{\mathrm{het}}$ or some related duality.

\subsection{T Duality}\label{ss:tdual}

As we discussed in section \ref{ss:heterotic}, there is a clear relation 
to the heterotic string for these type IIB compactifications.  In 
particular, with non 3-form fluxes, the $T^6/\Z$ compactifications are
dual to the heterotic string by T dualizing all six internal dimensions
and then S dualizing.  $K3\times T^2/\Z$ compactifications are dual to 
heterotic compactifications by T dualizing the $T^2$ and S dualizing.
Note, however, that there is no reason to
expect an effective heterotic description anywhere in the moduli space
for self-dual flux compactifications on $T^6/\Z$. 
For the $\N=4$ theories such a description holds when the IIB radii are
small and the ten-dimensional IIB coupling is large, but in the generic 
flux compactifications the latter coupling is always of order one.  
Also, because of the T duality rules for the NS-NS flux, taking six T 
dualities gives a metric that might be hard to interpret.
On the other hand, T duality is still a useful tool when only some of
the radii become small.  For example, supersymmetric self-dual flux 
compactifications on 
$K3\times T^2$ are dual to type C compactifications in type I string 
theory and type A models in the heterotic theory
\cite{Dasgupta:1999ss,Becker:2002sx}.  Below, I will stick with 
$T^6/\Z$ compactifications and illustrate the relations to other known
types of solutions.

\subsubsection{An Aside: ``$-1$'' T Dualities}\label{ss:MtypeB}

Before getting into the T dualities of the internal directions,
we should mention that the self-dual flux compactifications of type IIB
string theory are T dual on a \textit{noncompact} direction to 
M theory with three noncompact dimensions.  The supersymmetry conditions
for such compactifications were first studied in \cite{Becker:1996gj},
and the analysis is similar to the type IIB case (as presented in 
section \ref{ss:susy}).  Also, the equations of motion admit 
nonsupersymmetric solutions of the same form, just as in IIB SUGRA
\cite{Becker:2001pm}.   

The T duality to type IIB theory was first presented in 
\cite{Gukov:1999ya,Dasgupta:1999ss,Greene:2000gh}, as follows.  Start with
M theory on an 8D manifold with a shrinking $T^2$.  One of the directions
of the torus is the M direction, so we have type IIA string theory
with a shrinking circle.  These IIA compactifications have been studied
in detail, with attention paid to moduli fixing, on the $T^7/\Z$ 
orientifold \cite{Argurio:2002gv}.  Then we can T dualize on the type IIA
circle to get type IIB theory with 6 compact dimensions; the small circle
in IIA becomes noncompact in IIB string theory, and it has precisely the
right warp factor to preserve 4D Lorentz invariance \cite{Dasgupta:1999ss}.

One thing interesting to note is that, since the M theory compactifications
have only $SO(2,1)$ Lorentz invariance, they can have a smaller number
of supercharges than the IIB compactifications.  The 3D $\N =1$ 
compactifications were given in type IIA language in \cite{Argurio:2002gv},
and they correspond to M theory on a $\mathnormal{Spin}(7)$ manifold
with flux \cite{Becker:2000jc}.  The gauged supergravity of these
compactifications has been pursued in \cite{Berg:2002es}.

\subsubsection{One T Duality}\label{ss:IIA1}

For our discussion of T dualities and their action on self-dual flux
compactifications, I follow the discussion of \cite{Kachru:2002sk} and
use the background given by (\ref{n2fluxes}).  For simplicity, we consider
the initial background to have a diagonal metric
\bea
ds^2 &=& R_1^2 \left[(dx^4)^2 +|t_1|^2 (dx^7)^2\right]
+ R_2^2 \left[ (dx^5)^2 +|t_2|^2 (dx^8)^2\right]\nonumber\\
&&+R_3^2 \left[ (dx^6)^2+|t_3|^2 (dx^9)^2\right]\ ,\label{n2metric}\eea
remembering that we are also taking $\tau, t_i$ to be purely imaginary.
For the purposes of T duality, we take the coordinate radii to be
$2\pi\sqrt{\ap}$ for all internal coordinates.  We also take a gauge
\be\label{b2gauge}
B_2 = -\frac{2}{2\pi\sqrt{\ap}} \left( x^5 dx^4\wedge dx^6 +x^8
dx^7\wedge dx^6\right)\ .\ee
Also, we will ignore the warp factor because it introduces some subtleties
which we will discuss briefly at the conclusion of this section.

Using the T duality rules given for IIA/B string theories in
\cite{Bergshoeff:1995as}, \cite{Kachru:2002sk} showed that a single 
T duality along the $x^4$ direction gives a type IIA background of the
form
\bea
ds^2 &=& \frac{1}{R_1^2} \left(dx^4+\frac{2}{2\pi\sqrt{\ap}} x^5 dx^6
\right)^2+R_1^2|t_1|^2 (dx^7)^2 \nonumber\\
&&+R_2^2 \left[ (dx^5)^2 +|t_2|^2 (dx^8)^2\right]
+R_3^2 \left[ (dx^6)^2+|t_3|^2 (dx^9)^2\right]\nonumber\\
H_3&=& \frac{2}{2\pi\sqrt{\ap}} dx^7 \wedge dx^8\wedge dx^6\ 
\textnormal{or}\ B_2 = -\frac{2}{2\pi\sqrt{\ap}}x^8 dx^7\wedge dx^6
\nonumber\\
F_2 &=& \frac{2}{2\pi\sqrt{\ap}} dx^5 \wedge dx^9\ \textnormal{or}\
C_1 = \frac{2}{2\pi\sqrt{\ap}} x^5 dx^9\nonumber\\
F_4 &=& \frac{2}{2\pi\sqrt{\ap}} dx^4 \wedge dx^7\wedge dx^8 \wedge dx^9
\nonumber\\
e^\phi &=& \frac{1}{R_1 |\tau|}\ .\label{IIAback1}
\eea
Here the $F_p=dC_{p-1}$ are the IIA R-R field strengths and $H_3$ is the 
IIA NS-NS field strength.  D3-branes and O3-planes become D4-branes
and O4-planes wrapping the $x^4$ direction.

There are some very interesting features to this background.  First of
all, the metric is ``twisted,'' meaning that traversing the circle
$x^5$ changes the complex structure of the $x^4,x^6$ torus by two units.
Since integer shifts of the complex structure of a $T^2$ are dualities
of toroidal compactifications, these are compactifications of the 
Scherck-Schwarz type \cite{Scherk:1979zr,Dabholkar:2002sy}.  In fact,
the $x^{4,5,6}$ directions form a coset manifold 
\cite{Strominger:1986uh,Kachru:2002sk}, which turns out to be 
non-K\"ahler \cite{Kachru:2002sk}.  
The gauge invariant field strength $\t F_4 =F_4-C_1 H_3$ is also twisted.
In addition, we have lost one of the
components of the NS-NS flux (which has turned into the spin connection
of the metric \cite{Kachru:2002sk}).  

The moduli of the original background include $R_{1,2,3}$ and
$\tau$, $t_{1,2,3}$ subject to the constraints (\ref{t-taufix2}).  
Therefore, for this IIA background to be the good effective background,
we should be in a region where $R_1 <1$ and $|t_1|>1/R_1$ from geometrical
considerations.  Then (\ref{t-taufix2}) implies that $|\tau|<R_1$,
so the IIA string coupling is $e^\phi>1$.  In fact, if we want
$R_1$ very small, we find that the solution has very large coupling.
This problem is somewhat worse when $\tau,t_1$ are fixed independently,
because then it is impossible for both the $x^4$ and $x^7$ proper radii
to become large in IIA, and the string coupling still becomes very large
as the original $R_1$ becomes small.  

The previous paragraph suggests that we should probably think about these
IIA solutions in terms of M-theory on a 7D orbifold with M5-branes.
Because $F_2$ is the Kaluza-Klein field strength of M-theory, there should
be twists involving the M circle, as well.  Finally, there are also 
4-form fluxes.  While it has not been confirmed explicitly, these 
solutions must be of the type classified in \cite{Becker:2000rz}.

\subsubsection{Two T Dualities}\label{sss:IIB2}

Now take a second T duality, this time along the $x^7$ direction, so
the background returns to IIB string theory.  In geometric terms,
this should be a good description for $R_1 <\min (1,1/|t_1|)$.  Again
ignoring the warping, \cite{Kachru:2002sk} showed that the IIB background
becomes
\bea
ds^2 &=&\frac{1}{R_1^2} \left(dx^4+\frac{2}{2\pi\sqrt{\ap}} x^5 dx^6
\right)^2+\frac{1}{R_1^2|t_1|^2} \left(dx^7+\frac{2}{2\pi\sqrt{\ap}}
x^8 dx^6\right)^2 \nonumber\\
&&+R_2^2 \left[ (dx^5)^2 +|t_2|^2 (dx^8)^2\right]
+R_3^2 \left[ (dx^6)^2+|t_3|^2 (dx^9)^2\right]\nonumber\\
F_3&=& \frac{2}{2\pi\sqrt{\ap}} \left( (dx^7+2x^8 dx^6) dx^5 dx^9-
(dx^4 +2x^5 dx^6) dx^8 dx^9\right)\nonumber\\
\tau^\prime &=& iR_1^2 |t_1||\tau|\label{IIBback2}
\eea
with all other fluxes, including $H_3$, vanishing.  The prime on $\tau'$
is to distinguish it from the original dilaton-axion.  The original
D3-branes and O3-planes become D5-branes and O5-planes.

This type of background (modulo the warp factor) is just a type C background
as we discussed in section \ref{ss:susy}.  There should be relations between
the flux $F_3$ and the complex structure of the manifold, as in 
equation (\ref{Ffromtorsion}); however, we note as in \cite{Kachru:2002sk}
that the correct complex structure may not be obvious from the T duality.
One question that might arise is whether $H_3=0$ always after two T dualities,
as it should for type C.  Start with the original (type B)
model; section \ref{ss:susy} shows that $H_3$ is primitive.  For a diagonal
metric, primitivity implies that the indices of $H_3$ all fall in different
complex coordinates (for example, $H_{12\b 1}$ is not allowed).
In the complex
coordinates (\ref{complexgeneral}), therefore, each component of $H_3$ has
either an $x^4$ index or an $x^7$ index but not both, so $H_3$ does indeed
vanish after two T dualities.

We do into problems again with large string coupling for the same reasons
as after a single T duality.  Therefore, we should S dualize;
we end up with a type A solution \cite{Strominger:1986uh}.  
In the case of $K3\times T^2/\Z$,
\cite{Dasgupta:1999ss,Greene:2000gh,Becker:2002sx} made this realization,
although the S dual of those compactifications is heterotic.

\subsubsection{Three T Dualities}\label{sss:IIA3}

We have so far chosen our T dualities to be along directions along which
the background is independent.  If we continue in this vein, we can choose
to make a third T duality along $x^{6,9}$.  In addition, we want to 
T dualize in a direction with no cross-components in the metric in order to
prevent the reappearance of the $H_3$ flux; the possible T duality directions
are $x^{5,8,9}$.  Therefore, we will chose to T dualize along $x^9$, 
still following \cite{Kachru:2002sk}\footnote{However, as we discuss
in the subsection below, it is not necessary to T dualize along a direction
of which the background is independent, so it might be very interesting to
study T dualities along $x^{5,8}$.}.  We find that the metric is unchanged
up to $|t_3|R_3\to 1/|t_3|R_3$ and 
\bea
e^\phi &=& \frac{1}{R_1^2 R_3 |t_1||\tau||t_3|}\nonumber\\
F_2 &=& \frac{2}{2\pi\sqrt{\ap}}\left( (dx^7+2x^8 dx^6) dx^5
- (dx^4 +2x^5 dx^6) dx^8\right)\ .\label{IIAback3}\eea
There are now D6-branes and O6-planes.

When the radii of this
geometry are all large, the dilaton is large also, so this is really an
M-theory background.
Up to any details from the warp factor, this background is purely 
geometrical when lifted to M-theory \cite{Kachru:2002sk}.  Since it is a
4D $\N=2$ background, it must be M-theory on a CY 3-fold times a circle.
Performing a 9-11 flip, it could reduce to IIA on a CY 3-fold.  As
\cite{Kachru:2002sk} point out, this result is very interesting because
it shows flux compactifications on $T^6/\Z$ are connected to Calabi-Yau
compactifications.  One puzzle is that a Calabi-Yau compactification
typically does not have the same hierarchy of scales as a flux 
compactification and does not have a superpotential for moduli.  A possible
resolution of the paradox is that the modes frozen by the fluxes in the IIB
picture end up with masses of order the Kaluza-Klein scale on the M-theory
side of the duality.  We are working, after all, out of the large radius
limit of the original IIB theory.

Finally, as a bit of nomenclature, taking three T dualities gives a 
mirror symmetry.  Therefore, we could have started with a CY 3-fold
geometry in type IIB string theory and taken three T duals.  This more
general case has recently been studied in \cite{Gurrieri:2002wz}.

\subsubsection{The Warp Factor}\label{sss:warp}

Including the warp factor in analysis of T duality, 
with D-brane and O-plane sources, is problematic because it is not 
uniform along the directions of T duality.  Ignoring the point sources,
however, \cite{Becker:2002sx} showed that the warp factor and dilaton
behave precisely as required under two T dualities and S duality to turn
a $K3\times T^2/\Z$ compactification into a heterotic type A solution.  
Given the O7-plane boundary conditions, it is also easy to see that the
two T dualities remove the 5-form flux, as necessary for type A and C 
solutions \cite{Becker:2002sx}.  It is less obvious how to understand
the vanishing of the 5-form after two T dualities on $T^6/\Z$ because
naively there could be components $\t F_{4mnpq},\t F_{7mnpq}$.  If we
ignore brane sources, it should be possible to take $C_4$ gauge trivial on
the internal manifold,
in which case $\t F_5=-C_2H_3$; in this gauge, then, the fluxes 
(\ref{n2fluxes}) seem to require that $C_2H_3$ always have both $x^{4,7}$
indices.  This approach is similar to that of \cite{Becker:2002sx}.  
However, the fluxes (\ref{n2fluxes}) are nongeneric; a more generic approach
is to examine the determining equation (\ref{warppoisson}) for the
warp factor in a double scaling limit $R_1\to 0,R_{2,3}\to\infty$.  Then
(\ref{warppoisson}) becomes the Laplace equation on the $x^{4,7}$ torus,
and the boundary conditions at the O7-planes indicate that 
$\del_{4,7}A\to 0$.  This is not rigorous, though, and we would like to have
the appropriate solution even away from limits of moduli space.  Another
resolution is that the solutions after T duality are not actually type C
but the more general interpolating solutions of section \ref{ss:susy},
but this issue has not been resolved in detail.

We should also note that the warp factor has not been considered in the
cases of one or three T dualities.

\chapter{The Scalar Potential \& Nonperturbative Effects}\label{c:potential}
As we have seen, even understanding the classical vacuum of self-dual
flux compactifications is a difficult problem.  Not only are the vacua
themselves complicated due to the warp factor and the potential for
the moduli, but there is a rich array of vacua on a given geometry because
of the large number of possible fluxes.  However, there are many reasons
to look beyond the classical vacuum.  For example, evidence is increasingly
in favor of an inflationary phase in the early universe 
\cite{Bennett:2003bz} as well as a less rapid acceleration in the modern
era.  Therefore, we want to see whether self-dual flux compactifications
can accommodate an accelerating universe.  The answer, as we shall see,
depends on exactly what we consider.

In this chapter, I return to the large radius limit, in which 10D
and 4D SUGRA remains valid, to study the cosmology of self-dual
flux compactifications.  
Our first approach will be to find the cosmological evolution of the 
flux-induced potential.  Then we will review some interesting brane
cosmology that can be embedded in self-dual flux compactifications.  
Third, we will see how nonperturbative effects
can generate a cosmological constant and examine the consequences.

\section{Cosmology from Moduli Fixing}\label{s:fluxcosmo}

As we saw in section \ref{s:gvw}, the fluxes give a potential to many
of the moduli of the geometric compactification, freezing the scalars
to definite vacuum expectation values.  We can hope to get double-duty
out of this potential; perhaps it is shallow enough that it leads to an
accelerating expansion of the universe, either for cosmological inflation
or for quintessence in the current era.  In this section, we ask what 
the cosmology of the flux-generated potential could be and find that 
the potential does not give inflation.  For specificity,
we consider the case of compactification on $T^6/\Z$, but our answer
will clearly generalize to more complicated compactifications.  This
discussion follows \cite{Frey:2002qc}.

\subsection{Canonically Normalized Scalars}\label{ss:canonical}

It is easiest to study the cosmology of canonically normalized scalars; so
we will break down the geometric moduli. For simplicity we will consider 
only the factorized case $T^6=(T^2)^3$. We can then parameterize the metric 
on an individual 2-torus (say, the $(4-7)$ torus) as
\be
\label{t2complex}
\gamma^{mn}= e^{2\lambda} \left[ \begin{array}{cc} e^{-\zeta}+e^\zeta d^2 &
-e^\zeta d \\ -e^\zeta d & e^\zeta\end{array}\right]\ . 
\ee
We use the definition of $\gamma^{mn}$ as in sections \ref{ss:modmetric}
and \ref{ss:heterotic}.
Here, $\lambda$ gives the overall size of the $T^2$, $\zeta$ gives the 
relative length of the two sides, and $d$ controls the angle between the
two directions of periodicity. Then the $\gamma$ kinetic term becomes
\be
\label{gammakin}
S_{\mathrm{kin}} = -\frac{M_P^2}{16\pi} \int d^4 x \sqrt{-g_E} \sum_{i=1}^3
\left[ 2\del_\mu\lambda_i\del^\mu\lambda_i +
\frac{1}{2}\del_\mu\zeta_i\del^\mu
\zeta_i +\frac{1}{2}\del_\mu d_i \del^\mu d_i \right]\ .
\ee
Here and in the rest of the section, we use the unreduced Planck constant
$M_P^2=8\pi M_4^2$ for convenience of comparison to the cosmology 
literature.  
For canonical normalization, the coefficient of the kinetic terms should
simply be $-1/2$, so a further rescaling is necessary.

For reference, with $\sqrt{\det g_{mn}} = e^{6u}$, equation 
(\ref{moddensity}) gives $\sum_i \lambda_i = 6u-(3/2)\phi-\ln 8$.

\subsection{The Potential from Fluxes}\label{ss:potential}

The scalar potential comes from dimensional reduction of the background
$3$-form terms in the IIB action. After converting to our variables, 
the potential action for the bulk modes is, in generality,
\be\label{Vgeneral}
S_V= \comment{\frac{M_P^2}{4!\cdot (32\pi^{3/2})^2}  
\left(\det\gamma_{mn}\right)\gamma^{mq}\gamma^{nr}\gamma^{ps}
\left[e^\phi (F-CH)_{mnp}(F-CH)_{qrs}+e^{-\phi}H_{mnp}H_{qrs}\right]
\ee
&=&} -\frac{M_P^2}{4!\cdot 8\pi}\int d^4x\sqrt{-g_E}
\left(\det\gamma_{mn}\right)
\gamma^{mq}\gamma^{nr}\gamma^{ps}e^\phi G_{mnp}\bar G_{qrs}\ee
\comment{\label{Vgeneral}\eea}
along with an additional term that subtracts off the vacuum energy
\footnote{This comes from the D3/O3 tension, which must cancel the 
vacuum potential for string tadpole conditions to be satisfied to 
leading order in $\ap$.}. 
This potential was derived from dimensional 
reduction in \cite{Giddings:2001yu,Kachru:2002he}, from gauged 
supergravity in \cite{Tsokur:1996gr,D'Auria:2002tc}, and from the 
superpotential of \cite{Gukov:1999ya}, as we reviewed in chapter
\ref{c:effective}. One feature to note in this 
potential is that it always has (at least) three flat directions at 
the minimum, corresponding to the radii of factorization 
$T^6=T^2\times T^2\times T^2$.  Also, the $\beta$ moduli do not 
enter into the potential, although some become Goldstone bosons via 
the super Higgs effect
\cite{Tsokur:1996gr,Frey:2002hf,Andrianopoli:2002rm,Andrianopoli:2002mf}.

For cosmological purposes, we will need to have a more explicit form of
the potential in hand.  Since there are 23 scalars $\gamma^{mn}, \phi, C$,
writing the full potential for a given set of 3-form fluxes would be 
prohibitively complicated, but we can write down a few simple examples and
focus on the universal aspects.  In the following, a subscript $0$ indicates
the VEV.

The simplest case is to take the three $T^2$ to be square, so that the
geometric moduli are $\gamma^{44}=\gamma^{77} = e^{2\lambda_1}$, etc., with 
all others vanishing.  
Then, above a vacuum that satisfies (\ref{imsd}),
we can calculate the potential
\bea
V_{\mathrm{dil}}\!\!\!\!& = \!\!&\frac{h^2 M_P^4}{4(8\pi)^3}  
e^{-2\sum_i \lambda_i}\!\left[ 
e^{-\phi_{0}}\cosh\left(\phi-\phi_{0}\right)\! +\!\frac{1}{2}e^\phi
\left( C-C_{0}\right)^2\! -\!1 \right]\! ,\label{Vdilax}\\
h^2&=& \frac{1}{6}h_{mnp}h_{qrs}\delta^{mq}\delta^{nr}\delta^{ps}
\label{fluxno}
\eea
This potential was written explicitly in $SU(1,1)$ notation in 
\cite{D'Auria:2002tc} and is valid for any 3-form background. The most
important feature of this potential is that there is a vanishing vacuum 
energy, and, further, the radial moduli $u$ feel a potential only 
when the dilaton-axion system is excited.  Since this is the simplest
potential to write down, it will be our primary focus in section 
\ref{ss:evol}.  It is very interesting to note that the cosmology of this 
potential for the dilaton-axion has been discussed earlier in 
\cite{Fre:2002pd,Kallosh:2002wj,Kallosh:2002gf} from SUGRA. Importantly,
though, their models did not include the radial moduli or the 
negative term that subtracts off the cosmological constant.

Adding the complex structure is more complicated and more model-dependent.
The simplest possible case, for example, 
$f_{456}=-h_{789}$, is non generic in that (\ref{imsd}) is satisfied
at $\phi - \sum_i\zeta_i = C=d_i=0$, so the $\zeta_i$ give extra
moduli compared to other background fluxes (at the classical level).  
However, we still have $\phi - \sum_i\zeta_i$ fixed by a $\cosh$ potential
with a polynomial in $C,d_i$:
\bea
V_0 &=& \frac{h^2M_P^4}{4(8\pi)^3} e^{-2\sum_i \lambda_i}
\left\{\cosh\left(\phi - \sum_i\zeta_i\right) +\frac{1}{2}e^{\phi+
\sum_i\zeta_i} 
\left[ C^2-2C d_1 d_2 d_3\right. \right. \nonumber\\
&&\left. \left. +d_1^2 d_2^2 d_3^2 + e^{-2\zeta_3}
d_1^2 d_2^2 +e^{-2\zeta_2}d_1^2 d_3^2 + 
e^{-2\zeta_1}d_2^2 d_3^2\right. \right. \nonumber\\
&&\left. \left. +e^{-2\zeta_2-2\zeta_3} d_1^2 +e^{-2\zeta_1-2\zeta_3} d_2^2
+e^{-2\zeta_1-2\zeta_2} d_3^2 \frac{}{}\right] -1 \right\}\label{v0}
\eea
using again (\ref{fluxno}). It is straightforward but tedious to show 
that this potential is positive definite, and the only extremum is at 
$\phi - \sum_i\zeta_i = C=d_i=0$.  As this case is nonsupersymmetric, 
quantum mechanical corrections should lift the flat directions.

On the other end of the supersymmetry spectrum are the $\N = 3$ models
of \cite{Frey:2002hf}, which fix the dilaton as well as all the complex
structure.  If we ignore $C,d_i$ (set them to a vanishing vacuum value),
we find a potential
\bea
V_3 &=& \frac{h^2 M_P^4}{(8\pi)^3} e^{-2\sum_i \lambda_i} \left[\frac{}{}
\cosh \left(\phi-\zeta_1-\zeta_2-\zeta_3\right) 
+\cosh\left(\phi-\zeta_1 +\zeta_2+\zeta_3\right)\right.\nonumber\\ 
& & \left. +\cosh\left( \phi+\zeta_1-\zeta_2+\zeta_3\right)+
\cosh\left( \phi+\zeta_1+\zeta_2-\zeta_3\right)-4\right]\ .\label{v3}
\eea
This again has the same $\cosh$ structure for the dilaton; the only 
difference is a factor of $4$ due to the number of components of flux in
the background.

Including the non-Abelian coupling for the D3-brane scalars $\alpha^m_I$
introduces new terms in the potential (see \cite{Ferrara:2002bt} for
a supersymmetry based approach). In the absence of fluxes and even in 
the ground state of the bulk fields, 
this potential is monotonic and simply forces the 
$\alpha^m_I$ to commute.  Otherwise, the branes pick up a $5$-brane
dipole moment and become non-commuting, as discussed in \cite{Myers:1999ps}.
Writing the brane positions as $U(N)$ matrices, the potential is
\bea
V_{\mathrm{b}} &=& 2\pi M_P^4 \left[ 2\pi e^\phi \gamma_{mp}\gamma_{nq}\ \tr
\left( [\alpha^m ,\alpha^n][\alpha^q, \alpha^p]\right)\frac{}{}
\right.\nonumber\\
&&\left.
+\frac{i}{12} (\det \gamma_{mn})^{1/2} e^{\phi} \left(e^{-\phi}h-
\star_6(f-Ch)\right)_{mnp}\tr\left(\alpha^m\alpha^n\alpha^p\right)\right]
\ .\label{Vbrane}
\eea
To illustrate this potential, we take $f_{456}=-h_{789}$ as before, 
set $C=d_i=\zeta_i=0$, and consider $\alpha^{4,5,6}\propto I_N$ and 
$\alpha^{7,8,9}=\rho t^{1,2,3}$ with $t^i$ a representation of $SU(2)$.  
Then
\bea
V_{\mathrm{b}} &=& 2\pi M_P^4 \left[ \frac{}{}16\pi e^\phi
\left(e^{-2 \lambda_1-2\lambda_2}+
e^{-2\lambda_1-2\lambda_3} +e^{-2\lambda_2-2\lambda_3}\right)
\rho^4\right.\nonumber\\
&&\left.+\frac{h_{789}}{2}e^{-2\sum_i \lambda_i}e^{\phi}
\left(e^{-\phi}-1\right) \rho^3\right]\ .\label{braneexamp}
\eea
There are actually more terms in this potential as required by supersymmetry;
these are just the lowest order terms that appear in the D-brane
action given by \cite{Myers:1999ps}.  For example, the underlying $\N = 4$
supersymmetry gives a $\rho^6$ term \cite{D'Auria:2002th}, 
and there is also a $\rho^2$ term from 
gravitational backreaction that has been 
calculated in one case (see 
\cite{Polchinski:2000uf,Freedman:2000xb})\footnote{It is somewhat unclear
whether the $\rho^2$ term would arise in a background with self-dual flux.}; 
in any event,
there is a local maximum in the $\alpha^m_I$ direction.  Like the bulk
potential, this potential has exponential prefactors from the $\lambda$
moduli, and, if the bulk scalars are away from their minimum, there is
the same $\exp [-2\sum_i \lambda_i]$ factor.

The key point to take from this discussion of the potential is the
exponential prefactor that appears in all terms, whether bulk or brane
modes.

\subsection{Cosmological Evolution}\label{ss:evol}

We seek the cosmological evolution of the dilaton and 
the moduli fields $\lambda_i$
in our 4D effective theory.  We start by looking at the bulk fields
only.

\subsubsection{Bulk-driven Evolution}\label{sss:bulkevol}

It is sufficient to consider 
a toy model which illustrates the behavior of the potentials 
$V_{dil}, V_{0}$ and $V_3$ described in the earlier section: 
\begin{equation}
\label{pot1}
V\approx e^{-\sum_{i}\alpha_{i}\lambda_{i}} V(\phi)\, .
\end{equation}
Let us also assume that the above potential has a global minimum 
$\phi_0$ determined by $V(\phi)$. At $\phi_{0}$ the potential vanishes.
In the above, $\phi$ mimics the dilaton (and complex structure) 
and $\lambda_{i}$ play the role of 
moduli with various coefficients $\alpha_{i}$ that determine the slope 
of the potential. For generality we have assumed that there are any number
of moduli. In our potentials (\ref{Vdilax},\ref{v0},\ref{v3}), 
all the slopes are fixed at 
$\alpha_{i}=4\sqrt{\pi}/M_P$ (with normalized scalars).

It is 
interesting to note that the potential (\ref{pot1}) is quite 
adequate to determine the cosmological evolution if it dominates the 
energy density, which is fixed by the value $V(\phi)$ in our case. 
Given generic 
initial conditions for all the moduli $\phi,\lambda_{i}\sim M_{P}$ in the 
dimensionally reduced action, we hope that the rolling moduli could lead 
to the expansion of the universe. The equations of motion in a 
Robertson-Walker space-time metric with an expansion 
factor $a(t)$ ($t$ is proper time)  are (in Einstein frame)
\begin{eqnarray}
\label{eqm1}
\ddot \phi +3H\dot \phi +e^{-\sum_{i}\alpha_{i}\lambda_{i}}V^{\prime}
(\phi)=0\,, \\
\label{eqm2}
\ddot \lambda_{i}+3H\dot\lambda_{i}-\alpha_{i}e^{-\sum_{i}\alpha_{i}
\lambda_{i}}V(\phi)=0\,, \\
\label{eqm3}
H^2=\frac{8\pi}{3M_{\rm P}^2}\left[\frac{1}{2}\dot\phi^2+\frac{1}{2}\sum_{i}
\dot\lambda_{i}^2 +e^{-\sum_{i}\alpha_{i}\lambda_{i}} V(\phi)\right]\,.
\end{eqnarray}
The Hubble expansion is given by $H=\dot a/a$, an overdot denotes $\del_t$, 
and prime denotes $\del/\del\phi$.

To determine whether the expansion of the universe is inflationary or not,
we assume that the dilaton (and complex structure) rolls slowly.  Our
justification is that once the dilaton reaches its VEV, the $\lambda$ moduli
have no potential, so expansion will be radiation or matter dominated
(depending on what other ingredients might be in the universe).  Therefore,
we expect that long periods of accelerating expansion are more likely if
the dilaton spends a long time away from its vacuum.  Let us see what
happens in that case.

We argue that there exists an attractor solution for the moduli
with a power law solution $a(t)\propto t^{p}$ which, from equation 
(\ref{eqm3}), dimensionally satisfies 
$H^2 \propto t^{-2} \propto e^{-\sum_{i}\alpha_{i}\lambda_{i}} V(\phi) $.  
Hence we write  
\begin{eqnarray}   
\label{genevol}   
e^{\alpha_{i} \lambda_{i}} & = & \frac{k_{i}}{t^{c_{i}}} \,, \\    
\label{csj}  
\sum_{i=1}^{n} c_{i} & = & 2 \,,    
\end{eqnarray}    
where $k_{i}$ are dimensional and $c_{i}$ are dimensionless constants    
respectively.    
Equation (\ref{genevol}), coupled with the equations of motion (\ref{eqm2}),
results in    
\begin{equation}    
(3p-1) c_{i} = \alpha_{i}^2 V(\phi) \prod_{k=1}^{n} k_{k} \,,    
\end{equation}    
from which we find, using equations (\ref{csj},\ref{eqm2}),
\begin{eqnarray}     
\label{gensol2}    
V(\phi)\prod_{k=1}^{n} k_{k} & = & \frac{2 (3p-1)}{\sum_{i=1}^{n}\alpha_{i}^2}
\,, \nonumber \\   
\left(\frac{c_{i}}{\alpha_{i}} \right)^2 & = & \frac{4 \alpha_{i}^2}    
{\left(\sum_{k=1}^{n} \alpha_{k}^2 \right)^2} \,.     
\end{eqnarray}    
When substituted into (\ref{eqm3}) with $\dot\phi \ll \dot \lambda_{i}$,
we obtain the key result
\begin{equation}   
\label{slope1}    
a(t)\propto t^p\ ,\ p = \frac{16 \pi}{M_{P}^2}\frac{1}{\sum_{j=1}^{n}    
\alpha_{j}^2} \,.    
\end{equation}   
It is also easy to see that
\begin{equation}    
\label{scal1}    
\left(\frac{\dot{\lambda}_{i}}{\dot{\lambda}_{k}} \right)^2 =     
\left(\frac{\alpha_{i}}{\alpha_{k}}\right)^2 \,.    
\end{equation}     
In fact, \cite{Liddle:1998jc,Copeland:1999cs} showed that this solution is
a late time attractor for fields $\lambda_i$ in exponential potentials.

So far we have concentrated on the toy potential
(\ref{pot1}). Nevertheless, the situation remains unchanged for the
type of potentials we are interested in 
(\ref{Vdilax},\ref{v0},\ref{v3}), since the only difference is the
number of ``dilaton'' fields $\phi$. Note that the dynamical behavior
of the moduli will remain unchanged.  
By inspecting the potentials we find the corresponding slope 
of the three moduli, $\alpha_{i}=4\sqrt{\pi}/M_{P}$. 
Therefore, the moduli driven expansion of the universe leads to 
\begin{equation}
p =\frac{1}{3} < 1\ \textnormal{or}\ a(t) \propto t^{1/3}\,.
\end{equation}
This expansion is actually slower than even matter or radiation dominated
expansion, so it is clearly not a candidate either for inflation or
quintessence!  It could be an exit phase of inflation before or during
reheating, however.

Because even a slowly rolling dilaton (or complex structure) gives
a decelerating expansion, the dilaton's evolution is not particularly
informative.  Details were given in \cite{Frey:2002qc} based on 
the soft-inflation models of 
\cite{Berkin:1990ju,Berkin:1991nm,Mazumdar:1999tk}. 

\subsubsection{Brane-driven Evolution}\label{sss:braneevol}

Now we briefly comment on the brane-induced potential (\ref{braneexamp}).
Note, even if the dilaton is settled down the minimum with $e^{-\phi}=1$, 
the moduli fields $\lambda$ still contribute to the potential. 
In fact, we are most interested when the dilaton is in its vacuum because
the $\rho^3$ term simply mimics the toy potential (\ref{pot1}) above. 

So we are interested in the potential
\begin{equation}
\label{pot12}
V_{b}=32\pi^2 M_{P}^4\rho^4
\sum_{s=1}^{n}\exp\left(\sum_{j=1}^{m}\alpha_{sj}
\lambda_{j}\right)\,.
\end{equation}
This kind of potential has also been solved exactly without using
slow-roll conditions \cite{Copeland:1999cs}, including the 
possibility of $\alpha_{sj}=0$ for some $s,j$. 
As in section \ref{sss:bulkevol}, we demand that 
$\exp\left(\sum_{j=1}^{m}\alpha_{sj}\lambda_{j}\right)\propto 1/t^2$.
There is a late time attractor solution
with \cite{Copeland:1999cs}
\begin{equation}
\label{fresult}
\left(\frac{\dot\lambda_{j}}{\dot\lambda_{l}}\right)^2=
\left(\frac{\sum_{q=1}^{n}\alpha_{qj} B^{q}}{\sum_{r=1}^{n}\alpha_{rl}B^{r}}
\right)^2\,.
\end{equation}
In the above equation, 
$B\equiv\left(\sum_{j=1}^{m}\alpha_{sj}\alpha_{qj}\right)_{COF}^{T}$, 
where $T$ stands for transpose and $COF$ stands for the cofactor, and
$B^{s}\equiv \sum_{q=1}^{n}B_{sq}$ is the sum of elements in row $s$.
The power law solution $a(t)\propto t^{p}$ can be found to be 
\cite{Copeland:1999cs}
\begin{equation}
\label{fresult1}
p=\frac{16\pi}{M_{P}^2}\frac{\sum_{s}^{n}\sum_{q}^{n} B_{sq}}{\rm det~A}\,,
\end{equation}
where $A_{sq}=\sum_{j=1}^{m}\alpha_{sj}\alpha_{qj}$.

Now, we can read $\alpha_{sj}$ from the potential (\ref{braneexamp}). After
little calculation with the normalized $\alpha_{sj}$, we obtain the 
value of $p$ 
\begin{equation} 
p =\frac{3}{16}\ll 1\,.
\end{equation}
Again we find that there is no accelerated expansion. 

\subsubsection{Comments}\label{sss:cosmocomments}

In this section, we would like to comment on the conclusion that we 
cannot get power-law inflation (or quintessence) from the $3$-form 
induced potential.  The reason seems related to comments in 
\cite{Hellerman:2001yi,Fischler:2001yj}; exponential potentials consistent
with the constraints of supersymmetry are generically too steep.  Our
results, then, are consistent with a generalization to many fields of 
the work of \cite{Hellerman:2001yi,Fischler:2001yj} that a system cannot
simultaneously relax to a supersymmetric minimum and cause cosmological
acceleration.  Even though the models considered here do not necessarily
preserve supersymmetry, they are all classically of ``no-scale'' structure,
meaning that they all have vanishing cosmological constant and no potential
for the radial moduli.  So even the non-supersymmetric vacua have 
characteristics of supersymmetric cases.  Furthermore, the potential arises
from the supergravity Ward identity \cite{Tsokur:1996gr,D'Auria:2002tc},
which means it suffers from the same kind of constraints imposed by
the arguments of \cite{Hellerman:2001yi,Fischler:2001yj}.  Heuristically,
the vacua of our system give Minkowski space time, which is static, and
there is no way to accelerate into a static state.  In all our
examples we found that the moduli trajectories follow the late time attractor
towards Minkowski spacetime.

This sort of argument based on supersymmetry is readily generalized 
\cite{Giddings:2003zw}
to the Calabi-Yau models with 3-form fluxes that were studied in 
\cite{Giddings:2001yu}.  Indeed, the form of the bulk mode potential
(\ref{Vgeneral}) is identical, although the complex structure decomposition
of the metric will differ from case to case.  The key thing to note is 
that the overall scale of the internal manifold is always a modulus,
as if we set $\lambda_{1,2,3}=\lambda$.  In fact, it works out so that the
exponential prefactor gives the same $a\sim t^{1/3}$ evolution.  
The potential for brane modes should also be similar, at least for small
non-Abelian parts of the brane coordinates.
Considering a more complicated CY compactification is not the route to 
an accelerating universe. Again, this seems to be a feature of the 
broken supersymmetry.  

We should contrast this case to other work that does find inflationary 
physics in supergravity. In the 1980s, \cite{Kounnas:1985vh,Molera:1987ks}
found no-scale supergravities with inflation, but they specified the
potential to give slow-roll inflation.  The freedom to insist on inflation
does not exist in our string compactifications. 
More recently, other gauged supergravities have been 
found that can give at least a give few e-foldings of inflation
\cite{Kallosh:2001gr,Kallosh:2002gf,Kallosh:2002wj,Fre:2002pd}, but these
do not yet have a known embedding in string theory. These gauged 
supergravities are not of the no-scale type and have a cosmological 
constant. 

\section{Cosmology from Brane Motion}\label{s:branemotion}

As I mentioned in the last section, there is generally great interest
in embedding models of physics in string theory, simply because string
theory is a candidate theory of everything and has strict conditions for
self-consistency.  Indeed one of the main interests in self-dual flux
compactification has been the fact \cite{Chan:2000ms} that it gives a
natural string embedding for the warped compactifications of 
\cite{Randall:1999ee}.  
As it happens, there has been much work on cosmology due to brane physics
in braneworld and warped compactification models\footnote{The literature is
too extensive to list here; see \cite{Quevedo:2002xw} for a review.}.
In this section, I will give a
brief review of two models of brane cosmology which naturally embed in
self-dual flux compactifications.  The first is a model of a brane that
passes from a contracting to an expanding phase while avoiding a Big Crunch
singularity; the second is an inflationary scenario based on brane motion
in a warp factor.  

We should note that both of the scenarios described in this section
will be affected by nonperturbative physics.  In particular,
\cite{Kachru:2003sx} has shown that, when the volume modulus is fixed by
a nonperturbative superpotential (as discussed in \ref{sss:basicds} below),
the D3-brane positions will also develop a potential that could modify the
bounce of \ref{ss:bounce} or violate the slow-roll conditions in the model
of \ref{ss:pterm} reviewed below.

\subsection{A Big Bounce on a Conifold}\label{ss:bounce}

One of the challenges of cosmology has been developing an understanding of 
the Big Bang singularity; in a string theoretical context, the question 
would be whether string could propagate in such a singular background or 
if the singularity could be avoided by some mechanism.  
Within a braneworld compactification, we could imagine that brane motion
through a varying warp factor drives cosmology.  Then it is very easy to
guess that the brane could smoothly transition from a contracting phase to
an expanding one, completely avoiding a Big Bang singularity. 
Here we discuss a stringy implementation of this idea, due to 
\cite{Kachru:2002kx}.

The background is just a self-dual flux compactification on a CY 3-fold,
as in section \ref{ss:background}, either with O3-planes or O7-planes.
The details are fairly unimportant.  Now consider a conifold throat on 
the compact space, as in section \ref{ss:deformation}.  In the model of
\cite{Kachru:2002kx}, a D3-brane (which feels no force in a self-dual
flux compactification) moves into a conifold throat.  At first it feels
the metric (\ref{conifold}) and warp factor (\ref{throatwarp}), and the
induced metric on the brane shrinks due to the warp factor.  The motion
is therefore just what we describe below in \ref{p:d3migrate} and
\ref{aa:migration} (but with no potential).  
In terms of brane cosmology, the warp factor just becomes the scale factor
of the induced metric:
\be\label{inducedscale}
a(t) = e^A(r(t))\ .
\ee
This type of brane cosmology has been described in \cite{Kehagias:1999vr}.

If the metric (\ref{conifold}) persisted all the way through the throat,
the warp factor would reach a minimum and then diverge (see the trajectory
plotted in \cite{Kachru:2002kx}).  However, as we know from section
\ref{ss:deformation}, the conifold is deformed with a fixed deformation
parameter (\ref{deform}) and a fixed warp factor at the tip (\ref{tipwarp}).
At the tip, the radial coordinate $y$ is just like the radial coordinate of
a plane, and the D3-brane can smoothly roll through the tip and back
out of the throat.  The D3-brane trajectory was numerically integrated in
\cite{Kachru:2002kx}.  

One thing to note is that the metric that has a cosmological evolution
is the induced metric on the brane rather than the Einstein frame metric
(\ref{einstein4d}), which has the warp factor scaled out.  The reason
is that the masses of brane fields vary with the warp factor in the
Einstein frame (see (\ref{higgs2},\ref{higgs3})); \cite{Kachru:2002kx}
argue that brane observers should work in a frame with fixed particle
masses.  Rescaling to make the masses fixed just gives the induced 
metric, as demonstrated in a toy model \cite{Kachru:2002kx}.  In fact,
other than the addition of a conformal coupling for one matter field,
the toy effective Lagrangian of \cite{Kachru:2002kx} is just the
Einstein-Hilbert term plus two copies of the brane action
(\ref{higgs2}) with one modification: if our Higgs scalar $H$ is one
of the D3-brane positions, we should add a factor of 
$g_{mn}=e^{-2A}\tg_{mn}$ to its kinetic and mass terms.  Then, as
\cite{Kachru:2002kx}, we can rescale $g_{E,\mu\nu}$ to the induced metric
and the scalar to $\tilde H =e^{-A} H$.  These give fixed masses and a
time-varying Planck scale.

We should note that \cite{Kachru:2002kx} showed that several types of
backreaction are unimportant, such as the Jeans instability for energy on
the brane (during the time of the bounce) and the backreaction of
the D3-brane on the geometry.

\subsection{D7/D3-brane Cosmology}\label{ss:pterm}

Brane physics can also give inflation.  In this section, we consider a
shallow potential that can arise between D7-branes and D3-branes at a
distance, which, when coupled to gravity, can give inflation
\cite{Dasgupta:2002ew}.  This
model is an embedding of ``P-term'' inflation from $\N =2$ gauge
theory (see \cite{Kallosh:2001tm} and references therein).

Consider, as in \cite{Dasgupta:2002ew}, a D7-brane and a D3-brane
with three common spatial directions.  The two branes are mutually 
supersymmetric, so they experience no forces.  From the D7-brane WZ
action, there is a term 
\be\label{d3ind7}
\mu_7 \int C_4\wedge \mathcal{F}\wedge \mathcal{F}
\ ,\ee
so a D7 worldvolume instanton transverse to the D3-brane directions 
has the charge (and it turns out, the tension) of a D3-brane.  If the
worldvolume field strength is self-dual 
$F=\star F$ on the directions transverse
to the D3-brane, the entire configuration is supersymmetric (remember that
$\mathcal{F}=2\pi\ap F -B$).  As in \cite{Dasgupta:2002ew}, we take a
configuration that is not self-dual, which breaks SUSY and introduces
a force between the two branes.  This model is roughly dual to a 
D4/D6/NS5-brane model given in \cite{Herdeiro:2001zb}.

There are several ways to find the potential between the D-branes
\cite{Dasgupta:2002ew}.  For simplicity, we discuss only one.  In the
self-dual flux compactifications at the orientifold limit, D7-branes
are expected to be fixed at the O7-planes (because the dilaton-axion,
which would vary if the D7s and O7s separated,
is frozen \cite{Tripathy:2002qw}).  Therefore, we take
the point of view that the D3-brane moves in the field of the D7-brane,
which is at a fixed position.  The background includes a warp factor,
a nontrivial dilaton, and potential $C_8$ from the D7-brane, and
the worldvolume field strength sources $C_{2,4}$.  Putting everything 
together, the D3-brane experiences a potential at large distance
\be\label{d3ind7pot}
V =\mu_3 \left[ 1+c_7\ln\left(\frac{r}{r_0}\right)\right]\ ,\ee
where $c_7$ is a constant depending on the D7-brane 
worldvolume field strength and $r_0$ is a length scale introduced to
regularize the calculation \cite{Dasgupta:2002ew}.  This potential 
in fact reproduces the form of the one-loop potential in the gauge 
theory, as was noted in \cite{Dasgupta:2002ew}.  Also, when the D3-brane
gets within roughly a string length of the D7-brane, the open string 
spectrum develops a tachyon (in string perturbation theory).  Therefore
the slowly-rolling period of inflation would end in a steep potential.
The system would end as a D7/D3-bound state with supersymmetry
\cite{Dasgupta:2002ew}.

Of course, to discuss inflation, the D-brane worldvolume theories should
really be coupled to 4D gravity through a compactification.  Our self-dual
flux backgrounds are ideal candidates for compactifying this model,
as they can accommodate both D7-branes and D3-branes.  In fact, that is
the proposal of \cite{Dasgupta:2002ew}.  In the simplest case, take 
coincident D7-branes and O7-planes on a $K3\times T^2/\Z$ compactification.
Turning on $G_3$ flux is not absolutely necessary, but it is useful for
satisfying tadpole cancellation, fixing the complex structure of the 
compactification, and reducing the number of D3-branes.  As initial 
conditions for the inflation, \cite{Dasgupta:2002ew} take a D3-brane
in the center of the $T^2/\Z$ where all the D7-branes have a non-self-dual
field strength.  This is actually a de Sitter (or inflationary) state;
eventually quantum fluctuations push the D3-brane into the basin of
attraction of one group of D7-branes.  The D3 then rolls down the 
logarithmic potential as before.

\section{Nonperturbative Physics and de Sitter}\label{s:dS}

So far, I have only discussed the classical potential.  However,
we expect that the potential will have both $\ap$ and quantum mechanical
corrections.  In this section, we examine the consequences of 
nonperturbative physics; in particular, I will review the construction
of 4D de Sitter spacetime (dS) in self-dual flux compactifications
\cite{Kachru:2003aw}.  I will continue by analyzing how the cosmological
constant can decay in those models.

\subsection{Construction of dS}\label{ss:dsconstruct}

String models of dS have been difficult to find partly
because, as non-supersymmetric vacua, they are isolated points in
moduli space with all moduli stabilized.   There are also
no-go theorems against SUGRA compactifications to dS 
\cite{Maldacena:2000mw,deWit:1987xg}, so (like the self-dual flux
compactifications) we have to access purely string theoertic physics.
Notably, 
\cite{Silverstein:2001xn} constructed dS backgrounds by 
compactifying noncritial strings, and \cite{Berglund:2001aj} argued
that certain F-theoretic 7-branes should lead to de Sitter spacetimes. 
The first well-controlled dS compactification in critical string theory
was given by Kachru, Kallosh, Linde, \& Trivedi \cite{Kachru:2003aw}.
They asked whether nonperturbative corrections maintained the no-scale
potential of self-dual flux compactifications and whether supersymmetry
breaking could lead to a cosmological constant. Their compactification
has no unfixed moduli.  We review it here\footnote{Throughout this 
section, a dimensionful quantity given as a pure number should be
taken to be in string/Planck units.  For comparison to the literature,
we assume that the string and 4D Planck scale are the same.}.

\subsubsection{Basics of dS Construction}\label{sss:basicds}

To begin, take a general F-theory background in the orientifold
limit with 3-forms, as in section \ref{ss:background}.  In this case,
the integrated Bianchi identity (\ref{intbianchi}) can be written as
\be
\label{tadpole} 
0 = N_{\textnormal{D3}} - N_{\bd} + \frac{1}{2\kappa_{10}^2 \mu_3} 
\int_M H_3 \wedge  F_3 - \frac{\chi (X)}{24}\ . 
\ee
We will work in the orientifold limit of F-theory, so $X$ is a $T^2$ fibered
over an (orientifolded) CY 3-fold $M$.
The Euler number of
the CY fourfold $\chi(X)$ gives the effective negative D3-brane charge
in IIB of O7-planes and D7-branes wrapped  on 4-cycles of.  For typical
choices of $X$, $\chi(X)$ can be up to $O(10^5) $
\cite{Klemm:1998ts}. Start by choosing the fluxes to balance $\chi$ 
precisely, so that there are no D3- or \bd-branes.  
Although the story would not be altered significantly otherwise, take
$h^{1,1}(M)= 2$, so that $M$ has a single
K\"ahler modulus $\rho$ (with imaginary part a function of the volume
modulus $u$ as given in equation (\ref{rhodef})).  
We will use the superpotential formalism of section 
\ref{s:gvw} and the metric (\ref{warp2}) with the warp factor pulled out
of the internal metric as in that section.

To generate a nontrivial potential for $\rho$, as suggested in 
\cite{Giddings:2001yu}, \cite{Kachru:2003aw} considered
non-perturbative  corrections to the superpotential
(\ref{superpotential}).   Both wrapped Euclidean D3-branes 
(as in section \ref{ss:describe}) and gluino
condensation on the worldvolume of non-Abelian D7-branes generate
additional terms of the form 
\be\label{nonperturb} 
\delta W = A
e^{ia\rho}\ , \ee 
where the constants $A \sim O(1)$ and $a \sim
O(10^{-1})$.   For simplicity, take $\rho$ to be purely
imaginary,  $\rho = i \sigma$, and $A$, $a$ to be real.  
With a tree-level superpotential $\wo$, the
potential for $\sigma$ becomes 
\be
\label{nonpertV} 
V = \frac{aAe^{-a\sigma}}{2\sigma^2} \left\{  Ae^{-a\sigma}  \left(1+
\frac{a\sigma}{3} \right) + \wo \right\}, 
\ee
 and for suitable $\wo < 0$
there is a supersymmetric vacuum with $V_0<0$, implying that the
non-compact directions are AdS \cite{Kachru:2003aw}.  
For $|\wo| \ll 1$, the AdS
minimum lies at $\sigma_{cr} \gg 1$ where the SUGRA can be trusted and
$\ap$  corrections are small.  Note that these AdS vacua exhibit a 
nonperturbative restoration of SUSY since $\wo\neq 0$ breaks SUSY at
tree level.  Also, if $\wo =0$, the nonperturbative superpotential
breaks SUSY because the vacuum runs away to infinity.  
In fact, this nonperturbative superpotential seems to make
$\N=2\to 1$ in $K3\times T^2/\Z$ compactifications impossible, 
recalling the no-go theorem of \cite{Mayr:2000hh}.  Additional, more
complex, nonperturbative superpotentials have been considered in 
\cite{Escoda:2003fa}.

The final step in the construction is to add \bd-branes.  
The \bd-branes break supersymmetry and add some extra
energy \cite{Kachru:2002gs}, 
\be\label{deltaV} 
\delta V = \frac{Dp}{\sigma^2}; \ \ \ 
D = 2 \mu_3 g_s e^{4A}\sigma^{-1}\ ,  \ee   
where $\mu_3$ is the
brane charge as usual.\footnote{Actually, \cite{Kachru:2002gs,Kachru:2003aw} 
used a potential $\propto \sigma^{-3}$, which was corrected in 
\cite{Kachru:2003sx}.  The qualitative physics is not changed; however,
the results of figure \ref{f:disc1} and tables \ref{t:models} and 
\ref{t:tensions} were obtained using the original (incorrect) potential.  
Therefore, even though the conclusions are unchanged, the numerical results
are not quite correct.}
To minimize their energy, the \bd-branes migrate to a region of small
warp factor, a 
conifold tip, for example, 
so the energy density per \bd-brane depends,  through
equation (\ref{tipwarp}), on the fluxes.  Note, though, that the warp factor
at the tip is proportional to $\sigma$ so $D$ is independent of $\sigma$.  
For sufficiently fine-tuned
parameters, this additional term in the potential lifts the AdS global
minimum to a dS local minimum.  Note that global $\t F_5$ charge must
still be conserved via equation (\ref{tadpole}), 
so we have to adjust the fluxes
to balance the \bd-branes.  

We should note that this construction is given in the 4D effective theory
and not the 10D string theory.  In particular, the sources no longer
saturate the inequality (\ref{d3ineq}) and the 5-form no longer satisfies
(\ref{d3noforce}).  It would be very interesting to see how the 
nonperturbative energy densities would allow these compactifications to
satisfy the de Sitter generalization of equation (\ref{stringnogo}).

\subsubsection{Explicit dS Compactification}\label{sss:parameters}

Obtaining the vacua constructed
in \cite{Kachru:2003aw} requires fine tuning subject to several
constraints.  First, one must adjust the bulk fluxes so that $|\wo|
\ll 1$.   Moreover, finding a dS minimum requires fine tuning of the fluxes,
$K$ and $M$,  in the conifold throat to give an appropriate warp factor 
because a given value of $\wo$ tightly
constrains one's choice for $D p$ (see (\ref{deltaV})).  
For example, \cite{Kachru:2003aw} presented a
model with $\wo = -10^{-4}$ and an AdS minimum of $V_0=-2.00\times 10^{-15}$;
by adding one \bd-brane with $D = 3 \times 10^{-9}$, they achieved
a dS minimum of $V_0=1.77\times 10^{-17}$.  This is a
very special choice of fluxes indeed.  For $D p \lesssim 3 \times
10^{-9}$, the minimum is  at $V_0 < 0$ and is AdS, and, for $D p \gtrsim
7.5 \times 10^{-9}$, a local minimum no longer exists.  There are
additional constraints as well.  In \cite{Kachru:2002gs}  it is shown
that there is a classical instability of the \bd-branes 
if  $p/M \gtrsim 0.08$.
Furthermore, results from \cite{Giddings:2001yu} 
(see section \ref{ss:background}) rely on
approximations valid when  $K/(g_s M) \gtrsim 1/2$.

Taking into account that the tuning
parameters $K$ and $M$ are discrete, one might question if it is
possible to build such a model at all. Tuning would require the
existence of a ``discretuum''\footnote{The authors of
\cite{Bousso:2000xa} coined this term to refer to situations in which
a discrete spectrum is sufficiently dense to allow for an (almost)
arbitrarily fine tuning.  Our discretuum is not as finely spaced as
those in \cite{Bousso:2000xa}.}.  It is possible to do 
numerical searches in
order to map out the discrete landscape of dS vacua, as was first
reported in \cite{Frey:2003dm}.  Figure
\ref{f:disc1} shows the existence of a discretuum of states.
%
%
\begin{figure}
\begin{center}
\includegraphics{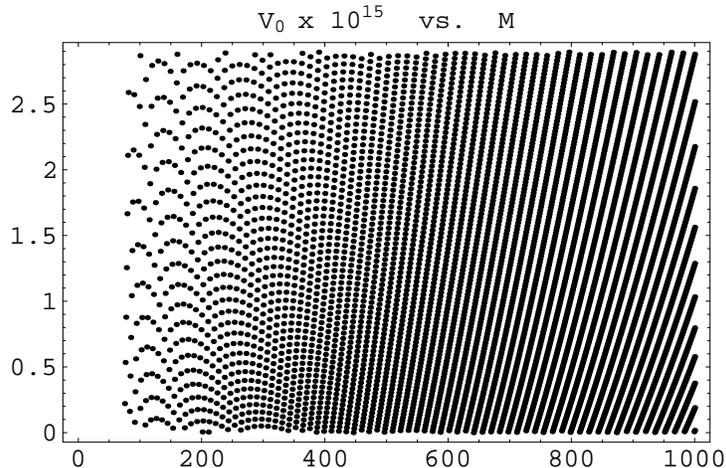}
\end{center}
\caption[Discretuum of dS States]{The possible dS vacua with 
$V_0$ for given $M$ illustrate the
density of states consistent with a discretuum.}
\label{f:disc1}
\end{figure} 
Here we have plotted the possible values $V_0$ that have a dS minimum
and can be achieved with integer fluxes for the parameters used in
\cite{Kachru:2003aw},  $\wo = -10^{-4},\  a = .1,\  A = 1,\ g_s=0.1,\ 
\k = \sqrt{\ap} =1$\footnote{In addition to tuning $V_0$ by varying 
the fluxes 
$M$ and $K$, one could, in principle, vary $\wo$ by adjusting the bulk 
fluxes.  While this would certainly increase the discretuum density, we 
leave $\wo$ constant as explicit calculation of $\wo$ in terms of bulk 
fluxes is prohibitively complicated.}.  
For each of the models studied
$K/g_s M > 1/2$.  Here we have allowed $M$ to
range from 75 to 1000.  The lower bound avoids the classical
instability (for $p \leq 6$).  As one goes to higher and higher values
of $M$, one must also increase the amount of induced D3-brane charge on
the D7-branes in order to satisfy (\ref{tadpole}).  This might require
adding more D7-branes and, thus, more degrees of freedom, which,
though massive, could cause problems when considering loop
corrections.  

The smallest possible value of $V_0$, for the parameters used in KKLT,
is $\mathcal{O}(10^{-20})$, a far cry from the desired
$\mathcal{O}(10^{-120})$.  In order to obtain a more realistic vacuum
energy, one must attempt to construct a background
with $|\wo| \sim \mathcal{O}(10^{-55})$.  While such a fine tuning seems 
improbable, with the number $b_3$ of 3-cycles sufficiently large, 
it is at least possible\footnote{One can estimate the smallest $|\wo|$ to 
have $\log(|\wo|) \sim -2b_3$.  
Thanks to S. Kachru for discussion on this point. This sort of estimate
is related to the statistics of string vacua as discussed in
\cite{Douglas:2003um,Ashok:2003gk}.}, 
if not particularly natural \cite{Bousso:2000xa}.

We have so far considered only a single conifold throat, as in 
\cite{Kachru:2003aw}.
However, a general CY 3-fold has many of them. In backgrounds
with multiple throats, the discretuum density increases 
dramatically.  One finds that (\ref{deltaV}) becomes,
\be\label{MdeltaV} 
\delta V = \sum_i \frac{D_i \ p_i}{\sigma^3}; 
\ \ \ D_i = 2\mu_3 e^{4A_i} \ .  
\ee
where $i$ labels the different
throats.  Clearly, by adjusting the fluxes in each individual throat,
one may tune $\delta V$ with greater accuracy.  
For a single throat we found $\mathcal{O}(10^3)$ 
configurations with a dS minimum.
Analogously, for 2 throats $(75 \leq M_1 \leq M_2,\ 75 \leq M_2
\leq 300)$ we find $\mathcal{O}(10^5)$ dS minima.   It is
easy to find configurations with $\mathcal{O}(10)$  conifold
throats\footnote{For example, in \cite{Greene:1996cy} a family of
quintics are constructed with 16 conifold singularities.}, leading to
an amazingly dense set of vacua.  The inclusion of a second throat
also lowers our minimum value  of $V_0$ by an order of magnitude.
Though this is nice, it does little good in helping build a model with
a realistic cosmological constant.  We suspect that even with the
addition of 10 or more throats the lofty goal of $V_0 \sim 10^{-120}$
would still be far out  of reach.

\subsection{Decays of Stringy dS}\label{ss:dsdecays}

De Sitter spacetime is a very unusual state of quantum gravity. 
The
observer-dependent horizon of dS, like a black hole horizon, yields a
thermal state with finite entropy.  Not only are the S-matrix
observables of string theory  precluded in $\Lambda > 0$ spaces
\cite{Hellerman:2001yi}, but, due to  the inevitable Poincar\'e
recurrences \cite{Dyson:2002pf}, all observables  are ill-defined
\cite{Banks:2002wr}.  
In fact,
it seems that dS cannot be a stable state in
any theory of quantum gravity.  The symmetries of dS are incommensurate with 
the discrete spectrum implied by finite entropy \cite{Goheer:2002vf}.  
Rather than a stable
vacuum, we should interpret dS instead as 
a metastable resonance whose lifetime, on
general entropic grounds, must be less than the recurrence time
\cite{Kachru:2003aw,Goheer:2002vf,Giddings:2003zw}.  In this section,
I address some of the decay modes of the dS string compactification
reviewed in \ref{ss:dsconstruct} above.  I will describe three decay
modes and then compare their decay rates.

\subsubsection{Decompactification Decays}\label{sss:decompact}

Generically, any string theoretic dS
compactification can decay and decompactify \cite{Dine:1985he,Giddings:2003zw}
because the 10D Poincar\'e invariant string vacuum is
supersymmetric and so has vanishing energy density.  In fact,
\cite{Kachru:2003aw} studied these decompactification decays within
the context of the self-dual flux compactifications.  We review
their results here, following the discussion in \cite{Frey:2003dm}.

Since the potential vanishes as $\sigma\to\infty$, the decompactification
decay can be described simply as barrier tunneling, which are well
understood even including gravitational back reaction
\cite{Coleman:1980aw}.
In terms of a
canonical scalar  field $\varphi =(\sqrt{3/2}\log\sigma)/\k$, the
Euclidean action is 
\bea S_E[\varphi] &=& \int d^4x \sqrt{g}
\left(-\frac{1}{2\k^2}R + \frac{1}{2}(\del  \varphi)^2 + V(\varphi)\right)
\nonumber\\ &=& -\int d^4x \sqrt{g} V(\varphi)\label{CDLaction} 
\eea
using Einstein's equations to get the second line.  The instanton
$\b\varphi$ is an O(4)-symmetric interpolation between the dS
vacuum at  $\varphi_{cr}$ and the supersymmetric vacuum at $\varphi =
\infty$.  When Wick rotated back to Minkowski signature, 
this gives an expanding
bubble of true vacuum inside the false dS vacuum.   The action of the
static dS vacuum is simply computed to give
\be\label{backgroundaction} 
S_0= 
-\frac{24 \pi^2}{\k^4 V_0} =
-\bm{S_0}\ , \ee 
where $\bm{S_0}$ is the entropy of the dS vacuum.  The
tunneling probability per unit volume
is given by the difference between the action
of the instanton solution  and the static dS vacuum \cite{Coleman:1980aw}:
\be\label{cdlprob1} P_{\textnormal{decay}} 
\sim e^{-S[\b\varphi] + S_0} \ .  
\ee
From eqn (\ref{CDLaction}), $S[\varphi] < 0$ for $V(\varphi) > 0$,
and the  resulting lifetime is exponentially less than the Poincar\'{e}
recurrence  time $t_r \sim e^{\bm{S_0}}$: 
\be\label{cdltime} 
t_{\textnormal{decay}} \sim e^{\bm{S_0} - |S[\varphi] |} < t_r 
\ee 
which is in line with the general arguments of
\cite{Goheer:2002vf,Susskind:2003kw,Giddings:2003zw}.

These tunneling instantons are often considered in the thin-wall
limit, in which there is a clear transition radius from one vacuum to
another.  For thin-wall instantons, we can say that there are 
well-defined ``bubble'' radius and  tension.  Appendix \ref{aa:grav}
shows how to calculate the thin-wall instanton action for given
initial and final cosmological constants and bubble tension, using the
formalisms of \cite{Coleman:1980aw,Banks:2002nm} and 
\cite{Brown:1987dd,Brown:1988kg}.  Of particular interest is the
existence of a critical tension (\ref{critical}) at which the bubble
fills half of the dS spacetime (the bubble radius is the dS radius).
Above the critical tension, the decay time rapidly approaches the
recurrence time; for example, if the decay has a final Minkowski vacuum,
the decay action is (see equation (\ref{kkltaction}) and \cite{Kachru:2003aw})
\be\label{Sdcmpt}
\Delta S_E = -\frac{S_0}{\left(1+T_c^2/T^2\right)^2} \ .
\ee
In fact, for decays, such as decompactification, that end in Minkowski
spacetime, 
\be\label{supercrit}
\frac{T}{T_c} = 
\frac{1}{\sqrt{4  V(\varphi_+) /3}} \int_{\varphi_+}^\infty d\varphi 
\sqrt{2 V(\varphi)} \
\geq \  1 \ ,
\ee
for any $V(\varphi)$ whose barrier width (in
string/Planck units for our normalization) is greater that $\sqrt{2/3}$.
Indeed, \cite{Kachru:2003aw} argue that their models have supercritical
decays to decompactification.

In addition, 
\cite{Kachru:2003aw} considered decompactification
decay via stochastic instantons
\cite{Hawking:1982fz}.  
(These are the opposite type of instantons when compared to the
thin-wall instantons that were considered in greatest detail in
\cite{Coleman:1980aw}.  However, it is clear \cite{Banks:2002nm} that
all types of instantons should contribute to the decay rate.)
Considering decays of general dS string
compactifications, \cite{Goheer:2002vf} and \cite{Giddings:2003zw}
also discussed thermal fluctuations using this reasoning. The thermal
instanton relies on thermal fluctuations to carry $\varphi$ to the top
of the potential, where it can then roll down the other side to the
true vacuum.  While the original discussions took this process to be 
homogeneous \cite{Hawking:1982fz}, \cite{Kachru:2003aw}
argued it should be interpreted as a horizon-sized fluctuation.  If
the potential has a broad, flat maximum at $\varphi_1$, the state
there is approximately dS with energy $V(\varphi_1) > V_0$ and entropy
$\bm{S_1}$.  The probability per unit volume for a thermal fluctuation
is given by the
difference in entropies between the fluctuation and equilibrium:
\be\label{HMprob}
P_{\textnormal{decay}} \sim e^{\bm{S_1} - \bm{S_0}}  \ . 
\ee
The decay time $t_{\textnormal{decay}} = (P_{\textnormal{decay}})^{-1}$ 
is again less than the recurrence
time $t_r$ and is also less than the thin-wall decay time (\ref{cdltime})
when the potential barrier is short and wide (precisely when the thin-wall
approximation is invalid).  In fact, the thin-wall tension is above
critical when the thermal decays begin to dominate the decay amplitude
\cite{Kachru:2003aw}.

\subsubsection{NS5-Brane Decays}\label{sss:ns5decay}

There are, of course, decay modes other than decompactification.  
For example, in any compactification
in which R-R 
fluxes contribute to the cosmological constant, D-brane instantons
change the fluxes and the cosmological constant.  This has been an
object of study in many papers, including
\cite{Brown:1987dd,Brown:1988kg,Bousso:2000xa,Feng:2000if,Maloney:2002rr}.
In the dS compactifications reviewed in \ref{ss:dsconstruct}, 
the $(0,3)$ part of the flux contributes to the cosmological constant in 
roughly that manner (although the situation is somewhat complicated due
to feedback into the minimum of $\sigma$).  We are interested in 
fluxes in a conifold throat.  These
do not contribute directly to the 4D cosmological constant, but, because
of tadpole cancellation (\ref{tadpole}), changing fluxes do change the
number of \bd-branes and therefore the cosmological constant.  

\paragraph{Conifolds and NS5-brane Instantons}\label{p:reviewkpv}

Remember from section \ref{s:gvw} 
that the geometry found at conifold points of 
self-dual flux compactifications was
first studied in the usual decoupling limit of string-gauge theory
dualities \cite{Klebanov:2000hb}.  The relevant gauge theory dual is a
duality cascade with an energy dependent effective number of
D3-branes; in the IR, most  of the D3-branes have been transformed
into 3-form fluxes.  The BPS domain wall that transforms the D3-branes
to fluxes was described by \cite{Kachru:2002gs}; 
it is a spherical NS5-brane that
carries D3-brane charges and bends over the $A$ cycle at the deformed
conifold tip \cite{Kachru:2002gs}.  As the NS5-brane moves over the
$A$ cycle, the D3-branes are absorbed into the background RR flux, and
the background NSNS flux jumps by a unit due to the NS5-brane charge.

Nonsupersymmetric gauge theories can be defined with $p$
\bd-branes at the tip of the conifold \cite{Kachru:2002gs}, 
as in the dS compactifications of \cite{Kachru:2003aw}.  Due to the
3-form flux background, the \bd-branes suffer a classical instability
to brane polarization (first discovered in \cite{Myers:1999ps}) as an
NS5-brane wrapping an $S^2$  in the $A$ cycle.  However, for $p\gtrsim
M/12$, the NS5-brane itself is unstable to collapse around the $A$
cycle, reducing the NS-NS flux and turning the \bd-branes into
supersymmetric D3-branes.  For smaller $p$, the decay of the NS5-brane
proceeds by tunneling; in Euclidean spacetime, the NS5-brane is
slightly polarized with \bd\ charge at infinity and bends around the
$A$ cycle to leave D3-branes at the origin \cite{Kachru:2002gs}.  This
process is illustrated in the top line of figure \ref{f:thinwall}.

For $p$ small enough, \cite{Kachru:2002gs} 
showed numerically that the thin-wall
approximation is very reasonable.  In that limit, the instanton
appears to be an NS5-brane wrapping the full $S^3$ of the $A$ cycle at
a fixed radius, as shown in the bottom line of figure
\ref{f:thinwall}.  The wrapped $F_3$ flux induces $M$ units of charge
in the NS5 worldvolume gauge theory which is canceled by the charge
carried by the ends of $M$ D3-branes.  The $p$ \bd-branes end on the
outside of the NS5-brane, and $M-p$ D3-branes end on the inside.  The
bubble tension in the effective theory  is just the NS5-brane tension
times the volume of the $A$ cycle. These instantons are clearly
related to the BPS domain walls.
%
%
\begin{figure}
\begin{center}
\includegraphics[scale=0.3]{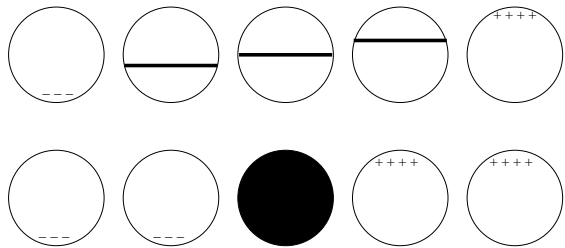}
\end{center}
\caption[Thin-wall NS5-brane Instantons]{\textbf{Top:} 
$p$ \bd-branes polarize
into an NS5-brane wrapping an $S^2$ on the $A$ cycle.  The NS5-brane
then slides to the opposite pole, becoming $M-p$
D3-branes. \textbf{Bottom:} In the thin-wall limit, the NS5-brane is
instead wraps the $A$ cycle at a particular Euclidean radius.}
\label{f:thinwall}
\end{figure} 

In the rest of this paper, we will focus on the thin-wall limit to
estimate the instanton bubble tension.  As has been argued strenuously
\cite{Banks:2002nm}, the thin-wall limit certainly does not describe
the full picture of the decays, but the other contributions to the
Euclidean path integral (such as thermal instantons at the other
extreme)  should only enhance the decay rate.  Therefore, we take the
point of view that the thin-wall limit estimates an upper limit for
the decay time.  As a consequence of the thin-wall limit, we may, as
in  \cite{Kachru:2003aw}, 
ignore the polarization of the \bd-branes in the initial
metastable state.  Before we turn to the modifications necessary for
including NS5-brane instantons in dS compactifications, let us also note
that our instantons are cousins of the supersymmetry-changing domain
wall bubbles found in \cite{Kachru:2002ns}, just as the AdS/CFT
instantons of \cite{Kachru:2002gs} are related to the BPS domain walls.
The following considerations were discussed in \cite{Frey:2003dm}.

\paragraph{Gravitational Backreaction}\label{p:backreact}

The first and most obvious correction due to the finite length of the 
conifold throat and overall compactification is that gravity  is no longer
decoupled, so we should include the effects of gravitation on the
decay time.  These effects are well known
\cite{Coleman:1980aw,Brown:1987dd,Brown:1988kg}; in appendix
\ref{aa:grav}, we work out the specific formula we need.  The decay
time, including gravity (but ignoring the large number of massive
fields in the  compactification), is $t_{\textnormal{decay}} 
\sim \exp[\Delta S_E]$,
where $\Delta S_E$ is the difference of the Euclidean actions for the
instanton and the initial background state as given in eqn
(\ref{bubbleaction}).  It depends only on the bubble tension, the
initial vacuum energy density, and the change in energy density.
Given two dS states from section \ref{sss:parameters}, we just need to
calculate the bubble tension and plug into (\ref{bubbleaction}).

\paragraph{NS5-brane Tension in Einstein Frame}\label{p:ns5einstein}

There are also modifications to the tension of the bubble.  The
easiest to calculate is an effect of working in the 4D Einstein frame.
Let us emphasize that we need to work in the 4D Einstein frame to use
the superpotential formalism of section \ref{s:gvw}, and this is also
the frame in which the potential has been calculated.  The Einstein
frame is also the frame used in calculating the instanton decay time.
It is easiest to get this by going to the NS5-brane action
\be\label{SeucNS5} 
S_E= \frac{\mu_5}{g_s^2} \int d^4x \sqrt{\det
g_{\mu\nu}} \delta \int d^3 x\sqrt{g_{S^3}}\ , 
\ee
where $g_{\mu\nu}$
is the 4D pullback of the 10D metric, $\delta$ is the delta
function at the radius of the bubble (with the determinant of the
metric included), and $g_{S^3}$ is the determinant of the metric on the 
$A$ cycle.  
With the 4D Einstein frame defined by equation (\ref{einstein4d}), the
NS5 action becomes 
\bea
S_E\!\!&=&2\pi^2 r^3 T_5 \ ,\nonumber\\
T_5 \!\!&\equiv& \mu_5 g_s e^{-9u} \left(\frac{z^{2/3}}{g_s
M}\right)^{3/2}( 2\pi^2) \left( b g_s M \ap\right)^{3/2} =
\frac{b^{3/2}z}{16\pi^3\ap{}^{3/2}g_s^{5/4}\sigma^{9/4}}.  
\label{NS5tension} \eea 
In
the first equality for the tension $T_5$, we have separated the
contribution from the conversion to Einstein frame, the warp factor,
and the volume of the $A$ cycle.  We have ignored the contribution to
the action from the NS-NS 6-form potential, which 
\cite{Kachru:2002gs} showed is
negligible in the thin-wall limit.  Heuristically this is because the
6-form potential only has two legs on the $A$ cycle and the 5-brane
fills the entire cycle, as shown in figure \ref{f:thinwall}.  However, the R-R
field strength $F_3$ gives a worldvolume anomaly that requires $M$
D3-branes to attach to the 5-brane.  Here, there are $p$ $\bd$-branes on the
outside and $M-p$ D3-branes on the inside.

\paragraph{Tension from Moduli}\label{p:modtension} 

The other correction we should make is due to the action for the moduli.
Since the moduli are fixed by the flux superpotential
(\ref{superpotential}), after the NS5-brane bubble changes the flux,
the VEVs of the moduli will change.  Therefore, we need to take into
account the rolling of the moduli to the new vacuum.  We will focus
on the deformation modulus $z$  of the conifold for the following
reasons.  First, it clearly changes  significantly when $K$ changes
(see (\ref{tipwarp})).  Also, for a noncompact conifold, $K$ does not
affect the dilaton or other moduli, so we would  expect that they
would be only minimally affected by a change of $K$ in the compact
case (the other moduli are typically fixed by fluxes on other cycles).
Also, \cite{Kachru:2003aw} 
have shown that the VEV of $\sigma$ does not change much
due to the presence of \bd-branes.  Therefore, since we expect $\tau$
and $\sigma$ to keep roughly the same values before and after the
decay, we expect that they will not roll much, and we will treat them
as constants.  There is actually a significant tree-level potential
for $g_s$ and $\sigma$ when $z$ is not at its VEV,  and we will consider its
effects below.  Nevertheless, we expect our estimate of the
contribution from $z$ not to be affected significantly by other moduli.  To
be conservative, one could multiply the contribution from $z$ by a
fudge factor, but we note that we are only making an estimate to begin
with, so we are not quite that careful.

To estimate the tension due to the rolling of $z$, we will assume that 
just inside 
the NS5-brane $z$
is in its original vacuum value outside of the NS5-brane and rolls
quickly to the new VEV inside.  This is probably not the exact
classical solution,  but we will use it and the thin-wall
approximation as an upper limit.  At tree level (where we are
working), we can write the action as 
\bea S_E(z) &=&
\frac{1}{\kappa_{4}^2}\int d^{4} x \sqrt{g_4} \left[ \mathcal{K}_{z\b
z}\del_\mu z\del^\mu \b z +\k^4 e^{\mathcal{K}}\mathcal{K}^{z\b z}
D_zW \b D_{\b z} \b W\right]\nonumber\\ &=&
\frac{2\pi^2}{\kappa_{4}^2}\int d\xi r^3\left[ \mathcal{K}^{z\b z}
\left( K_{z\b z} \del_\xi z- \k^2  e^{\mathcal{K}/2-i\omega}\b D_{\b
z}\b W\right) \left( K_{z\b z} \del_\xi \b z\right.\right.\nonumber\\
&&\left.\left.- \k^2
e^{\mathcal{K}/2+i\omega} D_{z} W\right) 
+ \k^2 e^{\mathcal{K}/2+i\omega}\del_\xi z  D_{z} W\right.\nonumber\\
&&\left. +\k^2
e^{\mathcal{K}/2-i\omega}\del_\xi \b z \b D_{\b z}\b W\right]\
,\label{SEz} 
\eea where $\omega$ is some phase (physically, we have to
take it so that the Euclidean action comes out positive because it
started positive definite).   As above, $r$ is the radius of curvature
of the bubble, while $\xi$ is the  radial coordinate corresponding to
proper distance.  This is clearly minimized when only the last two
terms contribute.  Taking the average K\"ahler potential in the
exponential, we get (up to numerical factors of order unity)
\be\label{zcontrib} T_z \approx 2\pi^2
e^{\langle\mathcal{K}\rangle/2} \left( |\Delta W| +|\Delta
\mathcal{K}||\langle W\rangle | \right) \ .\ee 
(This comes from the
definition of the covariant derivative and the chain rule.)  This
derivation is very similar to that of BPS domain walls and is also
used in \cite{Weinberg:1982id}.  Actually, it is easy to  generalize
this estimate to include other moduli, but we will only consider $z$
in the superpotential and K\"ahler potential.  We should note that
$\Delta W$ and $\Delta\mathcal{K}$ are calculated from the inside of
the NS5-brane (where $z$ is not in a vacuum state) to the new vacuum
on the interior of the instanton and not from the original vacuum to
the new vacuum.  Since we are just making an estimate,
$\langle\cdots\rangle$ will be an average value over the region of
variation of $z$.

The change in the superpotential is given entirely by the
superpotential of the conifold just inside the NS5-brane minus $W_0$.
This is because in the vacuum states, 
the $K$ and $M$ fluxes are $(2,1)$ 
forms and so do not contribute to the superpotential (see, for
example, \cite{Herzog:2001xk}).  Using the notation and conventions of
\cite{Giddings:2001yu,DeWolfe:2002nn}, we get 
\bea \Delta W &=& -W(z)
= -\frac{(2\pi)^2 \ap{}^{5/2}}{\kappa_4^8} \left( M
\mathcal{G}(z)-i\frac{K}{g_s} z\right)\nonumber\\ &\approx&
-\frac{(2\pi)^2 \ap{}^{5/2}}{\kappa_4^8} z \left( \frac{M}{2\pi i}\ln
z -i\frac{K}{g_s}\right) \approx  -i\frac{(2\pi)^2
\ap{}^{5/2}}{g_s\kappa_4^8}z\ ,\label{deltaW} \eea 
where $K$ is the
NS-NS flux on the inside of the bubble and $z$ is evaluated outside the
bubble.  This follows from the definitions (\ref{omegadefs})
and the relation from eqn (\ref{deform})
that $z(\mathnormal{outside}) \approx \exp(-2\pi/g_s M)
z(\mathnormal{inside})$.  To overestimate  $\langle W\rangle$, we will
take  \be\label{avw} |\langle W\rangle| \approx |\wo| +|\Delta W|\ .\ee
 
The K\"ahler potential is significantly more complicated, and, because
we are concerned with a modulus that lives at the bottom of a throat,
we need to take the warp factor into account.  We calculated this
effect already in section \ref{ss:deformation}, equation (\ref{kahlerwarp2}).
Then, using (\ref{kahlerkklt},\ref{kahlerwarp}) for $\rho,\tau$ and
assuming the complex structure gives small contributions to
$\mathcal{K}_c$,  we get roughly 
\bea
e^{\langle\mathcal{K}\rangle}\!\! &\approx& \frac{g_s}{16\sigma^3}\ ,\nonumber \\
\Delta\mathcal{K}\!\! &\approx& -\frac{\ap{}^3 (g_s M)^2}{2\pi\k^6}
|z|^{2/3} \left[ \frac{4\pi}{g_s M} e^{4\pi/3g_s M} +\log|z|^2 \left(
e^{4\pi/3 g_s M} - 1\right)\right]\! .\label{kdk}
\eea

The total bubble tension is therefore 
\bea T=T_5 +T_z
&\approx&\frac{b^{3/2}z}{16\pi^3\ap{}^{3/2}g_s^{5/4}\sigma^{9/4}}  
+\frac{2\pi^2 g_s^{1/2}}{4\sigma^{3/2}} 
\left\{ \frac{(2\pi)^2 \ap{}^{5/2}}{g_s\kappa_4^8}z
\right.\nonumber\\
&&\left.+\left( \wo + \frac{(2\pi)^2 \ap{}^{5/2}}{g_s\kappa_4^8}z\right)
\frac{\ap{}^3 (g_s M)^2}{2\pi\k^6}
|z|^{2/3} \left| \frac{4\pi}{g_s M} e^{4\pi/3g_s M}\right.\right.\nonumber\\
&&\left.\left. +\log|z|^2 \left(
e^{4\pi/3 g_s M} - 1\right)\right|\right\}\ .\label{tension} \eea
With the bubble tension in hand we are now in a position to calculate decay
rates.  However, before moving on we would like to take a closer look 
at subtle issues ignored in the above calculation.  

\paragraph{D3-brane Migration}\label{p:d3migrate}

The NS5-brane 
instanton bubble not only reduces the NS-NS flux $K$ and annihilates
\bd-branes, but it also leaves behind D3-branes.  If there are \bd-branes
in other throats, the D3-branes will feel an attraction and roll through the
bulk\footnote{We assume forces due to objects in the bulk, 
such as D7-branes with gluino condensation \cite{Ganor:1997pe}, 
can be ignored.  Thanks to S. Kachru and L. McAllister for discussion on 
this point.} and into the throat with the \bd-branes.  Eventually, they will 
annihilate
with the \bd-branes via tachyon condensation.  
If this migration is part of the instanton, then, in many cases, all the 
\bd-branes will be annihilated, leaving a Big Crunch spacetime with negative
energy density.  If there are more \bd-branes to start, the final state could
still be dS.

However, we argue that we should not consider the migration of the D3-branes
to be part of the instanton, but rather as a classical process that occurs
after the bubble nucleates.  Our logic is something like the discussion
of the bounce instanton of quantum mechanics in \cite{Coleman:1978ae};  
the instanton should only tunnel through the barrier to some energy slightly
lower than the initial state, and classical evolution should take over.
Typically, in the thin-wall limit, we just assume that the inside of the
instanton is just the final state.  In our case, though, we expect that
the D3 migration would not be well approximated at all by a thin-wall 
instanton because they are very far from the $\bd$-branes, so the potential is
very flat.  (Contrast this to the case for the $z$ modulus, where the 
gradient of the potential is Planck scale.)  This logic is consistent
with the discussion of thermal and related instantons in
\cite{Kachru:2003aw,Banks:2002nm}.

We expect the migration time to be similar to the bubble thickness
for the motion of the D3-branes in the Euclidean description of the 
instanton.  The migration times are larger than the bubble radius for the 
rest of the instanton, so we will treat the D3-brane
migration as a classical process.  In fact, the migration times are 
larger than the initial dS radius itself for the models we consider, 
which is the maximum bubble radius.

In appendix \ref{aa:migration}, we estimate the classical migration
time for a single D3-brane migrating from one tip to another.  For the
particular model we examine, $\Delta t_M \sim \mathcal{O}(10^{15})$ 
(in string
units).  As discussed below, decay times for the instantons we are
considering are much larger, $\mathcal{O}(\exp[10^9])$.  
Thus, in spite of the
fact that total migration will vary a great deal from model to model,
the total decay time is unaffected.

We should note that the classical D3-brane migration followed by 
D3/$\bd$ annihilation could leave a state with negative cosmological
constant.   In cosmology, if the spatial slices have
nonnegative curvature, the FRW constraint equation means that the 
universe cannot actually transition to a negative cosmological constant.
Instead, there is a Big Crunch singularity
\cite{Linde:2001ae,Felder:2002jk,Kallosh:2003mt}.  Though it is  
preferable to end in a dS state after the full decay, this is not
necessary as long as the initial instanton has a lifetime much longer
that the age of the universe.  Note that the instanton, however, ends
in a state of positive cosmological constant, so we avoid the concerns
raised by \cite{Banks:2002nm}.

As a diversion, let us note that 
D3/\bd-brane potential given in \ref{aa:migration} is quite flat, so these
instantons seem a good candidate for creating the initial conditions for 
slow-roll inflation.  In fact, \cite{Kachru:2003sx} argue that generally
D3/\bd-brane potentials with background warp factors do have inflationary
regimes (see \cite{Quevedo:2002xw} for a review of earlier work on
brane/anti-brane inflation).  However, the story is not that simple; the
K\"ahler potential (\ref{d3kahler}) in concert with the nonperturbative
superpotential generates a potential for the D3-brane moduli
\cite{Kachru:2003sx}.  Generically, the mass term from that potential
violates the slow-roll conditions of inflation, although a full analysis
would requite calculating the dependence of the nonperturbative superpotential
on the D3-brane positions \cite{Kachru:2003sx}.  As applied to the 
instantons, this nonperturbative potential could stabilize the D3-branes
from migrating to the \bd-branes.

\paragraph{Rolling Radius}\label{p:rollradius}

Now we should go back and examine the classical potential for $\sigma$
that arises because $z$ is away from its VEV.  We are in
a different regime from the studies of cosmology in section \ref{s:fluxcosmo}
and in \cite{Frey:2002qc} 
because we do not take $z$ to be slowly
rolling.  One point to address is that we cannot actually calculate
the K\"ahler potential  with warping for $z$ excited because it is not
clear if eqn (\ref{tipwarp})  would still hold as $z$ changes.
However, we will assume that it is valid since the starting and
ending points of our evolution are vacuum states for  some values of
the flux $K$.   The key point is that for instantons that go from dS
to dS, the boundary conditions on $\sigma$ mean it should not roll
much, so the following discussion does not apply.  What we are doing
here is comparing instantons with different boundary conditions, one
with $\sigma$ unchanged in the final state and one with
$\sigma\to\infty$ in the final state.

We make the comparison as follows.  The classical potential for
$\sigma$ and $z$ naturally pushes $\sigma$ to large radius  as long as
$z$ is not in its vacuum state (note that this potential is extremely
large compared to the nonperturbative potential (\ref{nonpertV}),  so we can
ignore the nonperturbative potential here).   We will make a very rough estimate
of the change in $\sigma$ while $z$ rolls to its vacuum.  If we
believe that $\sigma$ changes enough to get over the barrier of the
nonperturbaive potential before $z$ reaches its vacuum and the classical
potential vanishes, then we  expect dS to Minkowski decays -- mediated
by NS5-branes! -- will dominate over dS to dS decays.  This is
because the classical evolution should have a lower action.
Otherwise, the dS to dS decays will dominate,  at least in the
NS5-brane channel.  We will not say anything else about these dS to
Minkowski decays since they are less computationally tractable and are
somewhat redundant with other decays to large radius.

Now we can roughly estimate the potential for $\sigma$ and $z$.  As
in \cite{Giddings:2001yu}, we work assuming small $z$, which
implies that $\del_z W > (\del_z \mathcal{K}) W$, so we will consider
only the derivative of the superpotential.  As before, the $D_\rho W$ terms
cancel with $-3|W|^2$.  As a final approximation, we take
only the leading terms of the K\"ahler metric for $z$ small.  Thus, we
approximate the potential as \be\label{zpot} V = g_s^4 e^{-12u}
(2\pi)^5\frac{\ap{}^2}{\k^8 (g_s M)^2} \frac{|z|^{4/3}}{|\log |z|^2|}
\left| \frac{M}{2\pi}\log z+\frac{K}{g_s} \right|^2 \ .\ee We have
used  \be\label{kmetric} \mathcal{K}_{z\b z} = -\frac{(g_s
M)^2\ap{}^3}{18\pi\k^6} |z|^{-4/3}\log |z|^2 \ee as the K\"ahler
metric for $z$.  This is singular at $z=0$, but our evolution never takes
$z\to 0$.

To get a very rough estimate of the change in radial modulus $u$
(remember that $\sigma = e^{4u}/g_s$) while
$z$ changes, we approximate that the proper distance in the $u$
direction of moduli space is proportional to the proper distance moved
in the $z$ direction of moduli space.  The proportionality constant is
given by the directional derivative (in the moduli space orthonormal
frame) of the potential.  Using the K\"ahler metric (\ref{kmetric}) to
get the orthonormal frame, we find that \bea \Delta{u} &\approx&
\frac{\Del_{\hat u}V}{\Del_{\hat z}V}  \frac{\sqrt{\mathcal{K}_{z\b
z}} \Delta z}{\sqrt{12}}\label{deltau1}\\ \Delta u &\approx&
\frac{(g_s M)^2\ap{}^3}{18\pi\k^6} |z|^{2/3}  |\log |z|^2| \left(
e^{2\pi/g_s M}-1\right)\label{deltau} \eea up to factors of order
unity.  We have used $\sqrt{\mathcal{K}_{z\b z}}  \Delta z$ for the
proper distance in the $z$ direction.  The factor of $\sqrt{12}$ in
(\ref{deltau1}) comes from the normalization of $u$.

From the nonperturbative potential \cite{Kachru:2003aw}, we expect that $\Delta
u$  only needs to be $\gtrsim 0.1$ for the Minkowski decay to
predominate, which is achieved
by $z\gtrsim 10^{-3}$.   As it turns out, we will mainly be interested
in cases with smaller $z$, so we will not consider the 5-brane
mediated dS to Minkowski decays any further.

\paragraph{Thermal Enhancements}\label{p:thermal}

Due to the fact that dS has a temperature, we might expect that the
5-branes that make up our instantons should have some nonzero entropy.
Since the exponential of the entropy gives a density of states, the
decay time should be reduced by a factor
$\exp[-\bm{S}(\textnormal{NS5})]$.  This argument was first given in
\cite{Feng:2000if}.  There it was argued that the  brane instantons
probably are out of thermal equilibrium with any matter or radiation
in the cosmology, so they should have a temperature corresponding to
the dS temperature.  However, whether the temperature should be the
initial dS temperature, final dS temperature, or the geometric mean
was undetermined.  It is now clear \cite{Fabinger:2003gp} 
that the brane has a well-defined temperature because it corresponds to
accelerating observers in the two dS spacetimes.

We will, however, neglect this effect.  The bubble temperature
is just the inverse radius, 
$T=1/(2\pi r)$ \cite{Fabinger:2003gp}.
Therefore, the temperature is not high enough to excite the ``Kaluza-Klein''
modes of the bubble much, and the entropy would access only the zero-mode
quantum mechanics.  We expect that the enhancement factor would be
relatively weak, therefore.  

\subsubsection{Throat-to-throat Decays}\label{sss:t2t}

Another particularly simple decay mode occurs in models with multiple 
conifold throats.  The potential energy of a \bd-brane 
is proportional to $e^{4A}$, the inverse warp factor given by eqn 
(\ref{tipwarp}),
which is locally minimized at the tip of each throat.  However, the 
energy is lower still at the tip of other throats with smaller $z$.  
\bd-branes can therefore tunnel from one throat to another.  
On the other hand, $A \sim 0$ in the bulk, 
presenting a substantial potential barrier through which to tunnel. 
These instantons are similar to the glueball decays considered in 
\cite{Dimopoulos:2001qd,Dimopoulos:2001ui}.

As in previous examples, we consider models with two KS throats.  
The \bd-brane portion of the total potential is initially 
(cf. equation (\ref{MdeltaV}))
\be
\label{t2tdeltaV}
\delta V = \frac{2 \mu_3}{\sigma^3} e^{4A}(\t r_1) 
\ p_1 + \frac{2 \mu_3}{\sigma^3} e^{4A}(\t r_2) \ p_2 \ .
\ee 
After the tunneling occurs, the form of $\delta V$ is unchanged except for
$p_1 \to p_1 + 1$ and $p_2 \to p_2 - 1$.  These decays have little
effect on $\sigma$, and thus $\sigma$ will be treated as a 
constant throughout 
this calculation.  

To find the decay rate, we compute the instanton tension from the
Euclidean brane action in the thin-wall limit using \cite{Coleman:1980aw}:
\be\label{CDLtension}
T = (2\pi\sqrt{\sigma})^{3/2}g_s^{1/4}\ap\int_{\t r_1}^{\t r_2} dr \ 2
\sqrt{\frac{\mu_3}{\sigma^3} [e^{4A}(r)-e^{4A}(\t r_1)]} \ .  
\ee 
The prefactor is from the conversion between $r$ and a canonically 
normalized scalar in the 4D Einstein frame.
We then plug $T$ and $\Lambda_\pm = \kappa_4^2 (V + \delta V_\pm)$ into 
eqn (\ref{bubbleaction}) to obtain the ``throat-to-throat'' decay
time $(t_{\textnormal{decay}}^{T2T})$.  

\subsubsection{Calculated Decay Times}\label{sss:calculated}

Here we give some numerical results illustrating the relative 
speeds of the different decay modes.  Table \ref{t:models} lists
some specific models that we can use to compare decay rates; the models
have \bd-branes in two conifold throats.  The reader should note that
here $\Lambda_-$ is the cosmological constant for a vacuum with just the
\bd-branes in throat 2 and $|\Delta\Lambda|$ is the additional contribution
from the \bd-branes in throat 1.  For the NS5-brane decays, we take
the NS5-branes to form in throat 1 because the decay occurs faster for
smaller $z$.  For the throat-to-throat decays, the brane will tunnel 
from throat 2 to throat 1.

\begin{table}[t]
\begin{center}
\begin{tabular}{|c|c|cccc|cc|}
\hline 
Model & $p_1,p_2$ & $K_1$  &  $M_1$  &  $K_2$  &  $M_2$  
 &  $z_1 \times 10^{17}$  & $z_2 \times 10^{5}$  
 \\
\hline
\hline  
1 & 1,1 & 9  & 15 & 3 & 19 & 4.2 & 4.9 \\  
2 & 1,1 & 9  & 15 & 4 & 26 & 4.2 & 6.3  \\ 
3 & 1,1 & 9  & 15 & 9 & 69 & 4.2 & 28   \\ 
4 & 1,5 & 9  & 15 & 8 & 51 & 4.2 & 5.2   \\
5 & 1,5 & 9  & 15 & 13 & 91 & 4.2 & 13  \\ 
\hline
\end{tabular} \vspace{0.1in}
\begin{tabular}{|cc|}
\hline
$|\Delta \Lambda| \times 10^{31}$ &  $\Lambda_-  \times 10^{17}$  \\
\hline\hline
 3.9 & 69 \\  
 3.9 & 2.7 \\ 
 3.9 & 4.5 \\ 
 3.9 & 4.5  \\
 3.9 & 7.6 \\
\hline 
\end{tabular}
\end{center}
\caption[Some Models of dS]{\label{t:models}Models and Cosmological Constants}
\end{table}

The tensions (compared to critical) and decay times for the three types of
decays discussed in this section are given in table \ref{t:tensions}.  We
should note that the NS5-brane mediated decays occur much more rapidly
than the other types because they have subcritical tensions.  This fact is
due to the small values of the deformation parameters in throat 1, which
can be attributed to some extent to our choice of models.  If we consider
a model with \bd-branes in only one throat, this trend does not continue.
In particular, for 3
\bd-branes sitting at the tip of a throat with 
$K = 12,\ M=87$, one finds that the probability per unit volume for
NS5-brane mediated decay is $P \sim \exp (-10^{19})$.  Radial decays to
decompactification are much faster, $P \sim \exp (-10^{17})$. 
Moreover, as discussed in \cite{Banks:2002nm}, since all single throat
decays will have $\Lambda < 0$ in the final state, the instantons
mediating these decays might not exist.  It is for this reason we have
chosen to focus on models with 2 throats, which, after the initial
decay, have $\Lambda >0$.  

\begin{table}[t]
\begin{center}
\begin{tabular}{|c|ccc|ccc|}
\hline
Model & radial:& $T/T_c $ & $\ln (t_{\textnormal{decay}}) 
\times 10^{-18}$
      & NS5:& $T/T_c$  
      & $\ln (t_{\textnormal{decay}}) \times 10^{-9}$
      \\
\hline
\hline  
1 & & 1.8 & 0.32 & &0.163 & 0.66 \\
2 & & 7.7 & 8.9  & &0.164 & 86   \\
3 & & 5.9 & 5.2  & &0.164 & 40   \\
4 & & 2.9 & 1.1  & &0.163 & 3.7  \\
5 & & 4.6 & 3.1  & &0.164 & 18   \\
\hline
\end{tabular}\vspace{0.1in}
\begin{tabular}{|ccc|}
\hline
T2T:& $T/T_c $  
      & $\ln (t_{\textnormal{decay}}) \times 10^{-18}$ \\ 
\hline
\hline 
 & 24970 & 0.35 \\
 & 24257 & 8.9  \\
 & 16512 & 5.2  \\
 & 34054 & 1.1 \\ 
 & 33517 & 3.1  \\
\hline
\end{tabular}
\end{center}
\caption[Decay Times for 3 Decay Modes]{\label{t:tensions}Tensions and 
Decay Times}
\end{table}

\chapter{Future Directions}\label{c:future}
In conclusion, I would like to point out very briefly a few interesting
directions for future study.  There is no single theme to this chapter,
so we will jump into it.

\section{Relation to Other Compactifications}\label{s:othercompact}

As we pointed out in section \ref{ss:tdual}, T duality relates type IIB
self-dual flux compactifications to other interesting types of 
compactifications.  For example, a single T duality should give M theory
compactified to 4D with 4-form flux, which was discussed in 
\cite{Becker:2000rz}.  However, that relationship has not been confirmed
in detail.  Also, \cite{Kachru:2002sk} demonstrated that the mirror of
an $\N=2$ example is M-theory on a CY 3-fold times a circle.
Both of these dualities
build a bridge between IIB self-dual flux compactifications and 
compactifications of M-theory on manifolds of $G_2$ holonomy (the literature
is very extensive; see \cite{Acharya:1998pm,Acharya:2000gb,Atiyah:2001qf}
and \cite{Acharya:2002gu} for a review).  The case of one T duality would
seemingly make contact, at least in $\N=1$ cases, with the very interesting
case of fluxes on a $G_2$ manifold 
\cite{Acharya:2000ps,Beasley:2002db,Acharya:2002kv}.

The case of two T dualities is also of great interest.  We saw that the
T duals are type C compactifications, so the S duals are type A 
compactifications, either in type IIB or heterotic string theory.
Indeed, type A compactifications have been of great interest recently,
with \cite{Cardoso:2002hd,Becker:2003yv,Gauntlett:2003cy,Becker:2003gq,
Cardoso:2003af} studying both
geometric aspects and the effective theory.  In particular, 
\cite{Becker:2002jj,Becker:2003gq} have 
developed a superpotential analogous to the one
discussed in section \ref{s:gvw}.  

There is another intriguing aspect of these particular type A 
compactifications.  Starting with a self-dual flux compactification,
two T dualities and an S duality interchange the dilaton and the
volume modulus of the $T^2$ that was dualized \cite{Becker:2002sx}.
Since the dilaton is generically fixed in the original self-dual flux
compactification, this implies that the type A compactification has a
$T^2$ figure with a fixed volume.  The feature of a fixed-volume torus
resembles the ``nongeometric'' 
compactifications of \cite{Hellerman:2002ax}.  In those 
compactifications, a $T^2$ has monodromies in both $SL(2,\bm{\mathbbm{Z}})$
factors of its T duality group around different singularities.  These are
stringy implementations of Scherk-Schwarz compactifications
\cite{Scherk:1979zr,Dabholkar:2002sy}.  The monodromies in the
nongeometric models are actually the same twists as those
in the type A duals of self-dual flux compactifications, and the $H_3$ 
flux gives the monodromies in the other $SL(2,\bm{\mathbbm{Z}})$ factor.
The difference is that the geometric twists and $H_3$ flux 
in \cite{Hellerman:2002ax} both cover the same $T^2$, while they are 
on different tori in \cite{Becker:2002sx,Kachru:2002sk}.  Still, it would
be very interesting to see how closely the self-dual flux compactifications
are related to nongeometric compactifications.

The moduli space of nongeometric constructions 
also has a limit in which
they appear to be asymmetric orbifolds \cite{Narain:1987qm}.  
This correspondence was checked
by matching the SUGRA spectrum on the nongeometric compactification with
the perturbative spectrum of the asymmetric orbifold in two simple 
examples \cite{Hellerman:2002ax}.  This fact suggests mildly that the
self-dual flux compactifications (and their twisted T duals) may also be
related to asymmetric orbifolds, which would further link string vacua.
There is one further piece of evidence to suggest this link.  Perturbative
string theories with all degrees of supersymmetry were constructed in
\cite{Ferrara:1989nm} by an asymmetric orbifold of free fermions on the
worldsheet, and these exhibit properties similar to the self-dual flux
compactifications.  For example, the BPS spectrum also appears to have 
similarities to a particle in a magnetic field \cite{Kounnas:1998hi}. 
In addition, \cite{Kounnas:1998hi} showed that string dualities 
\cite{Sen:1995ff} map
their perturbative asymmetric orbifolds to models with a fixed dilaton.

\section{Phenomenology}\label{s:phenomenology}

There are many aspects of the phenomenology of self-dual flux models that
remain to be understood; no detailed models have yet been described.
Here we list some points that should be addressed as further progress is
made.
\begin{itemize}
\item As discussed in section \ref{s:eom}, the Kaluza-Klein decomposition
of the supergravity fields is nontrivial once the warp factor is included, 
even for the zero modes.  Therefore, the superpotential 
(\ref{superpotential}) and K\"ahler potential (\ref{kahlerwarp}) may not
be correct when we take into account the variation of the warp factor
with the geometric moduli.  Clearly, the full 10D equations of motion
need to be solved to address this thorny issue.  Some progress is 
currently being made in this direction \cite{gm}.

\item The worldvolume fields of D7-branes have not been studied carefully
so far.  We expect that some of the open string moduli, the D7-brane
positions, will be fixed by the flux \cite{Tripathy:2002qw}, and we also
anticipate that the gauge sector will have gluino condensation
\cite{Kachru:2003aw}, but the worldvolume theory needs to be worked out in
detail.

\item Where does the Standard Model live?  Originally, 
\cite{Giddings:2001yu} envisioned D3-branes at the bottom of a deep 
conifold throat, with the warp factor generating a large hierarchy
as in \cite{Randall:1999ee}.  The only difficulty is that the D3-branes
are free to move through the entire compact manifold.
However, \bd-branes do feel a potential and would be stuck at the bottom
of a throat, as in \cite{Kachru:2003aw}.  With \bd-branes, the problem
is to have large enough fluxes to allow enough \bd-branes for the
Standard Model.  Also, since \bd-branes break supersymmetry, they raise
the issue of the cosmological constant, as in section \ref{s:dS}.  Since the
hierarchies required are different, there would have to be different
throats for the cosmological constant and the Standard Model.
Embedding other types of singularities in self-dual flux compactifications
might also be interesting, such as the non-Abelian orbifold of
\cite{Berenstein:2001nk} that puts the Standard Model on a single D3-brane.
One example that has chiral fermions in the D3-brane spectrum has
been given by orbifolding the $T^6/\Z$ orientifold by an additional
$\Z\times \Z$ \cite{Cascales:2003zp}; chiral fermions also appear when
D9-branes are included \cite{Blumenhagen:2003vr}.

\item Stringy corrections to the 10D background were considered in 
\cite{Becker:2002nn}.  However, the full form of the corrections, even
at leading order in $\ap$, are not known in the presence of 3-form and
5-form flux.  Therefore, for our understanding of compactifications with
flux to advance to the same level as those without, we need to find the
full stringy corrections.  (See appendix \ref{aaa:corrections} for a 
discussion.)  In a related direction, loop and Kaluza-Klein
threshhold corrections are interesting for the same reasons.

\item Although \cite{DeWolfe:2002nn} calculated explicitly the gravitino 
masses, they discovered that the fluxes allowed by condition (\ref{imsd})
cannot break supersymmetry in the D3-brane gauge theory at tree level.
Indeed, \cite{Grana:2002nq} (which analyzed SUSY breaking in the
non-Abelian orbifold of \cite{Berenstein:2001nk}) showed that self-dual
fluxes, whether supersymmetric or not, do not affect the gauge theory
at tree level.  However, they would affect the gauge theory on \bd-branes.
A different approach (that avoids the explicit SUSY breaking of anti-branes)
is to consider loop and string corrections to the D3-brane action that
could communicate SUSY breaking from the bulk gravitational sector to the
brane Standard Model.  Although the corrections to the brane action are
not known completely (see \ref{aaa:corrections}), an analogy we discuss
below suggests that string corrections are as important as loops.

\end{itemize}

As the list above suggests, there are many investigations that could lead
to a better understanding of self-dual flux compactification phenomenology.
Below, we turn our attention briefly to a calculation that demonstrates
the importance of string corrections to the brane action.  Because the
corrections due to 3-forms are not known, we cannot see how the fluxes
break SUSY on D3-branes, but we briefly review an analogous communication
of symmetry breaking from the gravitational sector to the brane.

\subsection{An Analogy}\label{ss:analogy}

One of the interesting problems left to solve regarding self-dual flux
compactifications regards the communication of supersymmetry breaking
from the supergravity sector to fields on D3-branes.  No matter what
supersymmetries the 3-form fluxes preserve, as long as they satisfy
(\ref{imsd}), the D3-brane gauge theory retains the $\N =4$ structure
at the classical level \cite{DeWolfe:2002nn}.  Of course, we expect
that loops would mediate supersymmetry breaking in the brane sector,
but $\ap$ corrections to the action at string tree level could also.
Unfortunately, $\ap$ corrections to the D3-brane action involving the
3-forms are not known, so the necessary technology for studying that
effect is not available yet.

What we can do is look at an analogous problem in which symmetry breaking
occurs in the bulk but can be communicated to brane fields either by
loops or $\ap$ corrections.  The model is a braneworld, as before.
The following results were presented in \cite{Frey:2003jq}.

\subsubsection{Lorentz Violating Braneworlds}\label{sss:violating}

Despite the successes outlined in section \ref{s:dS}, there is still 
not a consensus on the explanation of the size of the cosmological
constant.
The review \cite{Weinberg:1989cp} considers many different
possible resolutions of the cosmological constant problem, including 
the possibility of ``tuning'' by a scalar field that reduces the effect of 
the vacuum energy density.  The conclusion is that such a method does not
work in models with standard 4D physics.  Nonetheless, self-tuning has 
enjoyed a revival of interest with the discovery by 
\cite{Arkani-Hamed:2000eg,Kachru:2000hf} that vacuum energy density on a
brane can be translated into curvature of a bulk scalar.  Because of 
singularities in the bulk, however, these models require fine-tuning to
reproduce 4D gravity and
therefore do not excape the no-go theorem stated above 
\cite{Csaki:2000wz}.

Seemingly, the way around this difficulty is to place the singularity 
behind an event horizon; additionally, the spacetime metric can be written
in a 5D Schwarzschild-like form
\be\label{schwarzschild}
ds^2 = -h(r) dt^2 +(kr)^2 d\vec x^2 +h^{-1}(r) dr^2 \ ,\ee
where $k$ is a natural mass scale of the 5D compactification 
(for $h(r) = (kr)^2$,
this is the same as \cite{Randall:1999ee,Randall:1999vf}.).  These types of 
solutions with a charged black hole background were first studied in 
\cite{Kraus:1999it,Csaki:2000dm,Csaki:2001mn,Csaki:2001yz} and generalized in 
\cite{Grojean:2001pv,Nojiri:2001ae}.  

Even though the full metric does not have $SO(3,1)$ symmetry, 
\cite{Kraus:1999it,Csaki:2000dm,Csaki:2001mn,Csaki:2001yz} 
argue that these asymmetrically
warped models evade limits on standard model Lorentz violation because
brane fields feel an $SO(3,1)$ symmetry as long as the brane stays at a 
fixed position.  The only effect at the field theory tree level 
would be to alter the speed of gravitons.  

Quantum mechanically, Lorentz violation
can be communicated to the Standard Model by loops \cite{Burgess:2002tb}
in a manner relatively independent of the model details.
As it turns out, any Lorentz violation in the Standard Model is subject
to tight limits, so the loop calculations can put limits on
gravitational Lorentz violation \cite{Burgess:2002tb}.
In particular, many 
unobserved effects, such as \v{C}erenkov radiation by charged particles 
in vacuum and photon decay, can occur if the 
maximum attainable velocities (MAVs -- the speed of light in Lorentz
invariant theories) of photons and fermions
are different \cite{Coleman:1998ti}\footnote{We do not consider interactions,
such as electromagnetic muon decay, that also violate lepton number.}.  
In fact, vacuum \v{C}erenkov radiation and photon decay are so efficient 
that we should observe no particles above threshhold for those interactions
(see, for example \cite{Jacobson:2001tu}).  These two interactions give
the constraint that the magnitude of the difference of photon and electron
MAVs should not be greater than $10^{-16}$ (in units with the speed of light
equal to unity).  A demonstration of this result, along with a review of 
many dispersion relation tests, is given in 
\cite{Jacobson:2002hd}\footnote{
There are more stringent constraints.  For example, ignoring the
parton structure, the proton MAV should differ from the photon MAV by no 
more than $10^{-22}$ \cite{Jacobson:2002hd}.  Atomic spectroscopy experiments
measuring spatial anisotropy of nuclear dipole and quadrupole couplings
give a similar (indirect) bound of $10^{-22}$ \cite{Lamoreaux:1986xz}; 
this is the limit used by \cite{Burgess:2002tb}.
Additionally, other Lorentz
(and even CPT) violating effects in QED have been considered by many authors
(see, for example, \cite{Kostelecky:2001mb,Bluhm:2001ms}).}.

To see how $\ap$ corrections can play the role of loops in communicating
Lorentz violation to the brane fields, we need a specific solution
of string theory that breaks 4D Lorentz invariance.  
We take an asymmetrically warped model as in
\cite{Kraus:1999it,Csaki:2000dm,Csaki:2001mn,Csaki:2001yz}:
\be\label{warpasym}
ds^2 = -e^{2A(x^m)} dt^2 +e^{2B(x^m)} d\vec x^2 + 
g_{mn}(x^m) dx^m dx^n
\ .\ee
Here the vector symbol $\vec{x}$ indicates just the spatial 
coordinates\footnote{In this section and appendix \ref{aa:lorentz}
only, I will also use $i,j$ indices
to indicate the three noncompact spatial directions.}.
An appropriate 5-form field strength can both solve the equations of
motion and stabilize the position of a D3-brane, which feels potential
$-\mu_3(e^{A+3B}/g_s -C_4)$ \cite{Frey:2003jq}, at least locally 
at the D3-brane position.  There are of course other ways to stabilize
the D3-brane position, but the only feature of concern is that the warp
factors have nonvanishing derivatives at the brane position.  That is,
taking our coordinates so that $A=B=0$ at the brane $x^m=0$,
\be\label{warpexpand}
A=a_m x^m +\frac{1}{2} a_{mn}x^m x^n+\cdots\, ,\ 
B=b_m x^m +\frac{1}{2} b_{mn}x^m x^n+\cdots\ .\ee

\subsubsection{Kinetic Terms at $\mathcal{O}(\ap{}^2)$}\label{sss:apkinetic}

Because the most stringent limits are from unobserved physics dependent
on differing MAVs for photons and fermions, we concentrate on finding
just the Lorentz violating kinetic terms.  Unfortunately, $\ap$
corrections are unknown for worldvolume fermions, but we can use the
worldvolume scalars to get an order of magnitude estimate.

To get the modified MAVs for the scalar fields on the brane, we just need
to calculate the Lorentz violating kinetic terms that arise in the 
corrected action (\ref{corrected}).  We can write these all as
$(\cdots)\vec{\del}X\cdot\vec{\del}X$ (because $\del_0 X\del_0 x$ differs
by a Lorentz invariant term).  Knowing this allows us to simplify our
calculation greatly; we need keep only terms up to $\mathcal{O}(\del X)^2$
and can ignore terms in which $X$ appears without a derivative as well as
$\del^2 X$ terms.  Also,
since the background would be Lorentz invariant if the warp factors
were equal, $A=B$, we can write $A=B+\Lambda$ and take only terms in which
$\Lambda$, the Lorentz violating function, appears.  To get the best limits
on Lorentz violation, we keep only terms that are linear in $\Lambda$.

The key to our results is that the extrinsic curvature has terms that are
zero-th order in the brane fluctuations that come from $\Gamma^m_{\mu\nu}$.  
In our background, the extrinsic curvature is
\bea
\Omega^m_{00} &=& a^m -2a_n\del_0 X^n\del_0 X^m+\cdots\ ,
\nonumber\\
\Omega^m_{0i} &=& -a_n\del_i X^n \del_0 X^m -b_n
\del_0 X^n \del_i X^m+\cdots\ ,\nonumber\\
\Omega^m_{ij} &=& - b^m \delta_{ij} -2b_n \del_{\left( i\right.}
X^n\del_{\left. j\right)} X^m+\cdots\ .\label{omega}
\eea
We use standard notation regarding symmetrization of indices with a weighting
of $1/2$.  Also, to reduce proliferation of indices, we replace the normal
bundle index $\hat a$ with the coordinate index $m$ since they are the
same at lowest order in perturbation theory.  We have still included all
the appropriate terms from the expansion of the normal bundle vielbein, 
however.  Additional zero-th order terms appear in the Riemann tensor part of
$\b R^{mn}$.  We leave those, along with the details of the rest of the
calculation to the appendix \ref{aa:lorentz}.

After much algebra, and carefully accounting for all terms, we find the
following Lorentz violating kinetic terms for the scalars to linear order
in $\Lambda$ \cite{Frey:2003jq}:
\bea
\delta S &=& \frac{\mu_3}{g_s} \int d^4 x \frac{(2\pi\ap)^2}{192}
\vec{\del} X^m\cdot \vec{\del} X^n \left[ 162 b^2 b_{\left( m\right.}
\lambda_{\left. n\right)} +90 b\cdot \lambda b_m b_n -4b^2
\lambda_{mn}\right.\nonumber\\
&&\left.+8b\cdot \lambda b_{mn}
-32\lambda^pb_{p\left(m\right.}b_{\left. n\right)} 
-32 b^pb_{p\left( m\right.}\lambda_{\left. n\right)} +188 
b^2 b\cdot \lambda
\delta_{mn}\right.\nonumber\\
&&\left. +16 b^{pq}\lambda_{pq}\delta_{mn}+32\lambda^{pq}b_p b_q
\delta_{mn}+160 b^{pq}\lambda_p b_q \delta_{mn}\right]\ .
\label{scalarmav}
\eea
These are probably not the only Lorentz violating
kinetic terms for the scalars because of the shortcomings of the corrected
action (\ref{corrected}) that we noted elsewhere.  
However, because Lorentz invariance is
broken by the background, there is no reason to suppose that the terms we
have calculated are precisely canceled by those that we have not.  Therefore,
we can take these as an estimate of the change in MAV for the scalars.
From (\ref{scalarmav}), we can see clearly why these terms modify the
MAV of the scalars; by a \textit{field dependent} redefinition of the
``speed of light,'' we can clearly combine all the time and space derivatives
of the fields into the usual form $\del_\mu X\del^\mu X$.  Each scalar
has a Lorentz invariant kinetic action, but the Lorentz groups are in 
general different for the different scalars.  This way of thinking about
our corrections is essentially the formalism of \cite{Coleman:1998ti}.

It is also possible to calculate the change to the photon MAV by using
a T duality argument, as in \cite{Frey:2003jq}.  
The result is that 
\bea
\delta S &=& \frac{\mu_3}{g_{s}} \int d^4x \frac{(2\pi\ap)^2}{192}
\mathcal{F}_{ij}\mathcal{F}^{ij} \left[ 44 b^2 b\cdot\lambda 
+12 b^{mn}\lambda_{mn} +24\lambda^{mn}b_mb_n\right.\nonumber\\
&&\left. +72b^{mn}\lambda_mb_n\right]
\ .\label{photmav}\eea
The T duality argument is reviewed in appendix (\ref{aa:lorentz}).

\subsubsection{Bounds on Lorentz Violations}\label{sss:bounds}

Given the limits on Lorentz violation in the Standard Model that we have
already discussed, \cite{Frey:2003jq} used the MAVs 
(\ref{scalarmav},\ref{photmav}) to put bounds on the Lorentz violating 
part of the warp factor, assuming that the Lorentz invariant warping was
about an order of magnitude lower than the string scale.  In comparison to
the loop calculations of \cite{Burgess:2002tb}, these string corrections
actually appear to be as important.  Additionally, they remain important
over a wider region of parameter space.  Therefore, in the context of
self-dual flux compactifications, we might expect that the $\ap$ 
corrections might play as much of a role in mediating supersymmetry breaking
on the brane as loops would.

\section{Conclusion}\label{s:conclusion}

It is my hope that this document proves a useful review for researchers
unfamiliar with self-dual flux compactifications.  I have tried to present
a comprehensive review, while highlighting my own contributions to the
subject.  The past few years have seen an expansion of our knowledge of
the vacua of string theory, and hopefully this document makes one type of
those vacua a bit more accessible.

\appendix
\chapter{Conventions}\label{a:conventions}
\section{Index and Tensor Notation}\label{aa:indices}

When dealing with compactification schemes, it is useful to have some type
of index notation that separates the compact (internal) dimensions from
the noncompact (spacetime or external) dimensions.  Here, I use capital
Roman indices $M,N,\cdots$ for all dimensions, Greek indices $\mu,\nu,\cdots$
for the spacetime dimensions (usually $0,1,2,3$), and lower case Roman
$m,n,\cdots$ for the internal dimensions.  Complex coordinates will be 
denoted $i,j,\cdots$ and $\bi,\bj,\cdots$.  Hats $(\, \hat{}\, )$ will denote 
orthonormal frame indices parallel to the corresponding coordinates.
(For example, $e^{\hat{m}}_m$ is the vielbein in the compact space.)  I will
use Greek indices from the beginning of the alphabet $\alpha,\beta,\cdots$
for pulled-back world-volume indices, and hatted Roman indices 
$\hat{a},\hat{b},\cdots$ for normal bundle indices for branes.  I also use
Roman indices from the start of the alphabet with and without dots for
spinor indices ($a,b,\cdots$ and $\dot{a},\dot{b},\cdots$); see 
\ref{aa:spinors} for spinor conventions.

Other conventions to note are the following.  I take the Levi-Civita
tensor to be a tensor as opposed to a tensor density; that is,
\be\label{lctensor}
\epsilon^{01\cdots D} = -\sqrt{-g}\ , \ee
although the value in an orthnormal frame is $\pm 1$ (as denoted with 
hats on the indices).  Symmetrization and antisymmetrization are denoted
by $(\cdots )$ and $[\cdots ]$ respectively and are weighted by $1/p!$,
where $p$ is the number of indices (anti)symmetrized.  Similarly,
forms are weighted,
\be\label{formweight}
F_p = \frac{1}{p!}F_{M_1\cdots M_p} dx^{M_1}\cdots dx^{M_p} \ ,
F_{(p,q)} = \frac{1}{p!q!}F_{i_1\cdots i_p \bi_1\cdots \bi_q} dz^{i_1} 
d\b z^{\bi_1}\cdots d\b z^{\bi_q} \ee
(the latter refers to forms in complex coordinates).  The square of a 
form is similarly weighted:
\be\label{formsquare}
|F_p|^2 = \frac{1}{p!} F_{M_1\cdots M_p} F^{M_1\cdots M_p}\ .\ee
Hodge stars are
defined by 
\be\label{hodge}
\star F_{M_1\cdots M_{D-p}} = \frac{1}{p!} \epsilon_{M_1\cdots M_{D-p}}
{}^{M_{D-p+1}\cdots M_D} F_{M_{D-p+1}\cdots M_D}\ , 
\star\star = \mp (-1)^{p(D-p)}
\ee
with the $\mp$ coming in Lorentzean/Euclidean signature.  These conventions
essentially follow \cite{Polchinski:1998rr}.  I will often drop the wedge
``$\wedge$'' in wedge products for notational convenience.

\section{Spinor Conventions}\label{aa:spinors}

We work with Dirac matrices that satisfy
\be\label{diracgamma}
\left\{\Gamma^M,\Gamma^N\right\} = 2g^{MN}\ .
\ee
With tangent frame indices, the metric is replaced by the flat metric
$\eta^{\hat M\hat N}$.  The 10D chirality matrix is 
$\Gamma_{(\widehat{10})} = \Gamma^{\hat 0}\cdots\Gamma^{\hat 9} =
(1/10!) \epsilon_{M_1\cdots M_{10}} \Gamma^{M_1\cdots M_{10}}$.
We use the notation that
\be\label{asymgamma}
\Gamma^{MN\cdots P} = \Gamma^{[M}\Gamma^N\cdots \Gamma^{p]}
\ .\ee

Since the 10D spacetime is broken into 4D and 6D parts, it is convenient
to take the Dirac matrices as tensor products of $SO(3,1)$ and $SO(6)$ parts:
\be\label{fourbysix}
\Gamma^\mu = \gamma^\mu\otimes 1, \ \Gamma^m = \gamma_{(\hat 4)}\otimes
\gamma^m\ .\ee
The 4D and 6D chirality matrices are
\be\label{chiralities}
\gamma_{(\hat 4)} = i\epsilon_{\mu\nu\lambda\rho}\gamma^{\mu\nu\lambda\rho},
\ \gamma_{(\hat 6)} =-i \epsilon_{mnpqrs}\gamma^{mnpqrs},\ 
\Gamma_{(\widehat{10})} = \gamma_{(\hat 4)}\gamma_{(\hat 6)}\ .\ee
Likewise, a 10D spinor $\varepsilon$ decomposes into $\zeta \otimes \chi$
with $\zeta$ an anticommuting 4D spinor and $\chi$ a commuting 6D spinor.

With complex coordinates, it is possible to define the spinors on the 
internal manifold in terms of spins.  The gamma matrices in those coordinates
are raising and lowering operators according to the algebra 
(\ref{diracgamma}), so we can define a ``spin-down'' spinor that is
annihilated by all antiholomorphic gamma matrices.  Then we define a 
spinor with spin vector $\bm{s}$ on an $n$-fold by 
\be\label{spinordef}
\ket{\bm{s}} = \left(\gamma^{n}\right)^{(s_n+1/2)} \cdots
\left(\gamma^{1}\right)^{(s_1+1/2)} \ket{--\cdots -}\ .\ee
External dimensions are similar; we just replace the first complex
coordinate with a lightcone coordinate.

In particular, we will make heavy use of an $SO(1,1)\times SO(8)$
decomposition of 10D spinors in section \ref{ss:counting}, following
complex coordinates (\ref{zcomplex}).  The Majorana-Weyl spinors
of $SO(8)$ are real linear combinations of the basis spinors
\bea  u_1=\ket{++++}+\ket{----} 
&,&v_1=\ket{+-++}-\ket{-+--} \nonumber\\
u_2=i\left(\ket{++++}-\ket{----}\right) &,
&v_2=i\left(\ket{+-++} +\ket{-+--}\right) \nonumber\\
u_3=\ket{++--}-\ket{--++} &,&v_3=\ket{-+++}+\ket{+---}
\nonumber\\ 
u_4=i\left(\ket{++--}+\ket{--++}\right) &,
&v_4=i\left(\ket{-+++} -\ket{+---}\right) \nonumber\\
u_5=\ket{+-+-}+\ket{-+-+} &,&v_5=\ket{+++-}-\ket{---+}
\nonumber\\ 
u_6=i\left(\ket{+-+-}-\ket{-+-+}\right)
&,&v_7=i\left(\ket{++-+} +\ket{--+-}\right) \nonumber\\
u_7=\ket{+--+}-\ket{-++-} &, &v_7=\ket{++-+}-\ket{--+-}
\nonumber\\ 
u_8=i\left(\ket{+--+}-\ket{-++-}\right)
&,&\!\!\! v_8\!=\!i\!\left(\ket{++-+}\! -\!\ket{--+-}\right)\!.
\label{so8basis}\eea
The $u_a$ ($v_{\dot a}$) form an $\mathbf{8}$ ($\mathbf{8}^\prime$).
To write a spinor from the $SO(3,1)\times SO(6)$ basis in this basis, 
we can simply find linear combinations of $u_a,v_{\dot a}$ that are
annihilated by the correct combinations of $SO(3,1)\times SO(6)$ 
gamma matrices converted to the complex coordinates (\ref{zcomplex}).

\section{String and Einstein Frames}\label{aa:frames}

Because the 10D string metric (the metric that the string worldsheet
feels) does not have a canonical Einstein-Hilbert action, many authors
use a Weyl rescaled Eintein frame metric even in 10D.  This metric is
\be\label{einstein10d}
g^{E}_{MN} = e^{-\phi/2} g_{MN}\ee
(see for example \cite{Polchinski:1998rr}).  
We should also note that this rescaling mixes the gravitino and dilatino
(and rescales the SUSY parameter) as \cite{Strominger:1986uh,Hassan:1999bv}
\be\label{gravdilmix}
\lambda^E = e^{\phi/8}\lambda\, ,\ \psi^E_M = e^{-\phi/8}\left( \psi_M 
+\frac{i}{4}  \lambda^* \right)\, , \ \varepsilon^E =e^{-\phi/8}
\varepsilon\ .\ee
Additionally, the rescaled metric rescales the definition of mass in 10D.
Consider a scalar mass.  The kinetic term is $e^{-\phi/2}g^{E,MN}\del_M
\psi\del_N\psi = g^{MN}\del_M\psi\del_N\psi$ while the potential term
is the same (up to overall scaling).  Therefore, the effective mass is
rescaled
\be\label{einsteinmass}
m^2_E = e^{\phi/2}m^2\ ,\ee
where $m$ is the mass in the string frame.

Also, the type IIB SUGRA supersymmetry transformations and equations of
motion were originally derived \cite{Schwarz:1983wa,Schwarz:1983qr} in
a formulation that is different from the usual string conventions.  
Conversions are given in multiple places in the literature, such as
\cite{Kehagias:1998gn,Grana:2001xn}.

\chapter{Actions and Equations of Motion}\label{a:actionsetc}

\section{String Frame IIB Actions}\label{aa:action}

\subsection{IIB SUGRA}\label{aaa:IIBsugra}

The action for type IIB supergravity is 
(\cite{Polchinski:1998rr,Johnson:2000ch} and references therein)
\bea
S&=& \frac{1}{2\kten^2} \int d^{10}x\sqrt{-g} \left\{ e^{-2\phi}
\left[ R + 4(\del\phi)^2 -\frac{1}{2} |H_3|^2\right]\right.\nonumber\\
&& \left. -\frac{1}{2} (\del C)^2 -\frac{1}{2} |\t F_3|^2 -\frac{1}{4}
|\t F_5|^2 \right\}\nonumber\\
&& + \frac{1}{4\kten^2} \int \left(C_4 -\frac{1}{2} B_2\wedge C_2\right)
\wedge F_3\wedge H_3\ .\label{IIBaction}\eea
The last term contributes only as a boundary term.  We have defined
$H_3=dB_2$, $F_3=dC_2$, $\t F_3 = F_3-C H_3$, $F_5 =dC_4$, and $\t F_5=F_5
-C_2 H_3$.  Note that $2 \kten^2 =(2\pi)^7\ap{}^4$.  Also, we commonly
write $\tau=C+ie^{-\phi}$.

\subsection{D-brane Actions}\label{aaa:dbranes}

The action for a D$p$-brane comes in two parts, the Dirac-Born-Infeld
part, and the Wess-Zumino part (again, see 
\cite{Polchinski:1998rr,Johnson:2000ch}).  These are
\be\label{DBIaction}
S_{\mathnormal{DBI}} = -\mu_p \int d^{p+1}\zeta e^{-\phi}
\sqrt{-\det \left( g_{\alpha\beta} +\mathcal{F}_{\alpha\beta}\right)}
\ ,\ee
where $\mathcal{F}=2\pi\ap F-B$ is a $U(1)$ field strength\footnote{The world 
volume gauge field
therefore transforms as $\delta A= \lambda_B/2\pi\ap$ under a SUGRA
gauge transformation $\delta B_2 = d\lambda_B$.}, and
\be\label{WZaction}
S_{\mathnormal{WZ}} = \mu_p \int e^{\mathcal{F}} \wedge \oplus_q C_q\ ,\ee
where the integral projects onto $p+1$ forms.  The D-brane charge is
$\mu_p = 1/(2\pi)^p \ap{}^{(p+1)/2}$.  The coordinates $\zeta^\alpha$
are the embedding coordinates of the D-brane.  Note that the spacetime
fields are pulled back to the world volume.

Multiple D-branes have a non-Abelian gauge theory and indeed non-Abelian
positions in the 10D spacetime.  The action in this case was found by
\cite{Myers:1999ps}.  Also, the action including worldvolume fermions
is given by $\kappa$-symmetry, which we discuss for a D-string in 
\ref{aa:kappaexpand}.

\subsection{String Corrections}\label{aaa:corrections}

Both the bulk and brane actions are corrected order by order in $\ap$.
However, these corrections are not very well known; mainly they are 
understood only for the metric even at lowest order in $\ap$.  The most
important corrections for us are those to the D7-brane action because they
give an induced D3-brane charge and tension.

The first study of $\ap$ corrections in the string action was CITE.
Recently, \cite{deHaro:2002vk} discussed the supersymmetric completion
of the $\ap{}^3$ terms; it also lists many references.  The reader can find
a discussion and review of corrections that include the other NS-NS fields
in \cite{Frolov:2001xr}, which also gives the metric contribution in
a particularly nice form:
\bea
\delta S &=& -\frac{\ap{}^3}{2\kten^2} \frac{\zeta(3)}{8}
\int d^{10}x\sqrt{-g} e^{-2\phi}\left[R^{TMNU}R_{PMNQ}R_T{}^{RSP}R^Q{}_{RSU}
\right.\nonumber\\
&&\left.+\frac{1}{2}R^{TUMN}R_{PQMN}R_T{}^{RSP}R^Q{}_{RSU}\right]\ .
\label{j0correct}\eea
These are the type of corrections examined in \cite{Becker:2002nn}.

Probably the most famous $\ap$ corrections to D-brane actions are from the
couplings to the RR potentials; Riemann curvature becomes lower-dimensional
brane charge.  The Wess-Zumino part of the action is
\be\label{aroof}
S_{WZ}=
\mu_p \int e^{\mathcal{F}} \wedge \oplus_q C_q\wedge 
\sqrt{\frac{\hat A\left( 4\pi^2
\ap R_T\right)}{\hat A \left(4\pi^2 \ap R_N\right)}}\ee
where $\hat A$ is the Dirac ``A-roof genus'' and $R_T,R_N$ are the tangent
and normal bundle Riemann tensors to be defined below
(see \cite{Cheung:1998az} for a full derivation and
\cite{Johnson:2000ch} for a review).  
These corrections are precisely those that give a D7-brane that is wrapped
on $K3$ a negative D3-brane charge, which arises from a term in the 
expansion of $\hat A$ proportional to $\tr R\wedge R$.

There are also corrections
to the DBI action, which have been studied in \cite{Bachas:1999um,
Wyllard:2000qe,Fotopoulos:2001pt,Wyllard:2001ye,Barabanschikov:2003fr}.  
These are responsible for modifying the tension of wrapped D7-branes,
for example.
The most complete 
results for curved backgrounds are in \cite{Fotopoulos:2001pt}, but only 
the geometry is considered and not other SUGRA fields.  The DBI action becomes
(considering the bosonic part only)
\bea
S_{\mathnormal{DBI}}&=& -\mu_p\int d^{p+1}\zeta\,  e^{-\phi}\sqrt{-\det \left(
g
+ \mathcal{F}\right)} \left[ 1 -\frac{(2\pi\ap)^2}{192}
\left( (R_T)_{\alpha\beta\gamma\delta}(R_T)^{\alpha\beta\gamma\delta}  
\right.\right.\nonumber\\
&&\left.\left. -2(R_T)_{\alpha\beta}(R_T)^{\alpha\beta}
-(R_N)_{\alpha\beta \hat a \hat b}(R_N)^{\alpha\beta \hat a \hat b} 
+2\b R_{\hat a \hat b}\b R^{\hat a \hat b}\right) \right]\label{corrected}
\eea
up to $\mathcal{O}(\ap)^2$.  There is an additional contribution at this 
order with an undetermined coefficient, but it vanishes on-shell, so it does
not affect S-matrix elements or dispersion relations \cite{Fotopoulos:2001pt}.
Here, $\hat a, \hat b$ are normal bundle 
indices in an orthonormal basis with vielbein $\xi^{\hat a}$.
Terms involving fermions are not known, although 
\cite{Barabanschikov:2003fr,Howe:2001wc} 
give steps in that direction.  

We now define the various Riemann and Ricci tensors above, as is discussed
in \cite{Bachas:1999um,Wyllard:2000qe,Fotopoulos:2001pt,Wyllard:2001ye}.
In the following, we use $P[\cdots]$ to denote the pullback to the 
worldvolume or pushforward to the normal bundle,
but for brevity we write $g_{\alpha\beta}\equiv P[g]_{\alpha\beta}$ and
$g^{\alpha\beta} \equiv (P[g]_{\alpha\beta})^{-1}$.
Start with the extrinsic curvature, or second fundamental form, of the
D-brane embedding
\be\label{extrinsic}
\Omega^{\hat a}_{\alpha\beta} = \xi^{\hat a}_M \left( \del_\alpha\del_\beta
X^M -(\Gamma_T)^\gamma_{\alpha\beta}\del_\gamma X^M +
\Gamma^M_{NP}\del_\alpha X^N\del_\beta X^P\right)\ ,\ee
where $(\Gamma_T)^\gamma_{\alpha\beta}$ is the Christoffel connection of
the pulled-back metric.  The tangent and normal bundle Riemann tensors 
can be shown to be
\bea
(R_T)_{\alpha\beta\gamma\delta}&=& P[R]_{\alpha\beta\gamma\delta}+
\delta_{\hat a\hat b} \left( \Omega^{\hat a}_{\alpha\gamma}
\Omega^{\hat b}_{\beta\delta}-\Omega^{\hat a}_{\alpha\delta}
\Omega^{\hat b}_{\beta\gamma}\right)\ ,\label{riemannt}\\
(R_N)_{\alpha\beta}{}^{\hat a \hat b} &=& P[R]_{\alpha\beta}{}^{\hat a\hat b}
+g^{\gamma\delta}\left( \Omega^{\hat a}_{\alpha\gamma}
\Omega^{\hat b}_{\beta\delta} -\Omega^{\hat b}_{\alpha\gamma}
\Omega^{\hat a}_{\beta\delta}\right)\ .\label{riemannn}\eea
Then $(R_T)_{\alpha\beta}$ is just the Ricci tensor associated with the 
tangent Riemann tensor, and
\be\label{rbar}
\b R^{\hat a\hat b} = g^{\alpha\beta}P[R]_\alpha{}^{\hat a\hat b}{}_\beta
+g^{\alpha\gamma}g^{\beta\delta}\Omega^{\hat a}_{\alpha\beta}
\Omega^{\hat b}_{\gamma\delta}\ .\ee

\subsection{$\kappa$-Symmetry and Superfield Expansions}\label{aa:kappaexpand}

As we saw in section \ref{sss:dqm}, the way to get the appropriate 
supersymetric action for a D-brane is to replace the spacetime fields by
spacetime superfields; in IIB string theory,
there are two Majorana-Weyl fermionic coordinates
$\Theta^1,\Theta^2$ in the superspace.  Therefore, the supersymmetry 
transformations take $\Theta^1\to\Theta^1+\varepsilon_1$, 
$\Theta^2\to\Theta^2+\varepsilon_2$.

D-branes have a local
$\kappa$-symmetry that can be
used to fix a gauge, keeping only 16 of those 32 degrees of
freedom. For a D-string, the $\kappa$-symmetry transformations are
\cite{Bergshoeff:1997kr,Kallosh:1998ky}
\be\label{kappastring}
\delta\left[\begin{array}{c} \Theta^1\\ \Theta^2\end{array}\right]
= \left( 1+\gamma_{(\hat 2)} \otimes \sigma^3 i\sigma^2\right)
\left[\begin{array}{c} \kappa^1\\ \kappa^2\end{array}\right]=
\left[\begin{array}{c} \kappa^1+\gamma_{(\hat 2)}\kappa^2\\ 
\kappa^2+\gamma_{(\hat 2)}\kappa^1\end{array}\right]\ .
\ee
It is clear from this formula that we can set $\Theta^2=0$ on the D-string,
so we take that gauge.  It is because of this gauge fixing that the
worldvolume supersymmetries are not simply the spacetime supersymmetries.
Henceforth we drop the superscript ``$1$'' on $\Theta$.

The expansions of the spacetime fields in terms of the world-volume
spinor $\Theta$ in this gauge are, for constant
dilaton-axion \cite{Grana:2002tu},
\begin{eqnarray}
\bm{e}_m^{\hat m}&=&e_m^{\hat m} + \frac{i}{8} \overline{\Theta} 
\Gamma^{\hat m\hat n\hat p}
\Theta w_{m\hat n \hat p}-  \frac{i}{16} \overline{\Theta} 
\Gamma^{\hat m np} \Theta H_{mnp} \nonumber\\
\bm{e}_{a}^{\hat m}&=& \frac{i}{2}
(\overline{\Theta}\Gamma^{\hat m})_{a} \nonumber \\    
\bm{e}_m^{a}&=& 0\nonumber \\ 
\bm{e}_{a}^{b}&=& \delta_{a}^{b} \nonumber\\  
\bm{\phi}&=& \phi- \frac{i}{48}  \overline{\Theta}
\Gamma^{pqr} \Theta H_{pqr} \nonumber\\ 
\bm{B}_{mn} &=& B_{mn}
+\frac{i}{4}\overline{\Theta} \Gamma^{\hat p\hat q}\,_{[m} \Theta 
\,w_{n]\hat p \hat q} -
\frac{i}{8}\overline{\Theta} \Gamma^{pq}\,_{[m} \Theta \,H_{n]pq}
\nonumber\\ 
\bm{B}_{ma}&=& \frac{i}{2} (\overline{\Theta}
\Gamma_m)_{a} \nonumber\\ 
\bm{C}_{mn}&=& C_{mn}-
\frac{i}{8}\overline{\Theta} \Gamma^{pq}\,_{[m} \Theta \,F'_{n]pq}+ C
B_{mn}|_{\Theta^2} \nonumber\\ 
\bm{C}_{ma}&=& C B_{ma}
\nonumber\\
\bm{C}&=& C -\frac{i}{48}\overline{\Theta}
\Gamma^{pqr} \Theta F'_{pqr}\ .
\label{expansionIIB} 
\end{eqnarray}
We have introduced a factor of $i$ in each fermion bilinear compared to 
the usual SUGRA notation, so that the
action (\ref{action1}) 
matches usual quantum field theory conventions and gives a real 
anticommutator.

\section{String Frame Equations of Motion}\label{aa:eom}

The equations of motion for type IIB supergravity, neglecting brane
fields, were derived in \cite{Schwarz:1983qr}.  To get them in string
frame in our conventions for the fields, we can convert from the
Einstein frame conventions of \cite{Polchinski:2000uf}.  We find
the following.

The equation of motion for the metric is
\bea
R_{MN}&=& \frac{1}{2} g_{MN}(\del\phi)^2 -2\Del_M\Del_N \phi
-\frac{1}{4} g_{MN}\Del^2\phi +\frac{1}{96}e^{2\phi}\t F_{MPQRS}
\t F_N{}^{PQRS}\nonumber\\
&&+\frac{1}{4} e^{2\phi} G_{(M}{}^{PQ}\b G_{N)PQ}
-\frac{1}{48} e^{2\phi}g_{MN} G_{PQR}\b G^{PQR} \nonumber\\
&& +\kten^2 e^{2\phi} \left( T'_{MN}-\frac{1}{8}g_{MN}T'^P_P\right)\ .
\label{einstein}\eea
Here $T'_{MM}$ is the stress tensor contriubtion from localized objects
such as D-branes and O-planes
defined by differentiating their action with respect to the metric. The
factor $e^{2\phi}$ comes from the noncanonical Einstein-Hilbert term.

The equation of motion for the scalars is
\be\label{dilaxeom}
\Del^2\tau =-ie^\phi|\del\tau|^2-\frac{i}{12}e^\phi G_{MNP}G^{MNP}\ .\ee
Note that the 3-form is \textit{not} multiplied by the complex conjugate
here and that $G_3$ includes the scalar $\tau$.  We are leaving off
brane sources for the dilaton.  There is also a 
Bianchi identity for these scalars that has 7-brane sources, but we
do not need it.

The 5-form satisfies ten-dimensional self-duality $\t F=\star\t F$, so
its equation of motion is given by its Bianchi identity
\be\label{bianchi5}
d\t F_5 = H_3\wedge F_3 +2\kten^2 \mu_3 \rho_3
\ .\ee
$\rho_3$ is the charge density 6-form of D3-branes, normalized to unity
for a single D3-brane.   (Thus, a normal O3-plane would have integrated
$\rho_3$ of $-1/4$.)  

The 3-forms are a bit more messy.  They have equations of motion
\bea
d\star \t F_3 &=& \t F_5\wedge H_3 + 2\kten^2 \mu_1 \rho_{\textnormal{D1}}
\nonumber\\
d\star \left(e^{-2\phi}H_3 -C\t F_3\right) &=& 
-\t F_5\wedge F_3 + 2\kten^2 \mu_1 \rho_{\textnormal{F1}}
\ .\label{3formeom}\eea
The charge densities, normalized to unity for a single string, are
$ \rho_{\textnormal{D1}}$ for a D-string and $\rho_{\textnormal{F1}}$
for an F-string.
The Bianchi identities are
\bea
dF_3 &=& 2\kten^2 \mu_5 \rho_{\textnormal{D5}}\nonumber\\
dH_3 &=& 2\kten^2 \mu_5 \rho_{\textnormal{NS5}}\ .\label{3formbianchi}
\eea
Again, $\rho_{\textnormal{D5}},\rho_{\textnormal{NS5}}$ are D5-brane
and NS5-brane charge densities.

\chapter{Auxilliary Material}\label{a:ancillary}

\section{Corrections to D-String QM}\label{aa:assumptions}

Here we will estimate a number of possible corrections to the Hamiltonian
given in section \ref{sss:dqm} and argue that they will not change
the counting of BPS states\footnote{The arguments of this appendix were
given in \cite{Grana:2002ti}.}.  
Because a single long multiplet could only be
formed from four short multiplets (of differing spins), we anticipate
that it would be difficult to lift the supersymmetric states given in
section \ref{sss:wavefunction}; this is essentially the argument of
\cite{Witten:1982df}.
We will mainly here be concerned if any
of the long multiplets could be made BPS by corrections to our 
Hamiltonian, eqn (\ref{hamil}).  We work in the framework of 
perturbation theory and concentrate on the bosonic terms.  We take all
the coordinate radii to be similar $\sim R$.

First we note the structure of the Hilbert space once we include excited 
states.  The center of mass modes considered in the text have excited
states with energy of the order of the magnetic field, $\sim \ap/R^3$.  Then
there are modes that move around the D-string, which form a tower of 
supersymmetric
harmonic oscillators with frequency $\sim 1/R$.  Each oscillator has
a unique supersymmetric ground state, and we don't expect perturbations to
the Hamiltonian to break the supersymmetry.\footnote{We should note that
the oscillator amplitudes are not periodic, unlike center of mass modes.
This is clear because these modes represent bending of the string, not
position.}

Just like the center of mass modes, the oscillators couple to the 3-form;
we treat this as a perturbation.  Because the perturbation couples different
oscillators, it appears only in second order perturbation theory.  Since
the magnetic field is of order $\ap/R^3$, the energy shift is of order 
$\ap{}^2/R^5$.  The 3-form flux $F_{mn7}$ also couples to the oscillators; this
perturbs the excited state energies by $\sim \ap{}^3/R^7$.  There are also
higher derivative terms in the D-brane action, coming from the
Born-Infeld determinant.  One typical term is $(\dot X \del_4 X)^2$.  This
can contribute at first order in perturbation theory, giving an energy
shift $\ap/R^2$ times the energy of the state.  

We also briefly consider bulk supergravity effects.  At large but finite 
radius, 
there is a metric warp factor; it's lowest order contribution is 
$\ap{}^2/R^4$ times the energy of the state.  The interaction with 
bulk modes is a little trickier to understand.  The massless moduli of
the compactification should not affect the BPS spectrum, since the same
supercharges are preserved in the spacetime.  The BPS particle mass and 
excited state energies will change, however.  The massive scalars of
the 4D effective theory, such as the dilaton, should also not change the
BPS spectrum.  One argument for this is that the dilaton enters into the
Hamiltonian only through the mass of the BPS particle (when one takes 
into account the normalization of the fermions), which affects only the 
excited states of the center of mass modes.  Another argument is that there
are some cancellations among different terms in the perturbation theory
for the dilaton.

Finally, we should note that the 4D gauge fields \emph{can} change the 
BPS spectrum rather violently.  For example, a BPS particle in the field
of an oppositely charged particle should have no supersymmetric ground
states. 

\section{Inclusion of Gravity in Bubble Nucleation}\label{aa:grav}

In this appendix, I compare two formalisms for calculating the
decay rate for thin-wall instantons\footnote{The calculations presented
here were given in \cite{Frey:2003dm}.}.
The method
of Coleman and De Luccia (CDL) \cite{Coleman:1980aw,Banks:2002nm} 
describes smooth
instantons in the limit of large radius of curvature; this formalism was
used by \cite{Kachru:2003aw} to argue that bubbles will always 
nucleate in a dS background before the recurrence time.  Alternately,
however, we could imagine that the bubble wall is truly an infinitesimally
thin membrane, such as a D-brane, with a delta function stress tensor.  
This type of configuration was studied by Brown and Teitelboim (BT)
\cite{Brown:1987dd,Brown:1988kg}.  In a description of dS solutions in 
noncritical string theory \cite{Silverstein:2001xn,Maloney:2002rr} use
the BT formalism to argue that there is a critical tension above which the
bubble occupies more than half the original de Sitter sphere and above 
which the 
decay time changes behavior as a function of the bubble tension.  
Additionally,
\cite{Maloney:2002rr} claims that the decay time is of order the 
recurrence time at the critical tension.  

In this appendix, we show that the CDL and BT formalisms actually agree; 
this is reassuring, since even D-branes should
be described as smooth objects in a complete version of string theory.
Our results show that the decay time is always less than the recurrence 
time, as in 
\cite{Goheer:2002vf,Kachru:2003aw,Giddings:2003zw,Susskind:2003kw}.  
Additionally, we confirm that more than half the original de Sitter 
sphere decays when the tension is above critical, but
we show that (due to some technical considerations) the decay time is 
actually a smooth function of the bubble tension.  We work in a 4D effective
theory throughout.

We begin by describing the two formalisms.  
In both CDL and BT, the nucleation/decay time is
given by exponentiating the difference of the (Euclidean) bubble and 
background actions.  Therefore, the exterior of the bubble, which is 
approximated by the background, contributes nothing in both formalisms.

At that point, CDL note that, since both bubble and background have the same
behavior at infinity, they can integrate some terms in the Ricci tensor
by parts.  After determining the bubble tension $T$ as a functional of
the potential, they find that the action is
\bea
\Delta S_E &=& 2\pi^2 r^3 T  +\frac{12\pi^2}{\kappa_4^2}\left\{
\frac{1}{\Lambda_-} \left[ \left( 1-\frac{\Lambda_-}{3}r^2\right)^{3/2}-1
\right]\right.\nonumber\\
&&\left.-\frac{1}{\Lambda_+}\left[ \left( 1-\frac{\Lambda_+}{3}r^2
\right)^{3/2}-1\right]\right\}\label{cdlaction}
\eea
where $\Lambda_\pm = \kappa_4^2 V_\pm$ are the potential outside and inside
the bubble respectively (this is a combination of eqns (3.11) and (3.13) 
from \cite{Coleman:1980aw}).  Minimizing this action with respect to the 
bubble
curvature radius $r$ gives the decay rate.

On the other hand, BT cannot use the same integration by parts because the
infinite stress of the bubble wall separates the interior and exterior 
regions.  Instead, the bubble action must include extrinsic curvature
terms; it is these terms that will explain the apparent contradiction between
CDL and BT formalisms.  The extrinsic curvatures of the interior and
exterior regions are
\be\label{curvature}
K_\pm = -3\sigma_\pm \left( \frac{1}{r^2} -\frac{\Lambda_\pm}{3}\right)^{1/2}
\ee
where $\sigma_\pm=1$ if the radius of curvature of the outside/inside
region of the bubble is increasing toward the exterior of the bubble and 
is $-1$ if the radius is decreasing.  Since the Ricci scalar in the bubble is
given by the cosmological constant, the action just becomes
\be
\Delta S_E = 2\pi^2 T r^3 + \frac{2\pi^2}{\k^2} r^3\left( K_- -K_+
\right) -\frac{1}{\k^2} \left( \Lambda_- \mathcal{V}_- -
\Lambda_+ \mathcal{V}_+\right)
\ . \label{btaction}
\ee
The interior volumes for the bubble and background are given by (for either
sign of the cosmological constant)
\bea
\mathcal{V}_\pm &=& 2\pi^2 \left(\frac{3}{\Lambda_\pm}\right)^2 \left\{
\frac{1}{3}\left[\sigma_\pm \left( 1-\frac{\Lambda_\pm}{3}r^2\right)^{3/2}
-1\right]\right.\nonumber\\
&&\left. -\left[ \sigma_\pm \left( 1-\frac{\Lambda_\pm}{3}r^2\right)^{1/2}
-1\right]\right\}  \ .\label{volumes}
\eea
It is algebraically simple to see that the extrinsic
curvature terms combine with the square root terms from the volume to give
exactly the CDL action (\ref{cdlaction}) up to the signs $\sigma_\pm$.
The reason \cite{Maloney:2002rr} found a different
action is that they omitted the extrinsic curvature terms.

Now we should see why the CDL result should actually have the 
signs $\sigma_\pm$.
For a de Sitter background, the terms in square brackets of eqn 
(\ref{cdlaction}) come from integrals
\be\label{integrals}
\int_0^{\xi (r)} d\xi\ r\left(1-\frac{\Lambda_\pm}{3} r^2\right)\ ,
\ee
which CDL evaluate by replacing $d\xi = dr (1-r^2 \Lambda_\pm/3)^{-1/2}$.
However, as they note, $r=\sqrt{3/\Lambda_\pm} \sin [\sqrt{\Lambda_\pm/3}\xi]$,
so $\xi (r)$ is double-valued.  In fact, the correct integral is
\be\label{newint}
\frac{3}{\Lambda_\pm} \int_{\sigma_\pm(1-r^2 \Lambda_\pm/3)^{1/2}}^1 dy\ y^2
\ee
which just introduces a factor of $\sigma_\pm$ in the $(\cdots)^{3/2}$ terms.
This precisely agrees with the BT results.  In this paper, we will be 
concerned only with the case $\sigma_- = 1$, and $\sigma_+ = -1$ only for 
$\Lambda_+ >0$ and tension above critical.

To find the radius of the bubble given the two cosmological constants and
the tension, we could minimize the action with respect to $r$.  However,
it is easier to use the Israel matching condition across the bubble wall, 
which has trace $K_+ - K_- = (3/2)\k^2T$.  The answer is given by 
\cite{Maloney:2002rr} and can be written as
\be\label{radius}
\frac{1}{r^2} = \left(\frac{\k^2T}{4}\right)^2 +\frac{\b\Lambda}{3}
+\left(\frac{\Delta\Lambda}{3\k^2T}\right)^2\ ,\ \ 
\b\Lambda= \frac{\Lambda_+ +\Lambda_-}{2}\ ,\ \Delta\Lambda = \Lambda_- -
\Lambda_+\ .\ee
It is tedious but straightforward to check that this matches the result
from minimizing the action.
The maximum radius occurs at critical tension
\be\label{critical}
\k^2T_c = \left( \frac{4}{3} |\Delta\Lambda|\right)^{1/2}\ee
and is $1/r^2 = \Lambda_+/3 = 1/R^2_{\textnormal{\scriptsize dS}}$ for positive
initial cosmological constant. Note that for vanishing initial vacuum energy,
gravity stabilizes the false vacuum for tension bigger than
critical, as in \cite{Coleman:1980aw}.  Also, for negative initial 
cosmological constant, the radius becomes infinite for tensions lower than
critical.  We will concern ourselves only with initial de Sitter spacetimes,
so we do not face some of the concerns raised by \cite{Banks:2002nm} about
decays of Minkowski and AdS spacetimes.  

We will finally write down the action for the bubbles:
\bea
\Delta S_E &=& 2\pi^2 r^3\left\{ T +\frac{6}{\k^2\Lambda_+ \Lambda_-}
\left[ \frac{\Delta\Lambda}{r^3} 
+\Lambda_+ \left(\left(\frac{\k^2T}{4}\right)^2 
-\frac{\Delta\Lambda}{6}
+\left(\frac{\Delta\Lambda}{3\k^2T}\right)^2\right)^{3/2}
\right.\right.\nonumber\\
&&\left.\left. -\sigma_+\Lambda_- \left(\left(\frac{\k^2T}{4}\right)^2 
+\frac{\Delta\Lambda}{6}
+\left(\frac{\Delta\Lambda}{3\k^2T}\right)^2\right)^{3/2}
\right]\right\}\label{bubble1}\\
&=& \frac{2\pi^2}{\left(\left(\frac{\k^2T}{4}\right)^2 +\frac{\Lambda_- 
-\Delta\Lambda/2}{3}
+\left(\frac{\Delta\Lambda}{3\k^2T}\right)^2\right)^{3/2}}
\left\{ T +\frac{6}{\k^2\left(\left(\Lambda_- -\frac{\Delta\Lambda}{2}
\right)^2
-\frac{\Delta\Lambda^2}{4}\right)}\right.\nonumber\\
&&\left.\times
\left[ \Delta\Lambda \left(\left(\frac{\k^2T}{4}\right)^2 +
\frac{\Lambda_- -\frac{\Delta \Lambda}{2}}{3}
+\left(\frac{\Delta \Lambda}{3\k^2T}\right)^2\right)^{3/2}
\right.\right.\nonumber\\
&&\left.\left. +\left(\Lambda_- - \Delta\Lambda\right)
\left(\frac{\k^2T}{4}-\frac{\Delta\Lambda}{3\k^2T}\right)^3
+\Lambda_-
\left(\frac{\k^2T}{4}+\frac{\Delta\Lambda}{3\k^2T}\right)^3\right]
\right\}\ .\label{bubbleaction}
\eea
While this is a mess, the reader should note that the sign of the last term
is independent of the tension.  That is because, for $T<T_c$, 
$\sigma_+=1$ but the quantity in the parentheses in the last term of 
(\ref{bubble1}) is the square of a negative number, so the square root 
introduces a sign.  For supercritical tension, that quantity is the square of
a positive number, but then $\sigma_+ = -1$.  We have chosen to write the
variables in this form in order to illuminate the dependence on the level
spacing.  Please see figure \ref{f:cdlbt} for the qualitative features of the
action $\Delta S$ as a function of $\Lambda_-,\Delta\Lambda$.  While in
some ways the physics depends more directly on the initial cosmological
constant $\Lambda_+$, in this paper we typically works with a fixed
final $\Lambda_-$, and $\Delta\Lambda$ depends on the same moduli that
control the bubble tension.

\begin{figure}[t]
\begin{centering}
\subfigure[Lines, from top to bottom, are at 
$dL =-0.1,-0.3,-0.55,-0.75,-1,-1.5$.]{
\includegraphics[scale=0.6]{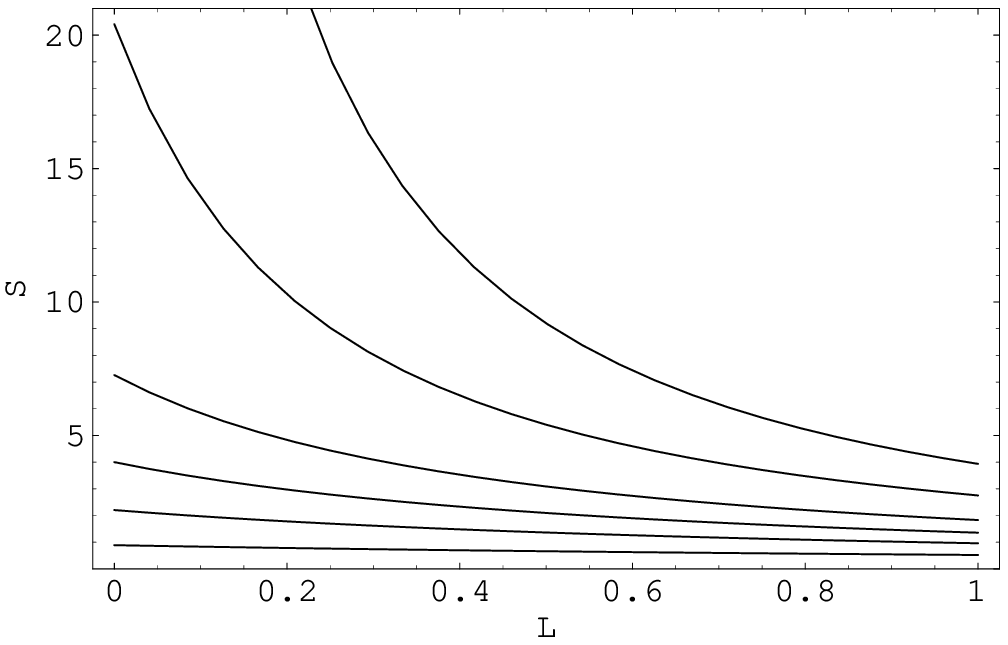}}
\subfigure[Dependence on both variables.]{\includegraphics[scale=0.6]
{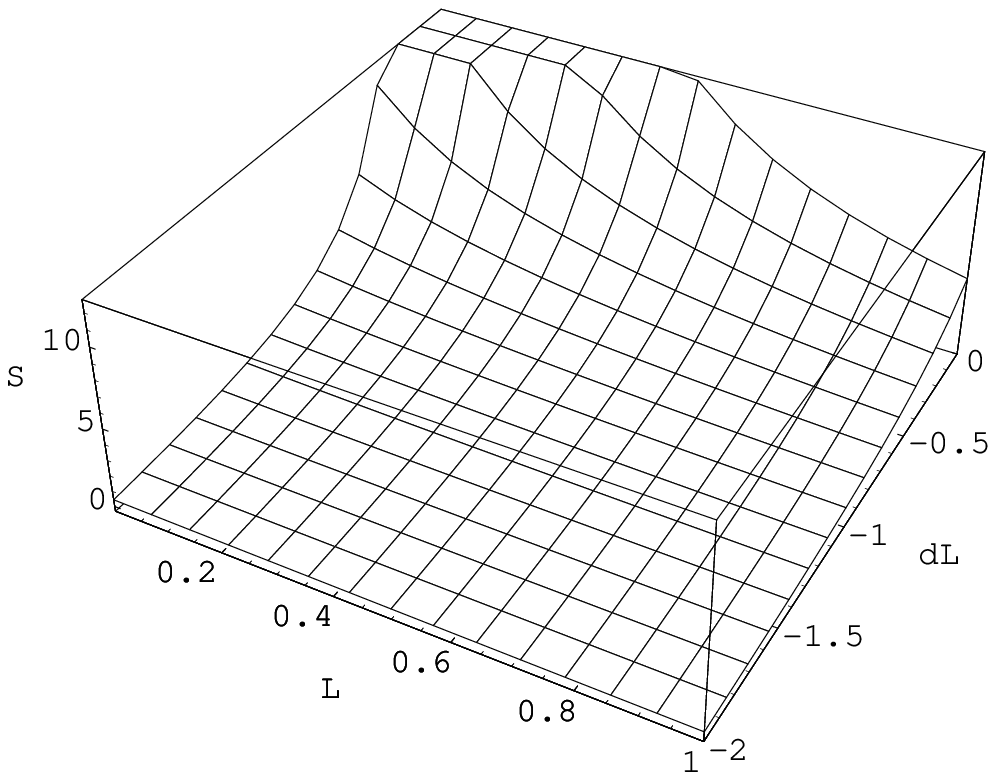}}
\caption[Instanton Decay Times]{\label{f:cdlbt}The 
bubble minus background action as a function of
the cosmological constants.  The variables are $S=\k^4T^2 \Delta S$,
$L=\Lambda_-/\k^4T^2$, and $dL=\Delta\Lambda/\k^4T^2$.}
\end{centering}
\end{figure}

We should note that this action reduces to the known formulae in special
cases.  In particular, the result of CDL as quoted in 
\cite{Kachru:2003aw},
\be\label{kkltaction}
\Delta S_E = -\frac{S_0}{\left(1+T_c^2/T^2\right)^2}\ee
is valid for all tensions when the final vacuum energy vanishes.  
A related result is that, for any $\Lambda_\pm \ge 0$, as the bubble 
tension goes to  infinity, the decay time goes to the recurrence time of 
the original dS.

As final comments, let us reemphasize, following 
\cite{Coleman:1980aw,Banks:2002nm}, that the final states are not maximally
symmetric spacetimes but rather cosmological ones.  In particular, decays
with a negative final cosmological constant lead not to AdS but to a Big
Crunch singularity within the bubble.  Additionally, as mentioned in
\cite{Banks:2002nm}, these instantons are technically different from 
instanton decays of inflationary spacetimes.  It seems reasonable that for
sufficiently small decay rates that treating the initial spacetime as dS 
is a good approximation, but it remains an interesting problem to study
decays of possibly more cosmologically relevant spacetimes.

\section{Migration of D3-branes}\label{aa:migration}

The NS5-brane decay 
leaves D3-branes at the tip of the  throat where it occurs.
In this appendix we analyze their subsequent classical motion in 
configurations with multiple throats\footnote{These calculations were 
given in \cite{Frey:2003dm}.}.  
The D3-branes, produced by a 
decay in one throat (Throat 1), are attracted by \bd-branes in another 
throat (Throat 2), migrate across the compact manifold $M$,
and eventually annihilate the $\bd$-branes.  
As discussed in section
\ref{p:d3migrate}, the decay times are so large that the
migration times have little effect.  
We assume that the back-reaction of the migrating D3-branes 
is negligible, the proper velocity of the branes remains small, and that the
majority of the travel time comes from the two throats (i.e. the time
through the bulk of the CY may be ignored).

We ignore the most of the details coming
from the deformation of the conifold since the motion is
assumed to be radial.  We will use the undeformed metric (\ref{conifold}) 
and, when
working near the tip, multiply the warp factor $e^{-4A}$ by an
overall constant $\sim 0.4$ to account for the deformation.
\footnote{One finds this correction by comparing  
the ``near tip'' warp factor found in \cite{Klebanov:2000hb} to 
the naive limit of the undeformed Klebanov-Tseytlin 
\cite{Klebanov:2000nc} geometry.}  
The warp factor, away from the tip, is
\bea\label{warpfactor}
e^{-4A}    &=& \frac{L^4}{r^4} \ln(r/r_s) \\
r_s  &=& r_0 \ \exp (-\frac{2 \pi(N+p)}{3 g_s M^2} - \frac{1}{4} ) \\
p  &\equiv& \# \textnormal{ of  \bd-branes} \ ; \ \  r_0^2 = 3/2^{5/3} \ \ . 
\eea
Due to the deformation of the conifold discussed above we will only be
interested in the region $\t r = r_s \exp (1/4) \leq r \leq r_0$.
Note that this avoids the naked singularity at $r = r_s$.   

The Ramond-Ramond potential $C_4$ depends only on the
radial distance $r$:
\be 
C_4 = \frac{f(r)}{g_s}  dt \wedge dx^1 \wedge dx^2 \wedge dx^3 
\ee
where $t \equiv x^0$.  Working in the
gauge $\xi^0 = \tau(t)$ (proper time) 
and $\xi^i = x^i$, the D3-brane Lagrangian becomes,
\be\label{fullL}
\mu_3^{-1} \mathcal{L}  = -  \frac{e^{4A}}{g_s}(\sqrt{\dot t^2-
e^{-4A}\dot r^2}) + f(r) \dot t .
\ee
Assuming that the proper velocity is small,
\be\label{Lag}
\mu_3^{-1} \mathcal{L}  \simeq \frac{1}{2} \frac{\dot r^2 \dot t^{-1}}{g_s}
                       - \frac{(e^{4A} - g_s f(r))}{g_s} \dot t 
\ee
It is easy to check that this a valid approximation. In particular, one
only needs to consider throat 2, since this is where the
D3-branes are moving fastest.  One can show that 
$p << (g_s M^2)/8$ will insure that (\ref{Lag}) is valid.
   
Since $t(\tau)$ is a cyclic variable, we know that 
$\partial \mathcal{L} / \del \dot t \equiv - E/g_s $ 
is constant, leaving us with
\be\label{Eeqn}
E = \frac{1}{2} (\partial_t r)^2 + V(r)\ ; \ \
V(r) \equiv (e^{4A} - f(r)) \ \ ,
\ee
and the travel time through a single throat is therefore,
\be\label{traveltime}
\Delta t = \pm \int_{r_0}^{\t r} \frac{dr}{\sqrt{2 (E - V(r))}} \ \ .
\ee
The $+/-$ corresponds to branes traveling into/out of the throat.

Note that the time here is a coordinate time, but we will see that it is
so small compared to decay times that we do not need to worry about
conversion to proper time in the 4D Einstein frame.

\subsection{Throat 1}
The D3-branes, produced at rest in throat 1, feel a slight gravitational
attraction to the \bd-branes at the bottom of throat 2.  
We see below that almost all the migration time does come in throat 1.

The Ramond-Ramond potential, $C_4 = e^{4A_1}/g_s$, 
is not effected by charge contained in throat 2.  Physically, 
this is due to the charge being screened; mathematically, we are
working on a compact manifold and may consider just the charge
enclosed in throat 1.  However, the geometry does, albeit slightly,
know about what is happening in throat 2.

In order to get an estimate of the migration time we will make the 
(perhaps bold) assumption that, as with most multi-pole solutions in
gravity, the warp factor is changed by an additive factor,
\be
e^{-4A_1}(r) \to e^{-4A_1}(r) + \delta(r); \ \ 
\delta(r) \equiv - \frac{27 \pi \ap}{(2 r_0 -r)^4} 
\left(N_2 + p + 3 g_s M^2 / 8 \pi \right) \ .
\ee
Here, the subscripts indicate the throat in which the fluxes are contained, 
$p$ is the number of $\bd$-branes in throat 2 (there are no $\bd$-branes
in throat 1).  Expanding to first order in $\delta$, we see that 
\be\label{throatpot1}
V(r) = - \frac{27 \pi \ap}{(2 r_0 -r)^4} 
\left(N_2 + p + 3 g_s M^2 / 8 \pi \right) e^{8A_1}(r) \ \ .
\ee 
%
%
\begin{figure}
\begin{center}
\subfigure[\label{f:v1plot} Throat 1]{\includegraphics[scale=0.6]{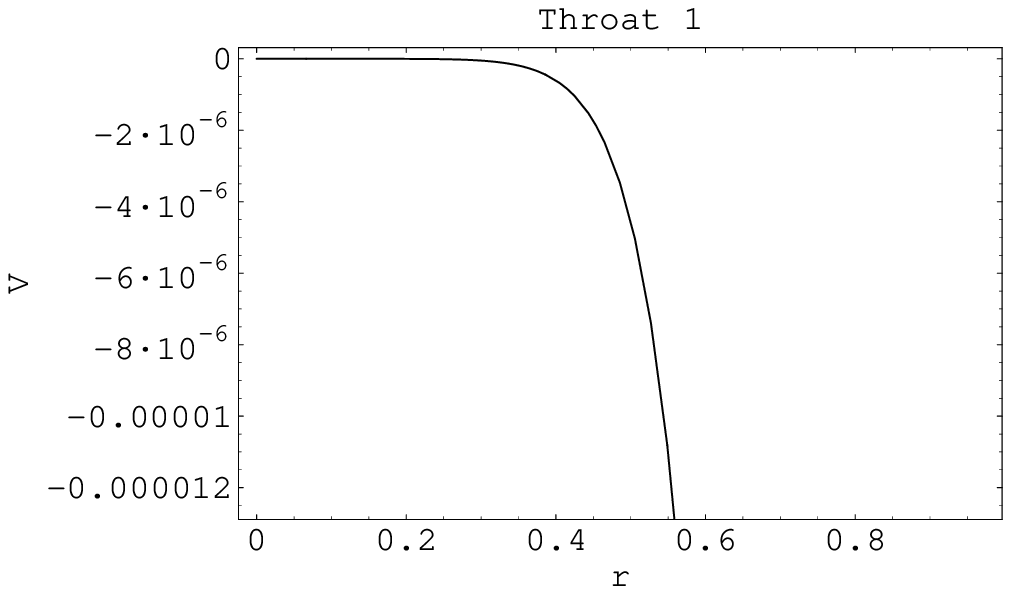}}
\subfigure[\label{f:v2plot} Throat 2]{\includegraphics[scale=0.6]{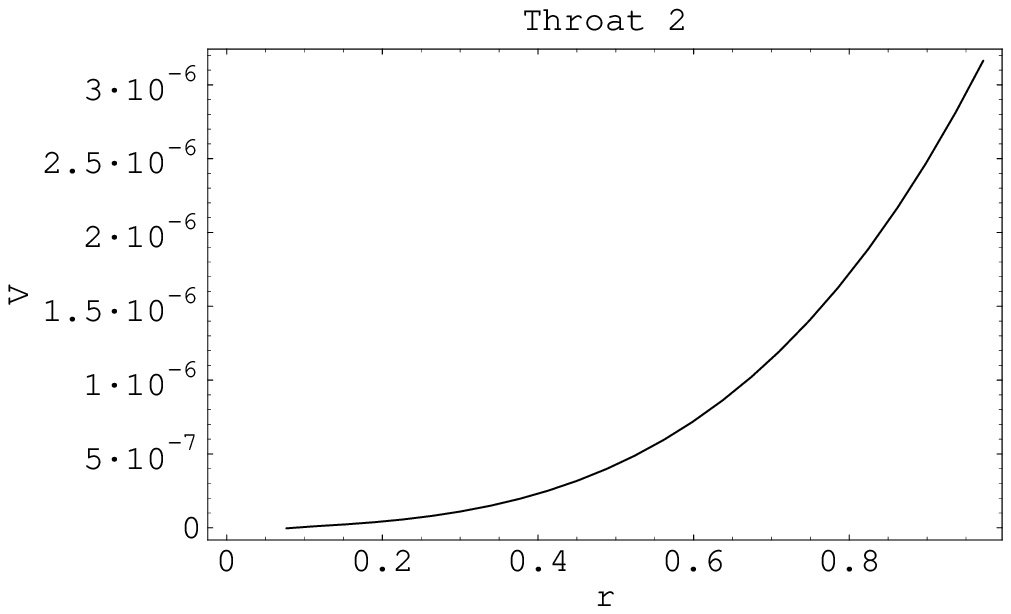}}
\end{center}
\caption[D3 in \bd\ Potential]{The potential for D3-branes in the
presenece of \bd-branes on the compact manifold.} 
\end{figure} 
Figure \ref{f:v1plot} shows the this potential.  Note that the
majority of the time will be spent at the tip of this throat.  We have
been unable to evaluate the integral (\ref{traveltime}) explicitly.
However, one may
gain control of the situation by linearizing the potential near
$r = \t r$ (where the branes are at rest) and integrate numerically.
It is in this limit that we multiply $e^{-4A}$ by the numerical
factor $\sim 0.4$ discussed above.
Once a safe distance away from $r=\t r$, which for technical reasons
coming from the linearization of (\ref{throatpot1}) we take to be 
$r \sim \t r + z r_0$, one may
evaluate the rest of the integral (\ref{traveltime}) numerically.
For model 1, one finds that
\be\label{time1}
\Delta t_1 = 1.5 \times 10^{16} \ .
\ee

\subsection{Throat 2}
For completeness, we will determine time spent in throat 2,
subsequently showing that it is of no importance.
In order to deduce $C_4$, recall that $\int \star d C_4 \sim
(N_{\textnormal{eff}}-p)$.  
Plugging in the appropriate constants, this leaves us
with
\be
r^5 e^{-8A} \partial_r f = (27 \pi {\ap}^2) (N_{\textnormal{eff}} - p) \ \ .
\ee
Using what we know from the geometry, we define $f = e^{4A} + V$,
where the potential, $V(r)$, satisfies
\be\label{Eeqn2}
\partial_r V = \frac{(27 \pi {\ap}^2 g_s) p}{L^5} 
\frac{r^3/L^3}{(\ln(r/r_{s+}))} \ \ .
\ee 
This can be integrated and gives a solution depending on exponential-
integral functions. 
The potential is shown in figure \ref{f:v2plot}.  Numerically
integrating (\ref{traveltime}) for throat 2 gives
\be\label{time2}
\Delta t_2 \approx 9.9 \  ,
\ee
which is clearly negligible compared to (\ref{time1}).

\section{$\ap$ Corrections in Asymmetric Warping}\label{aa:lorentz}

\subsection{Calculation of Scalar MAVs}\label{aaa:scalar}

Here we demonstrate how to calculate the $\ap$ corrections to the
D3-brane scalar kinetic terms in the presence of the asymmetrically warped
metric (\ref{warpasym}).  We keep only the Lorentz violating terms.

We start by specifying the tangent to the D-brane, which
we take as $\del_\alpha X^\mu = \delta^\mu_\alpha$ with perturbatively
small $\del_\alpha X^m$.  Then the normal bundle vielbein satisfies
\be\label{normviel}
\del_\alpha X^M \xi^{\hat a}_M = 0\ \mathnormal{or}\ \xi^{\hat a}_\mu
= -\delta^\alpha_\mu \del_\alpha X^m \xi^{\hat a}_m \ee
which has solution to second order
\be\label{normviel2}
\xi^{\hat a}_\mu = -\delta^\alpha_\mu \del_\alpha X^m e_m^{\hat m}
\delta^{\hat a}_{\hat m}\ ,\ \xi^{\hat a}_m = \delta^{\hat a}_{\hat m}
e^{\hat m}_m - \frac{1}{2} g^{\mu\nu}\delta_{\mu}^\alpha 
\delta_\nu^\beta \del_\alpha X^p \del_\beta X^n \hat e_n^{\hat n}
\delta_{\hat n}^{\hat a} g_{mp}\ . \ee
Here $e^{\hat m}_m$ is the vielbein of $g_{mn}$ and is trivial
at the position of the brane (we use Riemann normal coordinates on the
compact manifold at the brane position).  
Since its expansion would just give powers of
$X^m$ and not derivatives, we will ignore it from now on.  Also, for 
notational convenience, since the normal bundle indices $\hat a$ always
enter through Kronecker deltas with compact space indices, we will abuse
notation slightly and replace $\hat a$ with $m$ in future formulae.

The nonvanishing Christoffel symbols are
\be\label{christoffel}
\Gamma^0_{0m}=\del_m A\ ,\ \Gamma^m_{00}= 
g^{mn} e^A \del_n A\ ,\ \Gamma^i_{jm} = \del_m B\delta^i_j\
,\ \Gamma^m_{ij}= - g^{mn}e^B\del_n B \delta_{ij}\ee
on the spacetime and
\bea
(\Gamma_T)^0_{00}&=&\del_0 X^m \del_m A\ ,\ (\Gamma_T)^0_{0i}=
\del_i X^m \del_m A\ ,\ (\Gamma_T)^i_{00}=e^{A-B}
\delta^{ij} \del_j X^m\del_m A\nonumber\\
(\Gamma_T)^0_{ij} &=& e^{B-A}\delta_{ij}\del_0 X^m \del_m B\ ,\
(\Gamma_T)^i_{0j}=\delta^i_j \del_0X^m \del_m B\nonumber\\
(\Gamma_T)^i_{jk} &=& \left(\delta^i_k \del_j X^m+\delta^i_j
\del_k X^m - \delta_{jk}\delta^{il}\del_l X^m\right)\del_m B
\label{chrisT}
\eea
on the tangent space.  This is just the Christoffel symbol of the pulled-back
metric $g_{\alpha\beta}$.  
Plugging into eqn (\ref{extrinsic}), we can very 
easily get eqn (\ref{omega}) for the extrinsic curvature.  

Then, from eqn (\ref{riemannt}), the tangent Riemann components out to 
$\mathcal{O}(\del X^m)^2$ are
\bea
(R_T)_{0i0j} &=& -a^mb_m \delta_{ij} +2a_m b_n
\delta_{ij}\del_0 X^m \del_0 X^n -2 a_{(m} b_{n)}\del_i
X^m \del_j X^n \nonumber\\
&&-\left( a_{mn}+3 a_m a_n\right)
\del_i X^m\del_j X^n +\left(b_{mn}+b_mb_n\right)
\delta_{ij}\del_0 X^m\del_0 X^n\nonumber\\
(R_T)_{ijkl} &=& b^2\left(\delta_{ik}\delta_{jl}-\delta_{il}
\delta_{jk}\right)+\left(b_{mn}+3b_m b_n\right)
\left( \del_i X^m \del_k X^n \delta_{jl} \right.\nonumber\\
&&\left.+\del_j X^m \del_l X^n \delta_{ik}
-\del_i X^m \del_l X^n \delta_{jk} -\del_j X^m \del_k X^n \delta_{il}\right)
\ . \label{rtcomp}
\eea
The other components are only $\mathcal{O}(\del X^m)^2$, so they square to
$\mathcal{O}(\del X^m)^4$.  When squaring, we also need to take into 
account the second order part of the inverse metric pull-back.  Note that
the only terms involving $\Lambda$ come from $(R_T)_{0i0j}$.  To get
$(R_T)_{\alpha\beta}$ we just contract the Riemann tensor, being careful of
second order terms in the metric.

From eqn (\ref{riemannn}), the normal bundle Riemann tensor is always second
order in $\del X^m$, so it does not contribute to the kinetic terms.  That
leaves just $\b R^{mn}$.  We end up with
\bea
\b R^{mn} &=& -\left( a^{mn}+2a^m a^n\right) -3b^{mn}
+R_{p}{}^{mn}{}_q\del_\mu X^p\del^\mu X^q\nonumber\\
&& -2\left( a_p{}^{(m}+
3a_pa^{(m}\right)\del_0 X^{n)} \del_0 X^p -\left(b_p{}^{(m} -3b_p 
b^{(m}\right) \vec\del X^m\cdot \vec\del X^n\nonumber\\
&&- \left(a^{mn} +5a^m a^n\right)\delta_{pq}\del_0 X^p
\del_0 X^q \nonumber\\
&&+ \left(b^{mn}-b^m b^n \right)\delta_{pq}
\vec\del X^p\cdot \vec\del X^q\ .\label{rbarcomp}
\eea
Here $R_{mnpq}$ is the Riemann tensor of $g_{mn}$; note that
it enters in a Lorentz invariant fashion.
Getting the action (\ref{scalarmav}) is just a matter of squaring.

\subsection{T Duality for Photon MAV}\label{aaa:photlorentz}

Our T duality argument runs as follows.  Suppose we compactify one of the
noncompact spatial dimensions, say $x^3$, on a large circle.   Then
we perform T duality on $x^3$, giving a metric of 
\be\label{tdmetric}
ds^2 = -e^{2A} dt^2 +e^{2B} \sum_{i=1}^2 \left( dx^i\right)^2 +
e^{-2B}\left( dx^3
\right)^2 + g_{mn} dx^m dx^n \ee
with, as before, all components of the metric depending only on $x^m$.
Now $x^3$ is compactified on a small circle, and the braneworld is a D2-brane.
In fact, this metric is of the same form as our original metric 
(\ref{warpasym})
except that $x^3$ is not a Riemann normal coordinate.  The only difference 
this makes from the calculations of
scalar calculation is that we now have to keep in mind that there is a 
nonvanishing Christoffel symbol $\Gamma^m_{33}$ that contributes kinetic terms
to $\Omega^m_{\alpha\beta}$ ($\Gamma^3_{3m}$ will not 
matter because $\Omega^3_{\alpha\beta}$ will already be second order in 
$\del X$).  Otherwise some numerical factors differ.  If
we recalculate equation (\ref{scalarmav}) 
for fluctuations of $X^3$ in 2 dimensions,
we find
\bea
\delta S &=& \frac{\mu_2}{g_{s,2}} \int d^3x \frac{(2\pi\ap)^2}{192}
\left| \vec\del X^3 \right|^2 \left[ 44 b^2 b\cdot\lambda 
+12 b^{mn}\lambda_{mn} +24\lambda^{mn}b_mb_n\right.\nonumber\\
&&\left. +72b^{mn}\lambda_mb_n\right]
\ .\label{x3mav}\eea
Here 
$\mu_2,g_{s,2}$ are the appropriate tension and string coupling for the 
D2-brane case.  

If we then T-dualize back to the D3-brane and take the $x^3$ circle radius
to infinity, we get back the original tension and string coupling and take
$X^3\to (2\pi\ap)A_3$.  Then by gauge invariance, we must replace
$(2\pi\ap)\del_i A_3 \to \mathcal{F}_{i3}$, and isotropy requires that we
promote $\mathcal{F}_{i3}\mathcal{F}^{i3}\to \mathcal{F}_{ij}
\mathcal{F}^{ij}$.  Then we get the result (\ref{photmav}).

\bibliographystyle{utcaps}\bibliography{thesis}
\end{document}